\documentclass{aa}  

\usepackage{graphicx}
\usepackage{txfonts}
\usepackage{orcidlink}
\begin{document}

\title{Characterisation of starspot structure and differential rotation of Kepler-411}

\author{Mikko Tuomi\inst{1,2}\fnmsep\thanks{\email{mikko.tuomi@helsinki.fi}}\orcidlink{0000-0002-4164-4414} \and
Andr\'as Haris\inst{1}\orcidlink{0009-0008-5616-4633} \and
Thomas Hackman\inst{1}\orcidlink{0000-0002-1412-2610}
}

\institute{University of Helsinki, Department of Physics, PO Box 64, 00014 Helsinki, Finland\\
\and
University of Hertfordshire, Center for Astrophysics, College Lane Campus, Hatfield, Hertfordshire, UK, AL10 9AB\\
}

\date{Received ; accepted later}

\abstract
{Starspots and their movements on stellar surfaces enable investigating the mechanisms of stellar magnetic activity. Information on the spot distribution and differential rotation provide important constraints for the behaviour of stellar magnetic dynamos.}
{We analyse the Kepler photometry of Kepler-411, a known exoplanet host, to determine the distribution and properties of star spots on the stellar surface with two independent and complementary methods: modelling the photometric effect of rotation of spots on the stellar surface and mapping of spots by transiting planets.}
{By constructing a spot model accounting for geometry, differential rotation and spot evolution, we model the spots of the stellar surface giving rise to the observed brightness variations. We also search for evidence for occultations of starspots in high-cadence photometry.}
{Our spot models reproduce the observed photometric variations well and we are able to obtain information on the distribution and movement of spots on the stellar surface. We do not obtain evidence for differential rotation -- the rotational profile is consistent with rigid-body rotation with a period of 10.52$\pm$0.34 days. We detect three occultations of spots by planet c. The positions of these spots coincide well with the positions of larger spot structures identified by our modelling of the rotational modulation of the light curve.}
{}


\maketitle

\section{Introduction}

Young, active stars provide windows into the stellar interiors and physics by demonstrating a range of observable activity-induced phenomena such as starspots and active regions, their evolution, flares, coronal mass ejections, and differential rotation. Observations of these phenomena can enable assessing the mechanisms of stellar magnetic dynamos as well as their diversity and evolution. Yet, although we can readily study the nearest star, the Sun, obtaining information on the behaviour of other members of the stellar population can be painstakingly difficult.

There are limited observational approaches for obtaining information on stellar surface phenomena. For instance, Zeeman-Doppler imaging provides the means of generating rough brightness and magnetic field maps of inhomogeneous surfaces of active stars \citep[e.g.][]{willamo2022}, that can be interpreted as low-resolution spot maps of the stellar surfaces. Although, the precision of the resulting maps might enable studying differential rotation and spot evolution and/or lifetime in some cases, this has been very difficult in practice. Instead, detections of exoplanets around spotted stars can enable mapping the spot structures and even differential rotation on stellar surfaces given that planetary occultations of spots are detected during the transits \citep[e.g.][]{silva-valio2008,araujo2021a}.

While photometry can readily reveal information on flaring \citep[e.g.][]{gunther2020,ilin2021,vida2021,zeldes2021,howard2022}, photometric observations are also commonly applied to gain information on stellar rotation and the presence of spots. This can be clearly visible as periodic rotational modulation of the light curve \citep[e.g.][]{nielsen2013,reinhold2013,claytor2022,claytor2023}. Obtaining information on spot distribution has been attempted in such photometric data \citep{rodono1986,harmon2000,jarvinen2008,lanza2019,zhan2019,ioannidis2020,breton2024}, but robustly determining spot distributions based on photometry has been challenging. Details of the spot structure, distribution and evolution giving rise to such modulation are indeed difficult to reveal due to the ill-posedness of the resulting inverse problem of obtaining two-dimensional surface information from a one-dimensional light curve \citep{luo2019,basri2020,koen2021,luger2021,luger2021b}. This is particularly the case in the presence of differential rotation and spot evolution and movement \citep{basri2020} that are both observed for young stars \citep[e.g.][]{araujo2021a,basri2022}.

We study the star spots on the surface of a Kepler target star Kepler-411 to infer information on its differential rotation as well as spot distribution, evolution, and lifetimes \citep{xu2021}. However, we do not attempt to create comprehensive brightness maps of the stellar surface. Rather, we aim at identifying the most prominent spots on the stellar surface such that effects of spots on the light curve are well modelled with a minimal number of spots. We first obtain information on spots by modelling high-cadence Kepler photometry in order to characterise planetary occultations of star spots during transit events. Then, we model low-cadence observations around these occultation events in an attempt to obtain information on the same spot structures based on two independent approaches. These two approaches enable obtaining independent information on the star spots and cross-comparison between the two approaches is possible. We also estimate the differential rotation curve for Kepler-411 based on our results.

\section{Kepler-411}

The target star Kepler-411 (KOI 1781, KIC 11551692) is a spectral type K2 V star with a parallax of 6.5313$\pm$0.0080 mas \citep{gaia2021} implying a distance of 153.11$\pm$0.19 pc. It is a young and active star with an estimated age of 212$\pm$31 Myr \citep{sun2019} according to gyrochronological relations based on color index and the star's rotation. We have summarised relevant astrophysical parameters in Table \ref{tab:stellar_properties}.

\begin{table}
\caption{Observational and physical parameters of the target star Kepler-411.}\label{tab:stellar_properties}
\begin{center}
\begin{tabular}{lcc}
\hline \hline 
 Parameter & Estimate & Reference \\
\hline
Spectral Type & K2 V & \citet{sun2019} \\
$\pi$ (mas) & 6.531$\pm$0.008 & \citet{gaia2021} \\
$[\rm Fe/H]$ & 0.05$\pm$0.11 & \citet{morton2016} \\
$R_{\star}$ (R$_{\odot}$) & 0.820$\pm$0.018 & \citet{sun2019} \\
$R_{\star}$ (R$_{\odot}$) & 0.729$^{+0.016}_{-0.010}$ & \citet{gaia2018} \\
$R_{\star}$ (R$_{\odot}$) & 0.731$^{+0.068}_{-0.045}$ & \citet{stassun2018} \\
$T_{\rm eff}$ (K) & 4833$^{+32}_{-51}$ & \citet{gaia2018} \\
$T_{\rm eff}$ (K) & 4837$^{+150}_{-127}$ & \citet{stassun2018} \\
$M_{\star}$ (M$_{\odot}$) & 0.870$\pm$0.039 & \citet{sun2019} \\
$M_{\star}$ (M$_{\odot}$) & 0.785$^{+0.103}_{-0.085}$ & \citet{stassun2018} \\
$P_{\rm rot}$ (days) & 10.40$\pm$0.03 & \citet{sun2019} \\
$P_{\rm rot}$ (days) & 10.52$\pm$0.34 & This work \\
$i$ (deg) & 83.7$^{+6.3}_{-16.4}$ & \citet{hirano2014} \\
$i$ (deg) & 89.79$\pm$0.20 & This work \\
Age (Myr) & 212$\pm$31 & \citet{sun2019} \\
\hline \hline
\end{tabular}
\end{center}
\end{table}

Kepler-411 is a host to a system of four planets that are members of a nearly co-planar system with orbital inclinations ranging from 87.4$\pm$0.1$^{\circ}$ to 89.43$\pm$0.02$^{\circ}$ \citep{wang2014,morton2016,sun2019}. Three out of these four planets are transiting the star while the existence of the fourth one has been inferred based on transit timing variations (TTVs). A useful summary of the papers resulting in the detection of this system is provided by \citet{sun2019}. In their work, the TTVs were used to obtain information on the masses of the system of planets. This enabled estimating that the bulk densities of planets c and d are consistent with massive cores and significant fractions of volatiles whereas the innermost planet b has a high density indicating probable rocky composition. We have tabulated some parameters of the transiting planets relevant to the current work in Table \ref{tab:planetary_properties}.

\begin{table*}
\caption{Selected parameters of the planets around Kerpler-411. The parameter estimates for orbital period $P$, impact parameter $b$, planetary radius $R_{\rm p}$, semi-major axis and orbital inclination have been taken from \citet{sun2019}.}\label{tab:planetary_properties}
\begin{center}
\begin{tabular}{lcccc}
\hline \hline 
 & Planet b & Planet c & Planet d & Planet e \\
\hline
$P$ (days) & 3.005156$\pm$0.000002 & 7.834435$\pm$0.000002 & 58.02035$\pm$0.00056 & 31.509728$\pm$0.000085 \\
$b$ & 0.574$^{+0.012}_{-0.015}$ & 0.620$^{+0.007}_{-0.009}$ & 0.7193$^{+0.0014}_{-0.0012}$ & 1.688$\pm$0.006 \\
$R_{\rm p}$ (R$_{\oplus}$) & 2.401$\pm$0.053 & 4.421$\pm$0.062 & 3.319$\pm$0.104 & -- \\
$a$ (AU) & 0.0375$\pm$0.0008 & 0.0739$\pm$0.001 & 0.279$\pm$0.004 & 0.186$\pm$0.003 \\
$i$ (deg) & 87.4$\pm$0.1 & 88.61$\pm$0.04 & 89.43$\pm$0.02 & 88.04$\pm$0.02 \\
\hline \hline
\end{tabular}
\end{center}
\end{table*}

The planets have also been reported to occult as many as 198 star spots, allowing even constraining the stellar differential rotation \citep{araujo2021a}. Flares have also been reported for the star. \citet{araujo2021b} identified 65 flare events in the out-of-transit Kepler light curve.

Although the inclination of the star's axis of spin has not been characterised directly with, e.g., observations of the Rossiter-McLaughlin effect, the star has been reported to rotate in an edge-on orientation based on rotational velocities. \citet{hirano2014} reported a $\sin i$ value of 0.994$^{+0.077}_{-0.072}$ corresponding to an inclination of 83.7$^{+6.3}_{-16.4}$ deg.

We adopted the stellar radius of 0.820$\pm$0.018 R$_{\odot}$ \citep{sun2019} in our computations throughout the current work.

\section{Transit mapping of spots}\label{sec:occultation_modelling}

Information on stellar spot distribution can be inferred from the photometric time-series by identifying planetary transits of large starspots. Such transits have been reported for Kepler-411 \citep{araujo2021a} and it is possible to obtain information on projected spot contrasts and diameters as well as their temperatures and positions on the stellar disk. 

We modelled the spot transits in the Kepler photometry of Kepler-411 by adopting the statistical models and detection techniques of \citet{haris2024}. This methodology is also described here as applied.  We obtained the Pre-search Data Conditioning (PDC) Kepler observations of the target from the Mikulski Archive for Space Telescopes by using the \texttt{lightkurve} Python package \citep{lightkurve2018}. We restricted the analyses to observations taken with a 60s cadence. In total, the data contained the signatures of 46 transit events of planet c. As we only considered transit event such that they were at least by a factor of two deeper than the standard deviation of the local light curve variations there were only one and two suitable transits to search for spot occultations of planets b and d, respectively. We do not discuss these transits any further as they showed no evidence for spot occultations. 

Analyses of each transit event were conducted by first choosing subsets of data centered around transit mean times $t_{\rm c}$ such that data taken at $t_{i} \in [t_{\rm c} - \frac{3}{2}T_{14}, t_{\rm c} + \frac{3}{2}T_{14}]$ was selected. Here $T_{14}$ denotes the transit duration or time between contact points one and four. The rationale was to focus on data during transits but avoid the computational cost of evaluating the likelihood values for unnecessarily many points on both sides of the transit event.

Not all data subsets containing transit events of planet c were searched for signals of spot occultations. We neglected seven subsets during which transits of planet b occurred nearly simultaneously making calibration and/or spot identification less robust. Two of the subsets lacked data on either side of the transit event compromising the reliability of the normalization process. These subset were also removed because if the variations of the light curve around a given transit exceed a certain threshold, a second order polynomial may not produce a sufficiently good description of the out-of-transit parts. This could pose challenges for de-trending the entire light curve, leading to unrealistic estimates of transit depths and, consequently, the planetary properties. We were thus left with 37 subsets containing transit events with numbers of data ranging from 497 to 531.

The data subsets were normalised by choosing the points outside the transits and by removing second-order polynomial trends in the flux from each data subset, mostly due to rotational modulation of the light curve. This enables removing the rotation-induced variations from the observations without disturbing the transit signals too much. The resulting sets were then scaled according to the procedure of \citet{morris2017} to calibrate for differences caused by effects of rotational modulation and changing stellar brightness \citep{haris2024}.

Assuming co-planarity of the system, i.e., that the star is seen (almost) exactly edge-on, we can then estimate constraints for the spot positions on the stellar surface based on their transits \citep[e.g.][]{silva-valio2008}.

\subsection{Statistical modelling}

In order to detect occultations of starspots during transits, we compared two different models describing planetary transits with and without effects of star spots. For a model without spots we applied the transit model of \citet{mandel2002} with quadratic limb darkening. This is the reference model without spot occultations. The reference model was then compared to a model containing a spot occultation event corresponding to a temporary decrease in transit depth due to the planet transiting an area with lower brightness. This is modelled as a fractional decrease in transit depth of $h_{\rm s}$ in the interval $[t_{\rm in}, t_{\rm out}]$ and Gaussian decreases in the effect outside this interval with variance of $\sigma_{\rm spot}^{2}$ \citep{haris2024}.

In addition to these astrophysical effects, both models were assumed to have Gaussian random variable $\epsilon_{i}$ with zero mean and variance of $\sigma_{\rm S}^{2}+\sigma_{i}^{2}$ caused by stellar $(\sigma_{\rm S})$ and instrumental $(\sigma_i)$ noise sources. This forward model can be written as
\begin{equation}\label{eq:forward_model_tr}
  m_{i} = f_{\rm tr}(t_{i}) \bigg[ 1 - g_{\rm spot}(t_{i}) \bigg] + \epsilon_{i},
\end{equation}
where $f_{\rm tr}$ is the unaltered transit curve of \citet{mandel2002} and
\begin{eqnarray}\label{eq:spot_occultation}
  g_{\rm spot}(t_{i}) = \left\{ \begin{array}{ccc}
    h_{\rm s} \exp \bigg\{ -\frac{(t_{i} - t_{\rm in})^{2}}{2\sigma^{2}_{\rm spot}} \bigg\} & \mathrm{ if } & t_{i} < t_{\rm in} \\
    h_{\rm s} & \mathrm{ if } & t_{\rm in} \leq t_{i} \leq t_{\rm out} \\
    h_{\rm s} \exp \bigg\{ -\frac{(t_{i} - t_{\rm out})^{2}}{2\sigma^{2}_{\rm spot}} \bigg\} & \mathrm{ if } & t_{i} > t_{\rm out} \\
  \end{array} \right.
\end{eqnarray}

For consistency, we analysed $N$ data subsets containing transit events simultaneously in order to obtain information on the transit shape. The mean transit times were also allowed to vary from strict periodicity due to TTVs. To simplify the computational problem and enable faster posterior samplings, the orbital period, semi-major axis, and stellar radius were fixed to their published values from \citet{sun2019} and the orbits were assumed to be circular. We also set $N = 10$ at a time to reduce the computational cost of the samplings. These simplifications did not affect the results significantly \citep{haris2024}. As a result, there were four transit parameters (planetary radius, inclination, and two limb darkening coefficients), the Gaussian noise component, and one transit mean time parameter for each transit. The spot occultation component of the model contained additional four parameters for each transit: $h_{\rm s}$, $t_{\rm in}$, $t_{\rm out}$, and $\sigma_{\rm spot}$.

The parameter posterior densities of the models were sampled with the adaptive Metropolis (AM) algorithm of \citet{haario2001}, which is an adaptive version of the Metropolis-Hastings algorithm \citep{metropolis1953,hastings1970}. In this algorithm, the proposal density adapts to the information obtained from the sampling, and its shape approaches a multivariate Gaussian approximation of the target distribution. These samplings were performed such that the transit model without spot occultations was modelled for $N$ transit events. The obtained maximum likelihood values of each data subset were then used to compare models with and without spot occultations for each transit event. The priors were chosen uniform and uninformative such that we simply set them equal to unity for simplicity. Details on the procedure can be found in \citet{haris2024}.

Several samplings were conducted for each model. Typically, four samplings were used to investigate chain properties and calculate Gelman-Rubin statistics $R$ in order to estimate whether the chains demonstrated non-convergent behaviour \citep{gelman1992}. This statistics determines whether the variances within each chain are sufficiently large with respect to the variance between chains indicating that the different samplings produce similar results.  We required that this statistics was below 1.05 for all models. We found that the samplings were well-behaving such that the same solutions were identified consistently enabling us to conclude that the solutions were robust in terms of uniqueness.

We applied two different approaches to test the significance of spot occultation detections. We applied likelihood-ratio tests and Bayes factors calculated based on Bayesian information criterion (BIC) estimates \citep{liddle2007}. The latter statistics have been shown to yield robust results in related statistical problems in astrophysics \citep{feng2016}. We interpret the Bayes factors in accordance with the Jeffrey's scale of probabilities as discussed in \citet{kass1995}. The (natural) logarithms of Bayes factors in excess of 5.01 are considered to indicate very strong evidence for a spot occultation whereas values in the intervals of [1.10, 3.00] and [3.00, 5.01] are considered to represent positive and strong evidence, respectively. Using the likelihood-ratio test, a signal of a spot occultation can be stated to be significant when inclusion of the four-parameter spot occultation event in the model increases the logarithm of the maximum likelihood value by 13.28 (18.47) -- in such a case the null-hypothesis can be rejected with a probability of 99\% (99.9\%). Out of these two tests, Bayes factors based on BIC are more conservative in our applications.

\subsection{Detections of spot occultations}

\begin{figure*}
\center
\includegraphics[width=0.45\textwidth,clip]{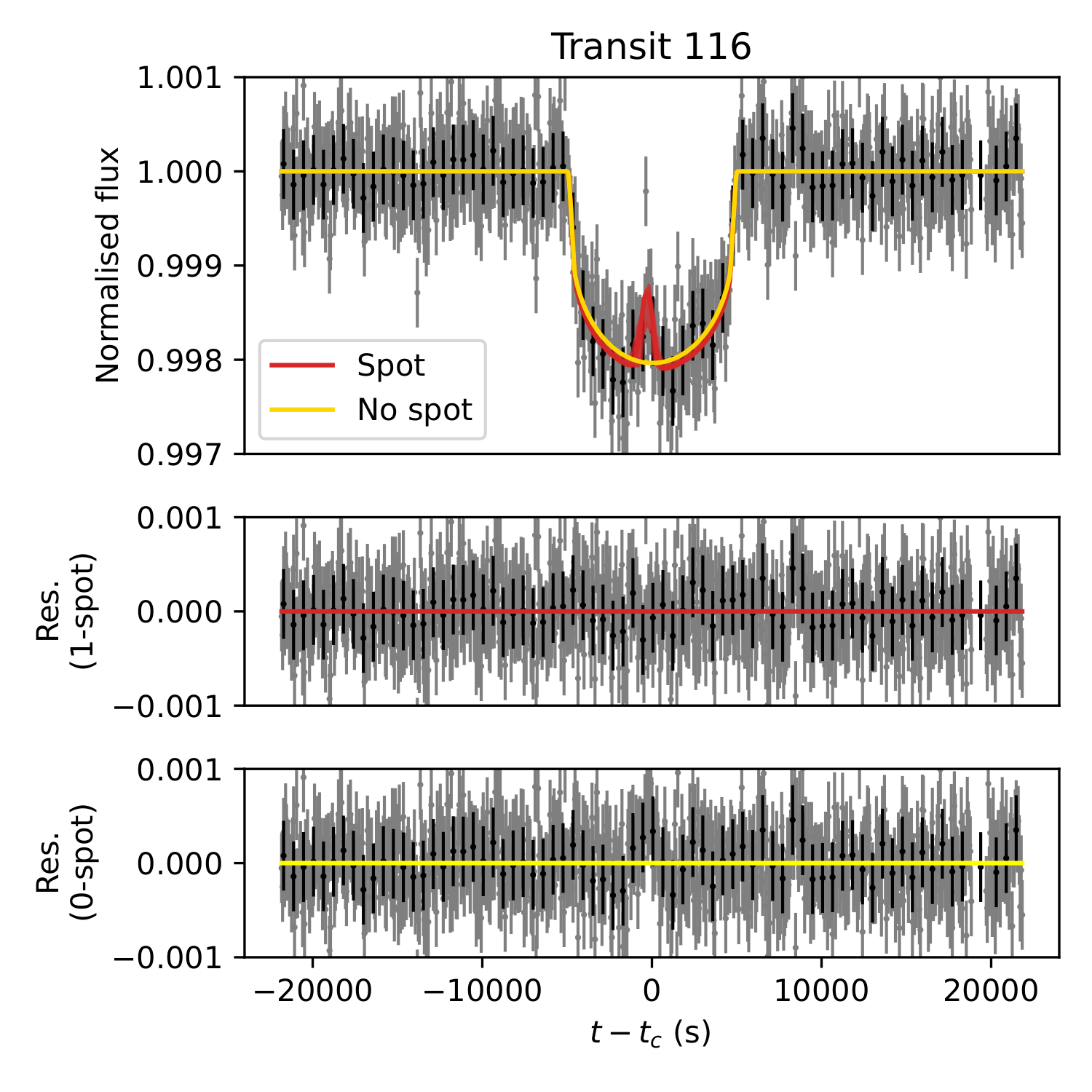}
\includegraphics[width=0.45\textwidth,clip]{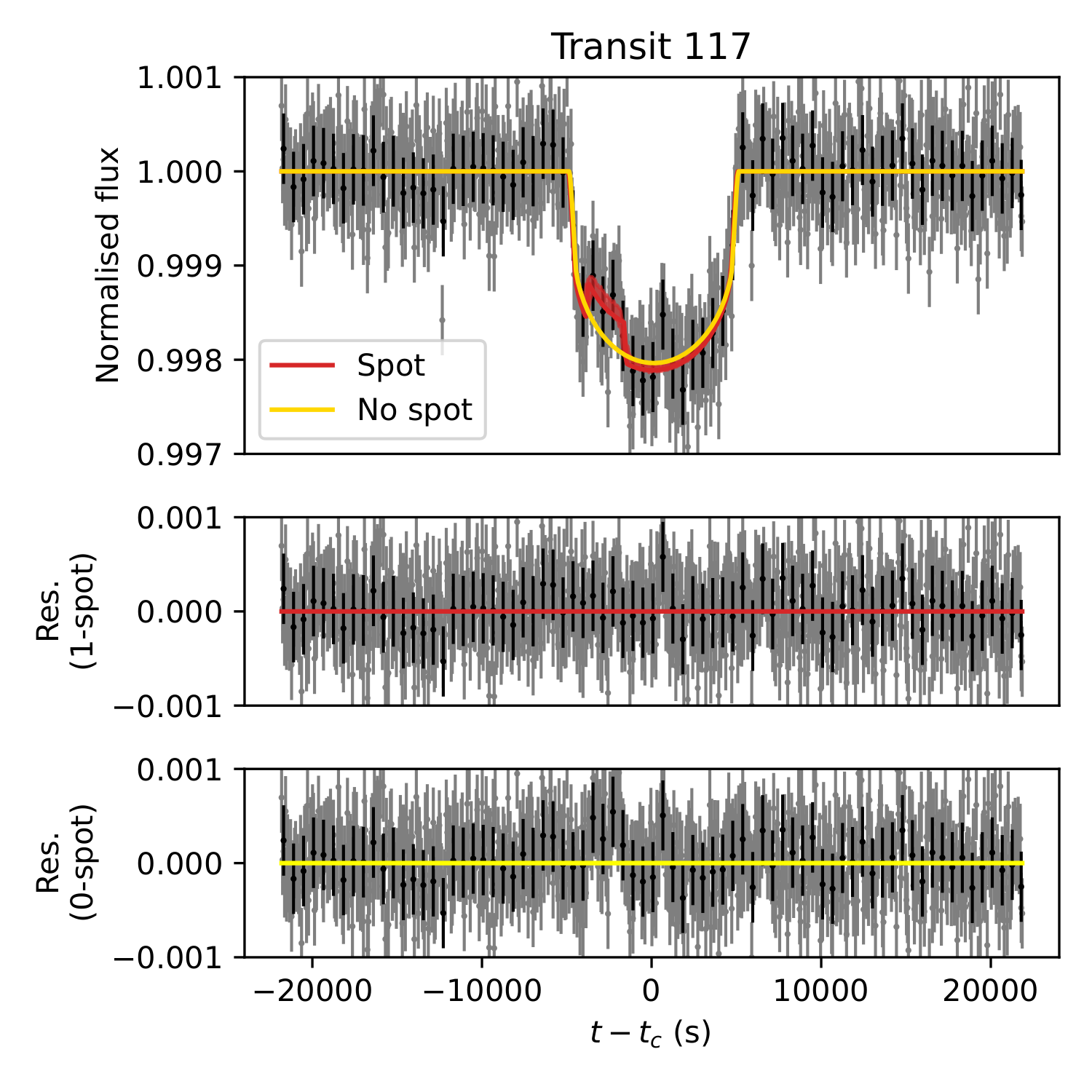}

\includegraphics[width=0.45\textwidth,clip]{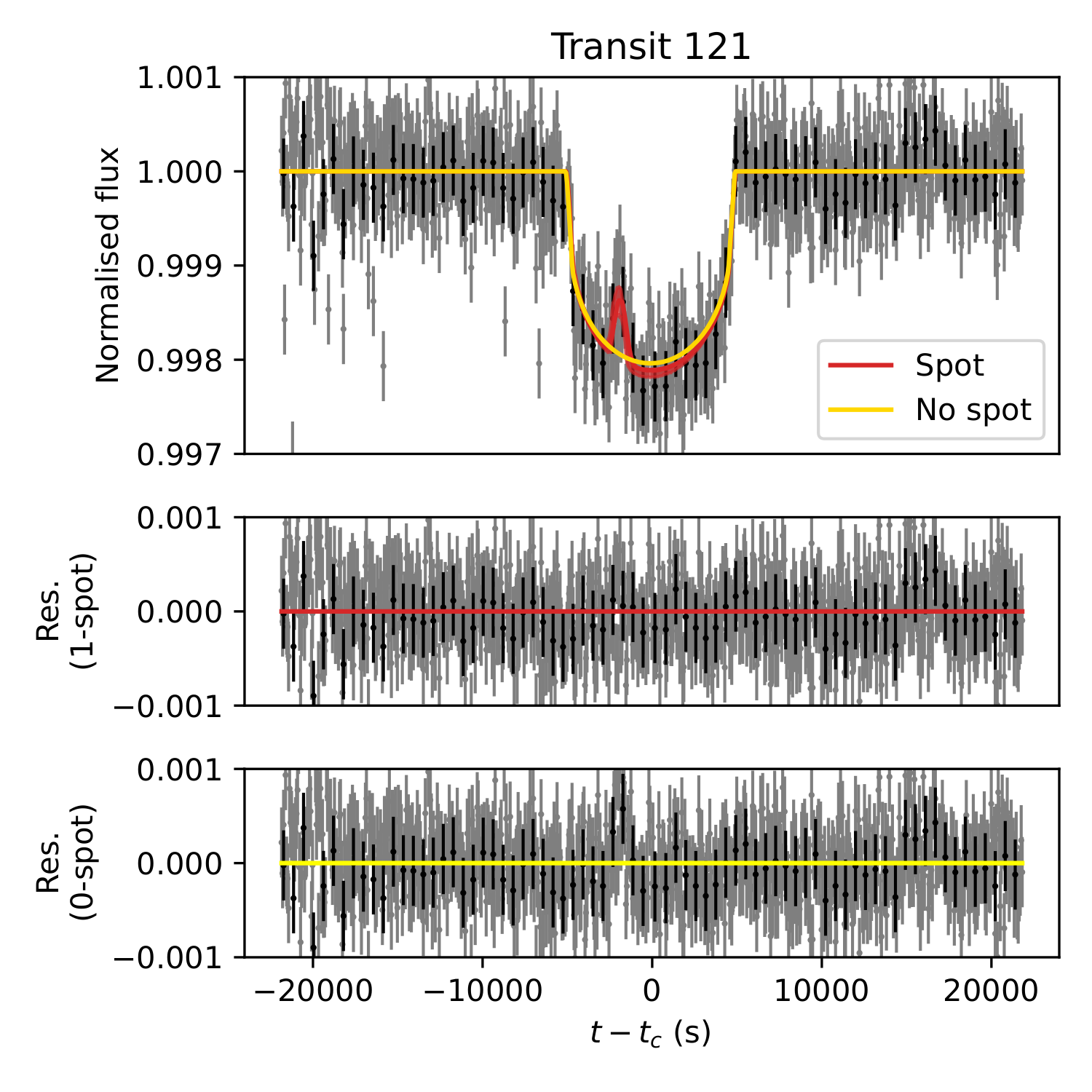}
\includegraphics[width=0.45\textwidth,clip]{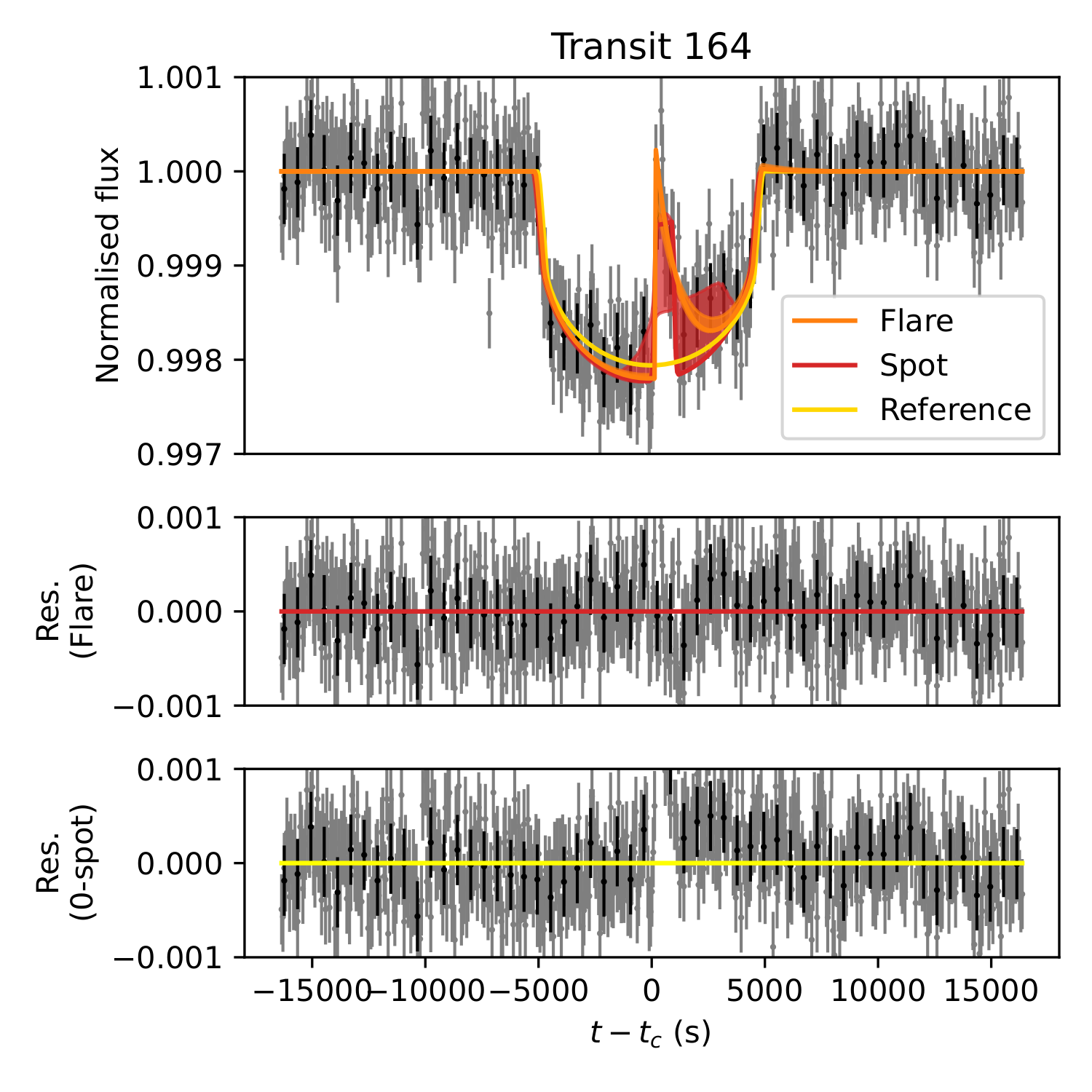}
\caption{Identified occultations of spots by Kepler-411 c during transit events 116, 117, and 121 and a flare-event occurring during the transit 164. The yellow, red and orange curves denote the reference model, a model with a spot occultation, and a model with a flare-event, respectively.}\label{fig:spot_transits}
\end{figure*}

We obtained very strong evidence for spot occultations during transit events 117 and 121 and strong evidence for an additional spot occultations during transit event 116. We have tabulated the significance statistics of these detections in Table \ref{tab:model_comparison_occ} and show our solutions in Fig. \ref{fig:spot_transits}. We have also tabulated our estimates for spot properties, angular diameter, contrast, and temperature in Table \ref{tab:model_comparison_occ} based on the equations of \citet{silva-valio2010}.

\begin{table*}
\caption{Model comparison statistics for selected candidate spot occultations in terms of logarithms of likelihood ratios $\Delta L_{1,0}$ and logarithms of Bayes factors $B_{1,0}$ in favour of model with one spot occultation $\mathcal{M}_{1}$ and against a model without them $\mathcal{M}_{0}$. The parameters reported here for each spot are the time of transit centre $t_{\rm c}$, the minimal angular diameter $\theta_{\rm min}$, spot contrast $c$, and temperature $T_{\rm spot}$.}\label{tab:model_comparison_occ}
\begin{center}
\begin{tabular}{lcccccccccc}
\hline \hline 
Transit & $\Delta L_{1,0}$ & $B_{1,0}$ &  $t_{\rm c}$ & $\theta_{\rm min}$ & $c$ & $T_{\rm spot}$ \\
 & & & (BJD-2450000) & (deg) & (-) & (K) \\
\hline
116 & 17.99 & 2.36 & 5877.01 & $2.1_{-1.5}^{+3.4}$ & $0.32_{-0.02}^{+0.04}$ & $4471_{-53}^{+26}$ \\
117 & 21.66 &  9.00 & 5884.85 & $15.1_{-4.6}^{+4.9}$ & $0.20_{-0.03}^{+0.03}$ & $4621_{-36}^{+35}$ \\
121 & 20.06 &  7.51 & 5916.18 & $2.9_{-1.7}^{+3.4}$ & $0.32_{-0.02}^{+0.03}$ & $4471_{-40}^{+26}$ \\
\hline \hline
\end{tabular}
\end{center}
\end{table*}

We accounted for the possibility that one or some of the detected spot occultations could be chance events caused by other astrophysical phenomena, small-scale brightness variations of the stellar surface or simply noise, and incorrectly interpreted as spot-induced effects. If this was indeed the case, such spot-resembling "bumps" should also be seen elsewhere in the Kepler light curve, outside the transit events. Therefore, we performed searches for such bumps in another 36 subsets of the Kepler data that were randomly selected but with the condition that they did not include any transit events. We did not obtain any evidence in favour of variations that could have been interpreted as spot occultations in any of these reference subsets had they occurred during transit events. This suggests that our methods are not very sensitive to false positives.

\begin{figure}
\center
\includegraphics[width=0.48\textwidth,clip]{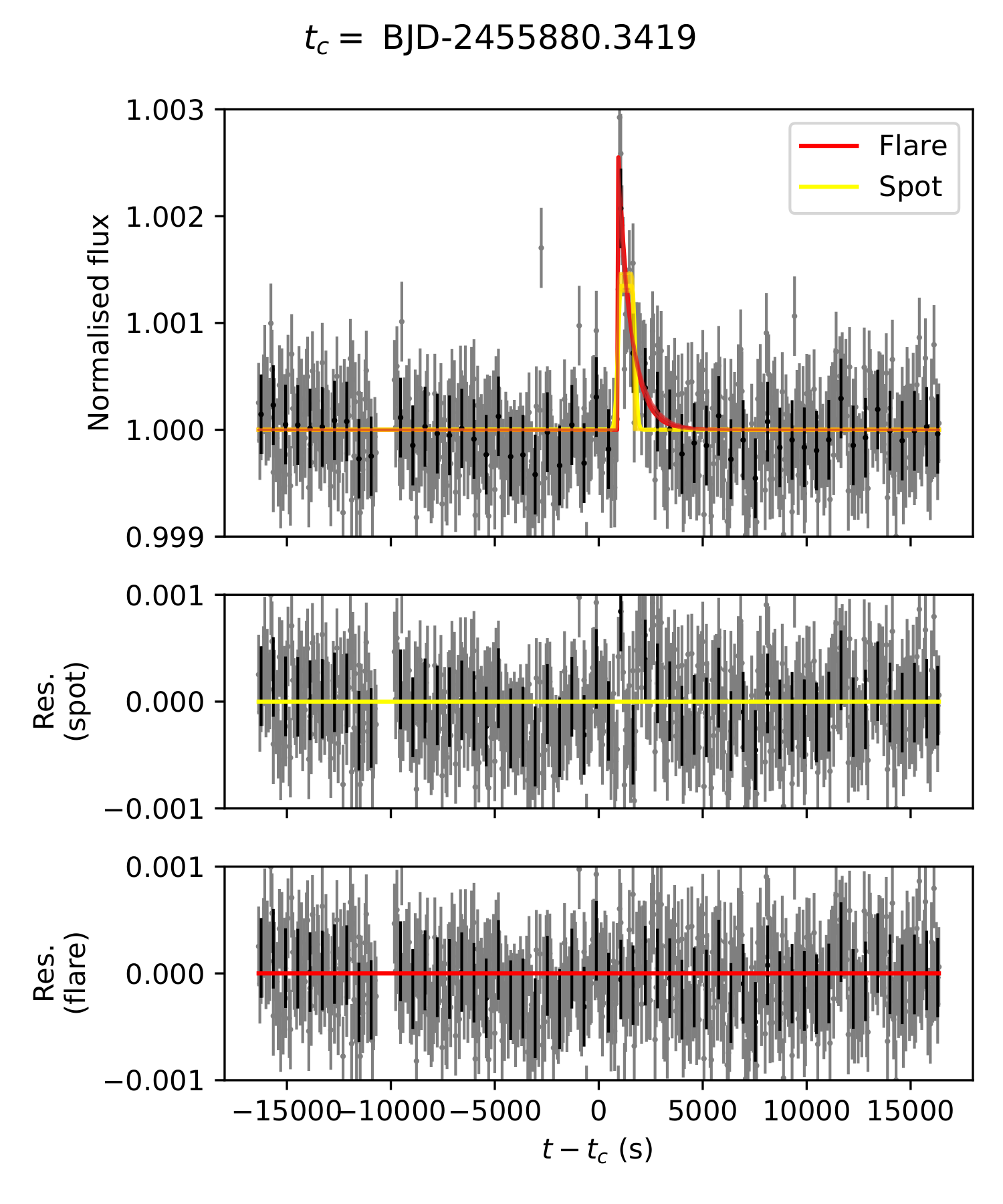}
\caption{The flare-event occurring outside transit events modelled for reference with both spot occultation model and a simple exponential flare morphology model. The latter model was favoured by our significance tests.}\label{fig:isolated_flare_event}
\end{figure}

Yet, in one of the data subsets outside transits we observed a flare event that was also visually determined to have a flare-like morphology \citep[see e.g.][]{howard2022}: a rapid increase followed by a slower decay. This was also confirmed by our model comparisons (Fig. \ref{fig:isolated_flare_event}). Applying a simple flare model with a discontinuity at time $t_{\rm f}$ such that the flux jumps above the baseline, followed by an exponential decay, we could improve the modelling of the event confirming that it indeed was better explained as a flare. 

During the transit event 164 there was a simultaneous flare-event (Fig. \ref{fig:spot_transits}). This event enabled us to see whether such an event could be confused with a spot occultation when occurring during a transit. Although our searches for spot occultation events yielded a positive detection, the flare model was again a better description of the variations.

\subsection{Physical properties of spots from occultations}

The occultations of spots by transiting planets can be used to estimate spot properties such as size, contrast and temperature. The contrast of the spot can be defined as a ratio between intensities of unspotted photosphere ($I_{\rm p}$) and spotted stellar surfaces ($I_{\rm s}$) \citep{morris2017}
\begin{equation}
c = 1 - \frac{I_{\rm s}}{I_{\rm p}} .
\end{equation}

For a spot larger than the planet's disk, the contrast can be approximated as $c=h_{\rm s}/\delta_{\rm p}$, where $h_{\rm s}$ is the amplitude parameter in Eq. (\ref{eq:spot_occultation}) and $\delta_{\rm p}$ is the transit depth at the location of the spot given an unspotted stellar surface. For smaller spots, the contrast is
\begin{equation}\label{eq:spot_contrast}
c=\frac{h_{\rm s}}{\delta_{\rm p}} \Bigg( \frac{R_{\rm p}}{R_{\rm s}} \Bigg)^{2} ,
\end{equation}
where $R_{\rm s}$ denotes the radius of the spot. \textrm{This is, during the transit the star's flux decreases by $\delta_{\rm p}$ only to increase by $h_{\rm s}$ during the spot occultation. Both of these parameters are thus defined in units of the unperturbed stellar flux.}

Under assumptions of blackbody radiation, it is possible to also estimate spot temperatures \citep{silva-valio2010}. The effective temperature of the spot $T_{\rm s}$ can be expressed as
\begin{equation}\label{eq:spot_temp}
T_{\rm s} = \frac{h \nu}{\kappa} \left \{ \log \left [ 1 + \frac{\exp \left ( \frac{h \nu}{\kappa T_{\rm eff}} \right ) - 1}{f} \right ]\right \}^{-1}
\end{equation}
where $\kappa$ and $h$ are the Boltzmann and Planck constants, respectively, $\nu$ is the frequency of the central wavelength of the observing instrument\footnote{The central wavelength of Kepler telescope is 640 nm.}, $f = 1-c$ is the spot intensity with respect to the central stellar intensity, and $T_{\rm eff}$ is the stellar effective temperature.

\section{Rotational modulation mapping of spots}\label{sec:rotational_modelling}

As the star rotates, spots come in and out of view leaving a distinctive pattern of variability modulated at or around the stellar rotation period. Simple geometric expressions for this modulation can be derived \citep{aigrain2012,kipping2012}, and it is possible to construct a model that describes the apparent rotation-induced variability of the star's brightness as a geometric consequence of $k$ spots co-rotating on the stellar surface.

We applied such modelling to the Kepler-411 PDC observations taken with 1800 s cadence. The modelling was applied to four subsets of the full Kepler light curve with a baseline of 1470 days. First, we selected a subset of the data around the two consecutive spot transits over transit numbers 116 and 117 at epochs 1044.016 and 1051.851 (BJD-2454833), respectively. We then limited the analysis to data taken ten days before the former transit and ten days after the latter one. This resulted in a data subset with a baseline of 29 days, roughly three times the rotation period of the star. We also selected three additional 29-day subsets beginning 200 days before and after the epoch 1034, and around the third spot occultation event during the transit 121 at epoch 1083.189 (BJD-2454833). The first two of these additional subsets were selected to see how much the estimated spot structure evolved over some twenty rotation periods of the star. We label these subsets as S834, S1034, S1069 and S1234, respectively, where the number denotes the starting epoch of each set.

We modelled the spot-induced effects on the Kepler flux as a superposition of the effects of several small spots. The rationale of this choice is that any irregularities in spot structure, or configuration of spot groups, can be approximated by several small spots as long as their effects are supported by data. Ultimately the whole stellar surface can be divided into a grid for which each surface element can be modelled independently. Yet, such modelling is well-known to correspond to an ill-posed statistical problem \citep{luo2019,luger2021,luger2021b}, and we thus attempt to minimise the number of spots while simultaneously describing the data as well as possible.
 
\subsection{Statistical modelling}

The brightness variations caused by spots co-rotating on the stellar surface can be modelled with a simple geometric model under the assumption that the spots are sufficiently small with respect to the star.

For the $i$th measurement $m_{i}$ we can write an expression for the stellar flux as
\begin{equation}\label{eq:forward_model}
  m_{i} = \gamma - \sum_{j=1}^{k}f_{j}(t_{i}) + \phi r_{i-1} + \epsilon_{i},
\end{equation} 
where the functions $f_{j}, j=1, ..., k$, denote the effects of $k$ spots on the stellar flux with respect to a flux from an unspotted surface $\gamma$. We apply a moving average (MA) model to account for intrinsic correlations in the data caused by e.g. astrophysical noise. In Eq. \ref{eq:forward_model} the parameter $\phi$, such that $|\phi| \leq 1$ to ensure stationarity, is a free parameter of the MA component and $r_{i}$ denotes the residual of the data point $m_{i}$ at time $t_{i}$. Finally, $\epsilon_{i}$ denotes a Gaussian random variable that has a zero mean and a variance of $\sigma^{2}_{i} + \sigma^{2}$ such that $\sigma$ is a free parameter. We note that the interpretation of parameter $\gamma$ is that of the flux coming from the unspotted stellar surface.

The effects of spots are assumed to be independent for each spot such that the function $f_{j}$ accounts for the projection of the $j$th spot on the stellar surface \citep{aigrain2012,kipping2012}. For a spot at a latitude $\psi_{j}$, with a phase $\omega_{j}$, and apparent rotation period $P_{j}$ around the star, the distance of the spot from the centre of the stellar disk is
\begin{equation}\label{eq:spot_position}
  \cos \delta_{j}(t_{i}) = \cos i \sin \psi_{j} + \sin i \cos \psi_{j} \cos \Bigg( \frac{2\pi t}{P_{j}} + \omega_{j} \Bigg) ,
\end{equation}
where $i$ is the stellar inclination. Accounting for limb darkening with a linear model, the observed flux is
\begin{equation}\label{eq:spot_model_full}
  f_{j}(t_{i}) = A_{j} \Big[ (1-\varepsilon) \mu_{j}(t_{i}) + \varepsilon \mu_{j}^{2}(t_{i}) \Big]
\end{equation}
where $A_{j}$ represents the magnitude of the dimming caused by the $j$th spot due to its temperature and size and we resort to the standard notation such that $\mu_{j} = \cos \delta_{j}$. We can then define $A_{j} = \gamma A'_{j}$ where $A'_{j}$ is a unitless number representing the fractional dimming caused by the $j$th spot if it were at the centre of the stellar disk. {The limb darkening parameter $\epsilon$ is a free parameter of the model.}

Spots evolve on a time-scale of one rotation period of the star \citep{ioannidis2020}, which is sometimes accounted for in models of photometric data \citep{kipping2012,basri2020}. We model the spot evolution by adopting a time-dependent amplitude $\hat{A}_{j}$ defined as\footnote{Coincidentally, this is essentially the same functional form as is used for the shape of the spot occultations in Eq. (\ref{eq:spot_occultation}).}
\begin{eqnarray}\label{eq:spot_model_evolution}
  \hat{A}_{j}(t_{i}) = \left\{ \begin{array}{ccc}
    A_{j} \exp \bigg\{ -\frac{(t_{i} - t_{j, \rm in})^{2}}{2\sigma^{2}_{j, \rm in}} \bigg\} & \mathrm{ if } & t_{i} < t_{j, \rm in} \\
    A_{j} & \mathrm{ if } & t_{j, \rm in} \leq t_{i} \leq t_{j, \rm out} \\
    A_{j} \exp \bigg\{ -\frac{(t_{i} - t_{j, \rm out})^{2}}{2\sigma^{2}_{j, \rm out}} \bigg\} & \mathrm{ if } & t_{i} > t_{j, \rm out} \\
  \end{array} \right.
\end{eqnarray}
According to this envelope model, spot-induced effects on the stellar flux are constant between $t_{j, \rm in}$ and $t_{j, \rm out}$ and outside those limits they appear and disappear in a Gaussian manner on time-scales of $\sigma_{j, \rm in}$ and $\sigma_{j, \rm out}$, respectively, for all spots $j = 1, ..., k$. The four numbers are also free parameters of the model resulting in a total of eight free parameters for each spot.

Assuming that the stellar axis of spin is approximately normal to the plane of the planetary system simplifies the expressions above, but we do not make such a general assumption. Instead, we compare models where the star is seen edge-on with a model such that inclination is a free parameter as it is possible to obtain information on stellar inclination from precision photometry \citep{walkowicz2013}. This corresponds to a standard comparison of two nested models. Yet, when searching for solutions we resort to simplifying the model by fixing the inclination to edge-on orientation. This is justified by the fact that the star is surrounded by a system of planets seen approximately edge-on, and a significant spin-orbit misalignment would be very difficult to explain.

We set uniform priors for the parameters with a notable exception. The period parameter should only be allowed to have a range of values available around some \emph{a priori} expected value corresponding to the overall stellar rotation period. We thus set constraints for the allowed differential rotation of the star. We choose a prior $\pi(P_{j}) = \mathcal{N}(\mu_{\rm P}, \sigma^{2}_{\rm P})$ such that $\mu_{\rm P} = 10.5$ days and $\sigma_{\rm P} = 2.0$ days, respectively. The former is based on our estimation of the star's rotation period and the latter on a maximum rotational shear for such stars. This choice is made in accordance with the results of \citet{reinhold2013}, where the maximum rotational shear for similar stars was estimated to be some 0.2 rad/day corresponding to variations in period of about 2 days. We thus choose the standard deviation of our prior to equal to maximal variability making it sufficiently conservative.

As for spot occultations, we applied Bayes factors and likelihood ratios to determine significances. For likelihood ratio tests, the model with one less signal and eight less parameters as a null-hypothesis can be rejected with 99\% (99.9\%) when the logarithm of the ratio is in excess of 20.09 (26.12). The only exception is the model with $k=1$ because it has nine more free parameters than the model with $k=0$. This is because the limb darkening parameter plays no role without signals. The corresponding 99\% (99.9\%) threshold is thus 21.67 (27.88).

\subsection{Identification of solutions}

We searched for solutions in the parameter space with the AM posterior sampling algorithm of \citet{haario2001}. But identification of solutions in the multidimensional parameter space remains a challenge, and although solution methods such as posterior samplings would ultimately enable identifying the global solution, given finite samples this is not guaranteed. We therefore keep this limitation in mind while interpreting the results but attempt to minimise the chances of identifying local solutions rather than global ones.

Because we set a prior for the period parameter, our modelling does not suffer from high multimodality of the posterior probability density in the period space as harmonics of the rotation period are ruled out \emph{a priori}. This removes one major obstacle in the search for solutions. We also assume that inclination of the star is consistent with edge-on orientation while seaching for solutions and allow the parameter to be a free parameter of the model when an optimal spot solution has been identified.

Our general algorithm for searching for solutions works as follows. Given that a solution for a model $\mathcal{M}_{k}$ with $k$ spot signals has been identified, we search for solutions for model $\mathcal{M}_{k+1}$ by first duplicating one of the signals in the model $\mathcal{M}_{k}$. Given that the $j$th signal, for some $j = 1, ..., k$, has parameters $\theta_{j}$, we set $\theta_{k+1} = \theta_{j}$ with one exception. The signal amplitude $A_{j}$ (Eq. \ref{eq:spot_model_full}) is replaced with $A_{j}/a$ and we set $A_{k+1} = A_{j}(1-a)$ for some $a \in (0,1)$. In this manner, we choose an initial state for the posterior samplings that matches the likelihood identified for the solution of model $\mathcal{M}_{k}$. If the model $\mathcal{M}_{k}$ is an attempt to model the superposition of many spots with only $k$ of them, at least one of the spot signals is a compromise between the simple model and the actual spotted surface of the star. Such a choice for an initial state enables reasonably rapid searches for solutions for model $\mathcal{M}_{k+1}$. In practice, we tested all signals $j = 1, ..., k$ and different values of $a$ for each. Typically, we selected values of $a$ ranging from 0.1 to 0.5 to enable searching for solutions where the superposition of one signal was caused by two spots with similar amplitude parameters and two spots with amplitudes differing by almost an order of magnitude.

When identifying solutions, we also tested whether the independent chains indeed yielded consistent solutions in the multidimensional parameter space. We calculated Gelman-Rubin statistics for testing whether there were differences in variances within and between chains. As in the previous section, we required that the Gelman-Rubin statistics $R$ was below 1.05 for our samplings.

\subsection{Solutions for S1034}\label{sec:section_1034}

\begin{figure*}
\center
\includegraphics[angle=270,width=0.24\textwidth,clip]{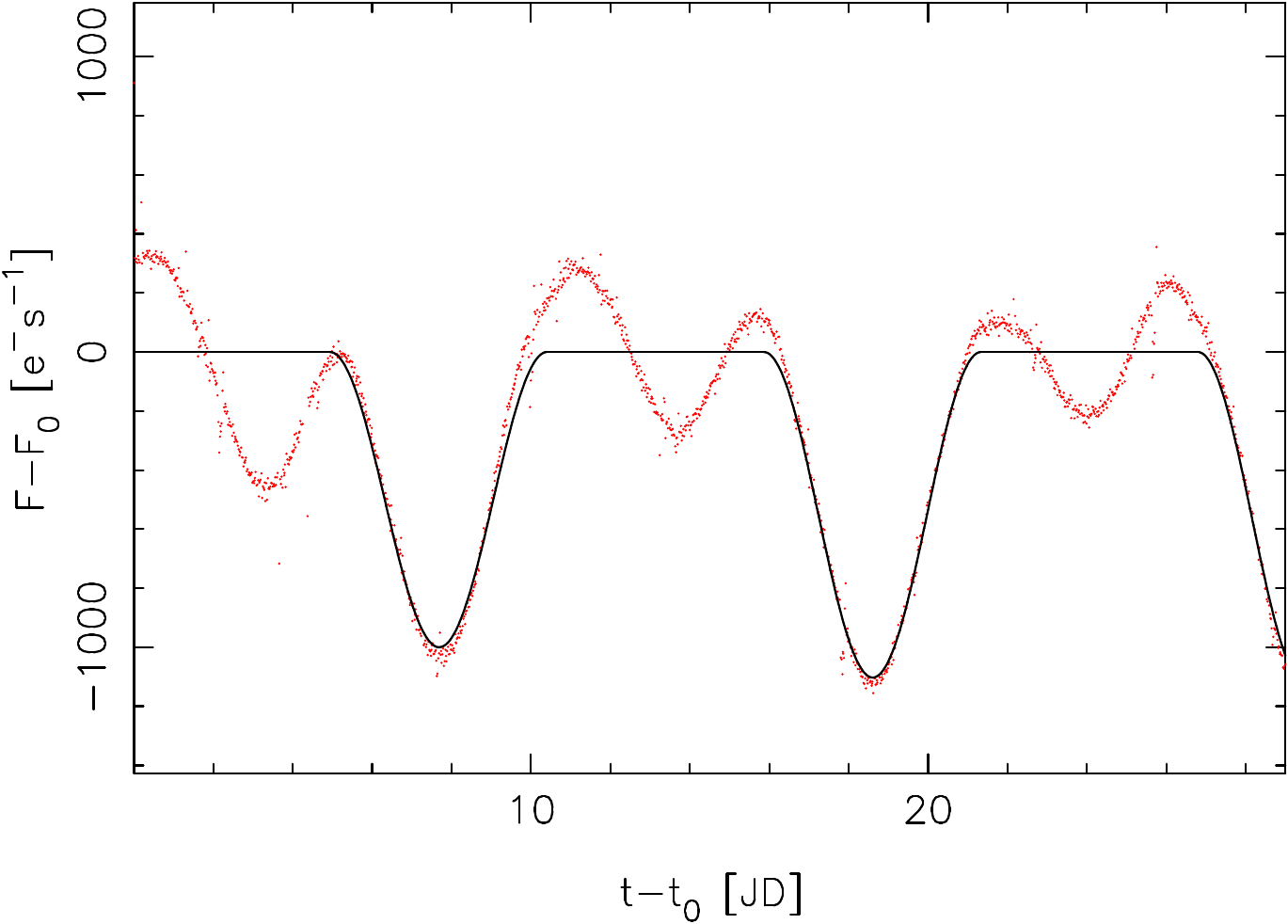}
\includegraphics[angle=270,width=0.24\textwidth,clip]{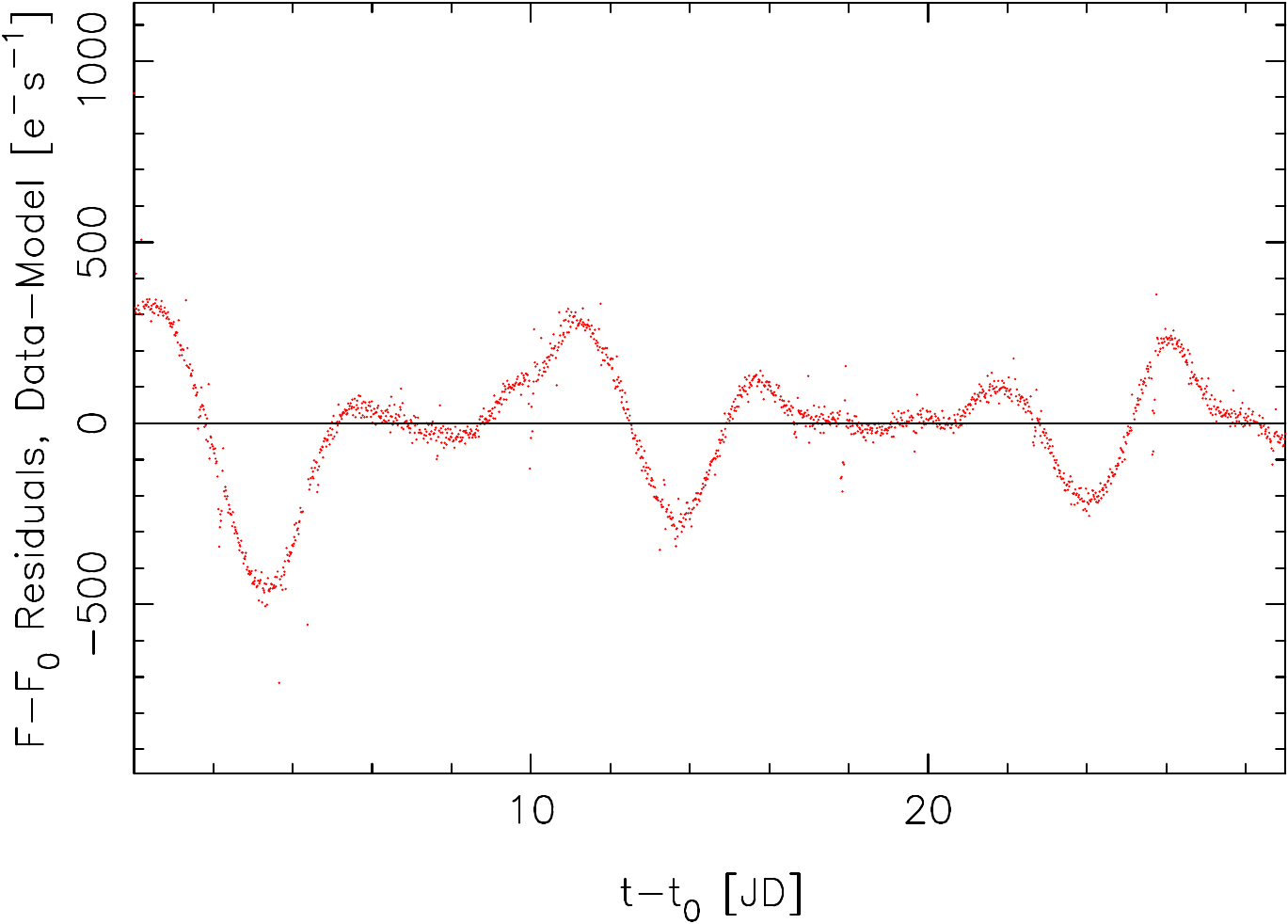}
\includegraphics[angle=270,width=0.24\textwidth,clip]{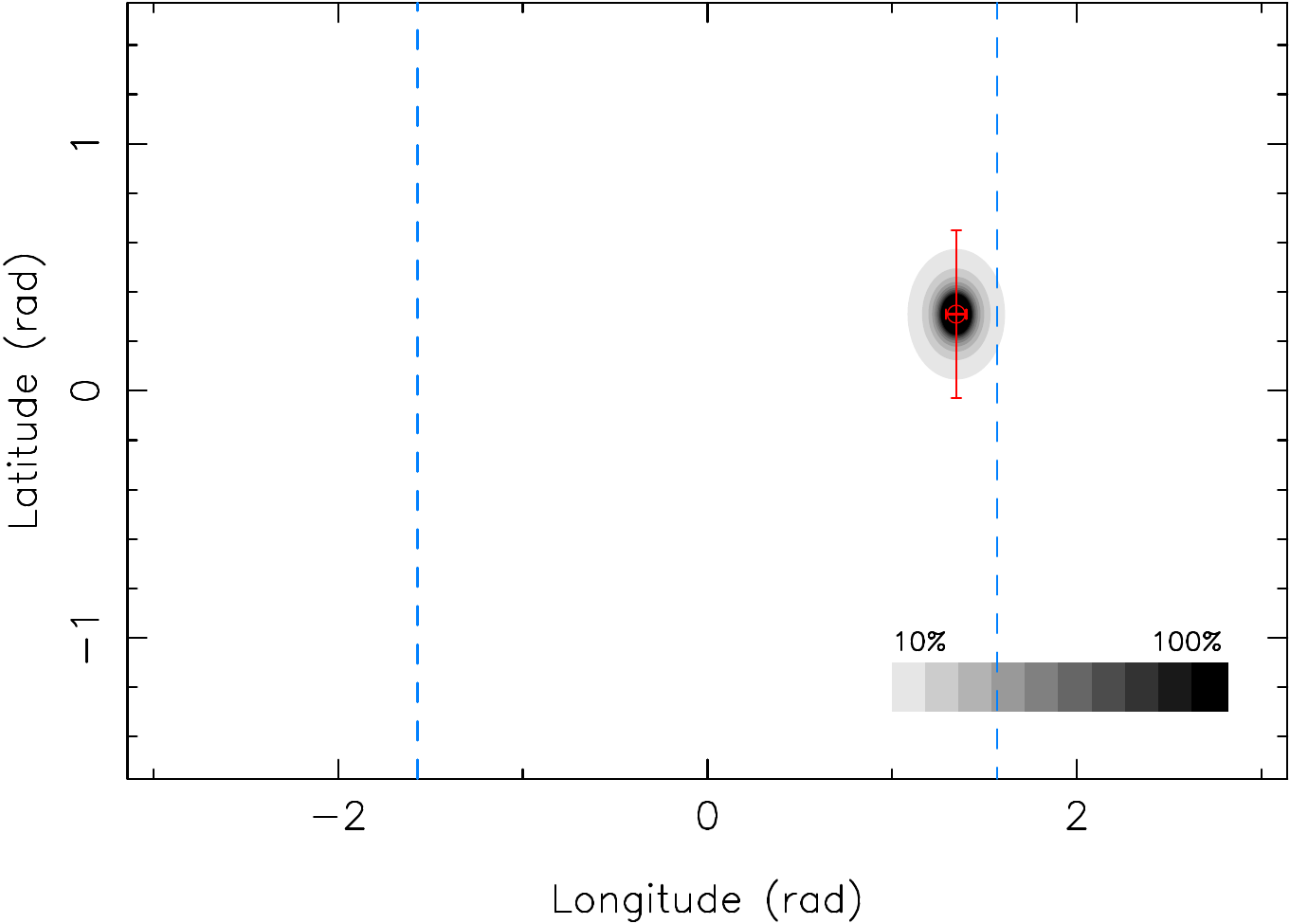}
\includegraphics[angle=270,width=0.24\textwidth,clip]{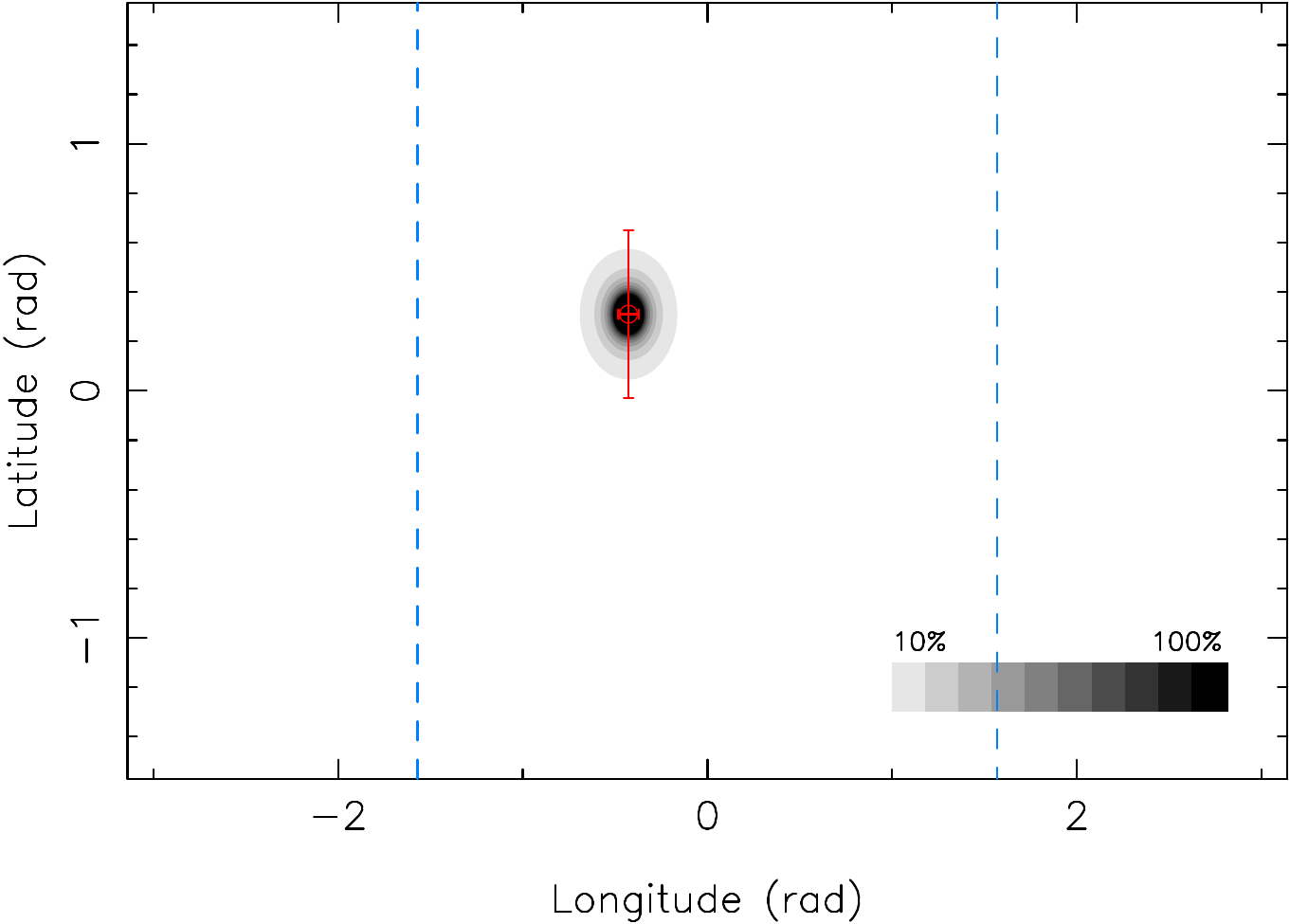}

\includegraphics[angle=270,width=0.24\textwidth,clip]{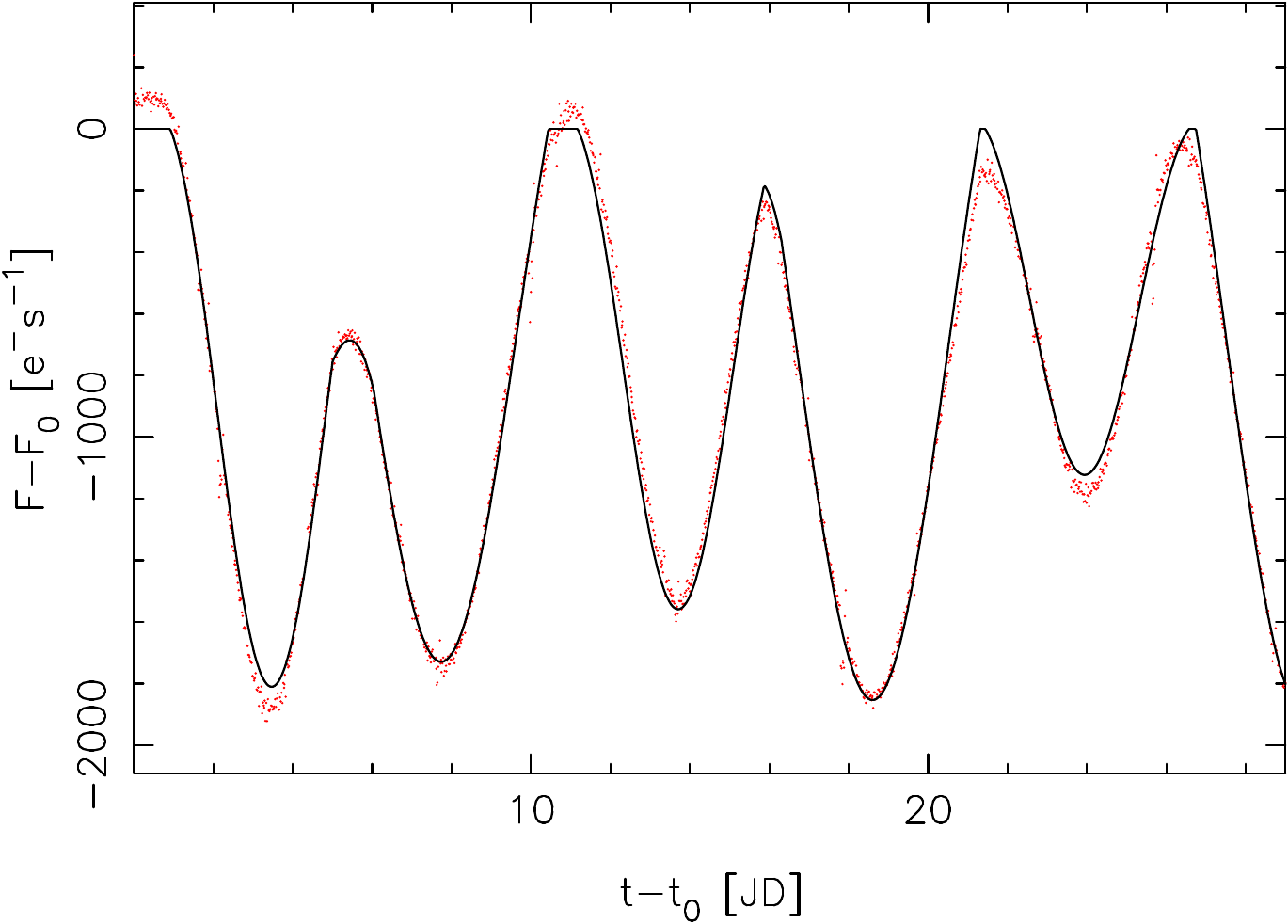}
\includegraphics[angle=270,width=0.24\textwidth,clip]{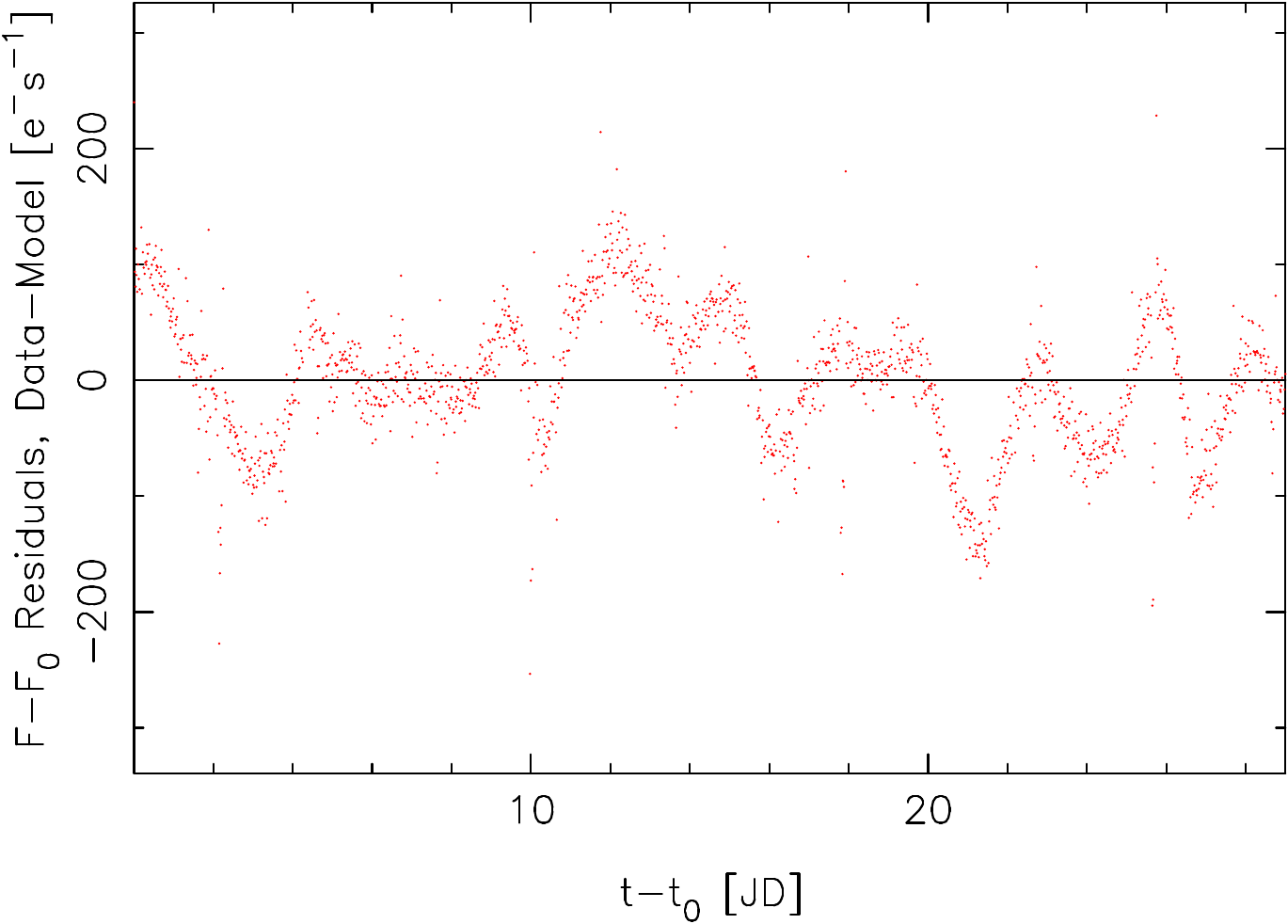}
\includegraphics[angle=270,width=0.24\textwidth,clip]{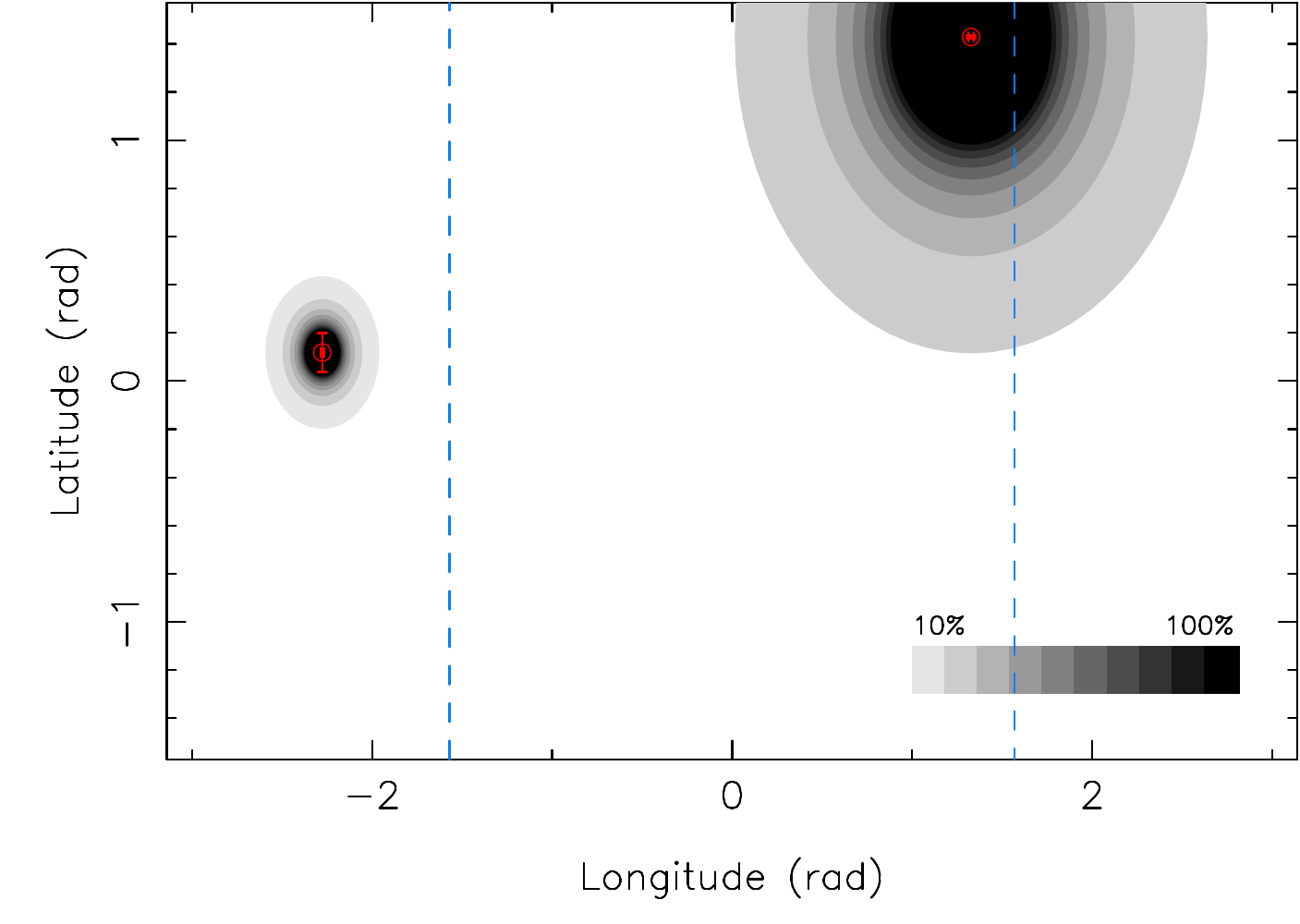}
\includegraphics[angle=270,width=0.24\textwidth,clip]{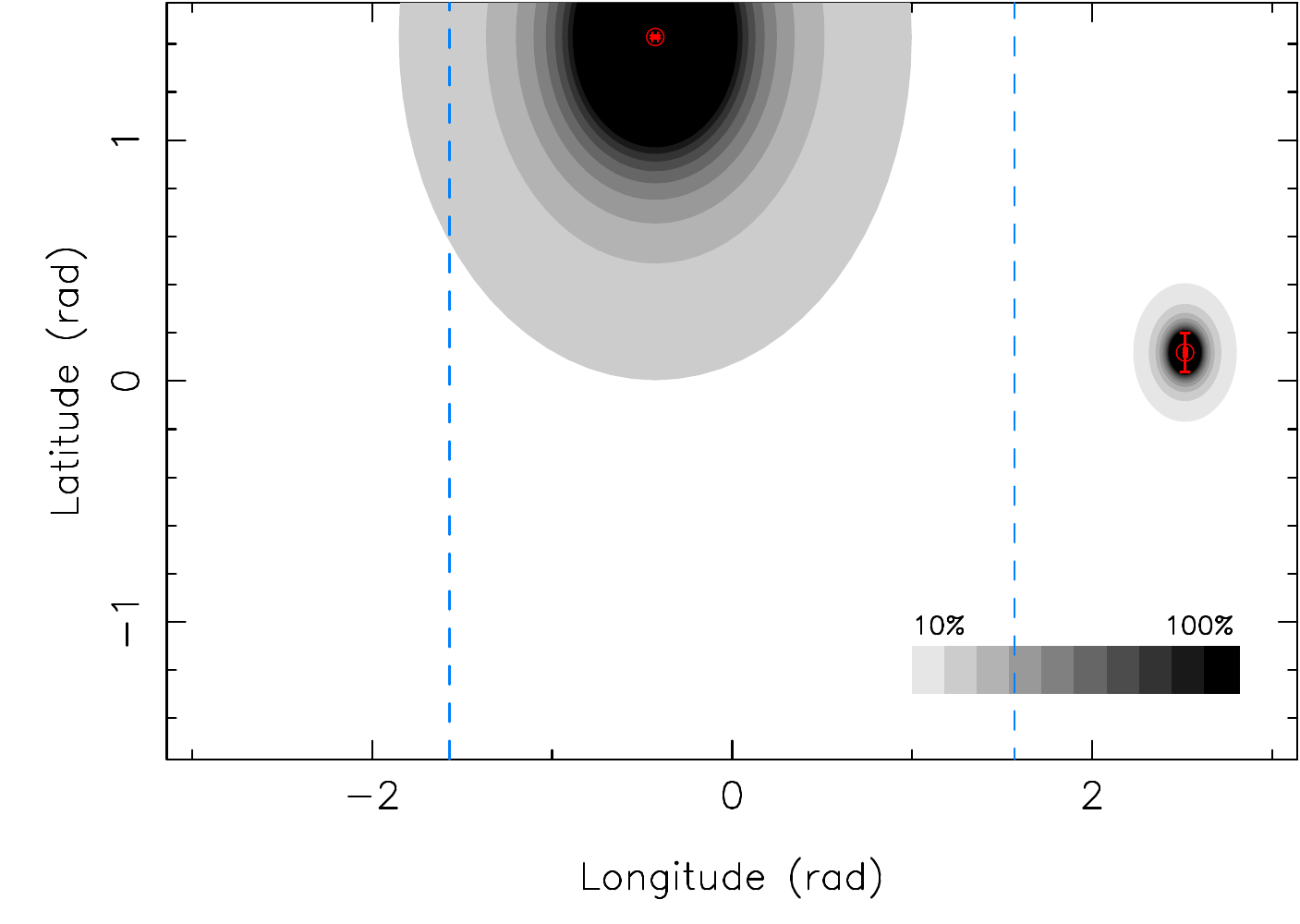}

\includegraphics[angle=270,width=0.24\textwidth,clip]{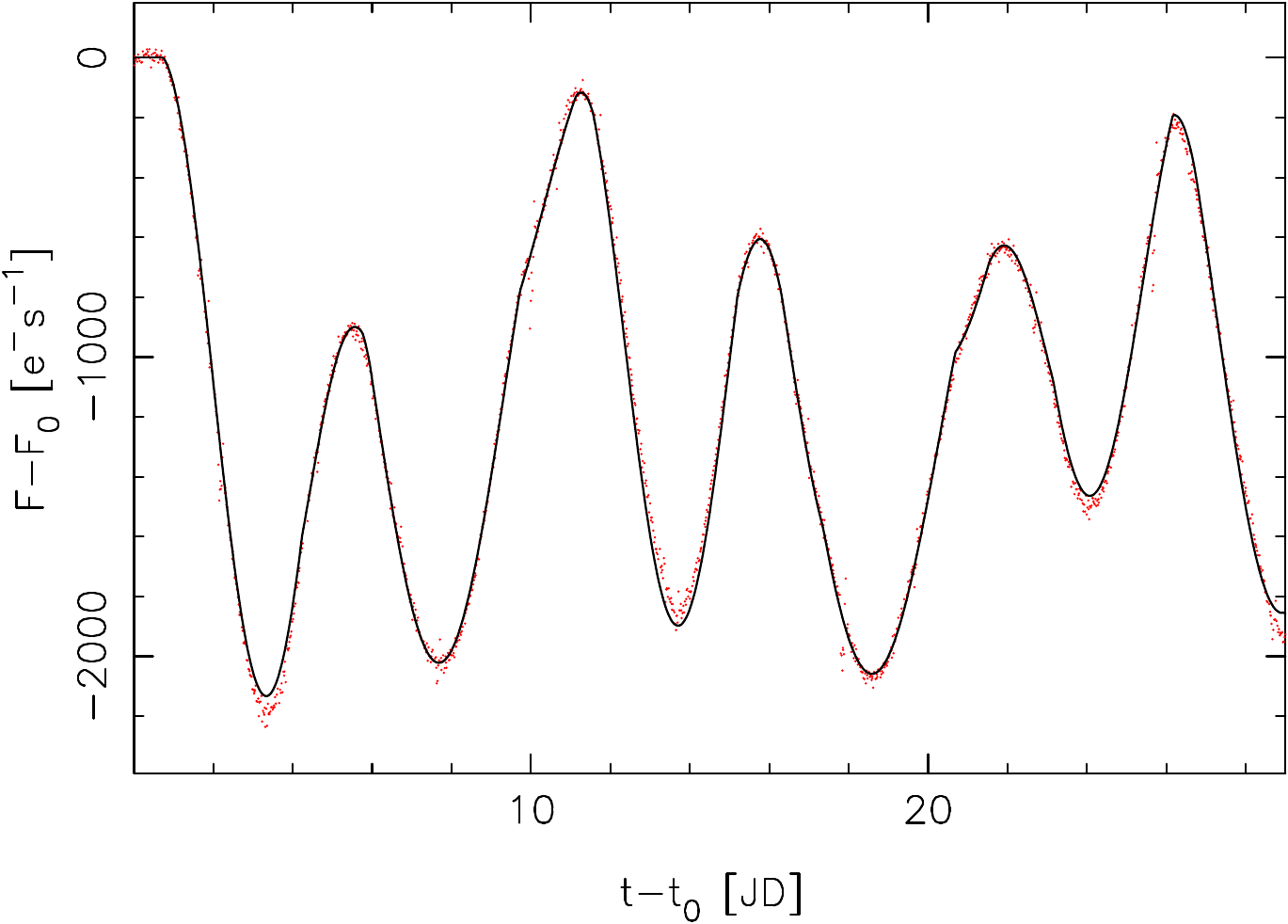}
\includegraphics[angle=270,width=0.24\textwidth,clip]{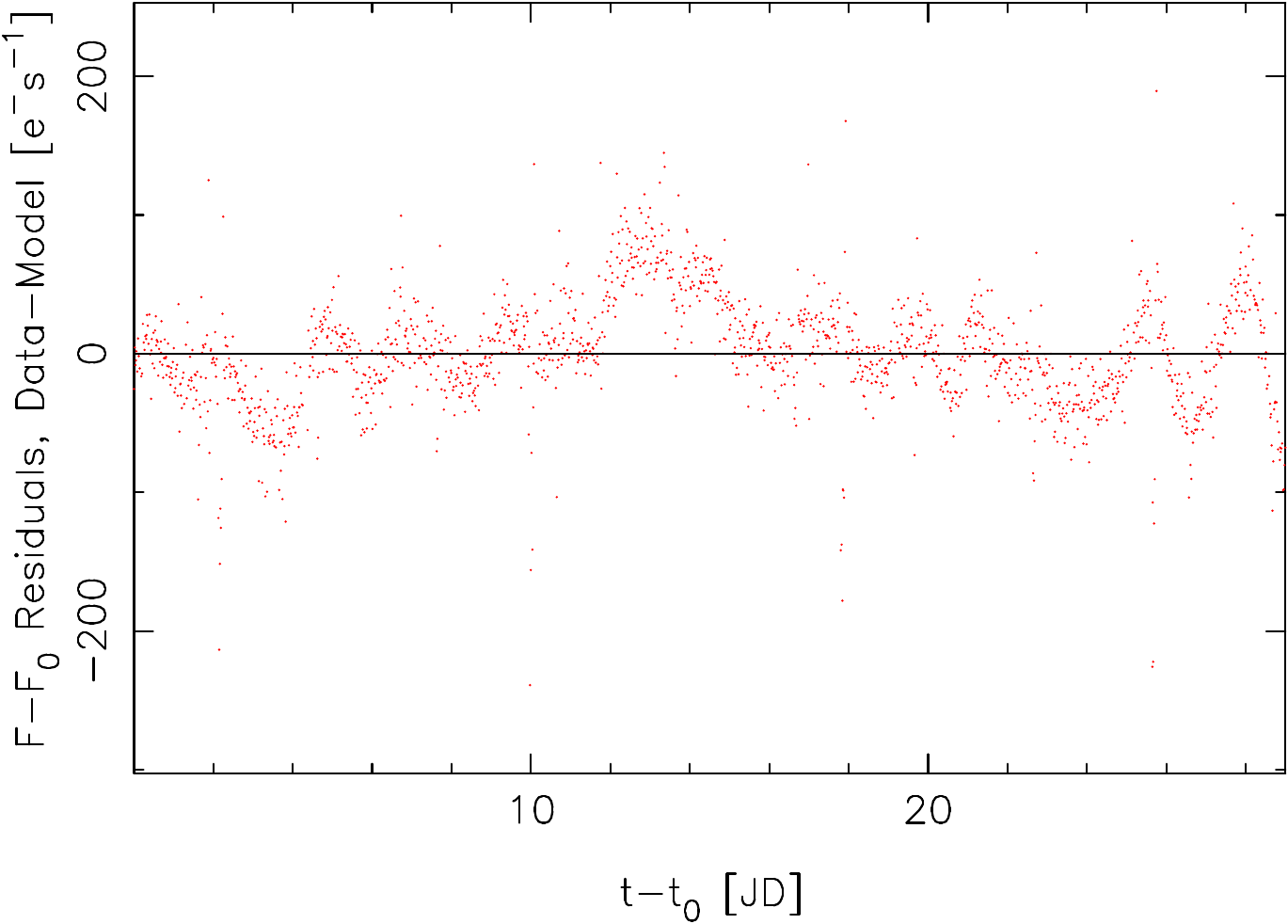}
\includegraphics[angle=270,width=0.24\textwidth,clip]{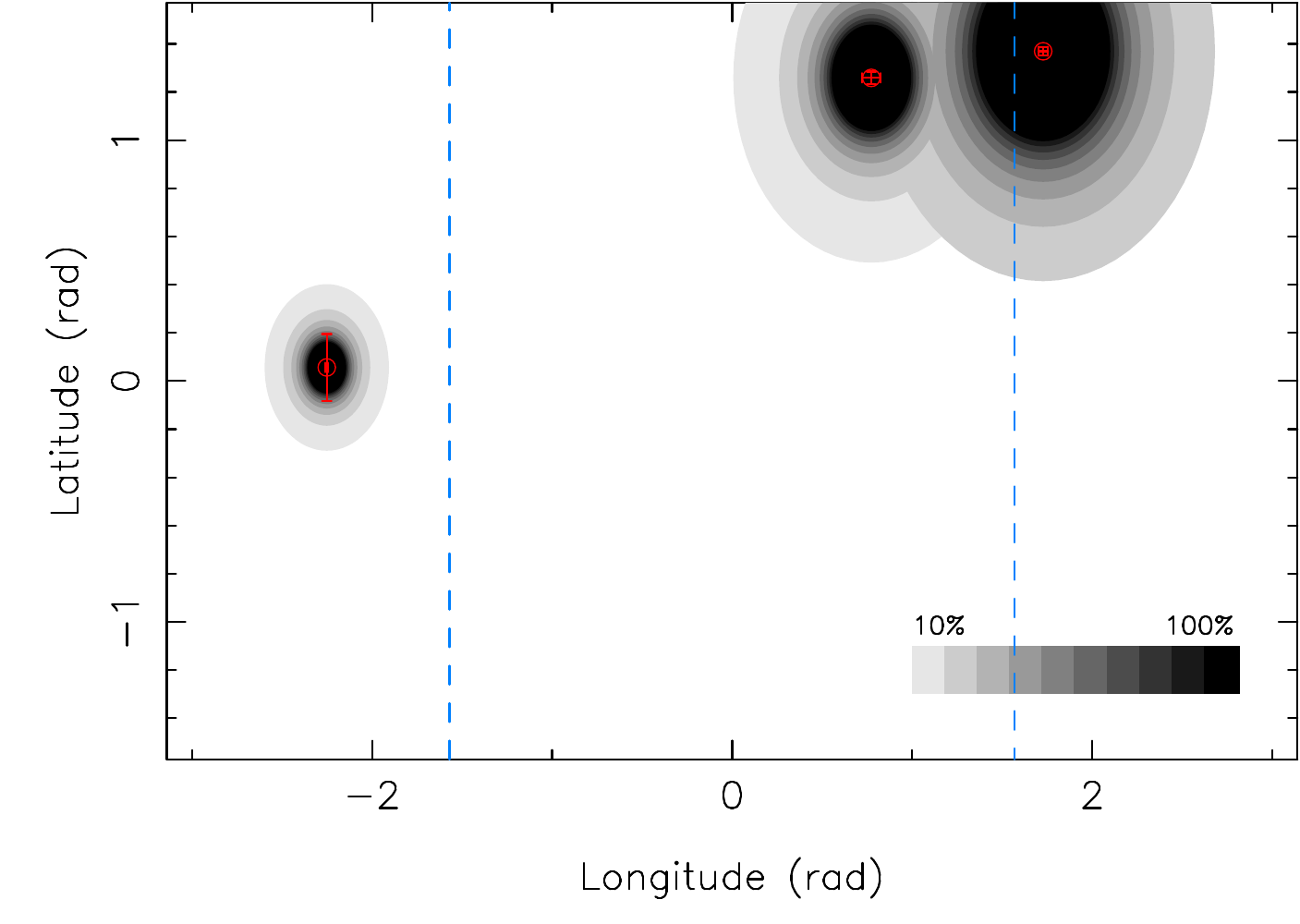}
\includegraphics[angle=270,width=0.24\textwidth,clip]{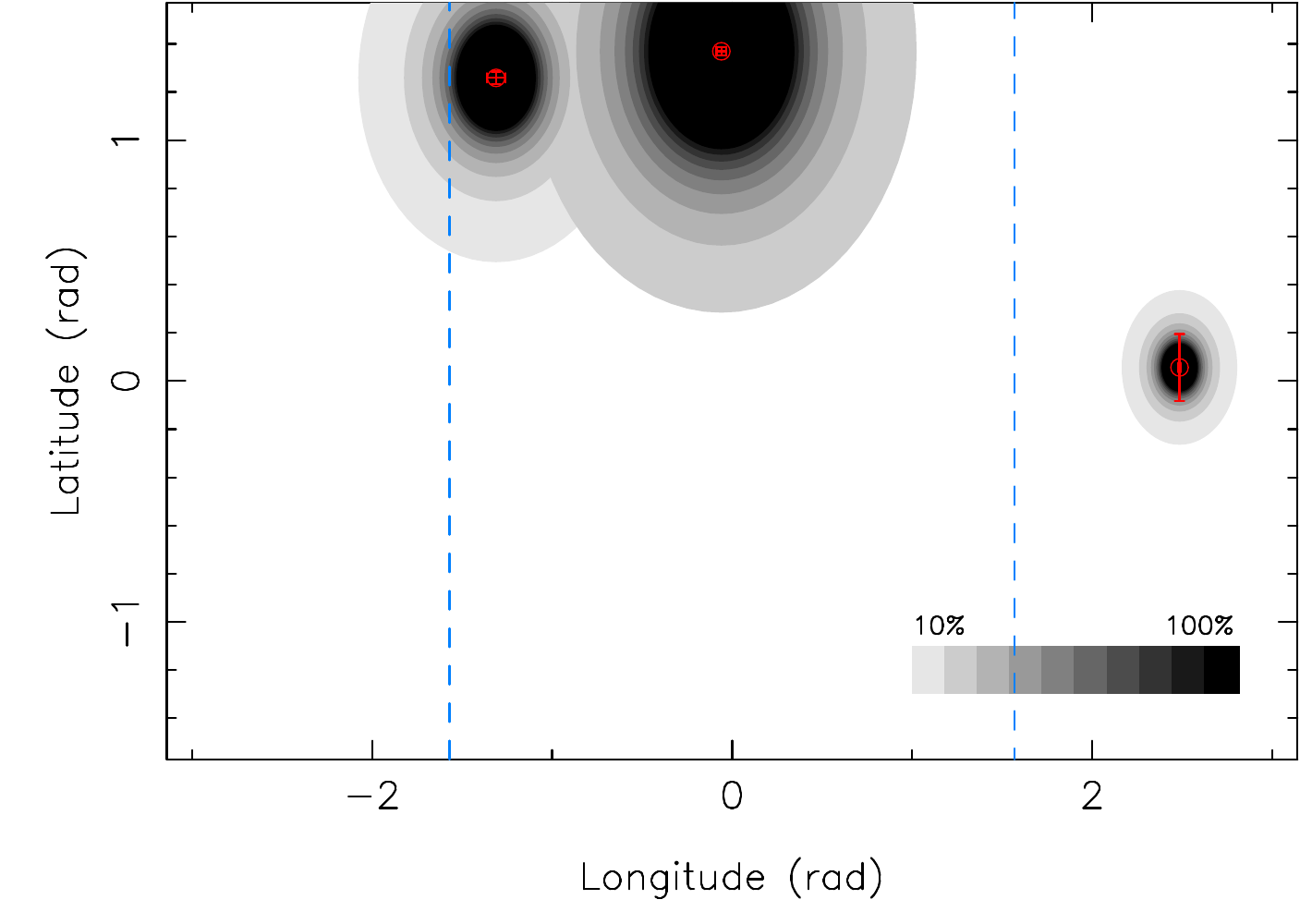}

\includegraphics[angle=270,width=0.24\textwidth,clip]{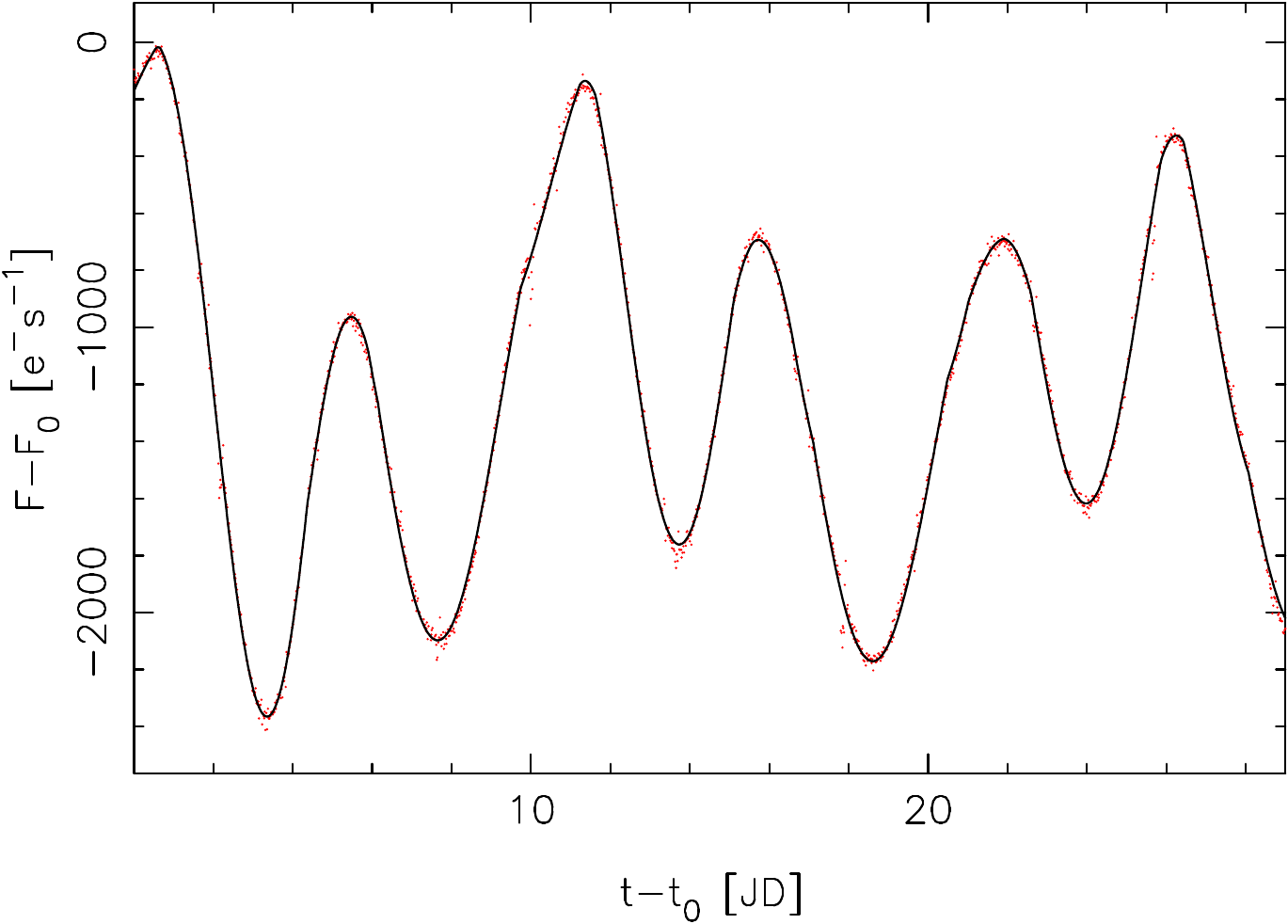}
\includegraphics[angle=270,width=0.24\textwidth,clip]{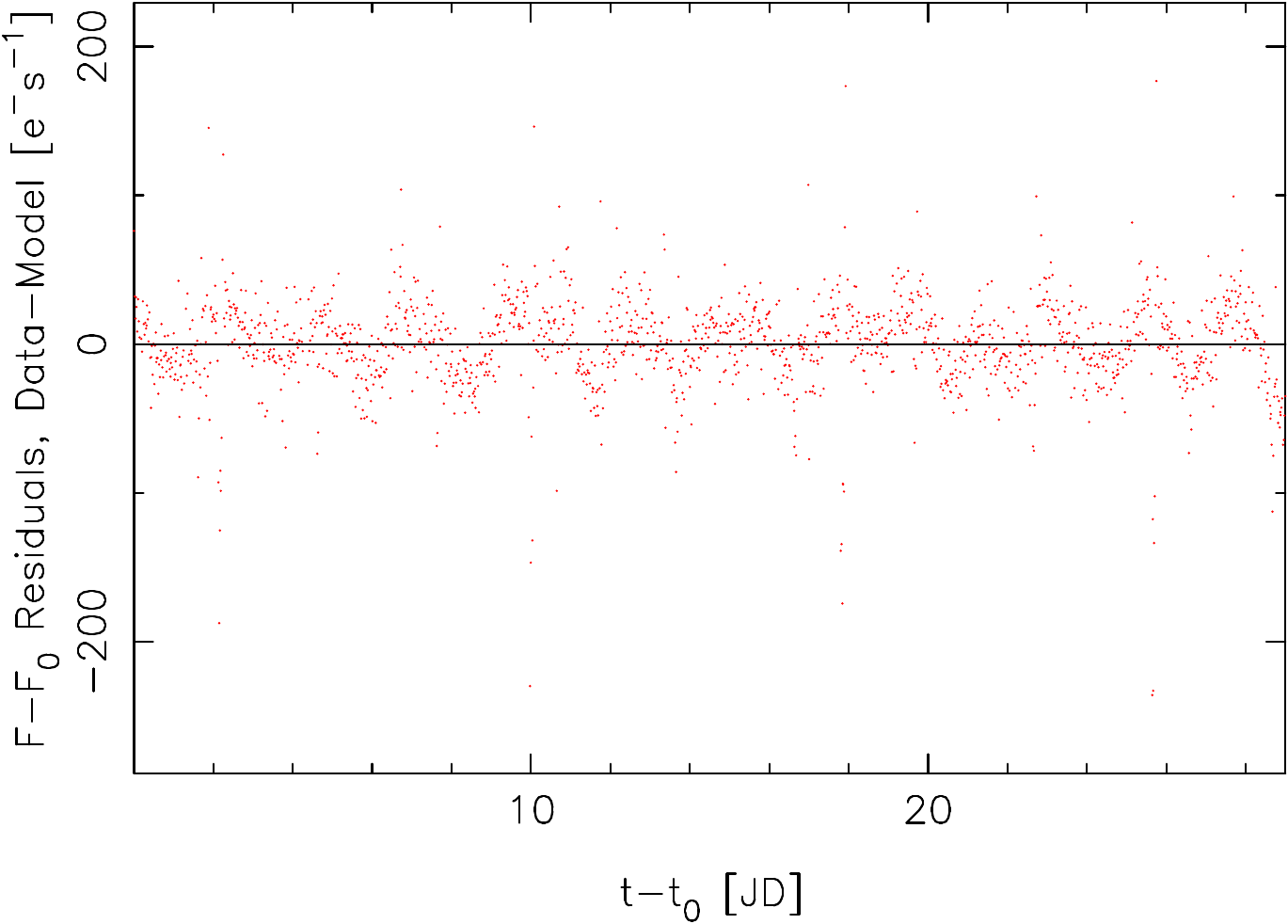}
\includegraphics[angle=270,width=0.24\textwidth,clip]{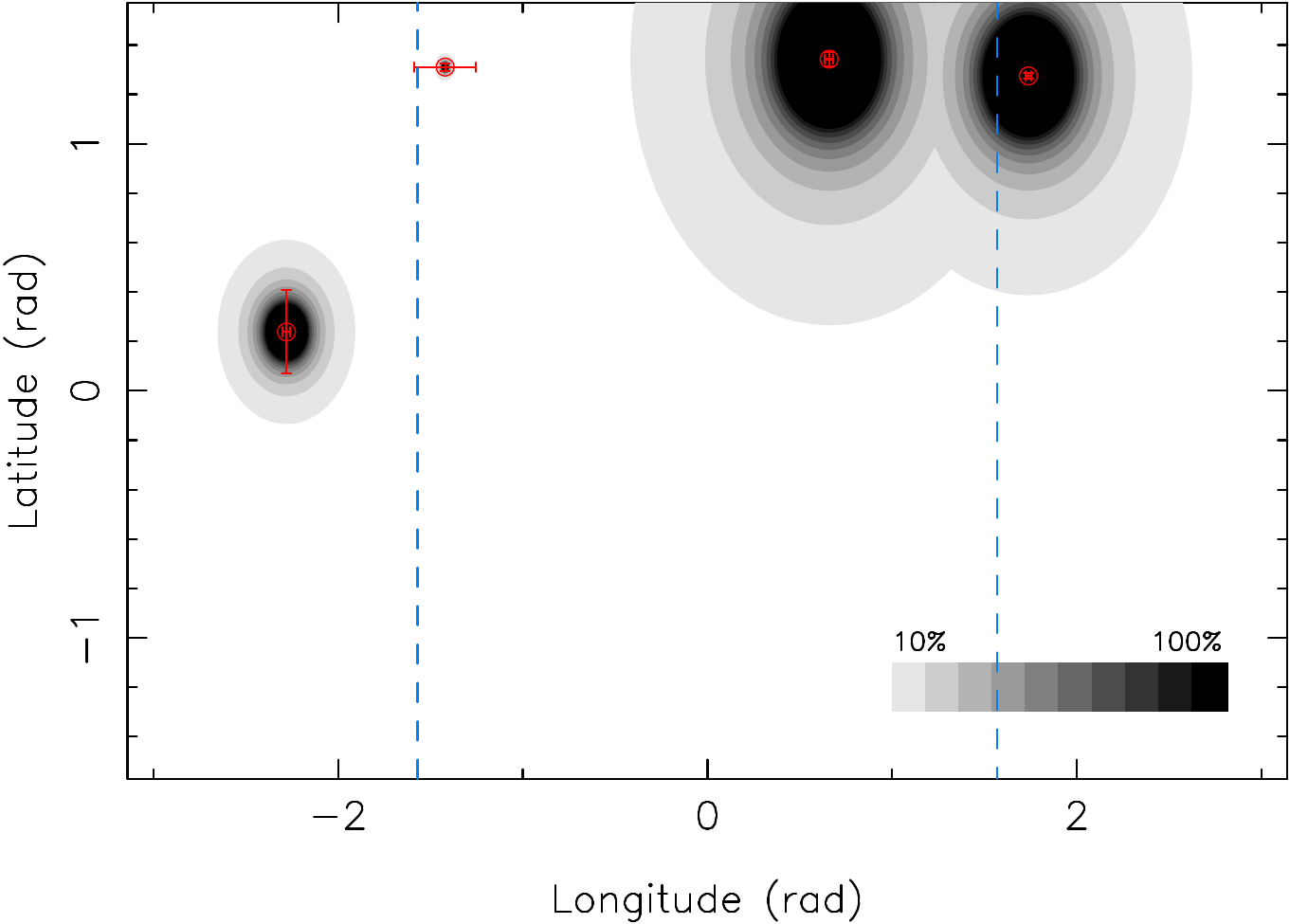}
\includegraphics[angle=270,width=0.24\textwidth,clip]{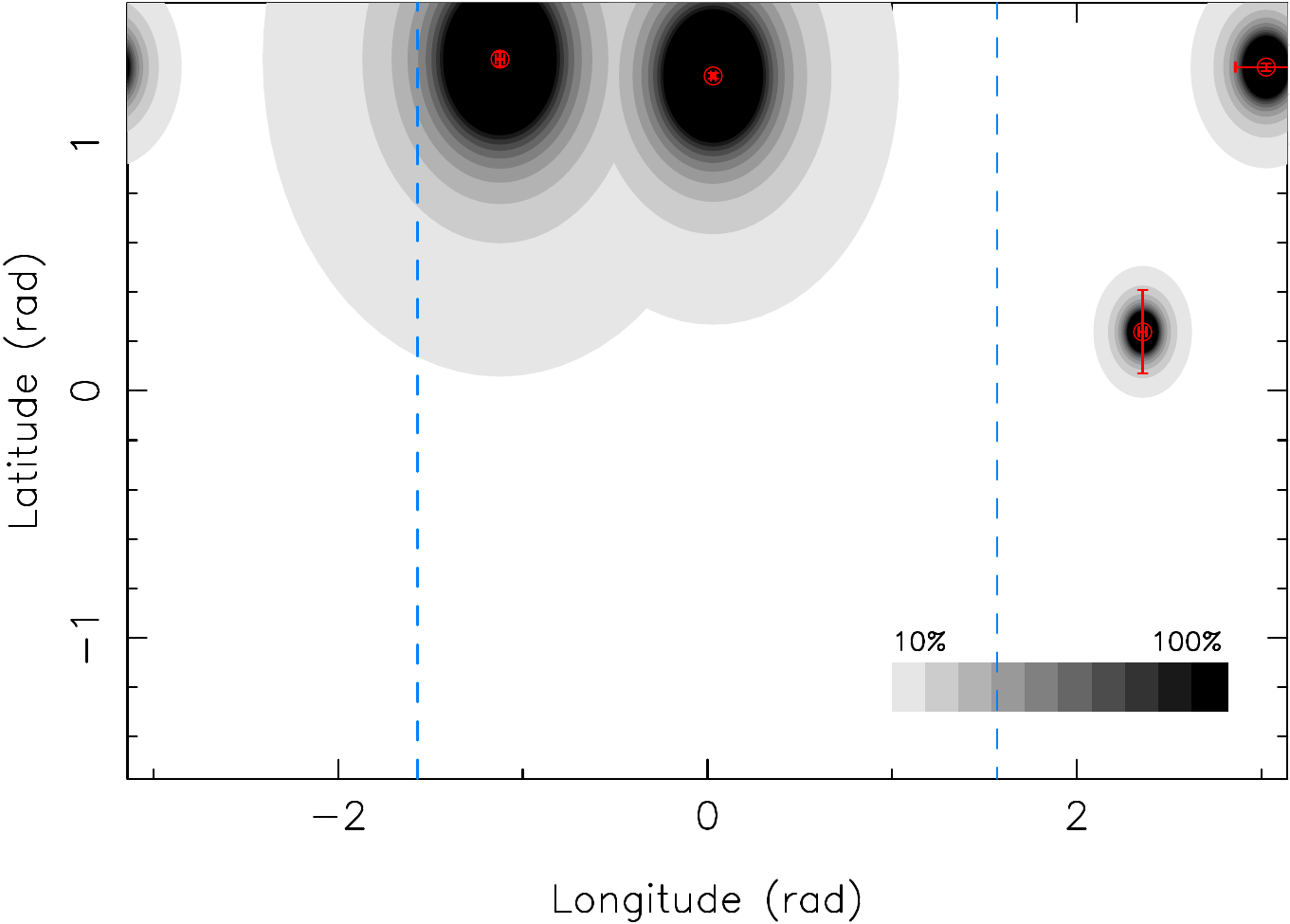}

\includegraphics[angle=270,width=0.24\textwidth,clip]{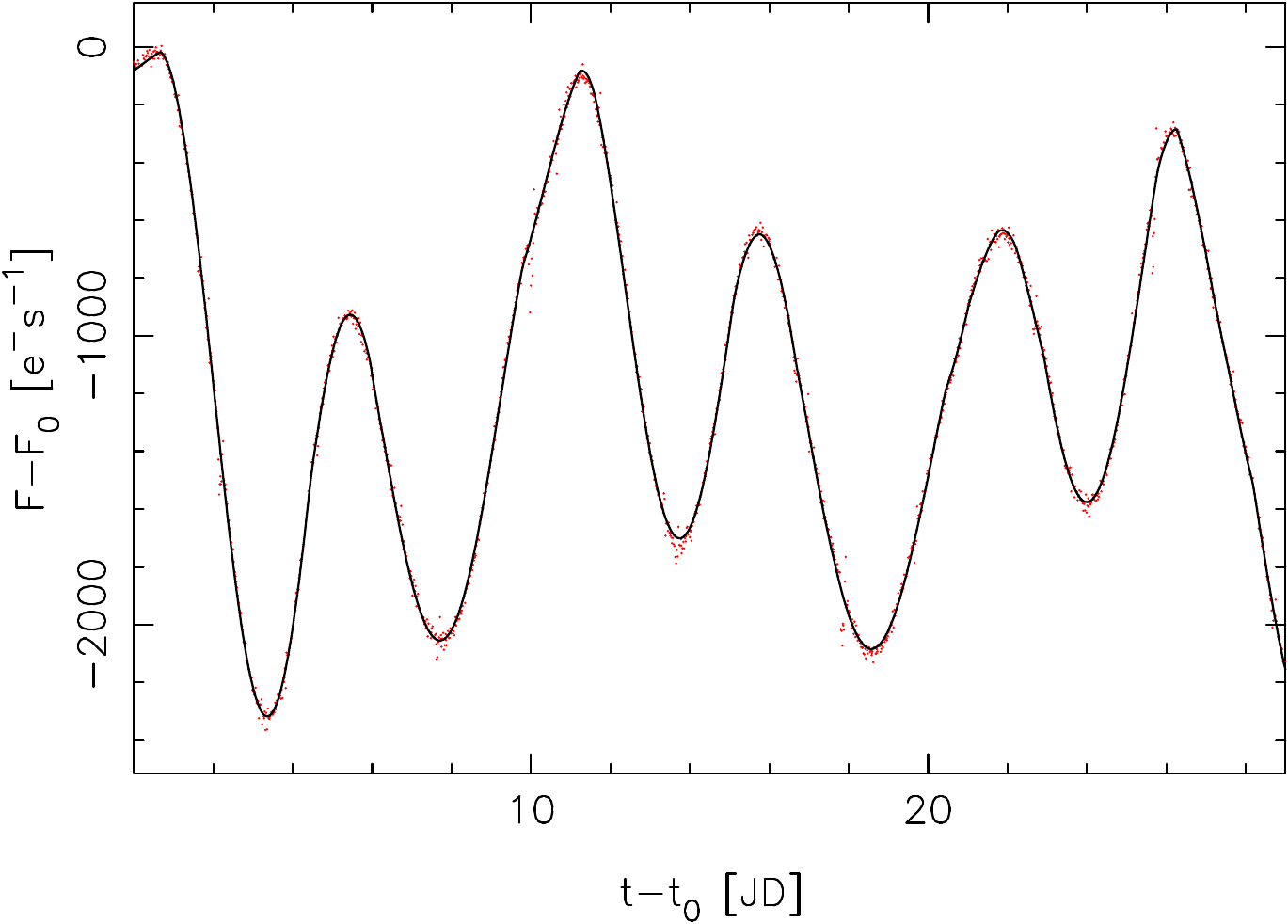}
\includegraphics[angle=270,width=0.24\textwidth,clip]{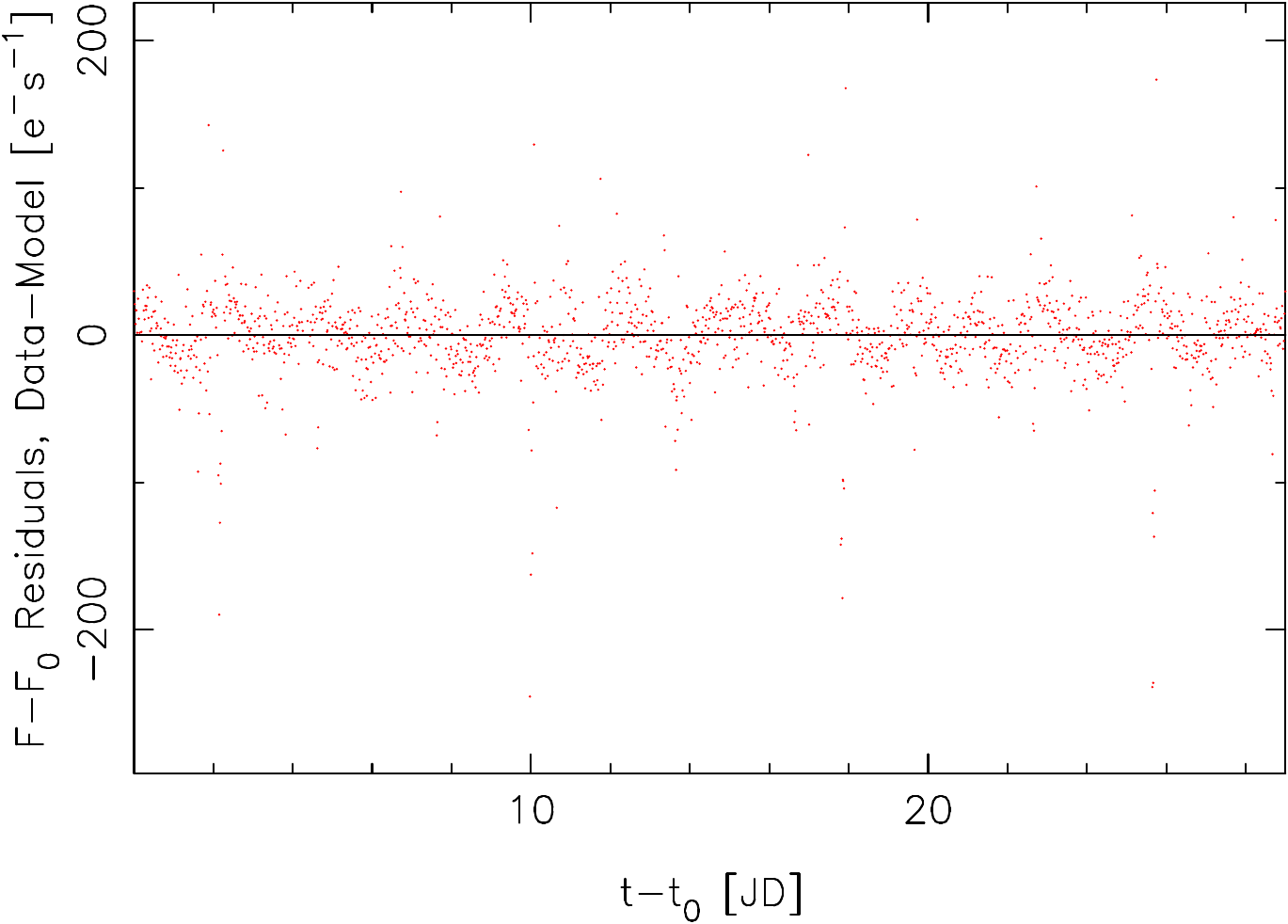}
\includegraphics[angle=270,width=0.24\textwidth,clip]{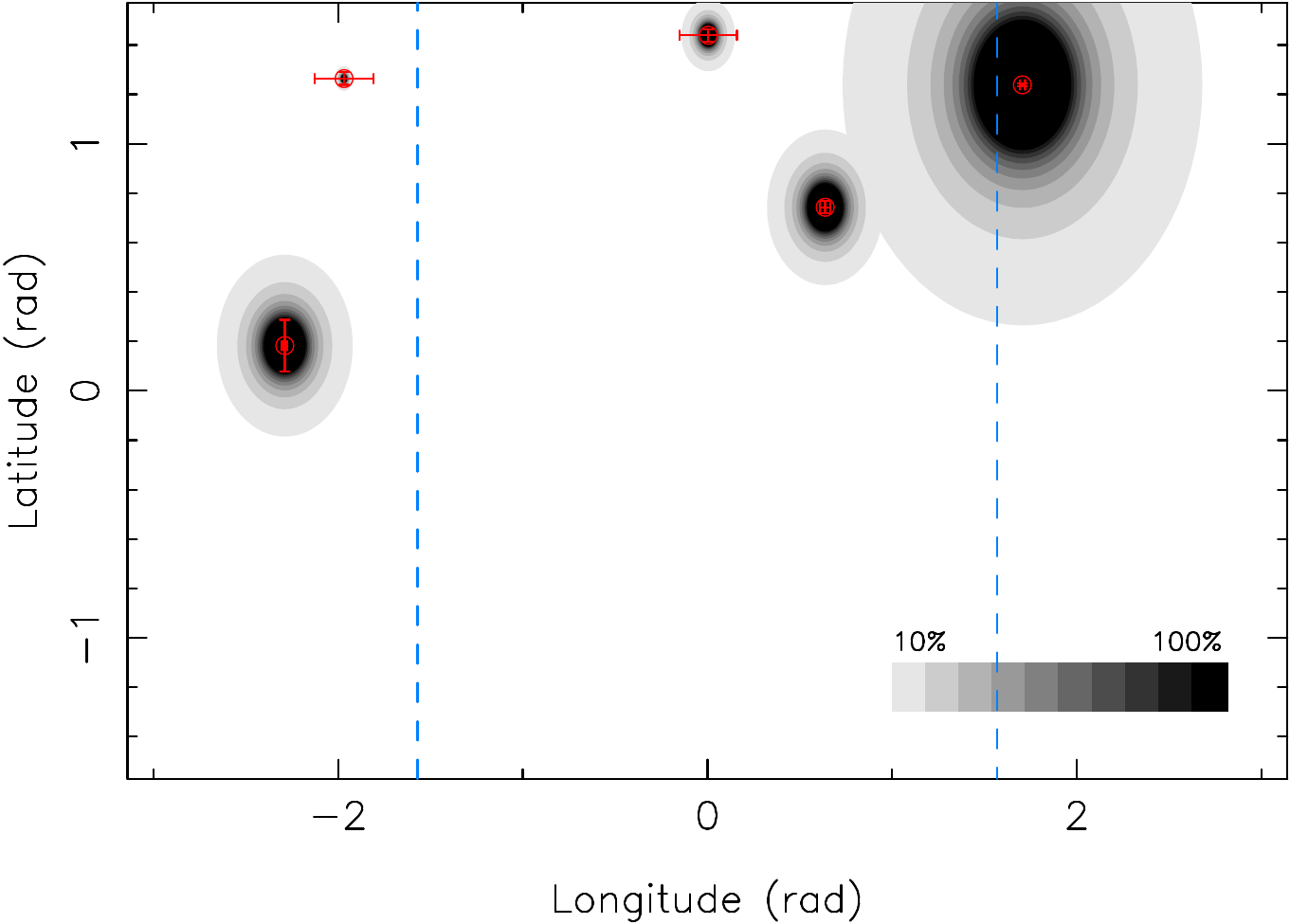}
\includegraphics[angle=270,width=0.24\textwidth,clip]{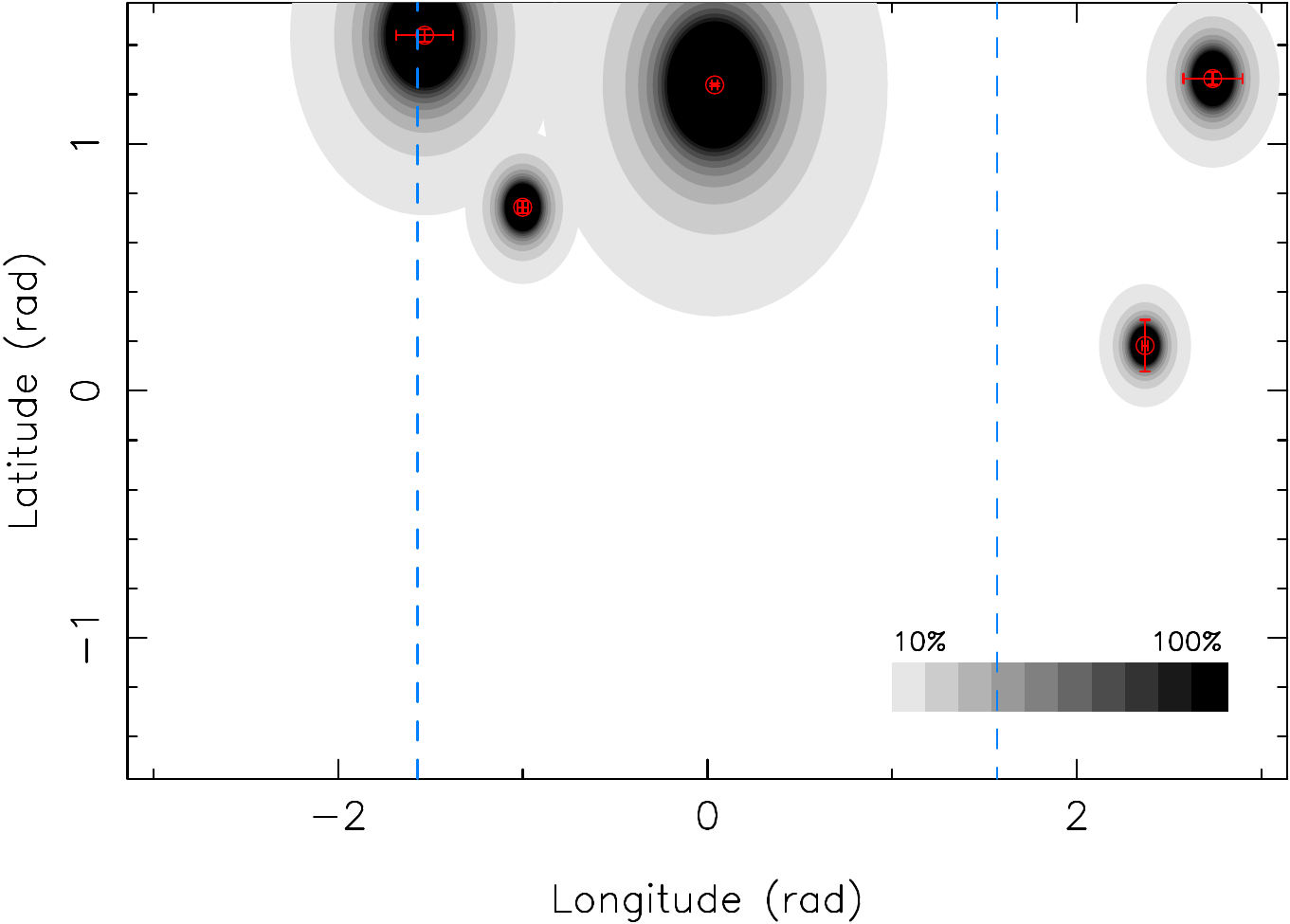}
\caption{Each row of panels contains Kepler-411 data (red dots) and maximum \emph{a posteriori} (MAP) solutions (black curves), model residuals, and estimated spot maps for epochs 1044.016 and 1051.851 corresponding to the mean times of transits 116 and 117. The rows from top to bottom are for models with $k = 1, ..., 5$. The blue vertical dashed lines denote the surface visible at those epochs. The greyscale denotes a range of spot sizes and contrasts as estimated based on the effect of the corresponding spot on the flux. The red circles and the corresponding uncertainty intervals denote MAP spot positions and standard errors. The reference time is $t_0 = 1034.0$ days. Only absolute values of latitude are known and all spots are placed on the same hemisphere.}\label{fig:S1034_solution}
\end{figure*}

We present in detail the analysis of the subset containing the two spot occultation events during transits 116 and 117, subset S1034. The maximum \emph{a posteriori} (MAP) solutions of this set are presented in Fig. \ref{fig:S1034_solution} for $k = 1, ..., 5$. As can be seen, the two-spot model already represents the observations rather well, and captures a majority of the flux variations. Yet, by adding more spots into the model improves the modelling very significantly for up to $k = 5$ (Table \ref{tab:model_comparison}). Uniquely identifying solutions for $k < 3$ was rather trivial, and we thus discuss only solutions for $k \geq 3$ in greater detail.

\begin{table*}
\caption{Model comparison statistics for models with $k$ spot signals and $n$ free parameters in terms of logarithms of likelihood ratios of models with $k$ and $k-1$ spots $L_{k}$ and logarithms of Bayes factors $B_{k}$ in favour of model $\mathcal{M}_{k}$ and against model $\mathcal{M}_{k-1}$.}\label{tab:model_comparison}
\begin{center}
\begin{tabular}{lccccccccc}
\hline \hline 
$k$ & $n$ & $B_{k}$ & $L_{k}$ & $B_{k}$ & $L_{k}$ & $B_{k}$ & $L_{k}$ & $B_{k}$ & $L_{k}$ \\
 & & S834 & S834 & S1034 & S1034 & S1069 & S1069 & S1234 & S1234 \\
\hline
1 & 12 &  622.29 &  654.80 &  276.03 &  308.64 & 1287.76 & 1319.98 &  969.90 & 1002.35 \\
2 & 20 & 1353.92 & 1382.82 & 1387.37 & 1416.35 &  709.10 &  737.75 & 1312.02 & 1340.86 \\
3 & 28 &  307.71 &  336.61 &  484.90 &  513.88 &  649.79 &  678.44 &  592.67 &  621.51 \\
4 & 36 &  187.62 &  216.53 &  347.54 &  376.52 &  211.85 &  240.49 &  116.30 &  145.14 \\
5 & 44 &  121.96 &  150.86 &   44.41 &   73.39 &  -10.53 &   18.11 &  117.42 &  146.26 \\
6 & 52 &   10.21 &   39.12 &  -15.55 &   13.43 &         &         &   -1.90 &   26.94 \\
\hline \hline
\end{tabular}
\end{center}
\end{table*}

\begin{figure*}
\center
\includegraphics[width=0.32\textwidth,clip]{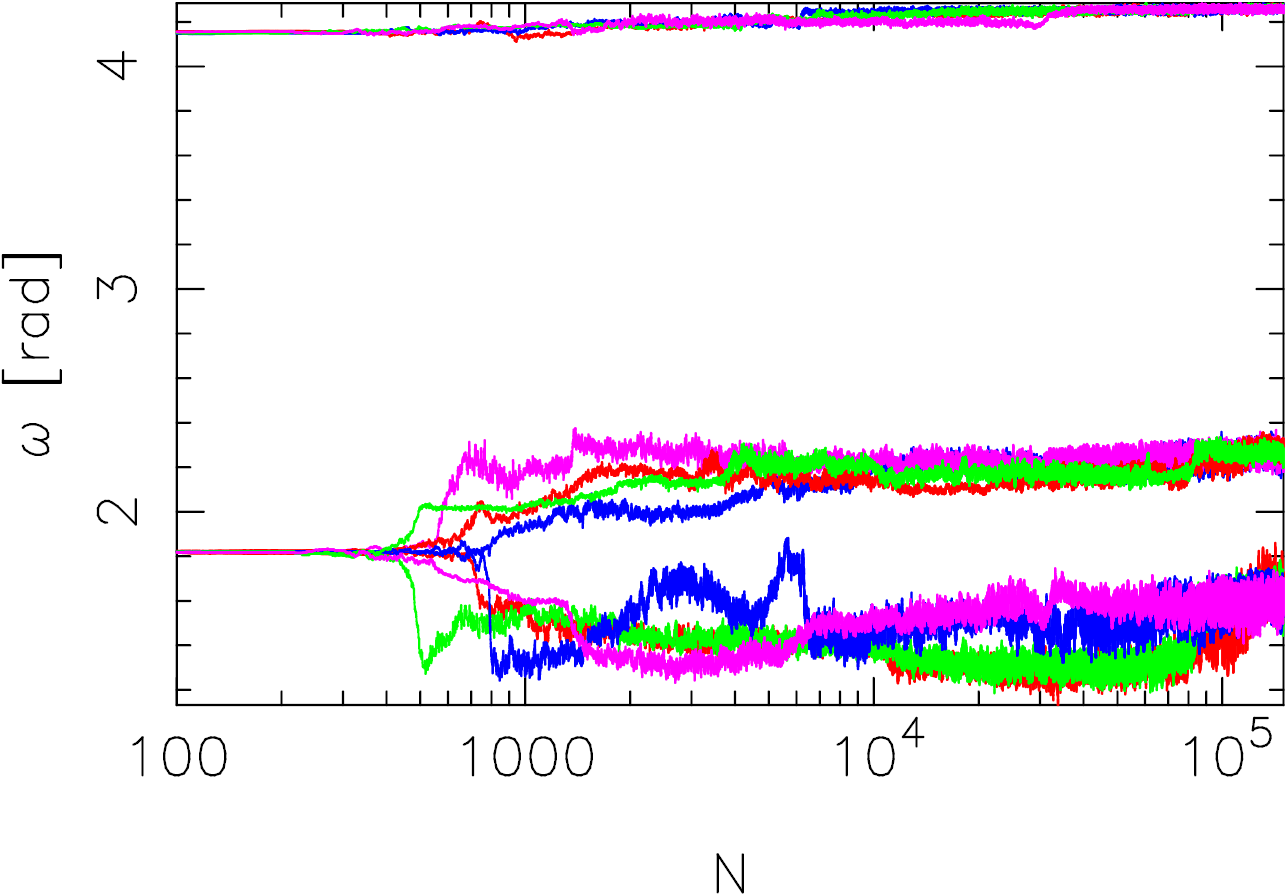}
\includegraphics[width=0.32\textwidth,clip]{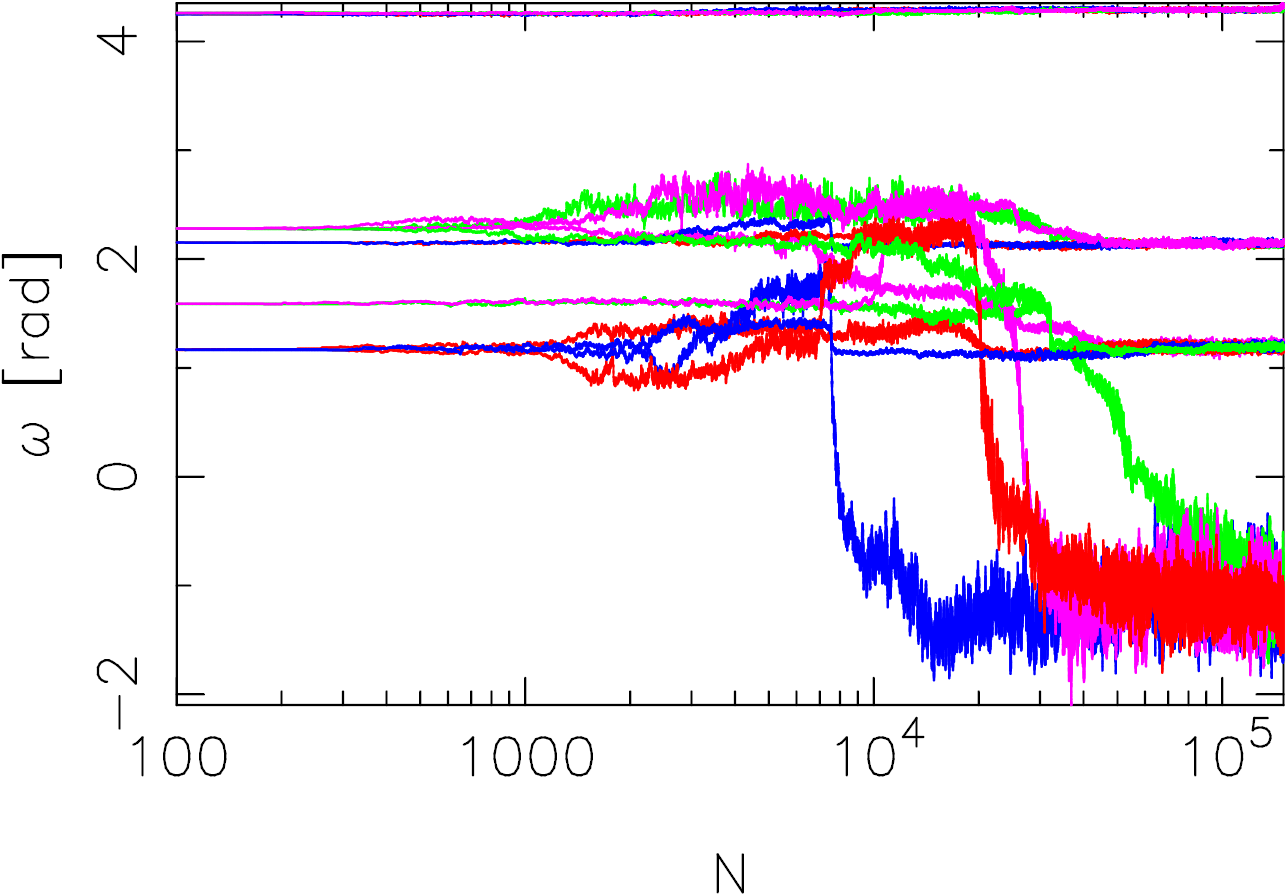}
\includegraphics[width=0.32\textwidth,clip]{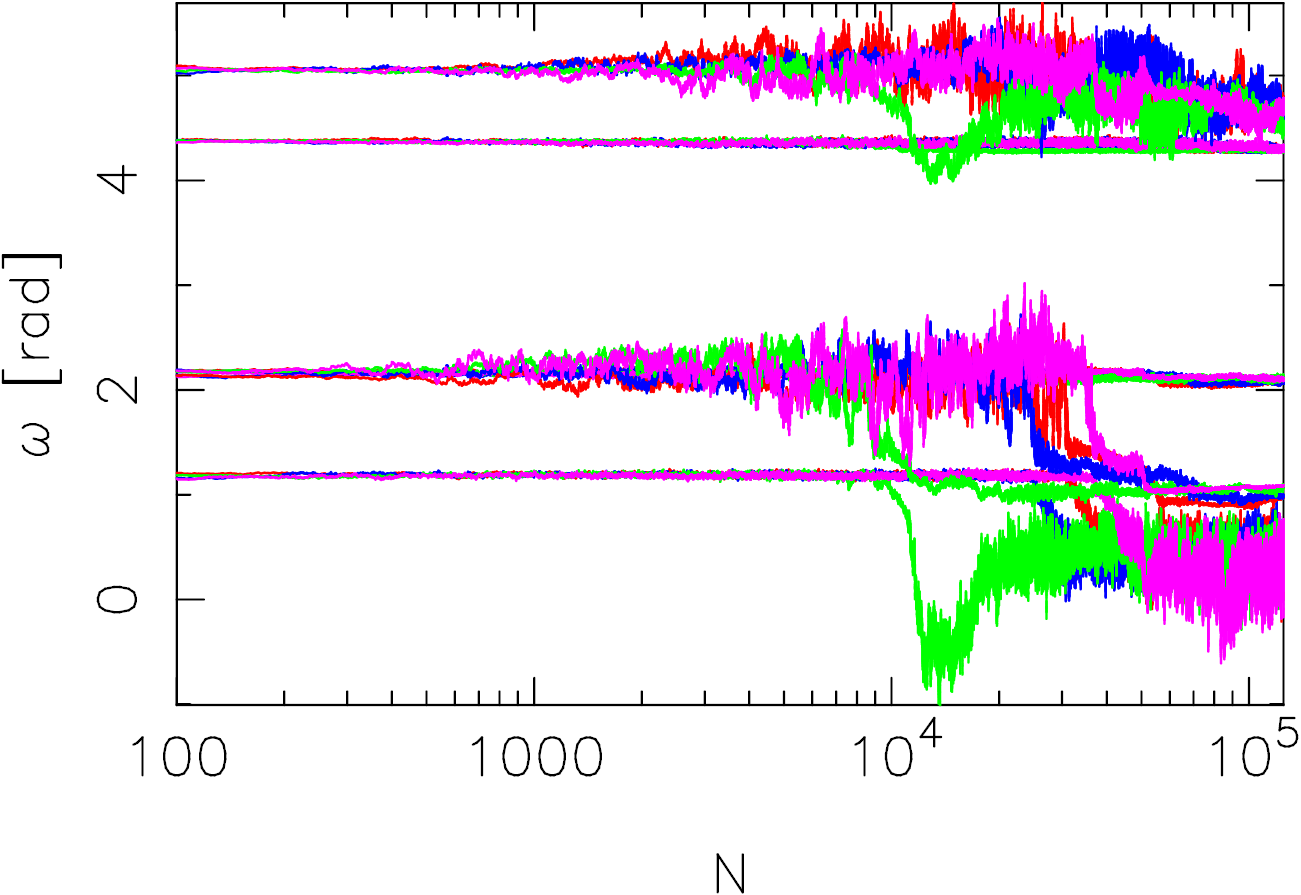}
\caption{Markov chains of the phase parameters of the signals identifying the solutions for models with $k=3, 4, 5$ (left to right) for the data subset S1034. The different colours denote independent samplings.}\label{fig:S1034_search3}
\end{figure*}

For $k=3$, our samplings identified a division of the polar spot in the model with $k=2$ into two nearby spot structures. Using the signal phase as a proxy for investigating the nature of the solutions we demonstrate how our samplings consistently identified the same solutions in Fig. \ref{fig:S1034_solution}. The improvements in the model goodness statistics (Table \ref{tab:model_comparison}) were very significant for up to $k=5$. In Fig. \ref{fig:S1034_search3} it can be seen that the phase ($\omega$) parameters demonstrate clear and consistent bifurcations for $k=3$ as the chains identify solutions.

Likewise, our samplings consistently identified a significant solution for the model with $k=4$. This solution corresponds to an emerging spot that appeared fully over the last rotation cycle of the star. As a polar spot we interpret this result as an indication of reconfiguration of the polar spot structure, for which only non-axisymmetric features are observable with our approach. Such a change would be interpreted by our model as the appearance of a spot out of nowhere when, in reality, it only resulted from non-axisymmetric changes in existing spot structure. 

Adding another spot into the model also resulted in a significant identification of a spot (Table \ref{tab:model_comparison}). For $k=5$, the pair of near-polar spots gave way for a solution with two smaller evolving spots at high latitudes and a third one cat mid latitudes (Fig. \ref{fig:S1034_solution}) and around the same longitudes. Again, we were able to identify the same solution with several samplings (Fig. \ref{fig:S1034_search3}), and could not identify any other solutions with comparable significance.

Allowing the stellar inclination to be a free parameter of the interval $[0, \pi/2]$ and obtaining a solution for $k=5$ based on samplings such that their initial state corresponded to the solution presented in Fig. \ref{fig:S1034_solution}, we were able to estimate the inclination angle. Our estimate for the inclination is 1.5671$\pm$0.0034 rad, which is consistent with $\pi/2$ and allowing the inclination to be a free parameter of the model did not improve the model in a statistically significant manner. We note that we are not fully confident that the inclination could really be determined this precisely. However, when searching for solutions by fixing the inclination to $\pi/2$ and letting it vary freely only after the optimal number of spots had been identified, probably plays a role in narrowing down its posterior probability density. It is a safe assumption to make for the current target that probably has an edge-on orientation given that it has a system of transiting planets supported by measurements of $v \sin i$ \citep{xu2021}.

We note that not all samplings enabled identifying the solutions presented in Fig. \ref{fig:S1034_solution}. We rejected, depending on $k$, some 10-50\% of Markov chains that did not succeed in identifying solutions with comparable statistics of goodness given fixed chain lengths. These chains only identified local solutions that are unlikely to represent the underlying spot structure of the target star.

Our solutions for subsets S834, S1069, and S1234 are presented in Appendix \ref{sec:all_solutions} and spot parameters are tabulated in Table \ref{tab:spot_parameters}. Although we identified a statistically significant solution containing six spots for S834, we were not able to replicate the result with several samplings as there appeared to be several almost equally probable local solutions. We thus rely on the solution with $k=5$.

\subsection{Stability of latitude estimates}

Obtaining latitude information from a photometric time-series seems challenging, especially, as the star is probably seen edge-on. However, the shapes of the spot-induced decreases in the flux indeed have different shapes for different latitudes. This is because limb darkening plays a larger role closer to the polar regions, effectively flattening the resulting light curves. It should therefore be expected that the latitude is the better constrained the closer the spots are to the polar regions. However, the whole light curve provides the constraints, and such considerations may be overtly simplistic in practice.

We subjected our solutions to sensitivity tests with respect to latitude. The modelled spot-induced effects are very similar for smaller spots at low latitudes and larger spots at high latitudes. We therefore truncated the identified solutions such that all spots were shifted to latitudes near the poles (equator) while simultaneously increasing (decreasing) their amplitudes. These truncated parameters were then used as initial states of posterior samplings in order to see whether there were corresponding high-probability areas in the parameter space indicating that latitudes could not be constrained as well as implied in Fig.\ref{fig:S1034_solution}. 

\begin{figure}
\center
\includegraphics[angle=270,clip,width=0.45\textwidth,clip]{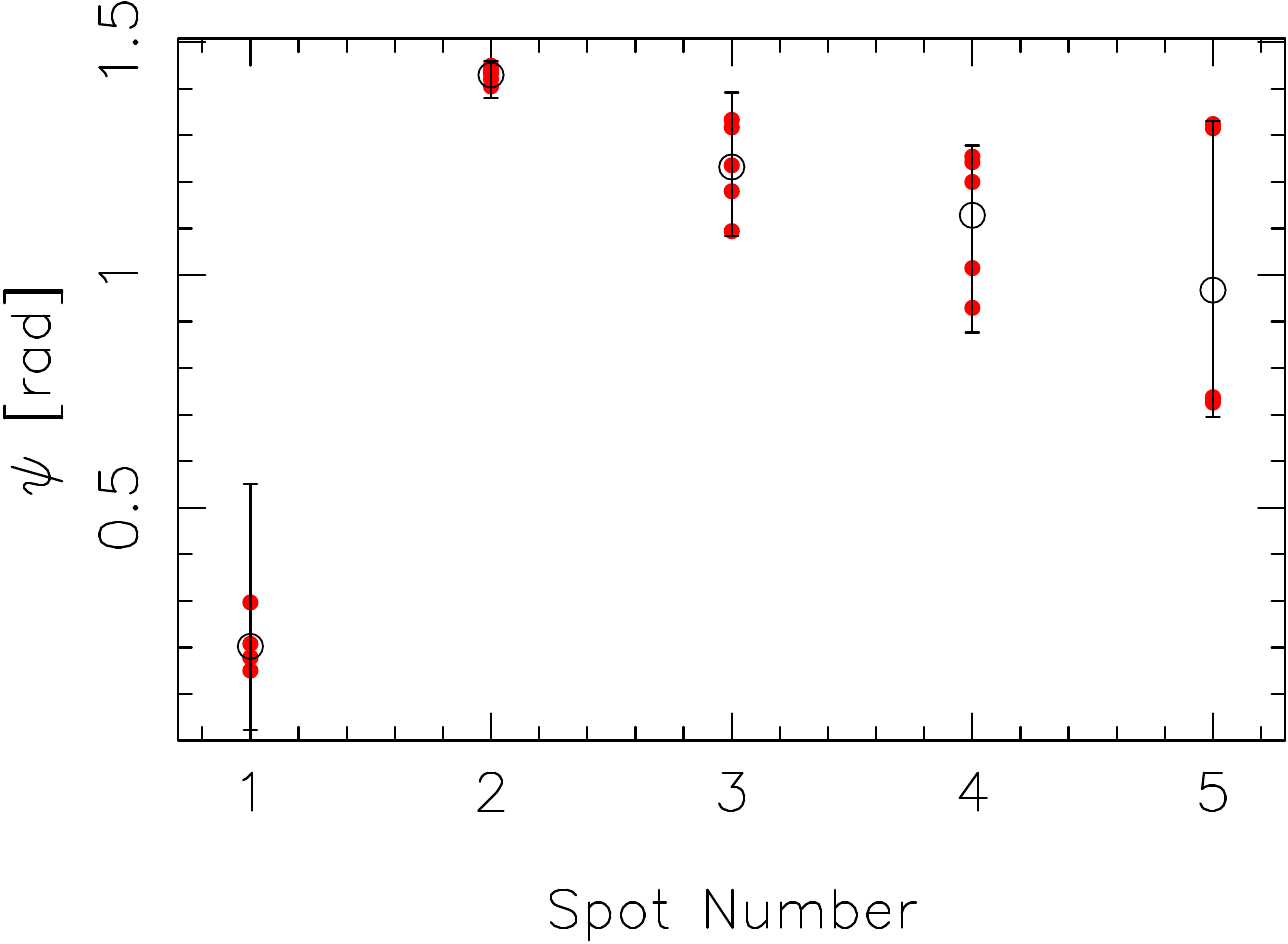}
\caption{Estimated latitudes (black circles) of the five spots identified in the data subset S1034. The red dots show where high posterior probabilities (above 1\% of the global maximum) were first identified by five samplings for which spots were randomly truncated to positions near the equator or the pole.}\label{fig:truncated_latitudes}
\end{figure}

According to our results, the latitudes are rather well constrained for all spots. We demonstrate this in Fig. \ref{fig:truncated_latitudes} where we show our solution together with results from the truncation tests for data subset S1034. The samplings with truncated initial states enabled us to confirm that the latitudes of all spots were constrained reliably, albeit the latitudinal uncertainties of some spots were rather large.

\section{Results}

We have presented our approach for spot occultation modelling in Section \ref{sec:occultation_modelling} and rotational modulation modelling in Section \ref{sec:rotational_modelling}. In all, we have detected three spot occultations by planet c in high-cadence data and were able to determine the positions of the respective spots on the stellar surface. We have also been able to obtain information on spot patterns on the stellar surface based on modelling of the rotational modulation in lower-cadence observations. Yet, given an edge-on orientation of the star, we cannot determine on which hemisphere the spots are and the planetary transits take place. With this caveat in mind, we present some further results based on our analyses.

\subsection{Spot estimates}

We cannot be absolutely certain that our solutions based on modelling of rotational modulation indeed are global and represent the physical reality of the stellar surface. Although such certainty is not even achievable in practice, we believe we have probed the parameter space carefully enough such that our solutions probably are global solutions. We have tabulated the parameters of identified spots in Table \ref{tab:spot_parameters} for all four data subsets. In this table, we have also denoted whether the spots are increasing, decreasing or approximately constant in terms of the behaviour of their envelope curve over the data baseline. If the envelope curve does not change more than 20\% over the data subset baseline we denote the spot as "constant", although actual constancy is strictly not the case for any of the spots.

\begin{table*}
\caption{Selected spot parameters. The status of the spot is denoted as increasing (I), constant(C) or decreasing (D), indicating at which phase of the envelope the spot is found. The status is considered "constant" if there are no changes in excess of 20\% in the spot amplitude over the data baseline. The longitudes are defined with respect to $t_{0}$ of each data subset and are thus not comparable for different subsets.}\label{tab:spot_parameters}
\begin{center}
\begin{tabular}{lcccccccccc}
\hline \hline 
Data & Spot & $P$ & $A'$ & $\psi$ & $\omega$ & Status \\
 & & (days) & ($10^{-3}$) & (rad) & (rad) & \\
\hline
S834 & 1  & 10.996$_{-0.163}^{+0.097}$ & 105.8$_{-75.0}^{+89.3}$  & 1.173$_{-0.376}^{+0.188}$ & 4.692$_{-0.057}^{+0.054}$ & D \\
S834 & 2  & 10.632$_{-0.239}^{+0.159}$ &  51.6$_{-58.2}^{+149.6}$ & 0.774$_{-0.660}^{+0.858}$ & 1.788$_{-0.268}^{+0.171}$ & I \\
S834 & 3  & 10.481$_{-0.047}^{+0.067}$ &  24.0$_{-12.4}^{+92.6}$  & 0.274$_{-0.262}^{+1.133}$ & 2.501$_{-0.030}^{+0.039}$ & C \\
S834 & 4  & 11.174$_{-0.304}^{+0.170}$ &  77.1$_{-63.5}^{+72.6}$  & 0.997$_{-0.747}^{+0.342}$ & 5.842$_{-0.089}^{+0.113}$ & C \\
S834 & 5  & 10.620$_{-0.225}^{+0.193}$ &  41.7$_{-34.1}^{+25.9}$  & 0.737$_{-0.717}^{+0.485}$ & 6.063$_{-0.272}^{+0.196}$ & I \\

S1034 & 1 & 10.584$_{-0.063}^{+0.072}$ &  15.0$_{-11.4}^{+16.5}$ & 0.202$_{-0.179}^{+0.348}$ & 4.313$_{-0.028}^{+0.043}$ & D \\
S1034 & 2 & 10.458$_{-0.138}^{+0.128}$ & 178.6$_{-38.5}^{+21.2}$ & 1.430$_{-0.049}^{+0.031}$ & 0.369$_{-0.213}^{+0.190}$ & I \\
S1034 & 3 & 10.679$_{-0.030}^{+0.046}$ &  82.9$_{-45.9}^{+97.8}$ & 1.232$_{-0.149}^{+0.160}$ & 2.100$_{-0.034}^{+0.038}$ & C \\
S1034 & 4 & 10.660$_{-0.196}^{+0.236}$ &  57.3$_{-31.4}^{+32.6}$ & 1.128$_{-0.252}^{+0.150}$ & 4.830$_{-0.256}^{+0.320}$ & I \\
S1034 & 5 & 10.669$_{-0.084}^{+0.130}$ &  32.2$_{-23.1}^{+35.9}$ & 0.968$_{-0.273}^{+0.363}$ & 1.033$_{-0.083}^{+0.090}$ & C \\

S1069 & 1 & 10.584$_{-0.063}^{+0.072}$ &  14.2$_{-1.3}^{+2.7}$   & 0.250$_{-0.197}^{+0.253}$ & 4.157$_{-0.021}^{+0.025}$ & C \\
S1069 & 2 & 10.361$_{-0.020}^{+0.021}$ &  90.0$_{-28.3}^{+36.9}$ & 1.286$_{-0.069}^{+0.060}$ & 0.475$_{-0.011}^{+0.010}$ & C \\
S1069 & 3 & 10.308$_{-0.012}^{+0.011}$ & 110.9$_{-35.2}^{+52.6}$ & 1.389$_{-0.055}^{+0.050}$ & 2.160$_{-0.089}^{+0.093}$ & I \\
S1069 & 4 &  9.919$_{-0.050}^{+0.055}$ & 137.4$_{-34.4}^{+22.0}$ & 1.361$_{-0.036}^{+0.027}$ & 2.747$_{-0.035}^{+0.035}$ & C \\

S1234 & 1 & 10.677$_{-0.049}^{+0.054}$ &   7.0$_{-4.1}^{+13.4}$   & 0.584$_{-0.418}^{+0.517}$ & 5.611$_{-0.056}^{+0.052}$ & C \\
S1234 & 2 &  9.885$_{-0.049}^{+0.045}$ &  10.9$_{-3.7}^{+5.2}$    & 0.431$_{-0.410}^{+0.344}$ & 0.821$_{-0.037}^{+0.043}$ & D \\
S1234 & 3 & 11.154$_{-0.033}^{+0.036}$ &  56.0$_{-45.8}^{+124.4}$ & 0.811$_{-0.781}^{+0.520}$ & 2.800$_{-0.028}^{+0.030}$ & C \\
S1234 & 4 & 10.275$_{-0.043}^{+0.051}$ & 104.4$_{-92.3}^{+55.4}$  & 1.041$_{-0.962}^{+0.259}$ & 2.534$_{-0.048}^{+0.052}$ & I \\
S1234 & 5 & 10.275$_{-0.020}^{+0.017}$ & 106.5$_{-92.1}^{+23.6}$  & 1.089$_{-0.701}^{+0.173}$ & 3.925$_{-0.022}^{+0.021}$ & C \\
\hline \hline
\end{tabular}
\end{center}
\end{table*}

We observe a total of six increasing spots whereas three spots are decreasing in terms of their envelope curves over the data baselines. The time-scale of such spot appearance and disappearance, as determined by parameters $\sigma_{j, \rm in}$ and $\sigma_{j, \rm out}$, respectively, varies from 5.02$\pm$0.11 to 24.84$\pm$0.35 days for appearing spots and from 8.89$\pm$0.23 to 11.87$\pm$0.32 days for the disappearing ones, respectively. This demonstrates that the stellar spot structure evolves significantly over a timescale of one rotation period of the star and spots can seem to appear from nowhere or disappear without a trace over a single rotation cycle. Although some spots remain virtually unchanged over roughly three rotation periods of the star, it seems clear that such modelling of spots is not possible without accounting for their evolution.

The effective sizes of the starspots range, in terms of parameter $A'$, from 7.0$_{-4.1}^{+13.4}$ to 178.6$_{-38.5}^{+21.2}$ parts per thousand (Table \ref{tab:spot_parameters}). For geometric reasons, the largest spots have high latitudes whereas the smallest spots can only be seen at or around the stellar equator. That latitudes can be constrained at all is dependent on whether there are spots at nearby longitudes and on limb darkening, which ensures that the shapes of the spot-induced curves are flattened near polar regions. Due to this flattening, there is a higher latitudinal resolution near poles than near equator (Table \ref{tab:spot_parameters}, Fig. \ref{fig:differential_rotation}). We note that the instrumental noise floor of the observations was  0.063 parts per thousand for all analysed data sets and the excess noise was approximately three times greater with a typical estimate of 0.19 parts per thousand. This demonstrates that the excess noise probably contained signatures of weak spots and noise from other astrophysical phenomena of the stellar surface.

\begin{figure}
\center
\includegraphics[clip,width=0.45\textwidth,clip]{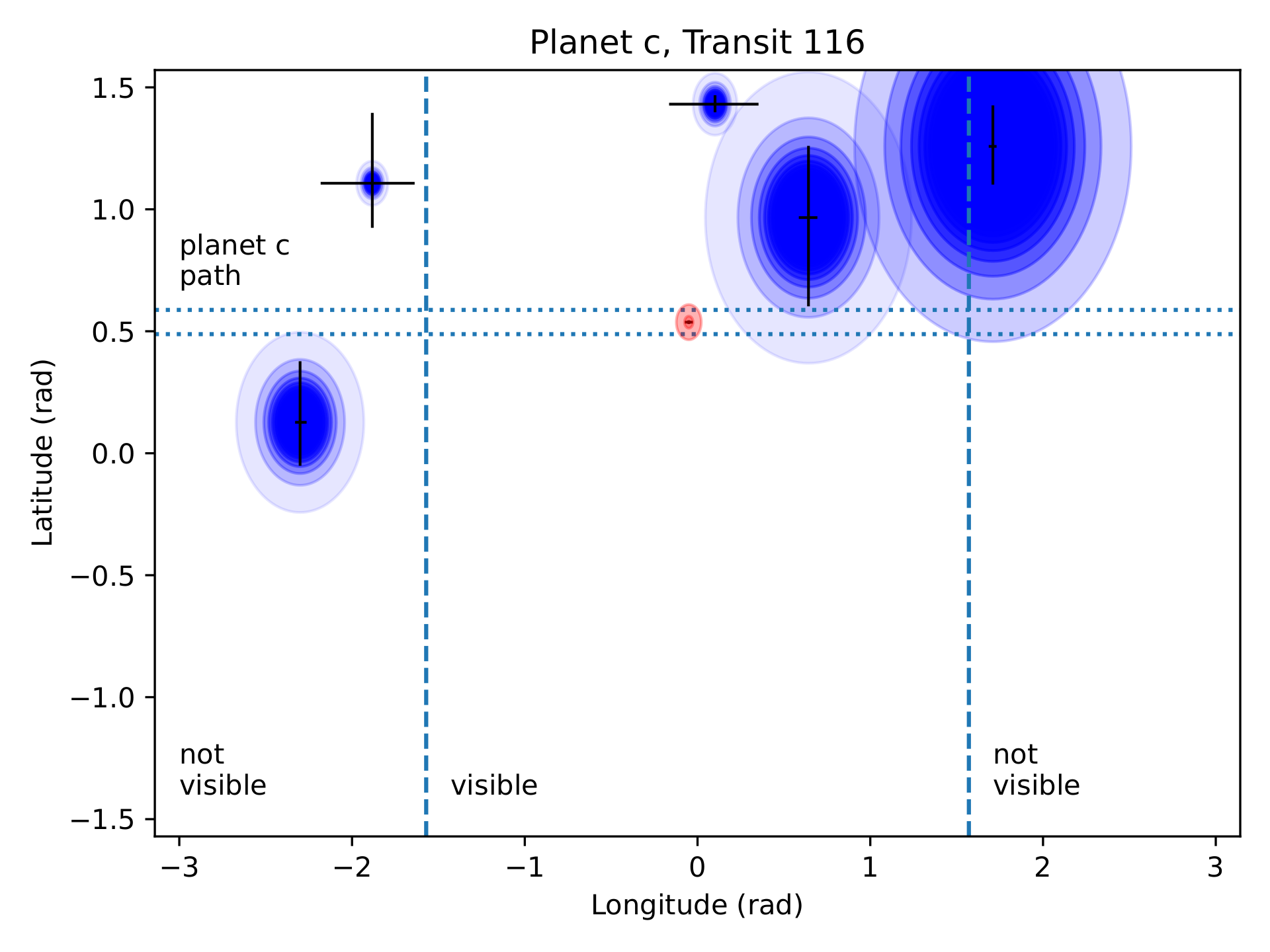}

\includegraphics[clip,width=0.45\textwidth,clip]{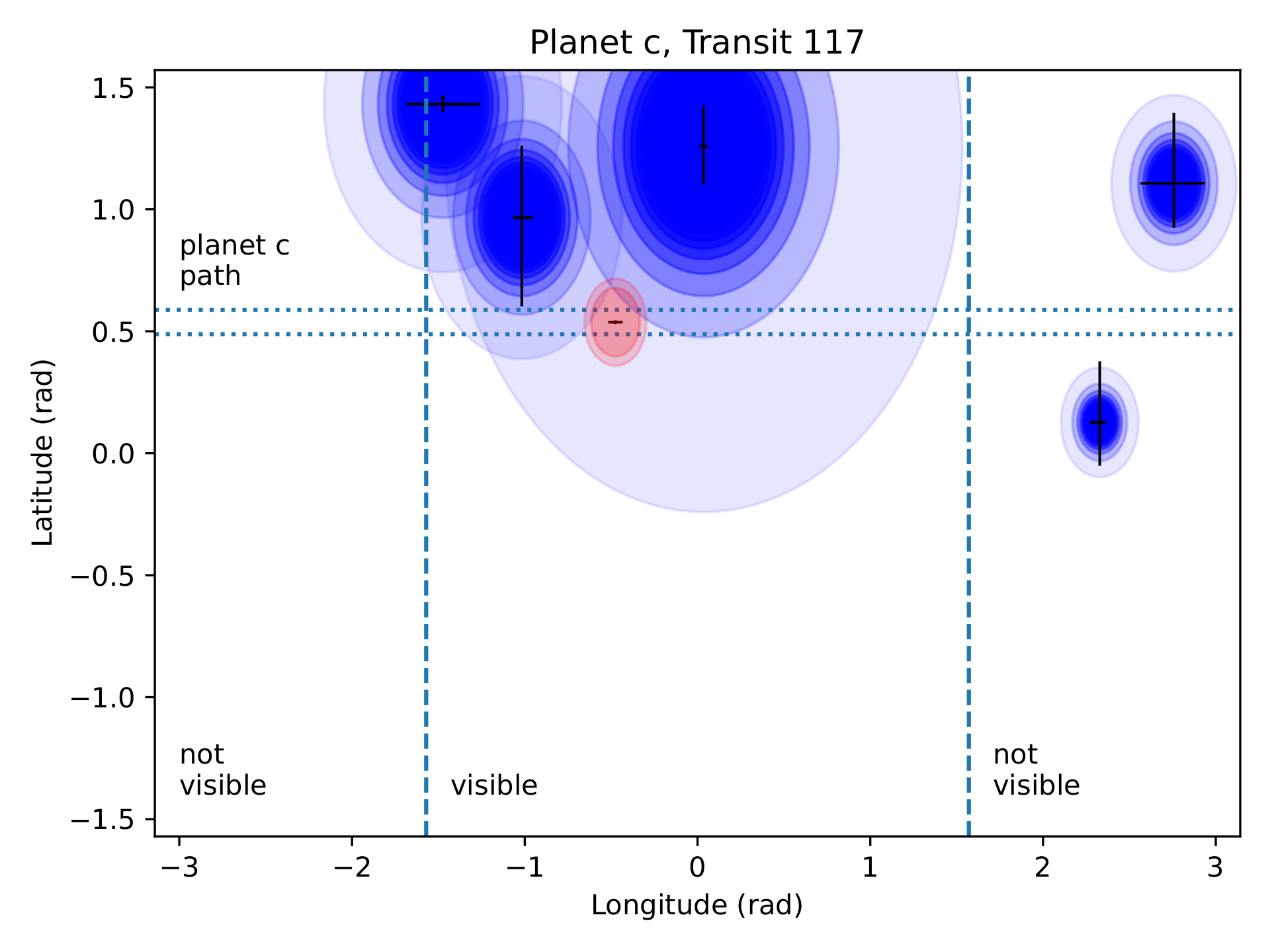}

\includegraphics[clip,width=0.45\textwidth,clip]{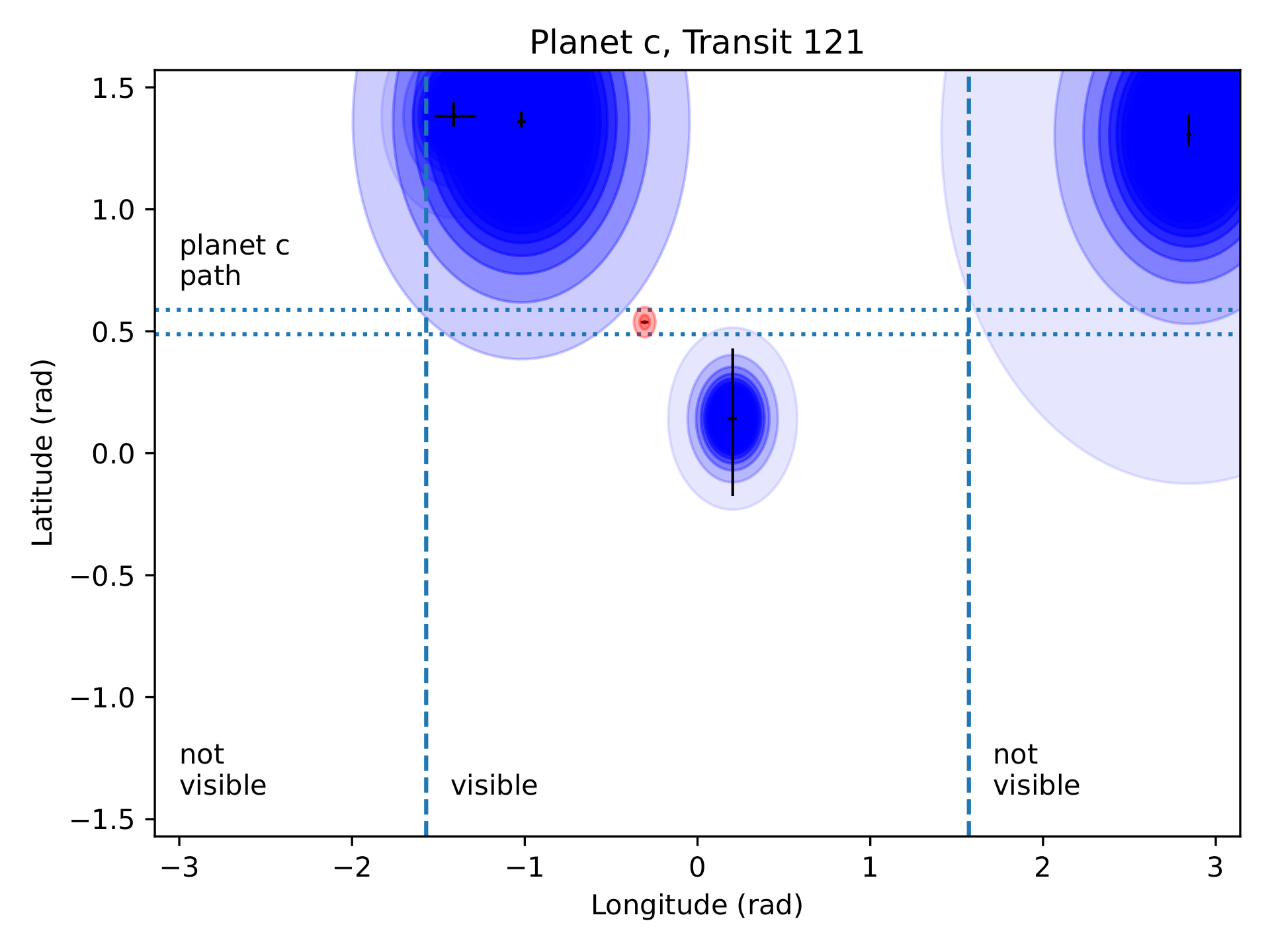}
\caption{Comparison of estimated spot patterns on the surface of Kepler-411 based on spot occultation results (red) and rotational modulation results (blue) for transit events 116, 117, and 121 (top to bottom). The different shades of blue (from dark to light) denote a range of spot sizes and their contrasts from 1.0 to 0.1 with intervals of 0.1.}\label{fig:comparison}
\end{figure}

The ability to identify spot occultations during transit events and estimation of properties of spots contributing to the rotational modulation provide independent information on the spot patterns of the stellar surface. It is therefore possible to perform cross-comparisons between the two. For this purpose, we have plotted the estimated star spots based on these two methods in Fig. \ref{fig:comparison}.

We note that we also considered a second order limb darkening model in an attempt to obtain more reliable results. However, it did not improve the modelling significantly. We suspect this is caused by the fact that rotational modulation caused by star spots is the dominating source of variability in the data. Because the variations caused by additional small spots cannot be distinguished, they also make the more accurate limb darkening models redundant.

We have also plotted the two-dimensional projections of posterior probability densities in the Appendix \ref{sec:all_solutions}. In Fig. \ref{fig:S1034_solution_contours} we show an example of the correlations between the parameters of a given spot. The amplitude parameter $A$ and spot latitude $\psi$ are clearly correlated in a nonlinear way, but otherwise the correlations are at most modest and/or linear and thus well-handled with the AM algorithm. We also show examples of correlations between parameters of two spots (Fig. \ref{fig:S1034_solution_contours2}) and some spot parameters and stellar parameters such as inclination and limb darkening (Figs. \ref{fig:S1034_solution_contours3} and \ref{fig:S1034_solution_contours4}).

\subsection{Differential rotation}

Based on our results regarding estimated spot latitudes and their respective independent angular velocities, we can study the differential rotation of the target star. We have collected the parameters of the estimated spots for all the four data subsets -- a total of 19 spots -- and investigate the relationship between spot latitude and angular velocity.

\begin{figure}
\center
\includegraphics[clip,angle=270,width=0.48\textwidth,clip]{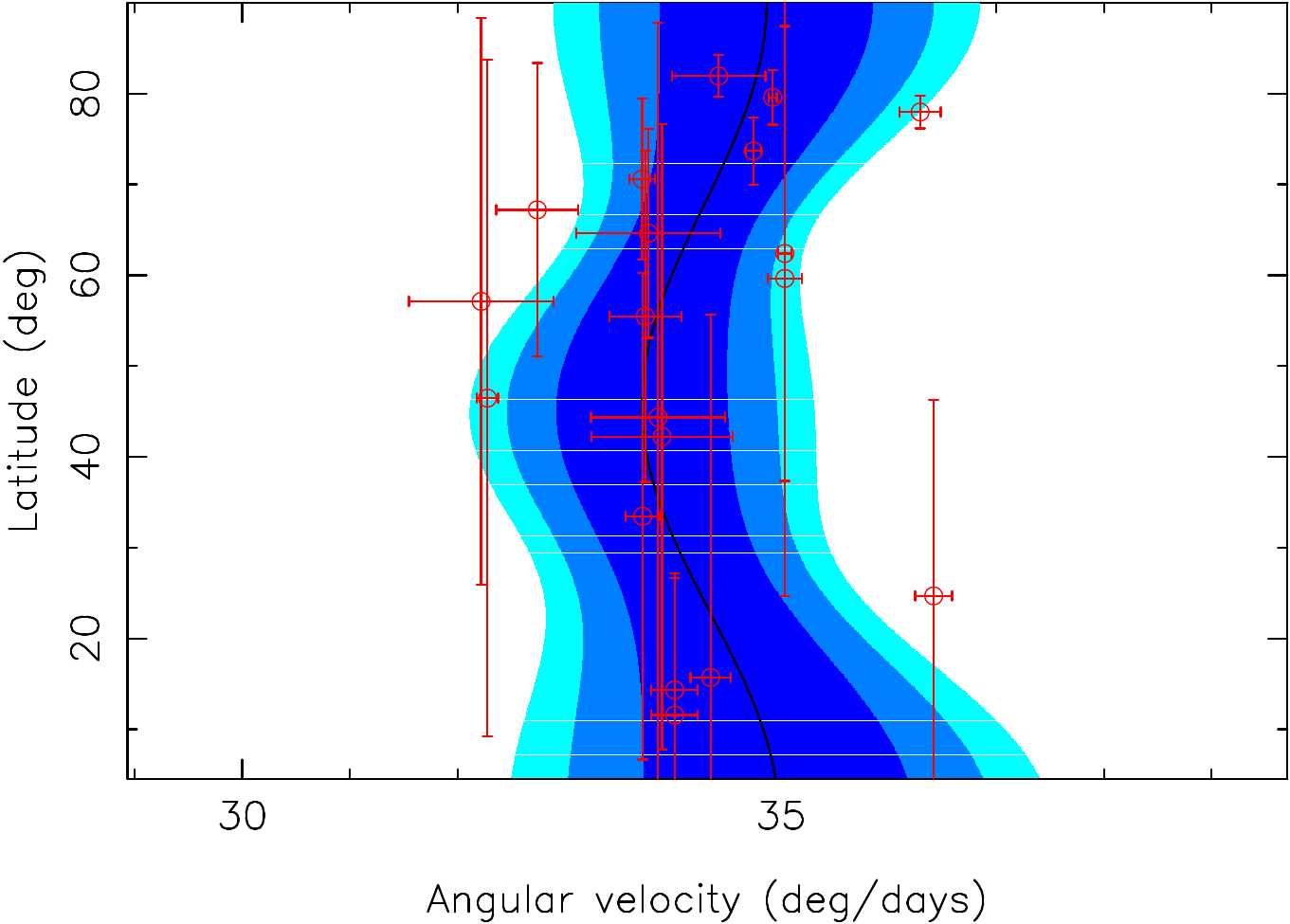}
\caption{Estimated differential rotation curve ($n=2$) for Kepler-411 based on spot latitudes and angular velocities from modelling of rotational modulation. The red circles denote each spot, the black curve represents the maximum-likelihood model curve and the shaded blue areas denote one, two, and three-$\sigma$ uncertainty intervals.}\label{fig:differential_rotation}
\end{figure}

We modelled the differential rotation with a simple sinusoidal model \citep[e.g.][]{wohl2010,lamb2017} such that
\begin{equation}
  \Omega(\psi) = \sum_{i=0}^{n} a_{i} \sin^{2i} \psi ,
\end{equation}
where $a_{0}$ denotes the equatorial angular velocity and only even powers of sine are allowed in order to achieve a symmetry with respect to equator. We modelled the angular velocity as a function of latitude such that we assumed Gaussian white noise with a variance of $\sigma_{i}^{2} + \sigma_{\omega}^{2}$, where $\sigma_{i}$ represent the uncertainties in the parameters of the $i$th spot and $\sigma_{\omega}$ is a free parameter. We compared models for $n \leq 3$ and found that $n=0$ corresponded to the preferred model. There was no evidence for any change in the angular velocity as a function of latitude, and it thus seems appropriate to conclude that differential rotation appears to be below our detection threshold. Illustration of our results from this modelling is shown in Fig. \ref{fig:differential_rotation} for a model with $n=2$ demonstrating visually that there indeed is no evidence for differential rotation. There are differences in the angular velocities experienced by individual spots but these velocities seem to have similar distributions for both high and low latitudes (Fig. \ref{fig:differential_rotation}), and for many of them the uncertainties are rather large.

We obtain an angular velocity of 0.597$\pm$0.019 rad/day that translates to an average rotation period of 10.52$\pm$0.34 days. We note that the individual periods were far more precisely constrained by the data than the prior probability density.

\section{Conclusions and discussion}

We obtained evidence for three separate spot occultations in the full Kepler high-cadence data. Two rather clear occultations by planet c at epochs of 1051.851 and 1083.189 (transit events 117 and 121, respectively) satisfied our detection thresholds for very strong evidence. Another spot transit at epoch 1044.016, corresponding to transit event 116, was also found to have strong evidence in favour of it (Fig. \ref{fig:spot_transits}.

The spot occultation events during transits 116, 117, and 121 correspond to spots with minimal angular diameters of $2.1_{-1.5}^{+3.4}$ deg, $15.1_{-4.6}^{+4.9}$ deg, and $2.9_{-1.7}^{+3.4}$ deg, respectively, on the surface of the star. Based on our modelling, and assuming circular spots, we estimate the spot contrasts to be $0.32_{-0.02}^{+0.04}$, $0.20_{-0.03}^{+0.03}$, and $0.32_{-0.02}^{+0.03}$, respectively. It is then also possible to estimate the spot temperatures based on the considerations of \citet{silva-valio2010}. We obtain values of $4471_{-53}^{+26}$ K, $4621_{-36}^{+35}$ K, and  $4471_{-40}^{+26}$ K for the spots detected during transit events 116, 117, and 121, respectively. These values are also tabulated in Table \ref{tab:model_comparison_occ}. Adopting the stellar effective temperature of 4833$^{+32}_{-51}$ K \citep{gaia2018} indicates that the spots are some 210 to 360 K cooler than the unspotted photosphere. We note that the occultations detected during transits 117 and 121 could correspond to the same spot given their separation by 31.3 days that is roughly three times the star's rotation period.

The obtained spot sizes and temperatures can be contrasted with observed spots on the solar surface. The largest spot group observed for the Sun was of diameter 9.0 deg \citep{aulanier2012} and is comparable to the size of the spots detected by transit mapping method for Kepler-411. Temperatures of Sun spots are typically between 3000-4500 K, which indicates that the spots on KEpler-411 would roughly correspond to some of the hottest spots on the solar surface.

Flares could, given their diverse nature, yield false positive detections that might be interpreted as spot occultations if they occurred during transits. However, because flares follow distinct asymmetric morphologies \citep{howard2022} they can typically be differentiated from more symmetric spot occultation events. This was found to be the case for Kepler-411 flares, that did not yield any false positives in our tests. We also searched for features resembling spot occultation events outside the transits in the data. However, we did not detect any candidates for such astrophysical false positives as we could not find any events that could have been interpreted as spot occultation events had they appeared during transits in our analyses. This suggests that the three detected spot occultation events are probably genuine signatures of starspots.  

There is no one-to-one match between the spot properties observed by modelling spot-induced photometric modulation and transit mapping. This is rather unsurprising as spots are the most likely not circular, homogeneous features but asymmetric, complex, and dynamical entities as is the case for large sunspots \citep[e.g.][]{aulanier2012}. All the spots identified with transit mapping are close to the modelled spots based on the photometric modulation. However, we cannot distinguish between the two hemispheres with either detection method, and it is not guaranteed that the planetary orbital plane coincides with the plane defined by the stellar rotation axis. This means that the planet paths highlighted in Fig. \ref{fig:comparison} are not necessarily horizontal, but can also cross the stellar equator.

We could not replicate the detections of nearly two hundred spots as claimed by \citet{araujo2021a} and \citet{araujo2023}. This is probably due to the fact that our detection thresholds based on Bayes factors are rather conservative. The far greater number of spot candidates detected by \citet{araujo2021a} can be explained by the fact that they seem not to have accounted for excess noise caused by the stellar surface but accepted all variations in excess of $3-\sigma$ as spot signatures.

In an attempt to visualise these differences, we present our modelling for the Kepler epoch 1083.189 corresponding to the transit event 121 in Fig \ref{fig:spot_transits}. This transit is the same as shown in Fig. 1. of \citet{araujo2021a} that they interpreted to show evidence for the occultation of a group of three small spots. In our analysis, we indeed obtained evidence for a spot. However, we could not identify any additional spot signatures during this transit event with our methods. When the white noise model suffices to explain the vast majority of the spot candidates identified by \citet{araujo2021a}, there is no need to postulate a multitude of spots on the stellar surface, although their existence cannot be ruled out either. We estimate that these qualitative differences in the results regarding spot detections are mostly due to the selected detection criteria. However, some of it is also caused by the fact that we accounted for the statistical excess noise in the data and treated it as a free parameter $\sigma_{\rm S}$ in our analyses. However, the instrumental noise was still the dominating source and the excess noise was less than 20\% of the instrumental noise during the transit events 116, 117, and 121, respectively.

We have presented a detailed analysis of the modelling results regarding rotational modulation of the light curve for data subset S1034 in Section \ref{sec:section_1034}. The results for the other subsets have been summarised in Table \ref{tab:model_comparison} and Appendix \ref{sec:all_solutions}. These results indicate that the differential rotation curve is indistinguishable from a flat one corresponding to a rigid-body rotation (Fig. \ref{fig:differential_rotation}).

A curious feature of the detected spot occultations is the fact that they were identified only at three out of six successive transit events but not during any other transits outside this space of approximately 40 days. This suggests that the star may have undergone an active phase with a larger number of spots on its surface during that period of time. However, we could not confirm this hypothesis based on rotational modulation modelling.

\section*{Acknowledgements}

The authors acknowledge Research Council of Finland project SOLSTICE (decision No. 324161). MT and AH acknowledge support from the Jenny and Antti Wihuri Foundation. The authors acknowledge CSC, IT Center for Science, Finland, for computational resources and support. This work has also made use of the University of Hertfordshire's high-performance computing facility.

\appendix

\section{Rotational modulation solutions}\label{sec:all_solutions}

\begin{figure}
\center
\includegraphics[angle=270,width=0.24\textwidth,clip]{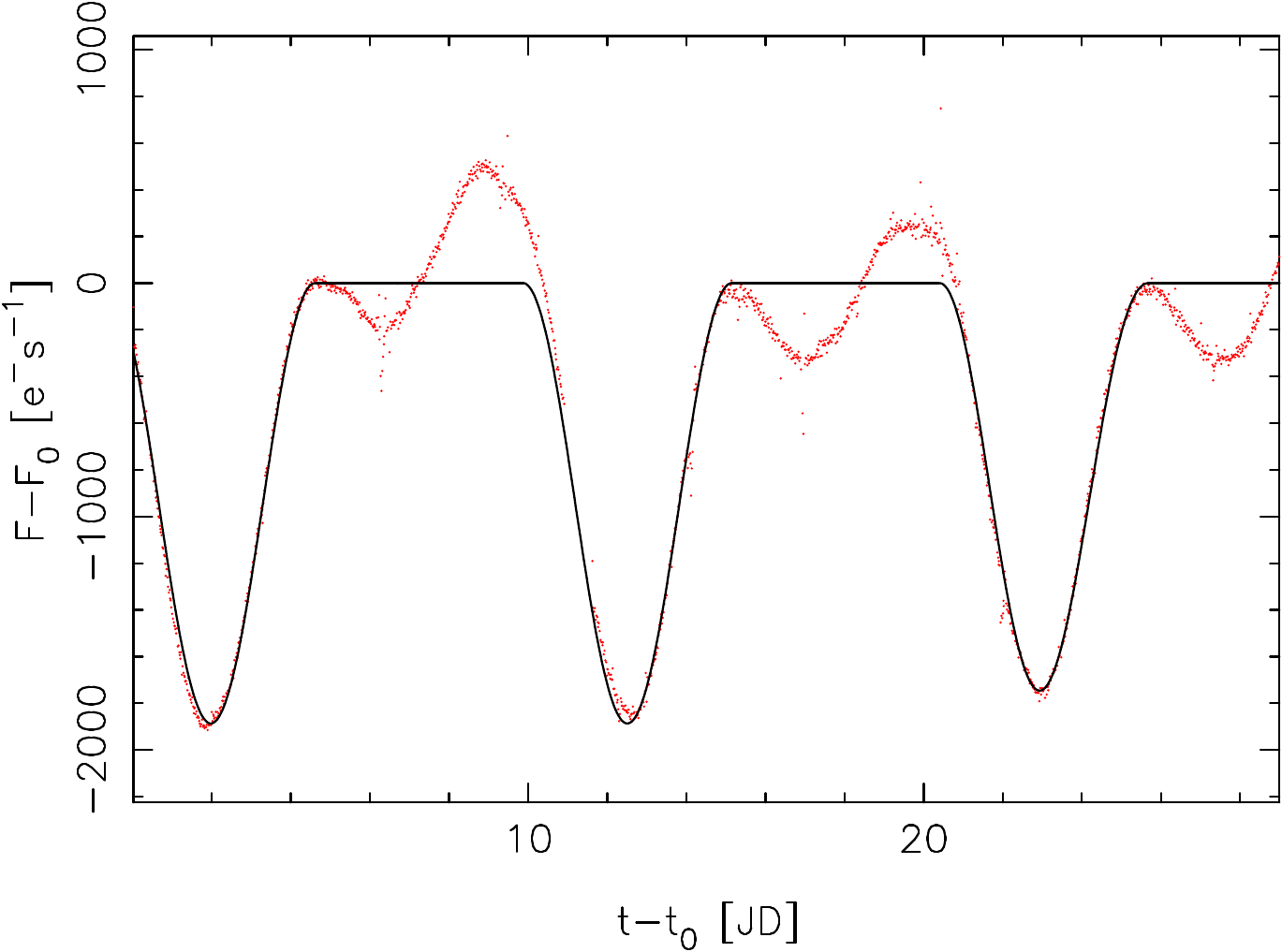}
\includegraphics[angle=270,width=0.24\textwidth,clip]{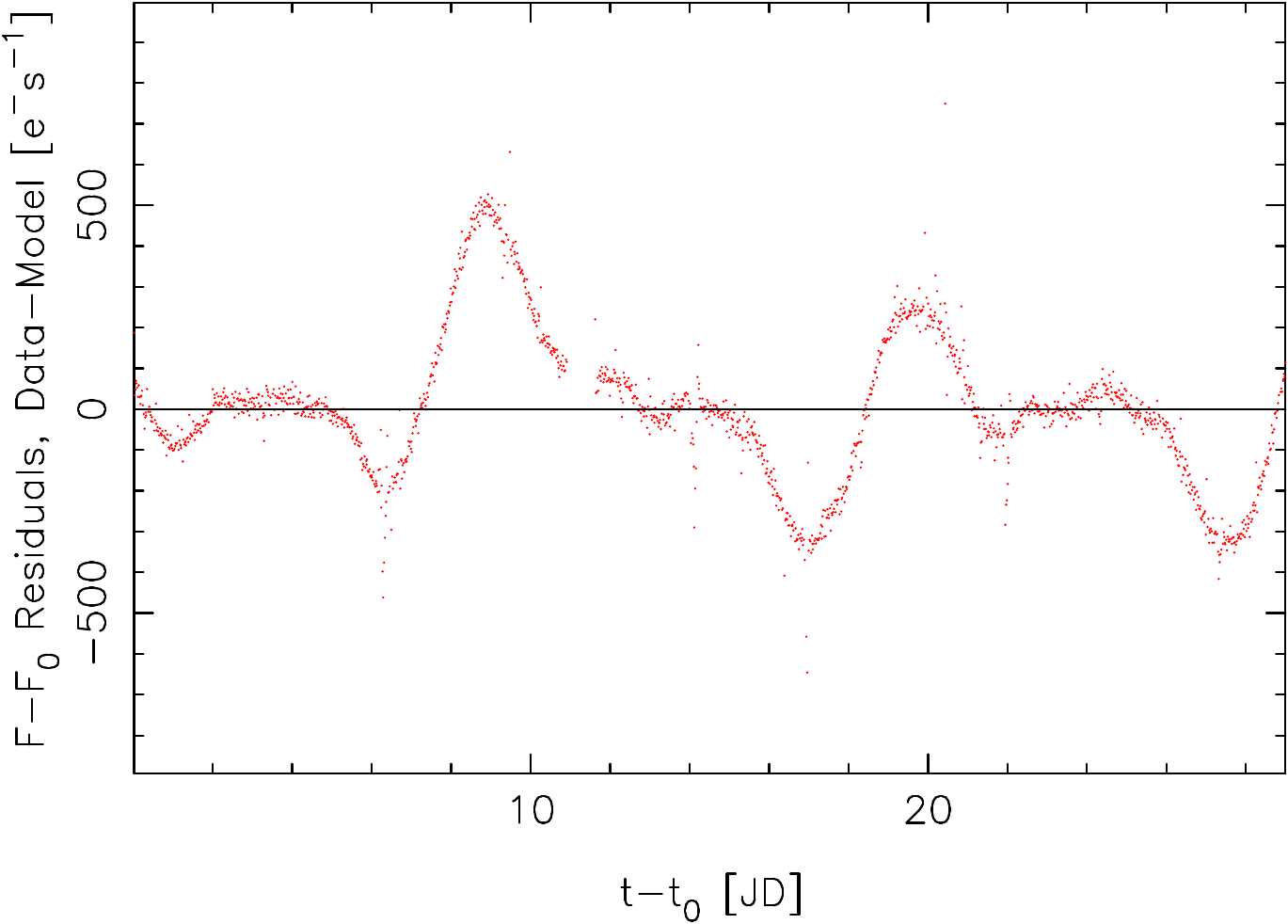}

\includegraphics[angle=270,width=0.24\textwidth,clip]{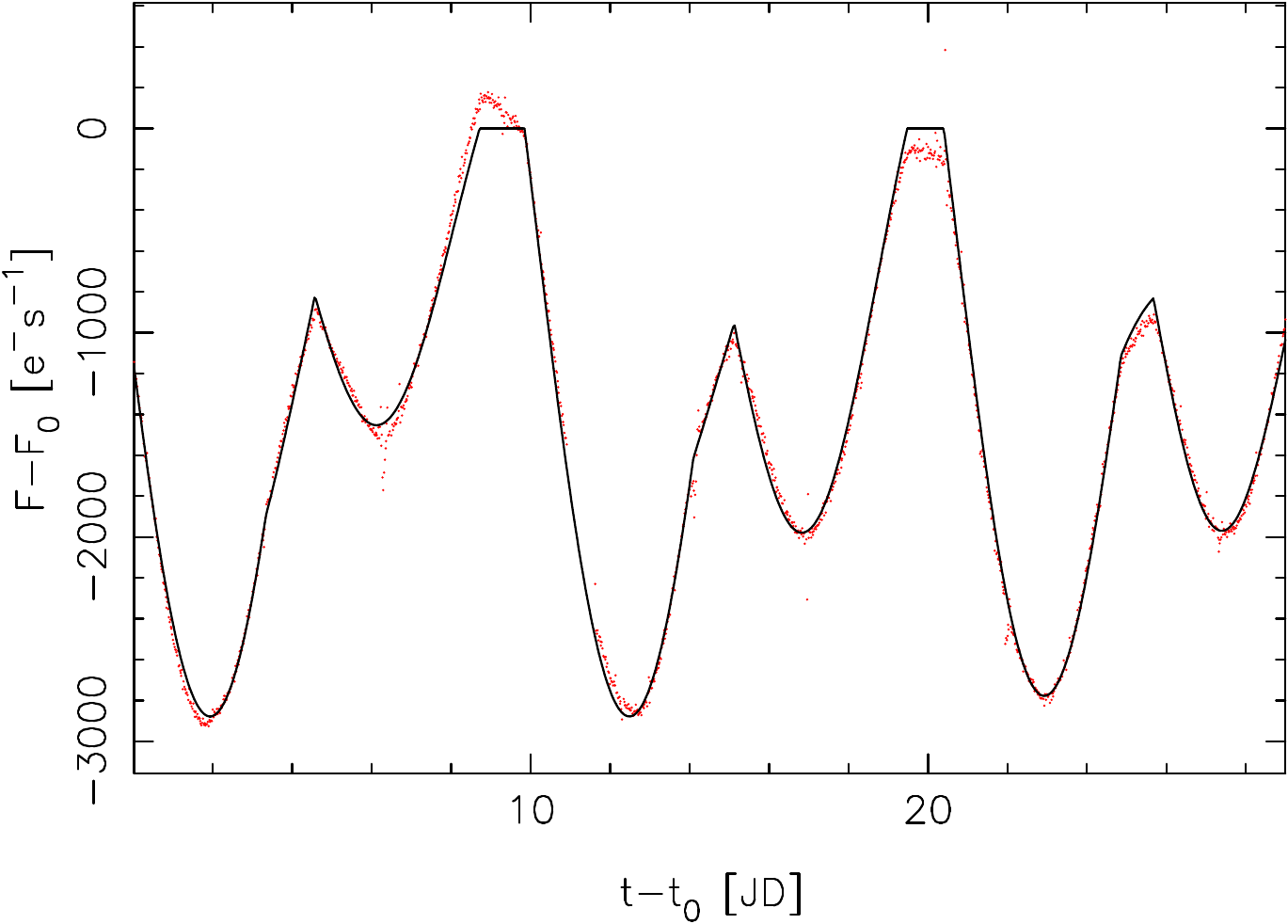}
\includegraphics[angle=270,width=0.24\textwidth,clip]{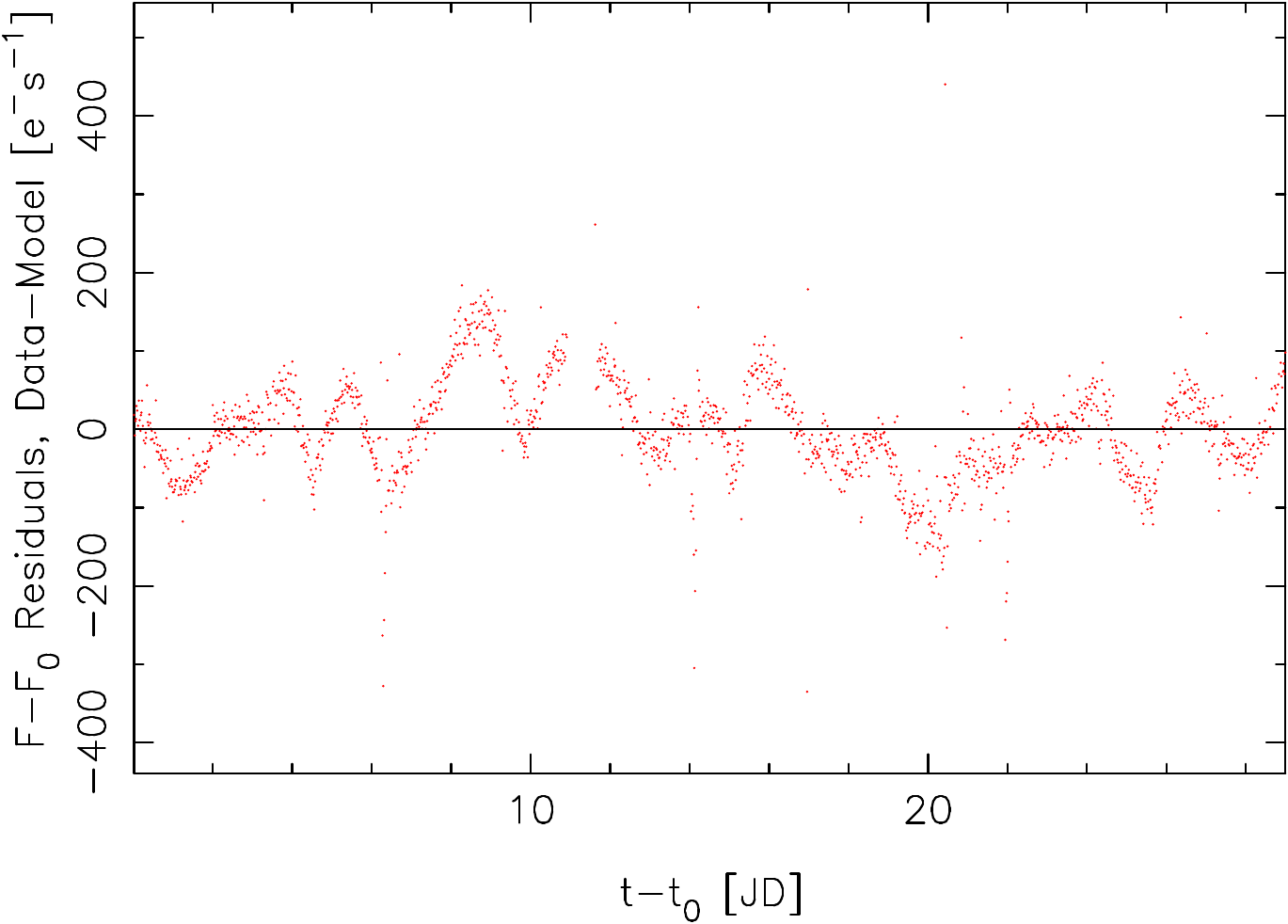}

\includegraphics[angle=270,width=0.24\textwidth,clip]{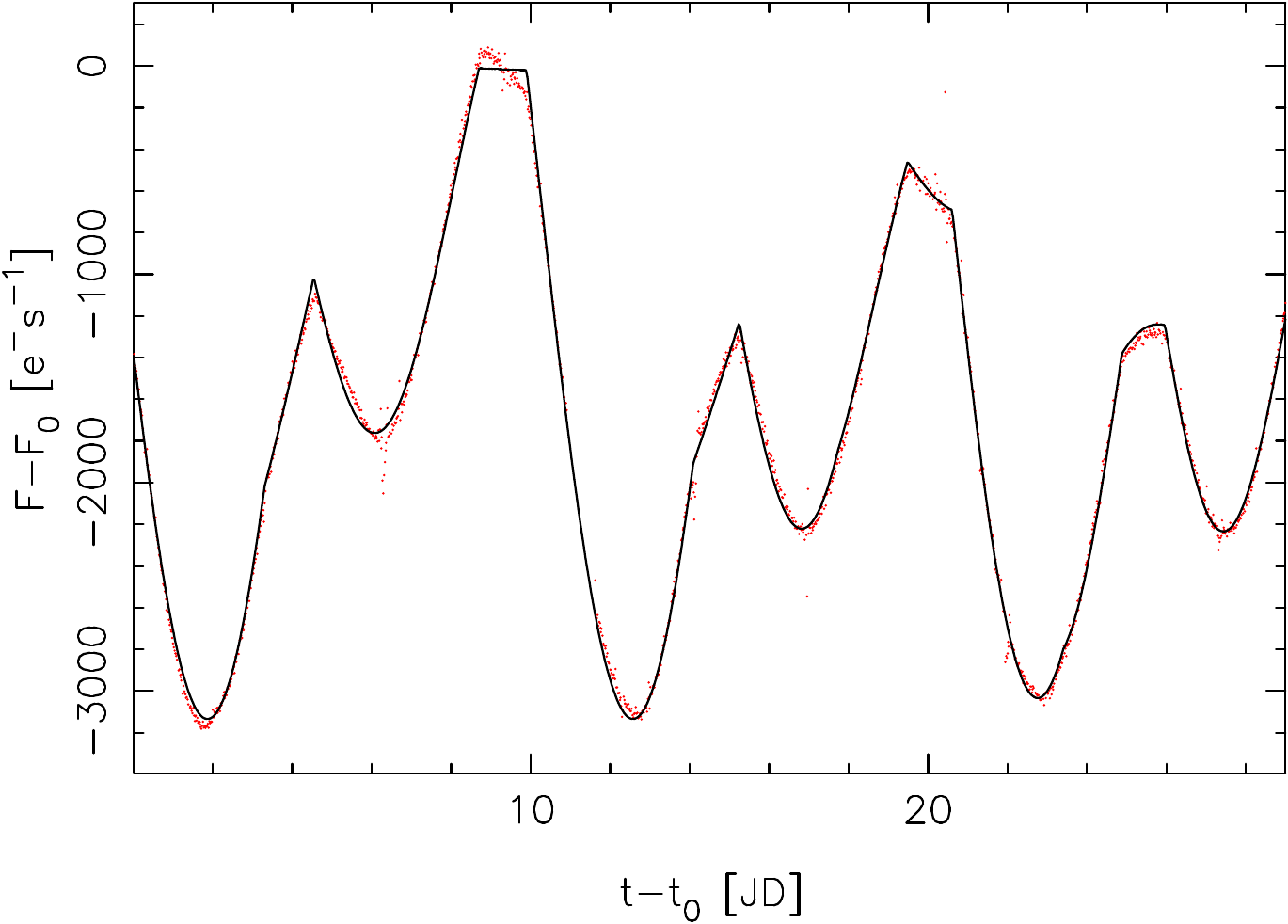}
\includegraphics[angle=270,width=0.24\textwidth,clip]{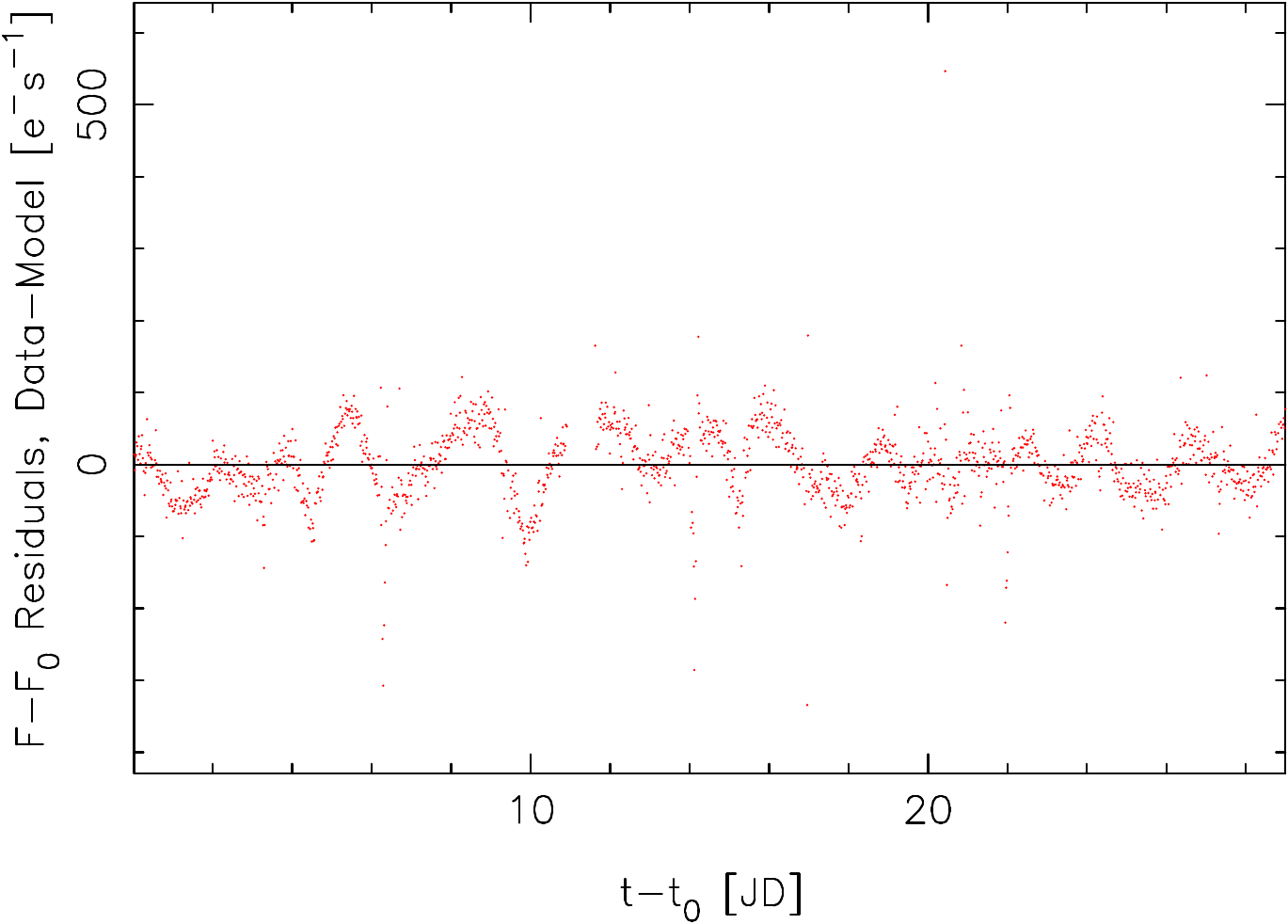}

\includegraphics[angle=270,width=0.24\textwidth,clip]{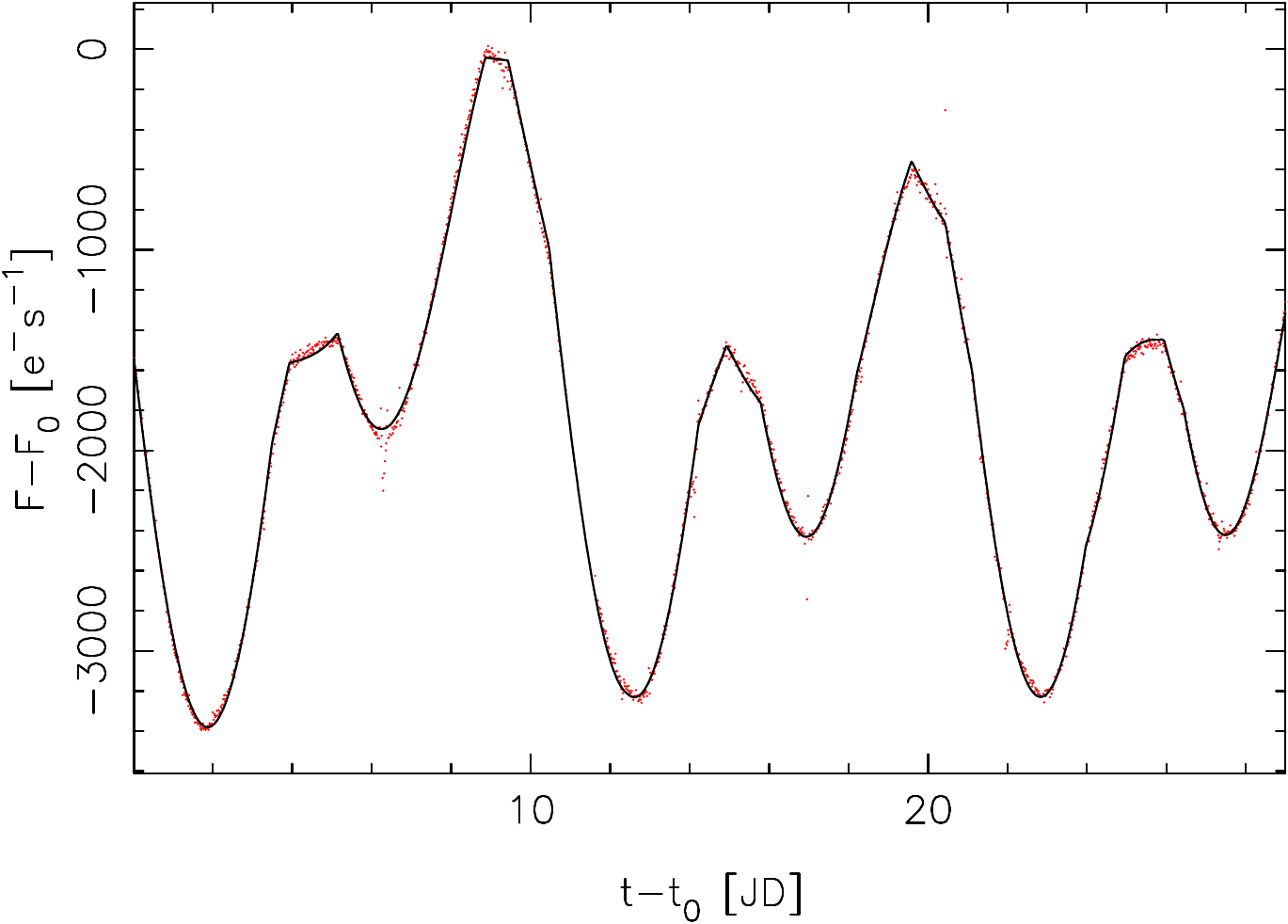}
\includegraphics[angle=270,width=0.24\textwidth,clip]{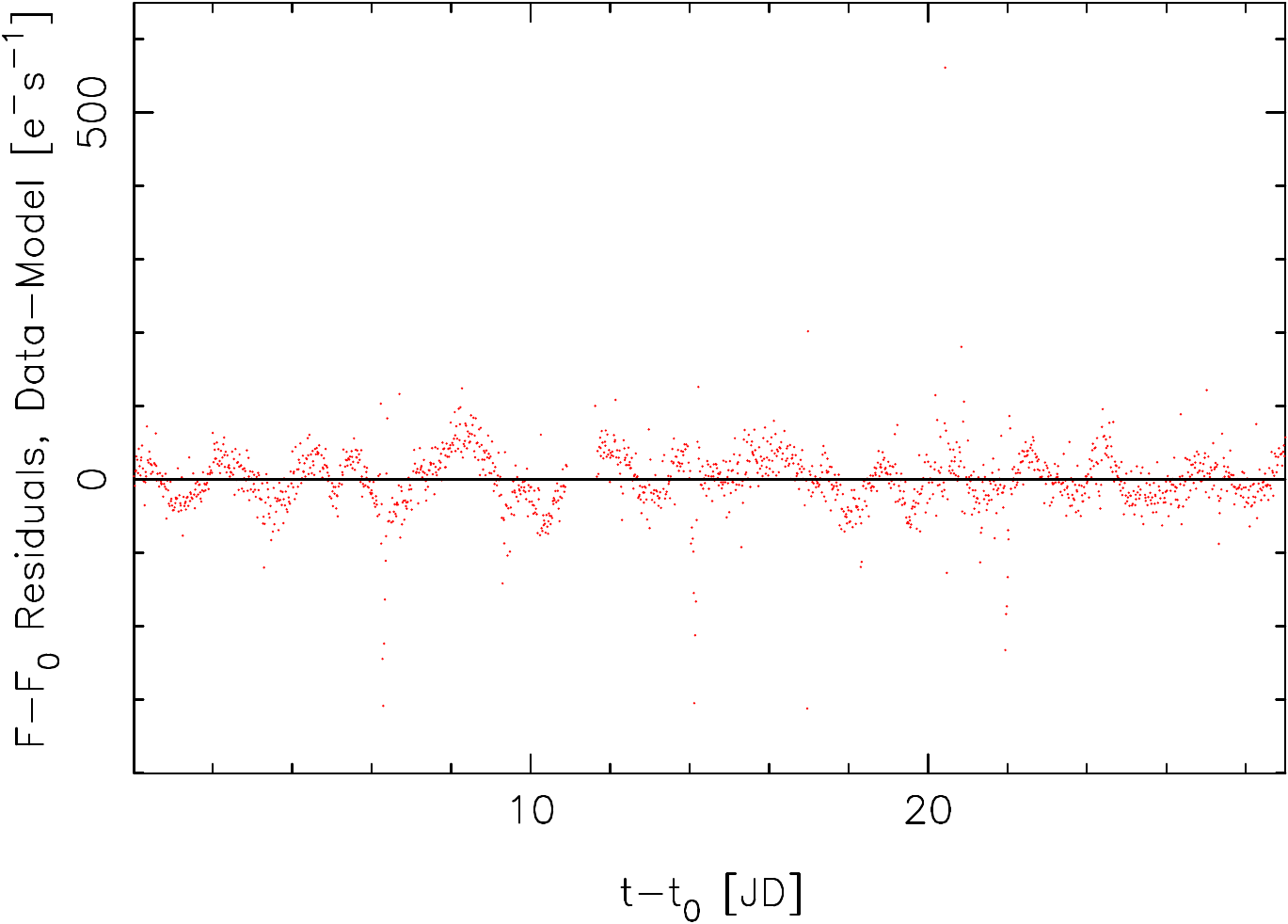}

\includegraphics[angle=270,width=0.24\textwidth,clip]{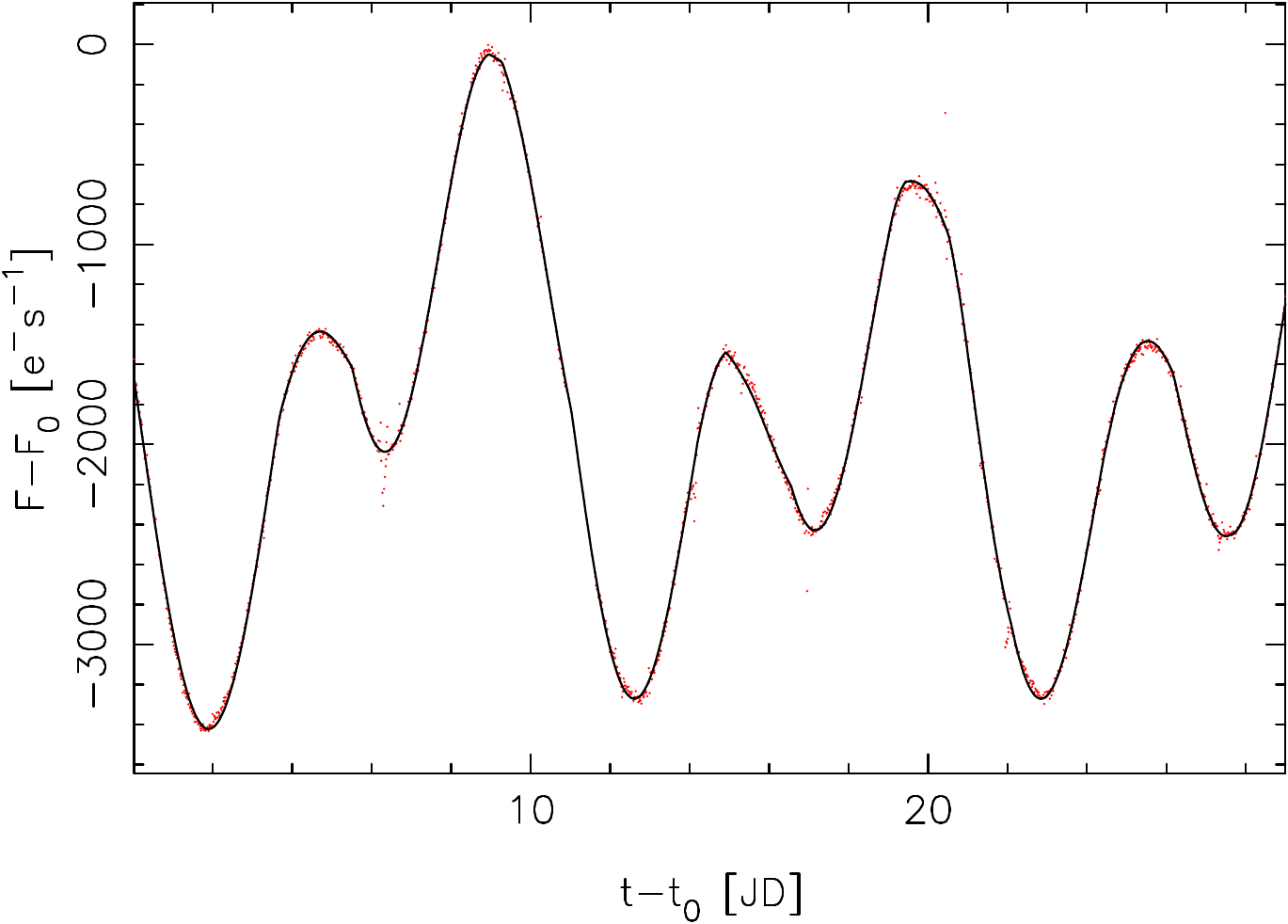}
\includegraphics[angle=270,width=0.24\textwidth,clip]{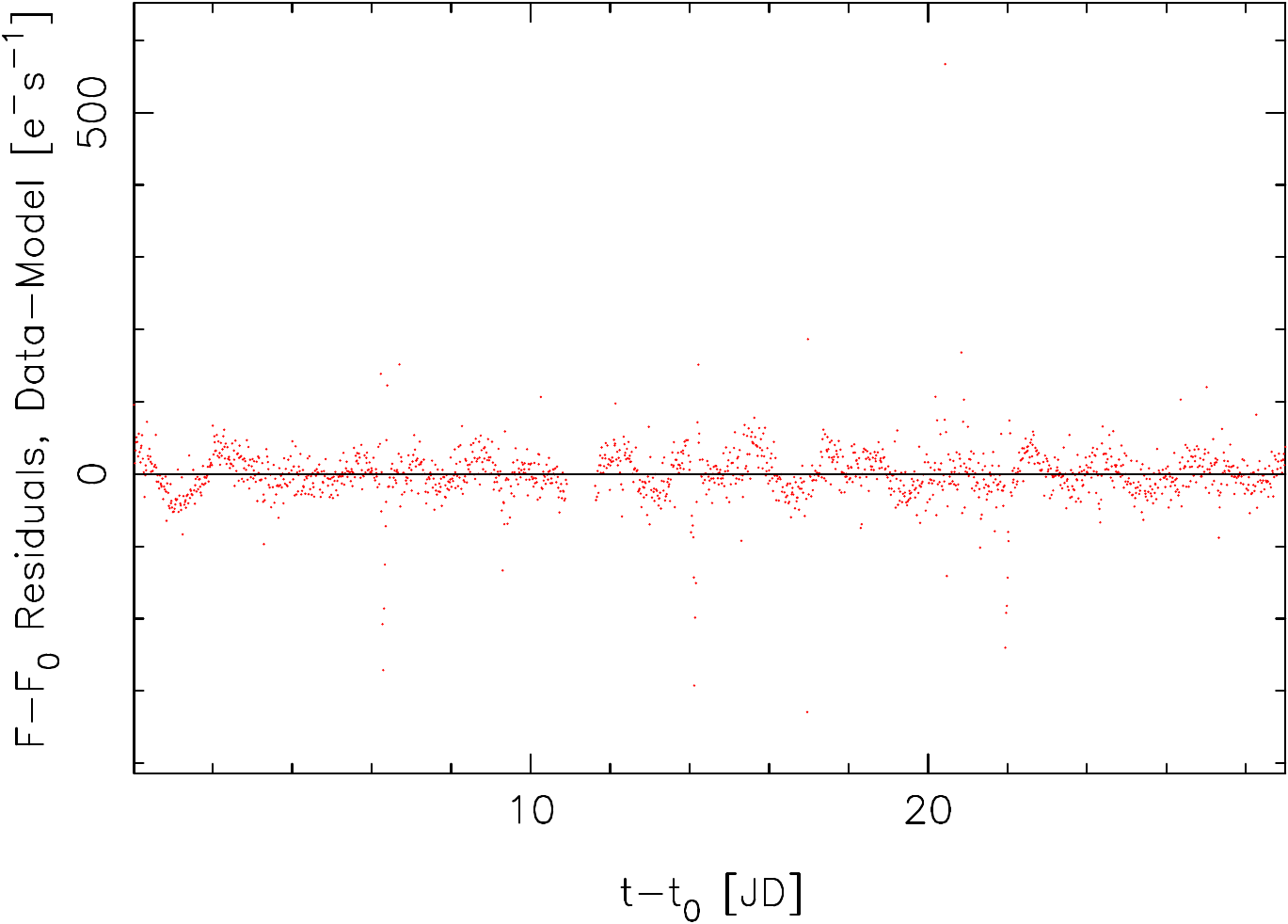}
\caption{Data (red dots) and maximum \emph{a posteriori} (MAP) solutions (black curves) as well as model residuals for data set S834 for which the reference time is $t_{0} = 834.0$. }\label{fig:S834_solution}
\end{figure}

\begin{figure}
\center
\includegraphics[angle=270,width=0.24\textwidth,clip]{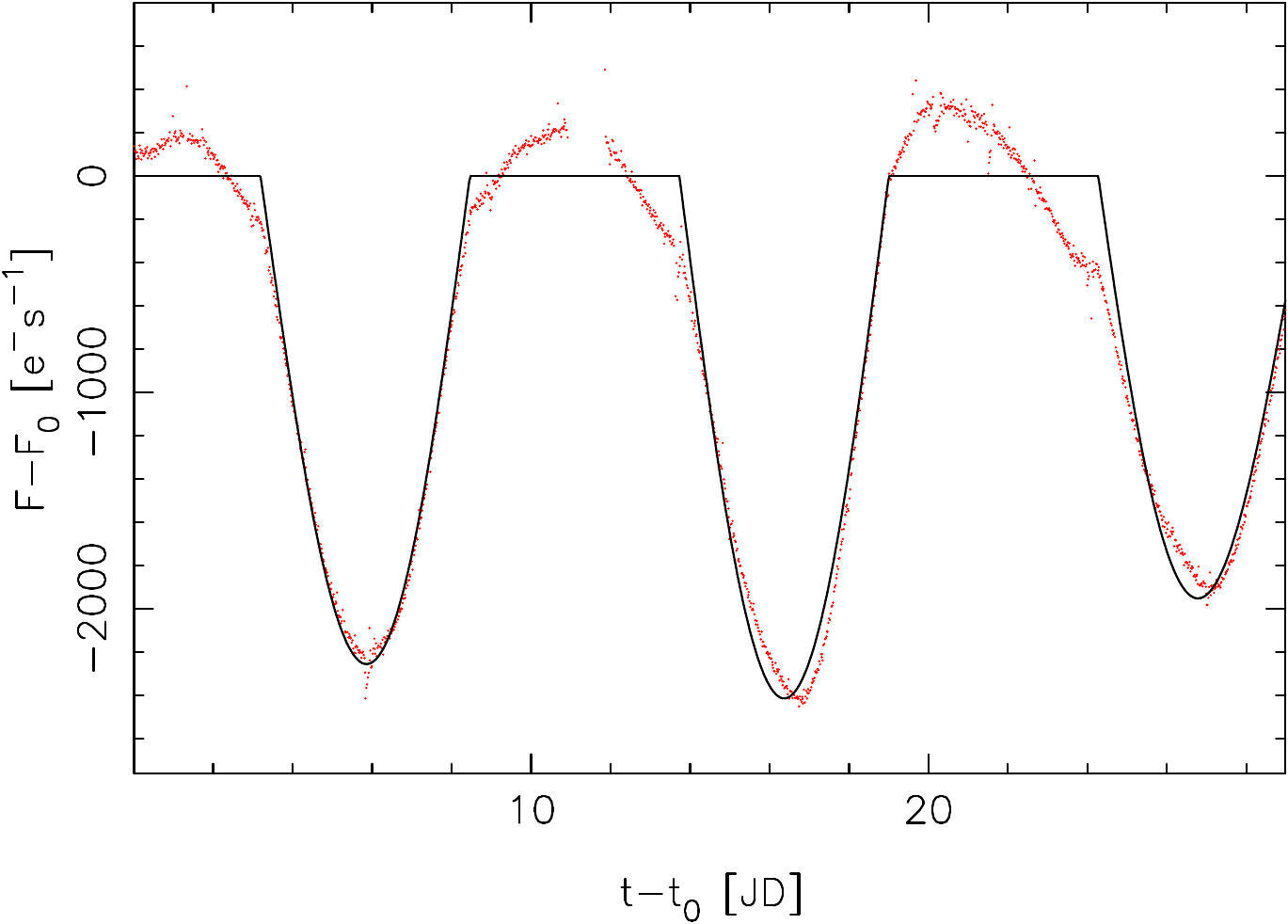}
\includegraphics[angle=270,width=0.24\textwidth,clip]{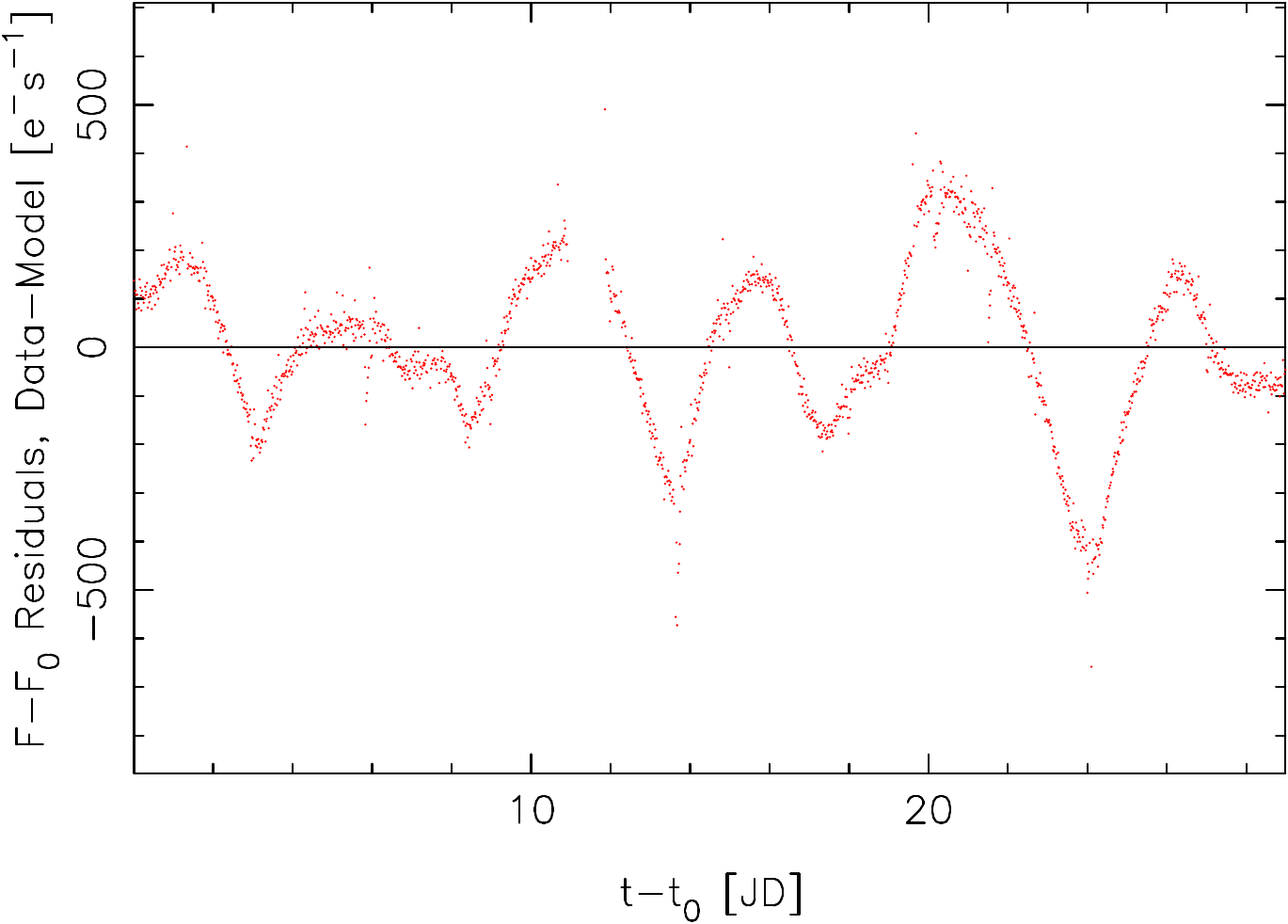}

\includegraphics[angle=270,width=0.24\textwidth,clip]{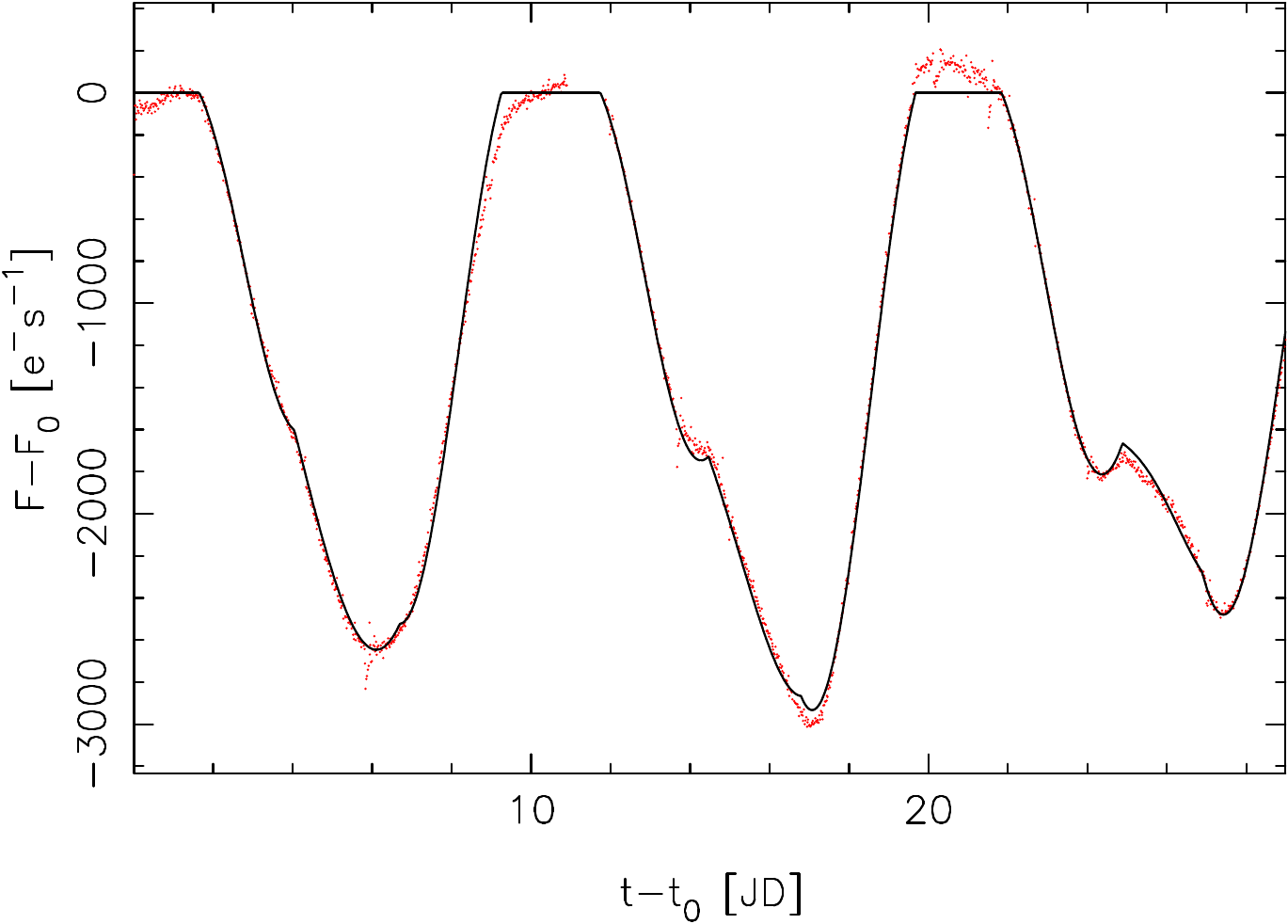}
\includegraphics[angle=270,width=0.24\textwidth,clip]{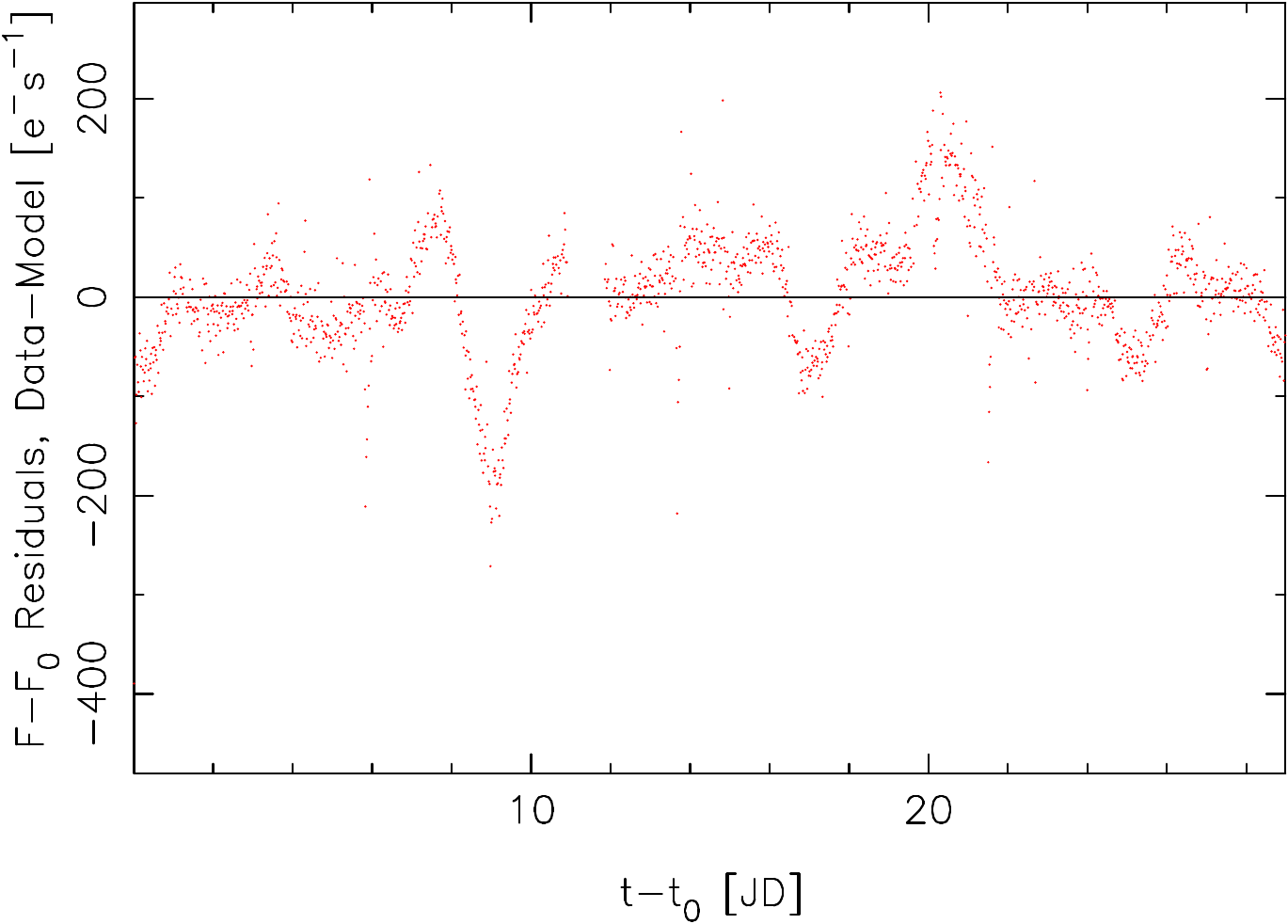}

\includegraphics[angle=270,width=0.24\textwidth,clip]{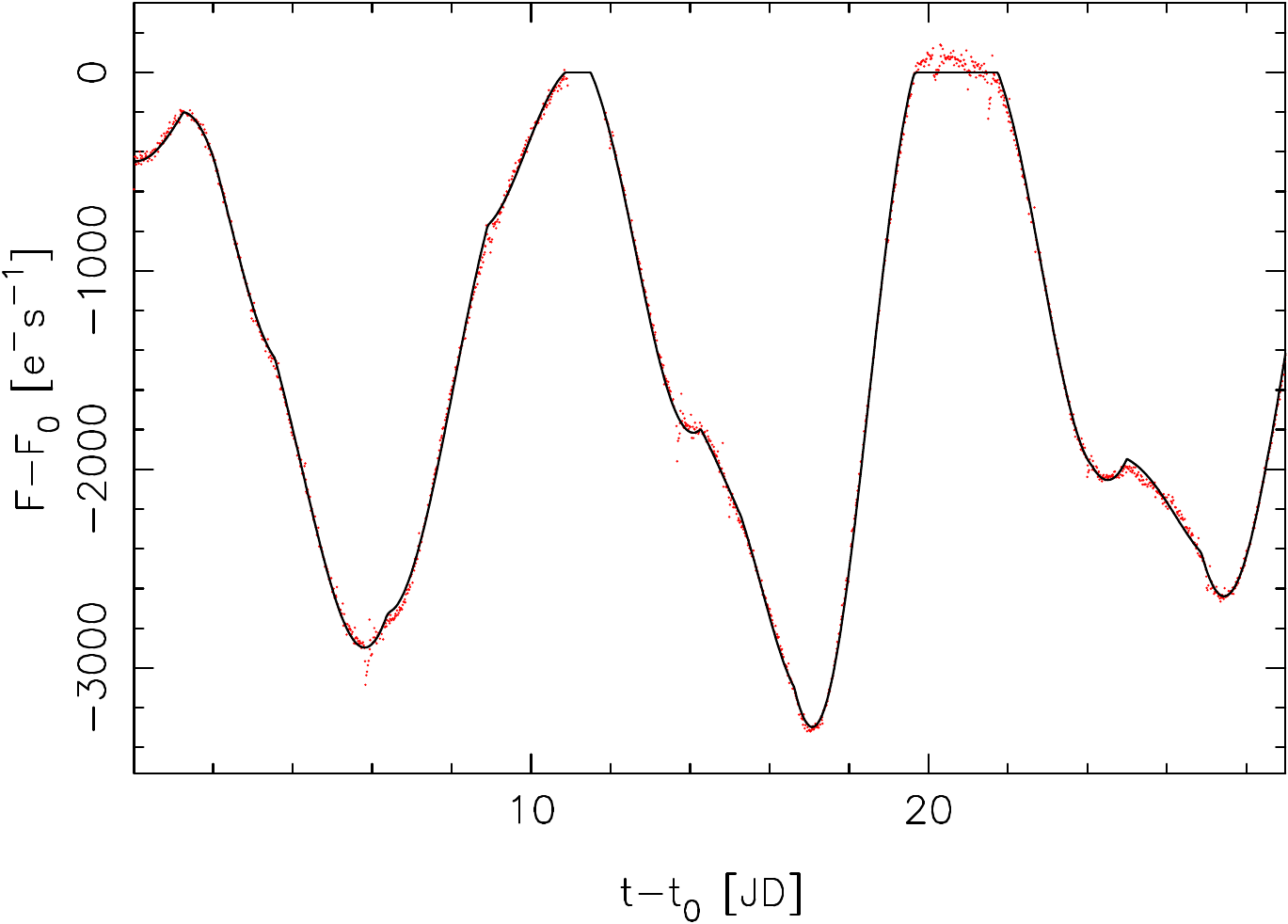}
\includegraphics[angle=270,width=0.24\textwidth,clip]{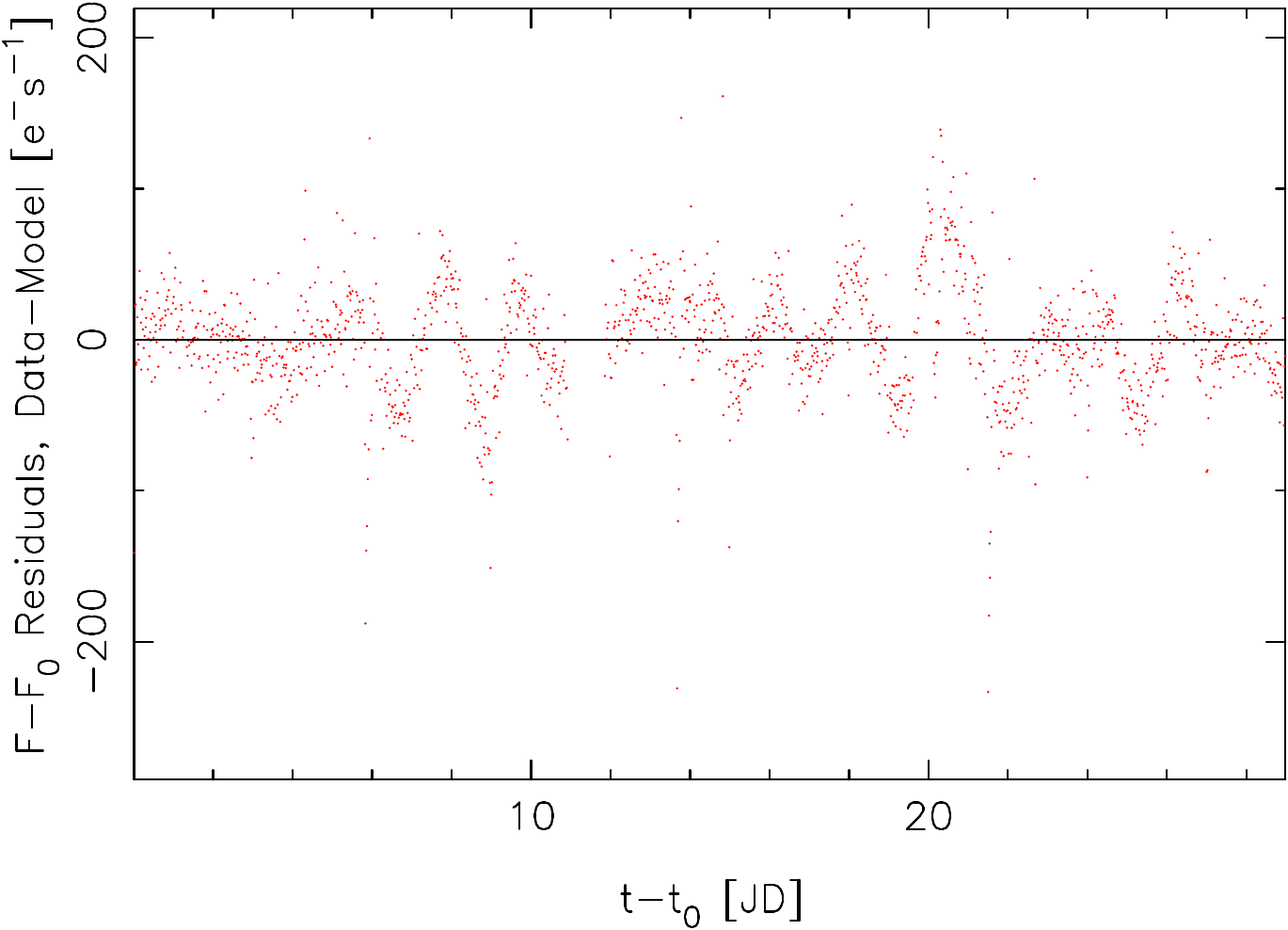}

\includegraphics[angle=270,width=0.24\textwidth,clip]{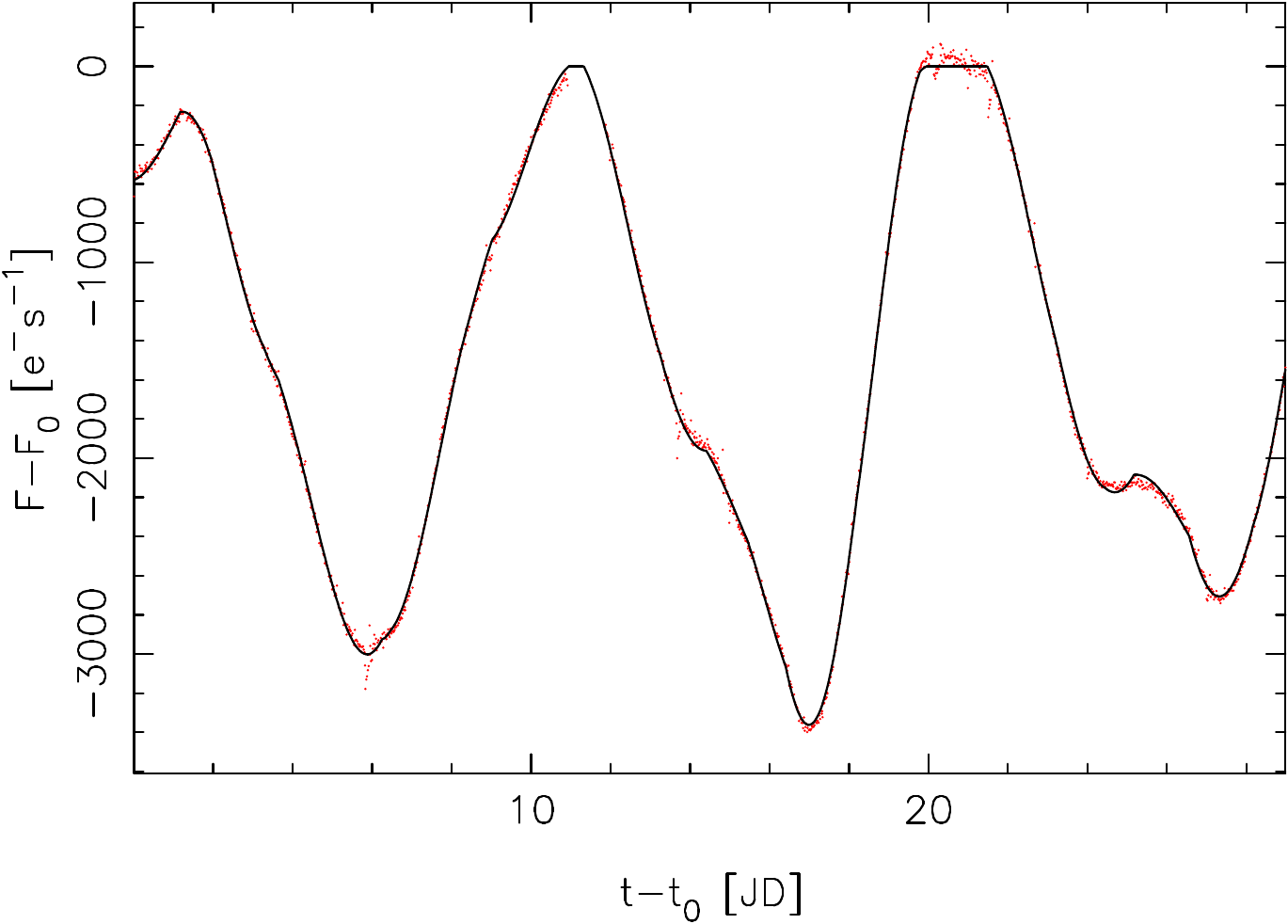}
\includegraphics[angle=270,width=0.24\textwidth,clip]{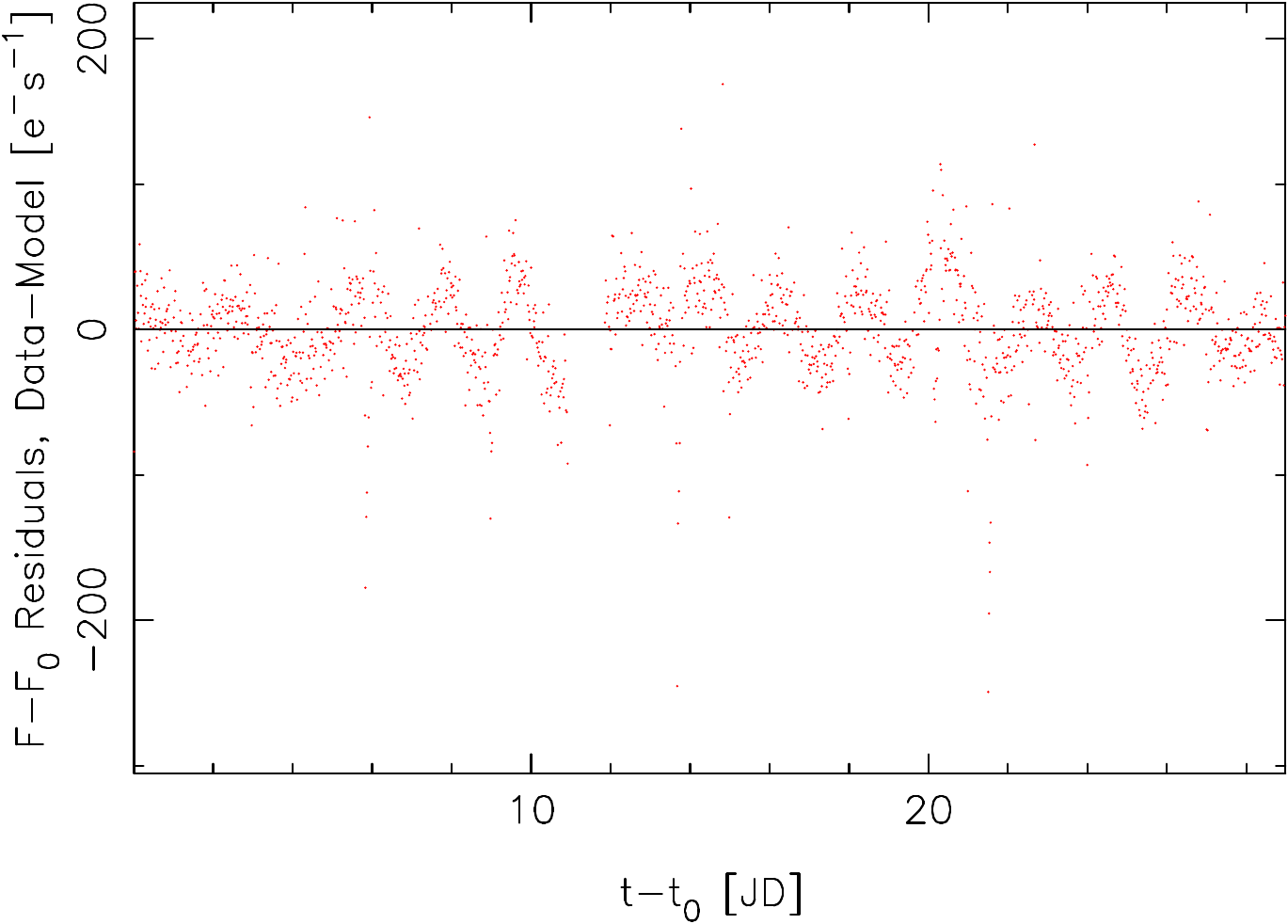}

\includegraphics[angle=270,width=0.24\textwidth,clip]{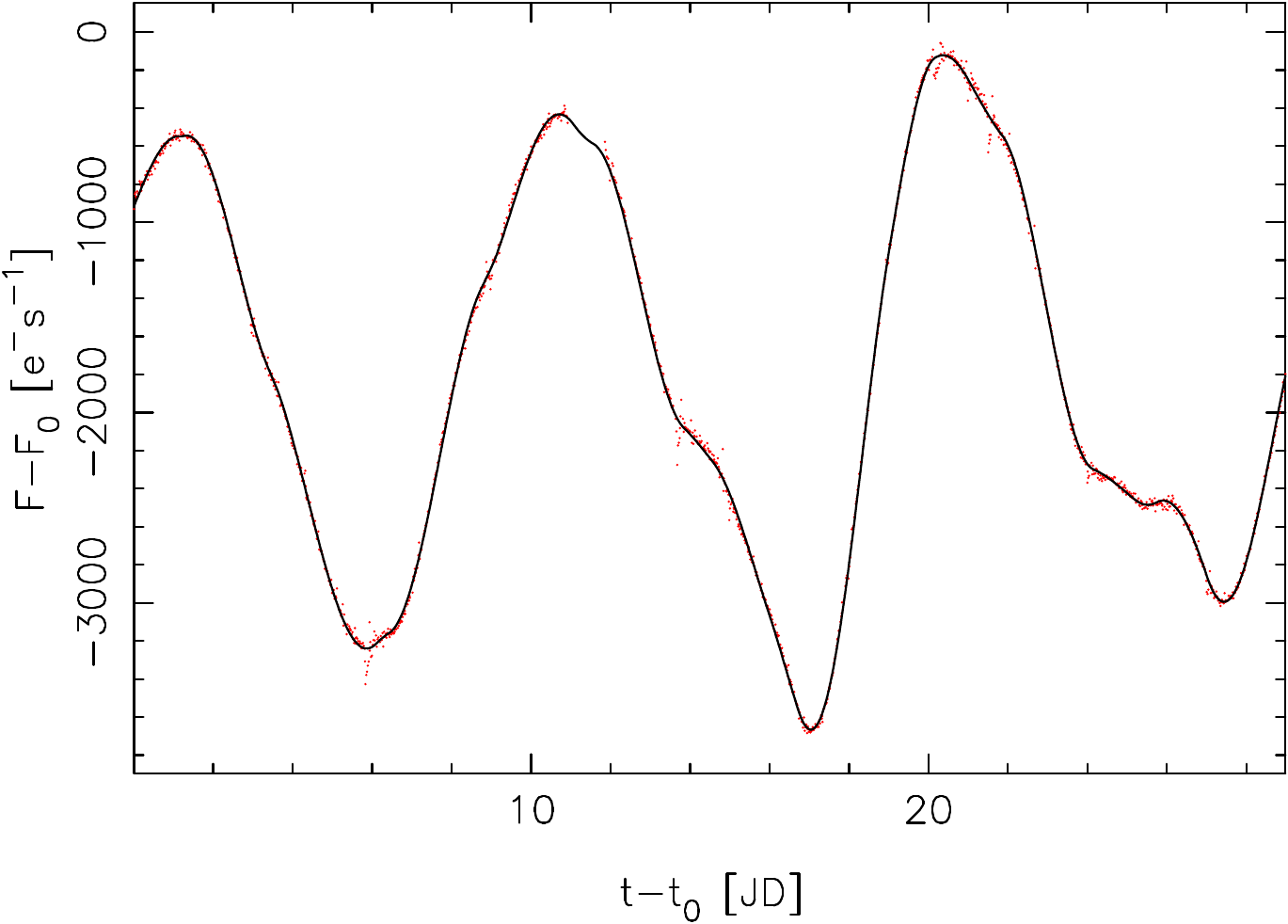}
\includegraphics[angle=270,width=0.24\textwidth,clip]{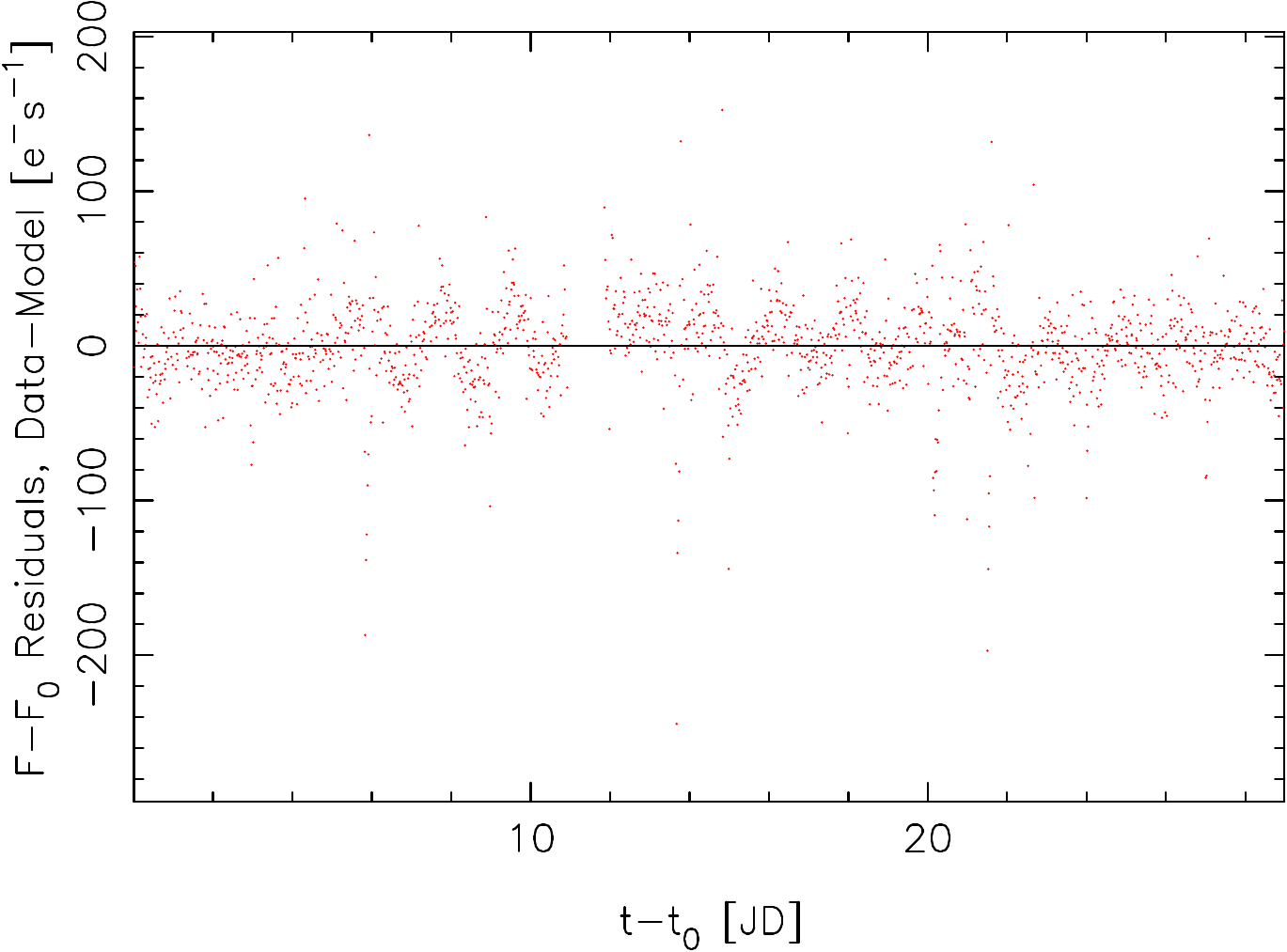}
\caption{As in Fig. \ref{fig:S834_solution} but for data set S1234 for which the reference time is $t_{0} = 1234.0$.}\label{fig:S1234_solution}
\end{figure}

\begin{figure}
\center
\includegraphics[angle=270,width=0.24\textwidth,clip]{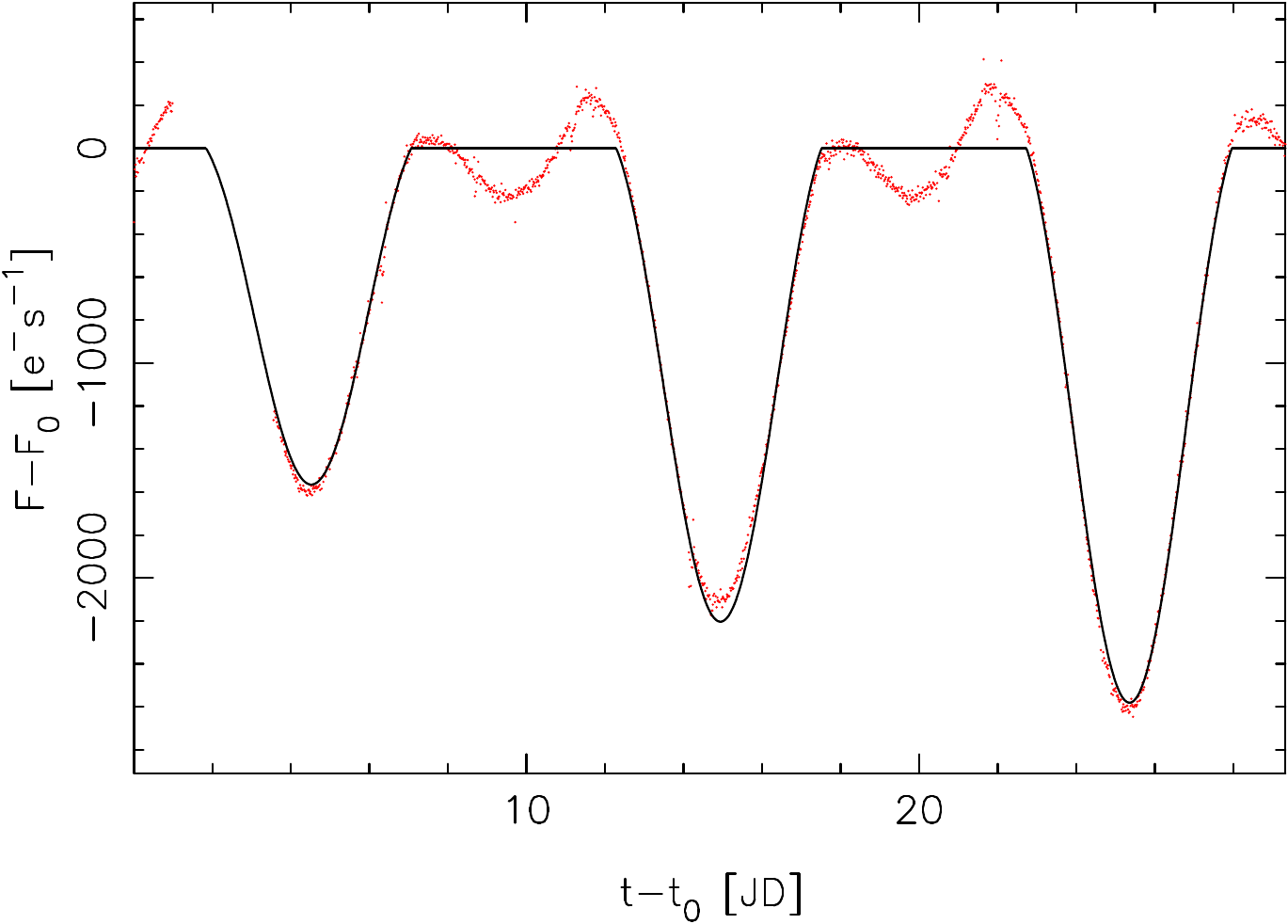}
\includegraphics[angle=270,width=0.24\textwidth,clip]{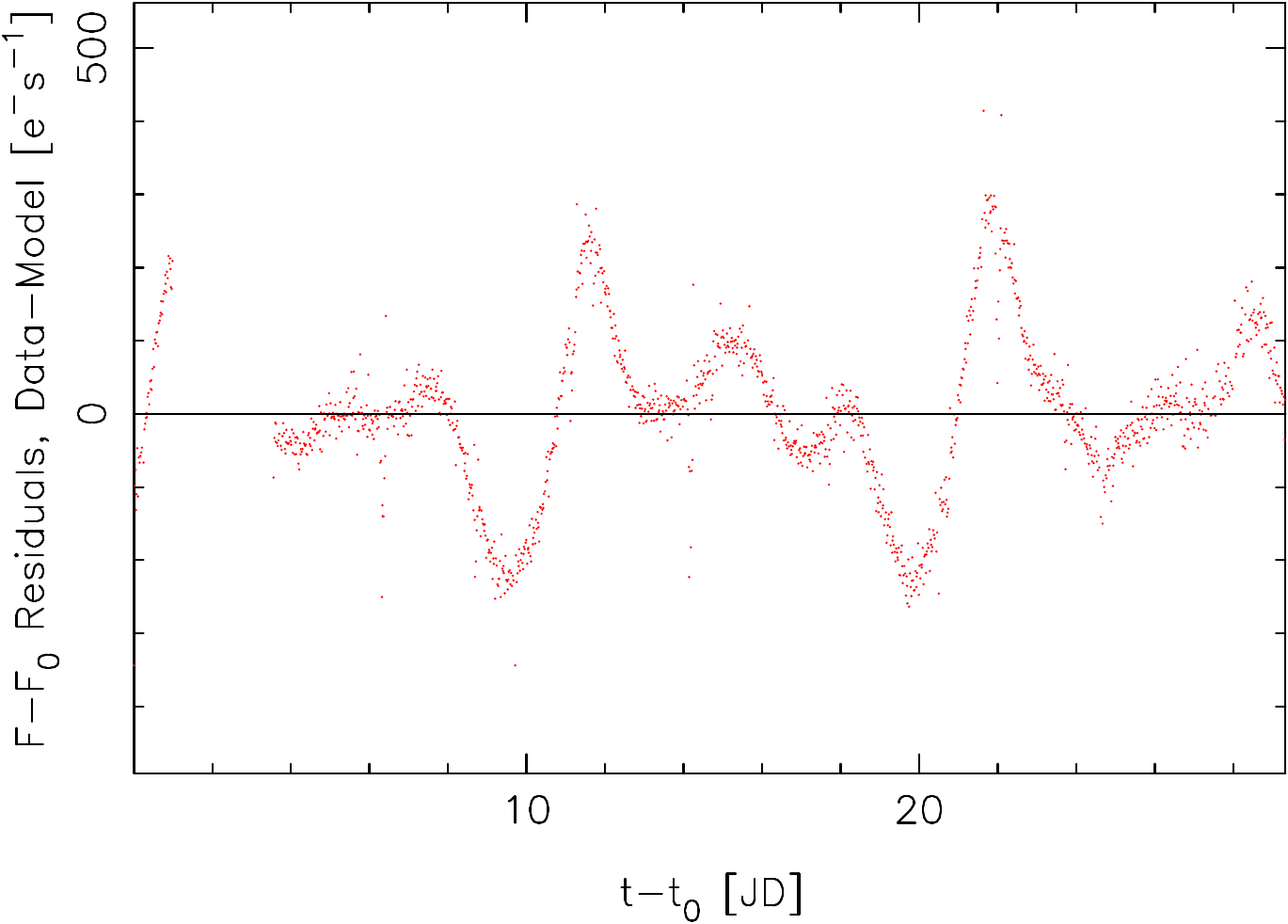}

\includegraphics[angle=270,width=0.24\textwidth,clip]{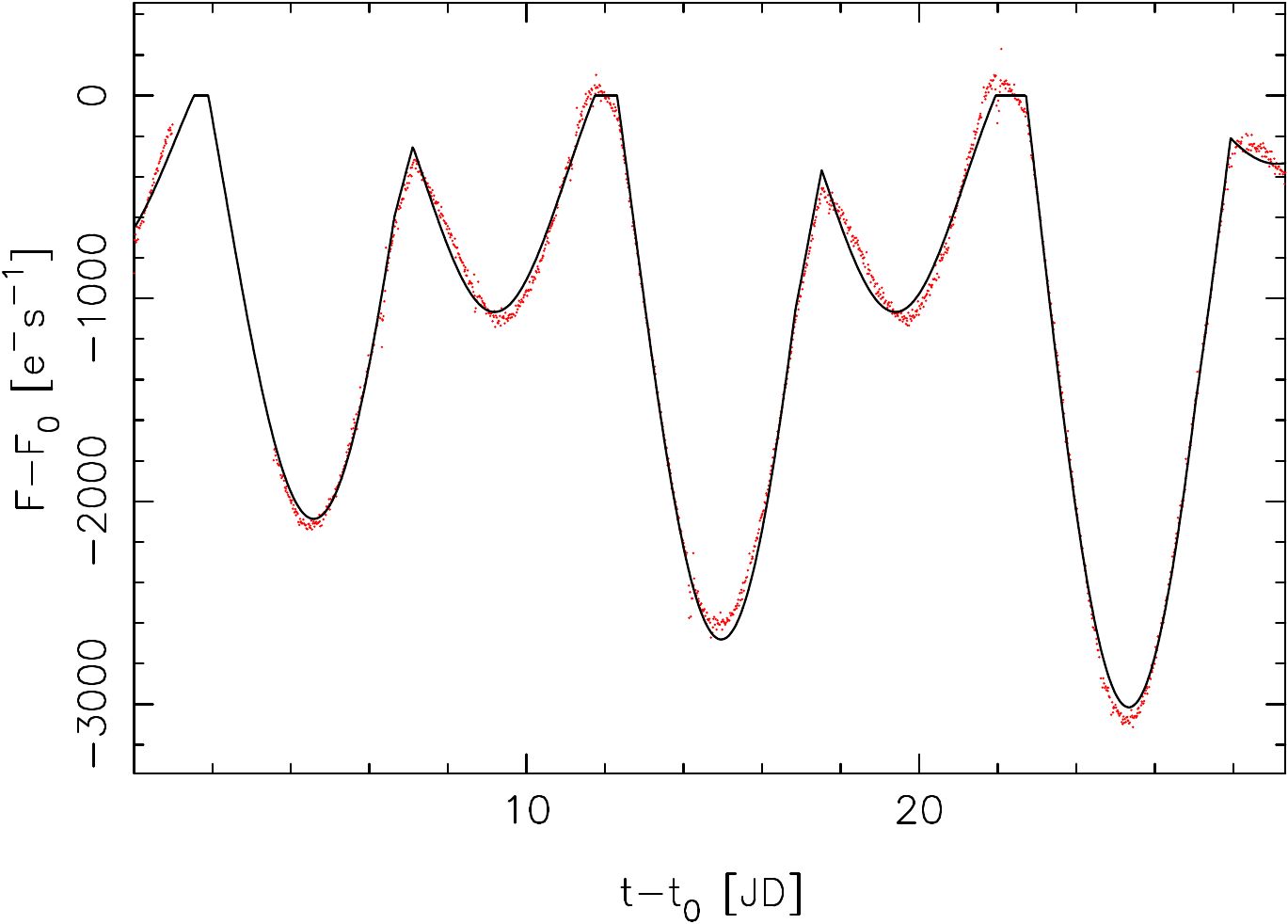}
\includegraphics[angle=270,width=0.24\textwidth,clip]{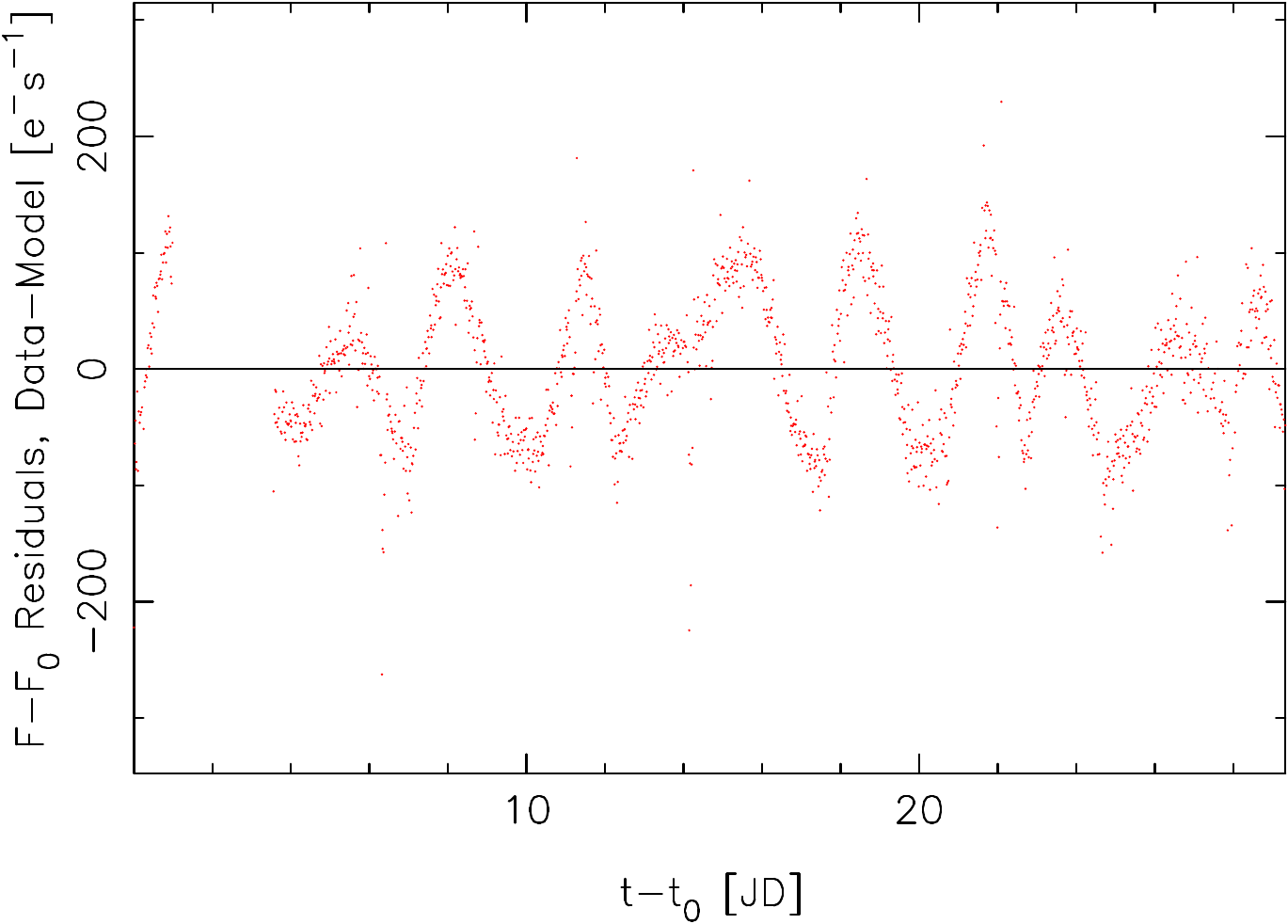}

\includegraphics[angle=270,width=0.24\textwidth,clip]{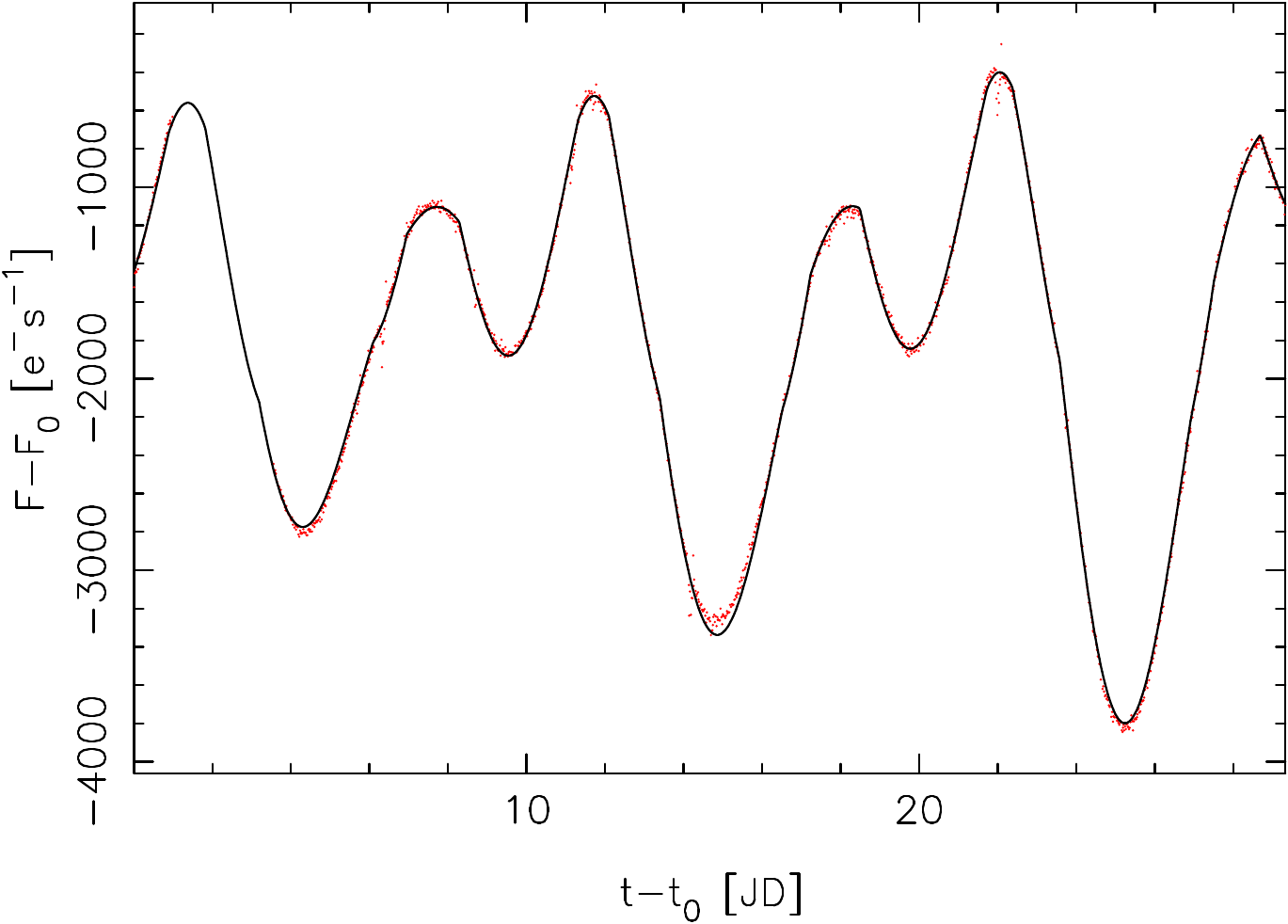}
\includegraphics[angle=270,width=0.24\textwidth,clip]{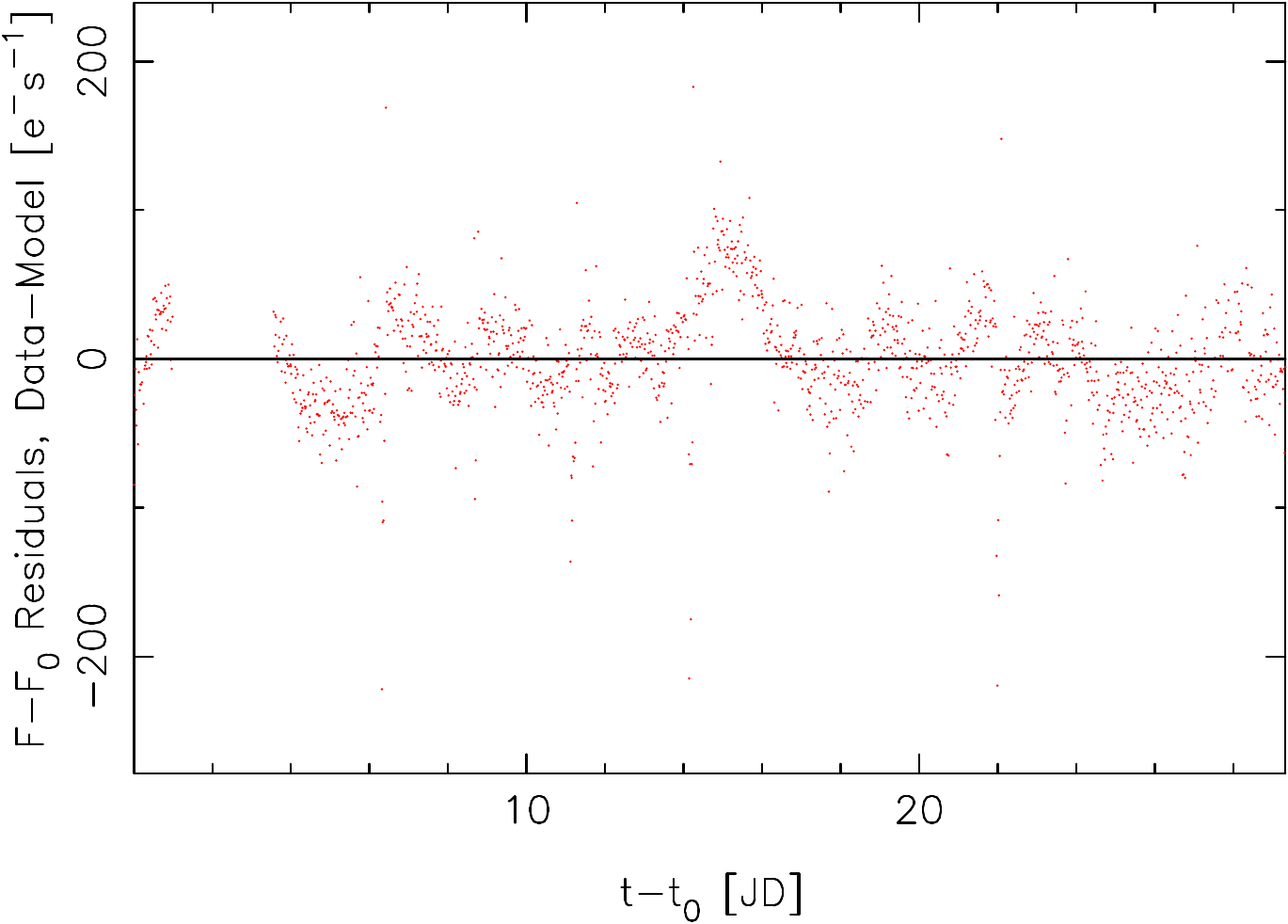}

\includegraphics[angle=270,width=0.24\textwidth,clip]{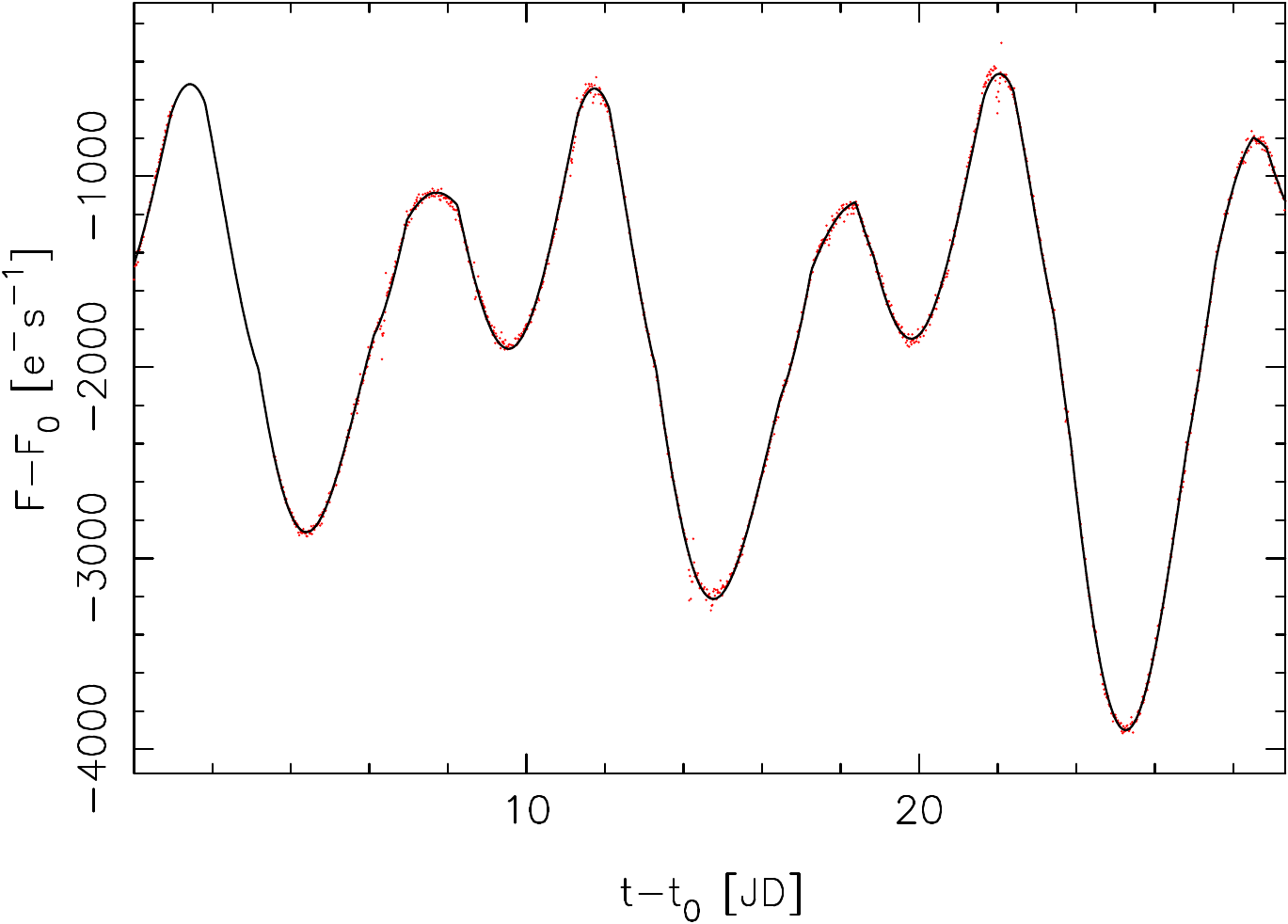}
\includegraphics[angle=270,width=0.24\textwidth,clip]{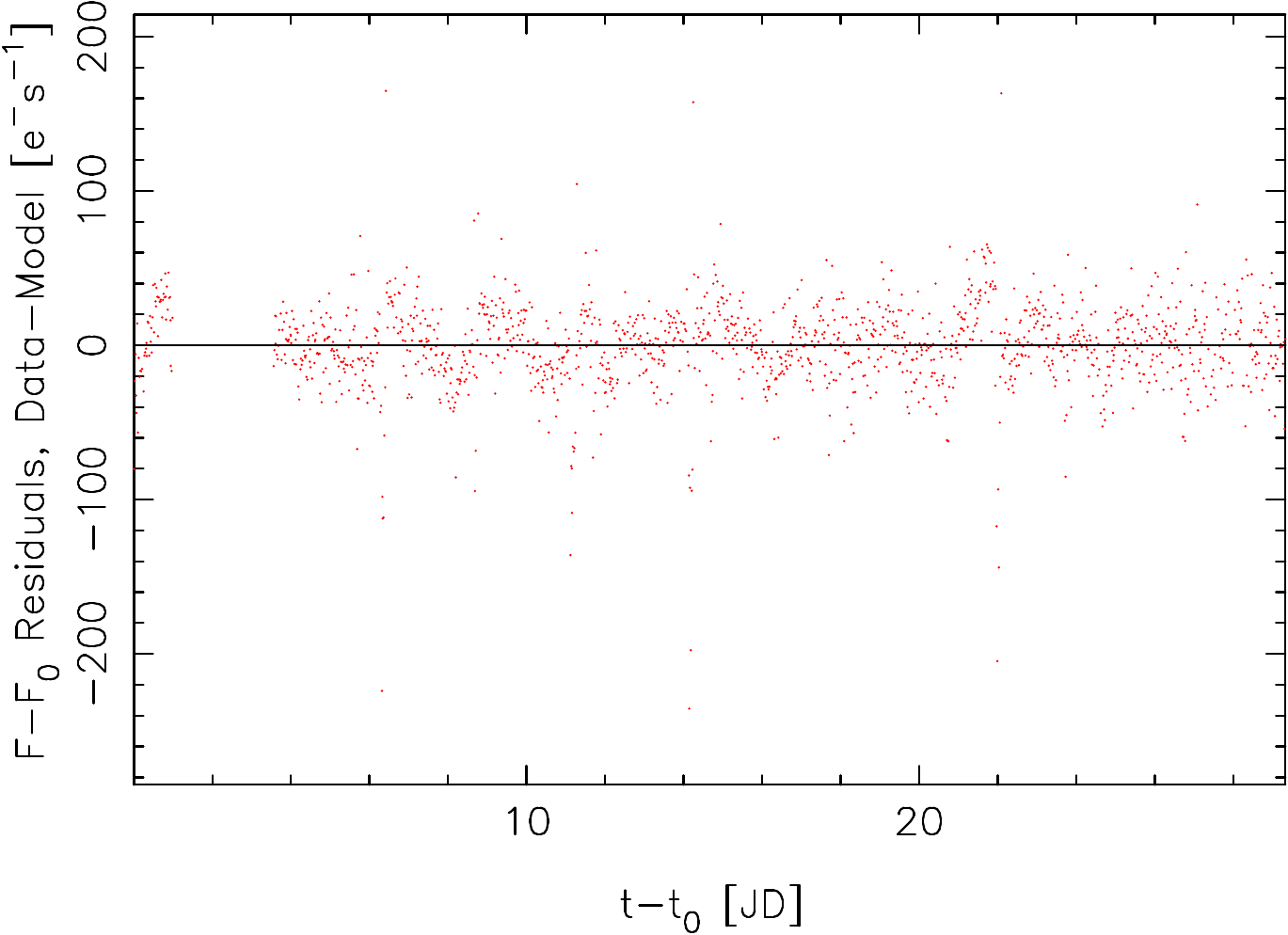}
\caption{As in Fig. \ref{fig:S834_solution} but for data set S1069 for which the reference time is $t_{0} = 1069.0$.}\label{fig:S1069_solution}
\end{figure}

\begin{figure*}
\includegraphics[angle=270,width=0.135\textwidth,clip]{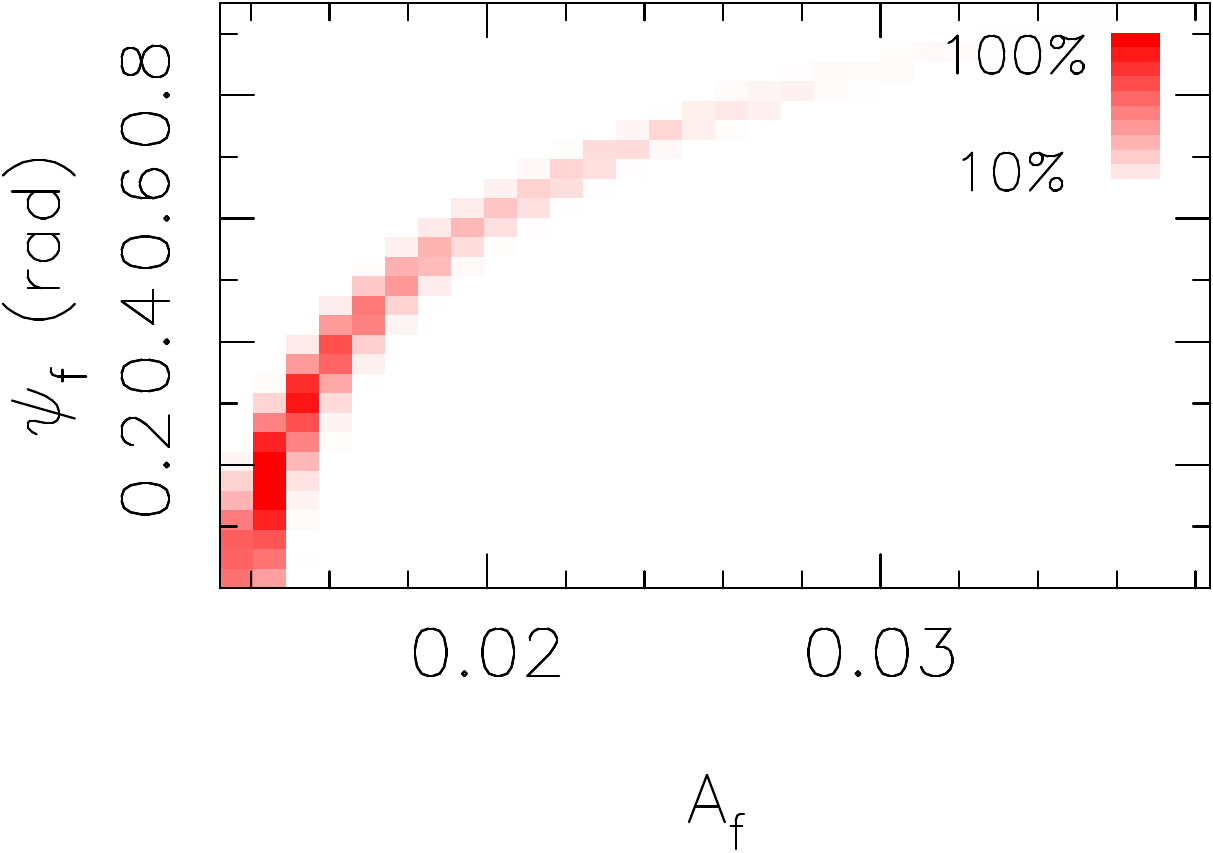}

\includegraphics[angle=270,width=0.135\textwidth,clip]{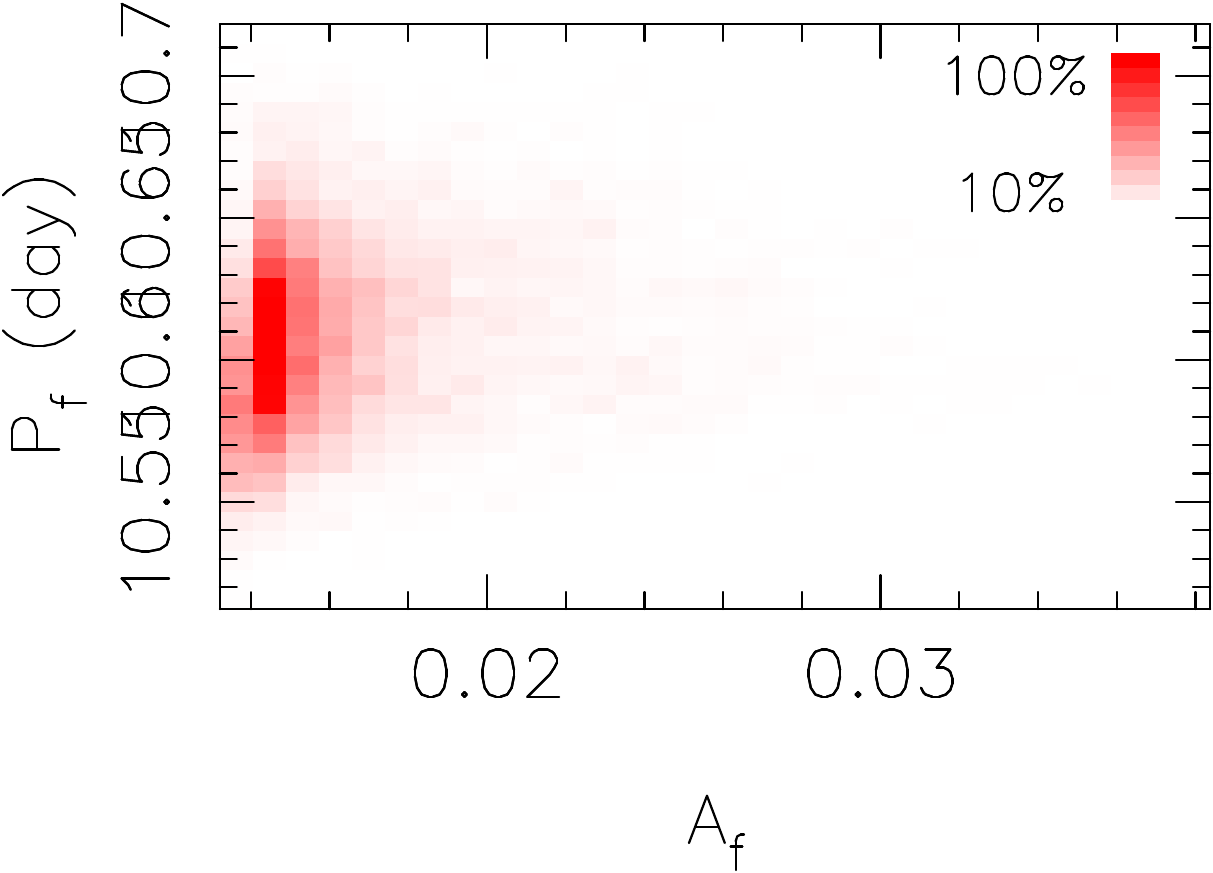}
\includegraphics[angle=270,width=0.135\textwidth,clip]{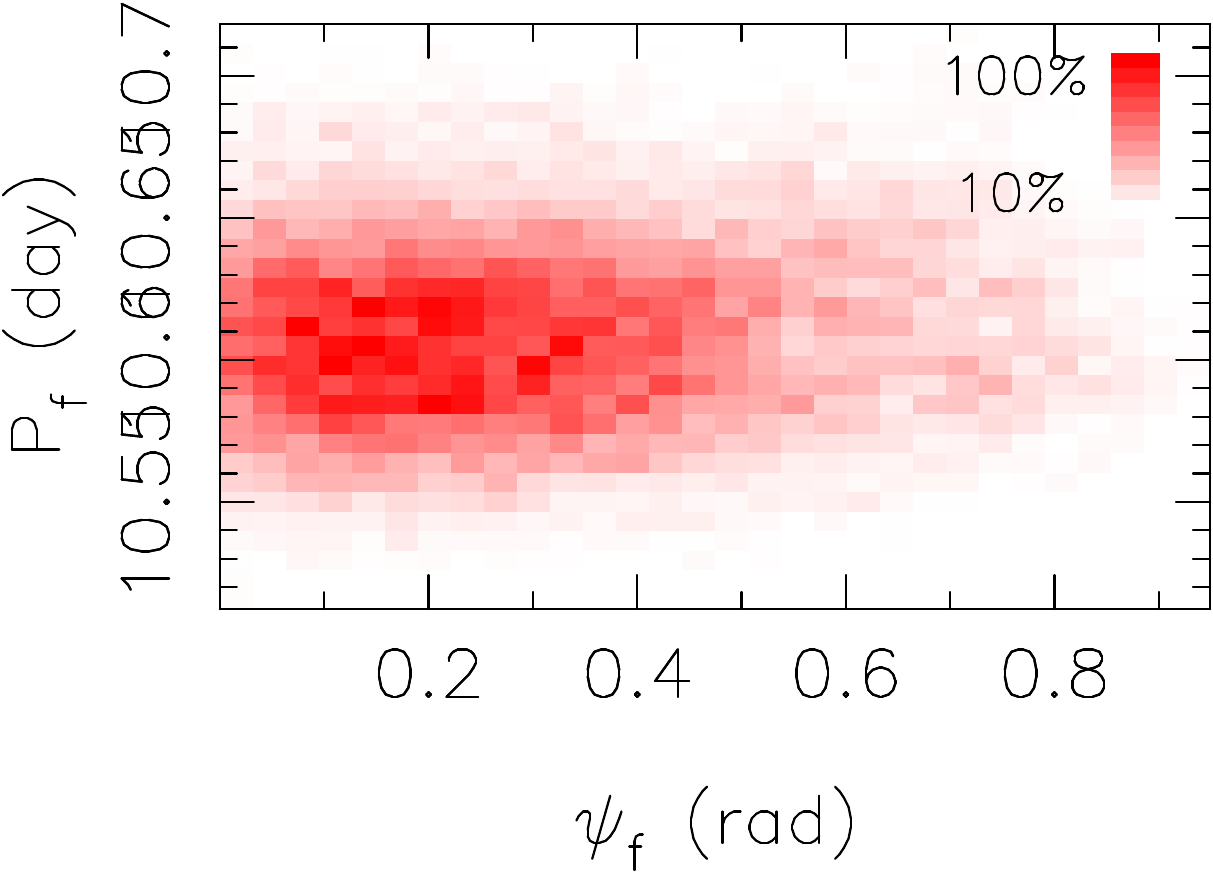}

\includegraphics[angle=270,width=0.135\textwidth,clip]{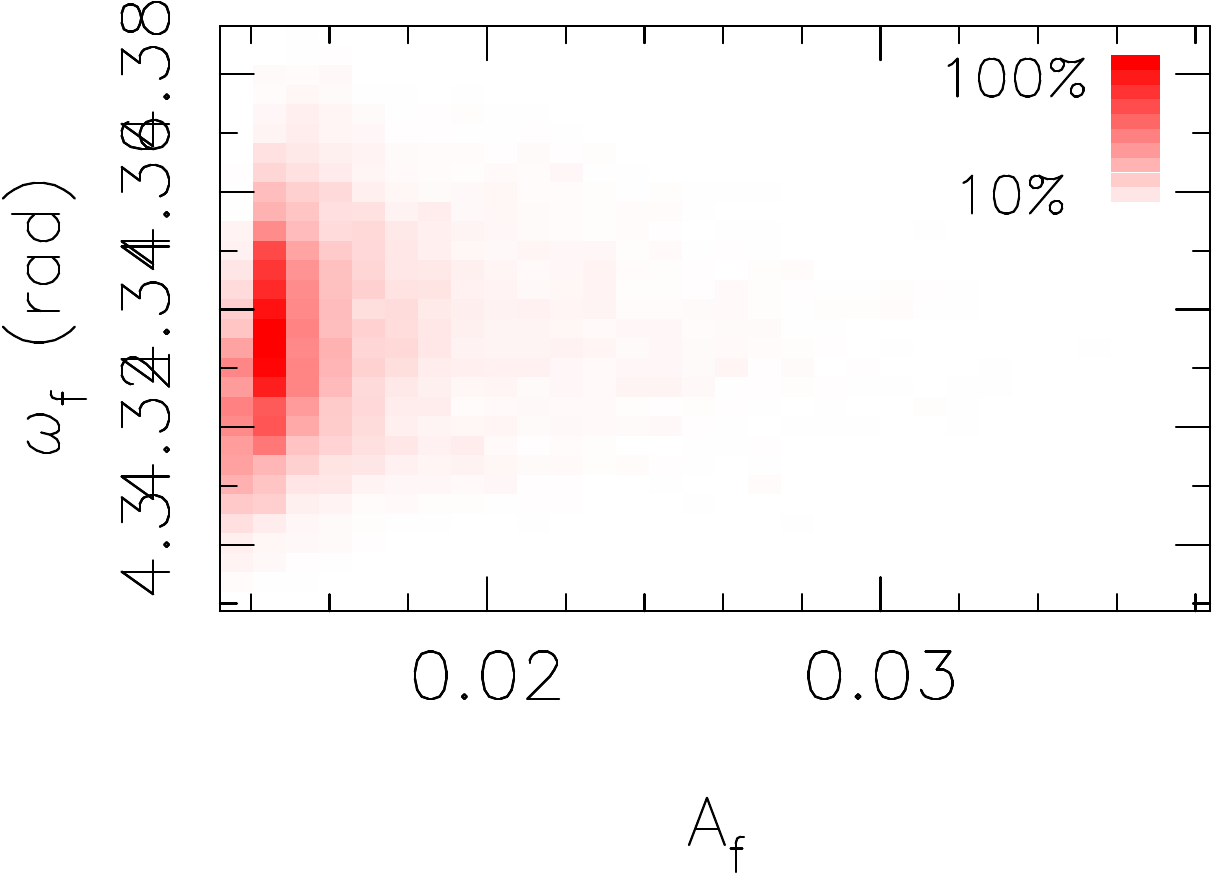}
\includegraphics[angle=270,width=0.135\textwidth,clip]{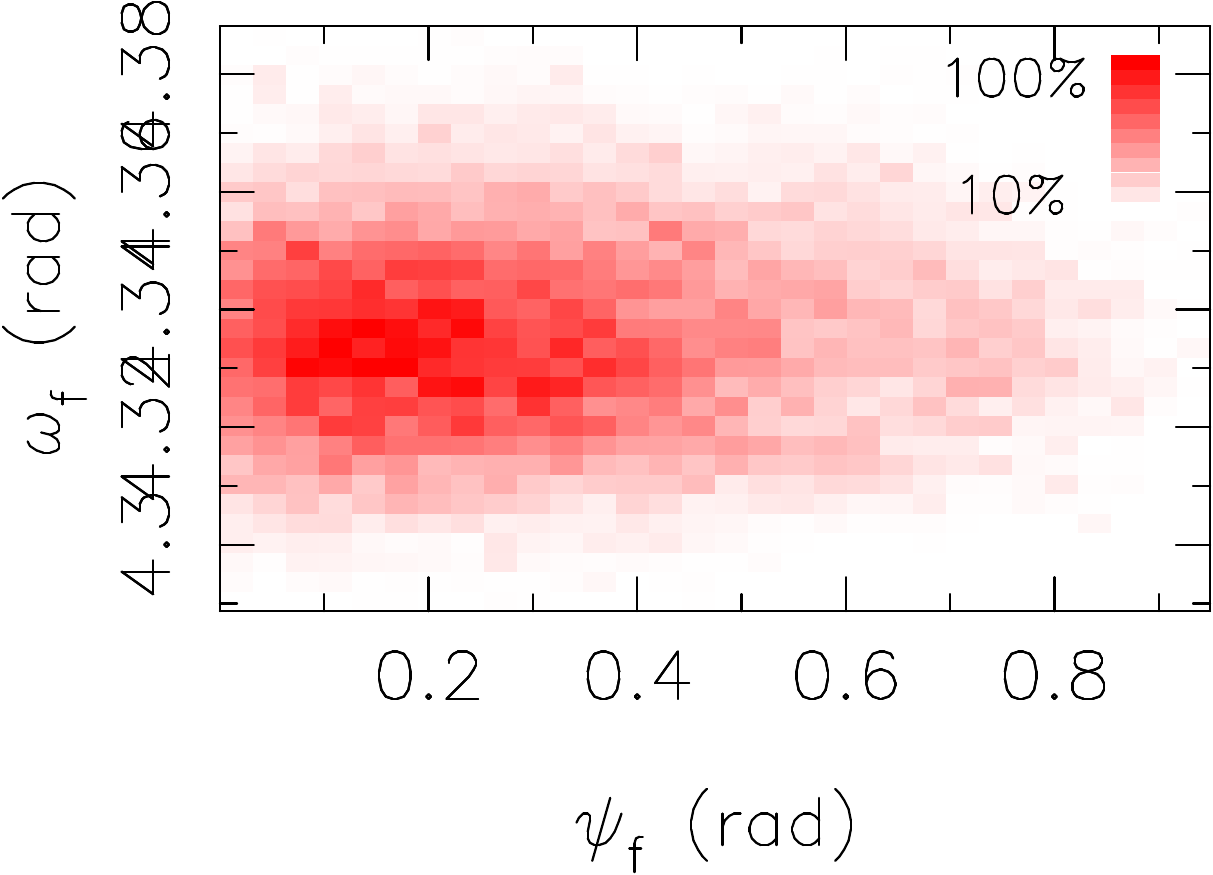}
\includegraphics[angle=270,width=0.135\textwidth,clip]{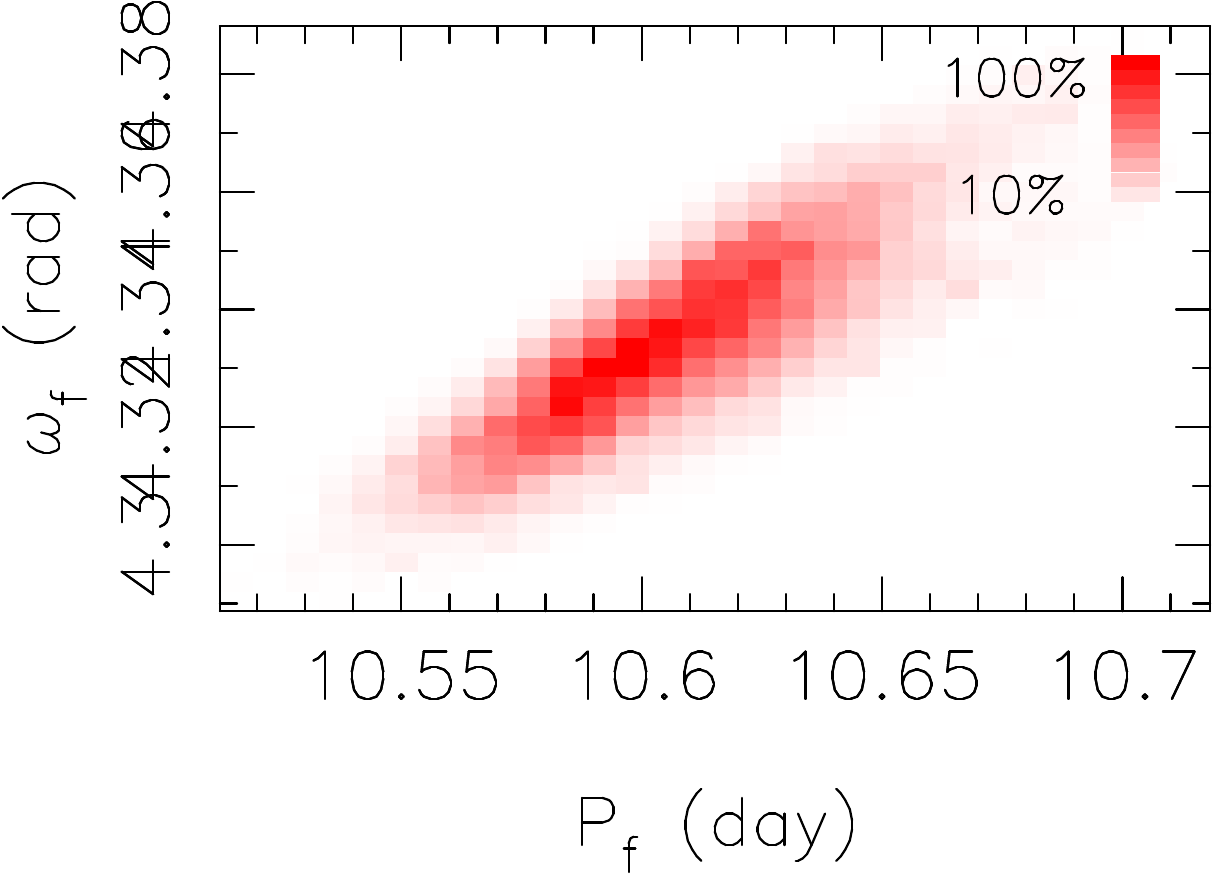}

\includegraphics[angle=270,width=0.135\textwidth,clip]{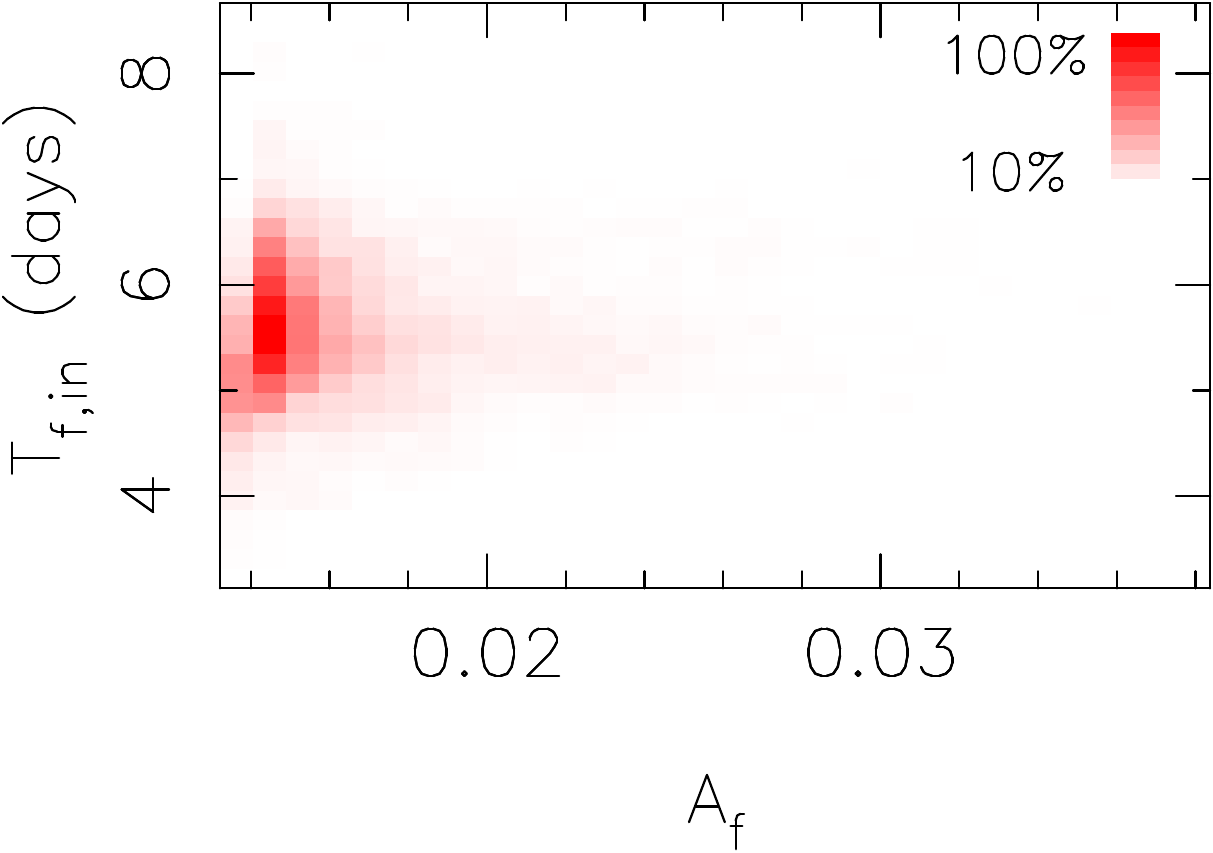}
\includegraphics[angle=270,width=0.135\textwidth,clip]{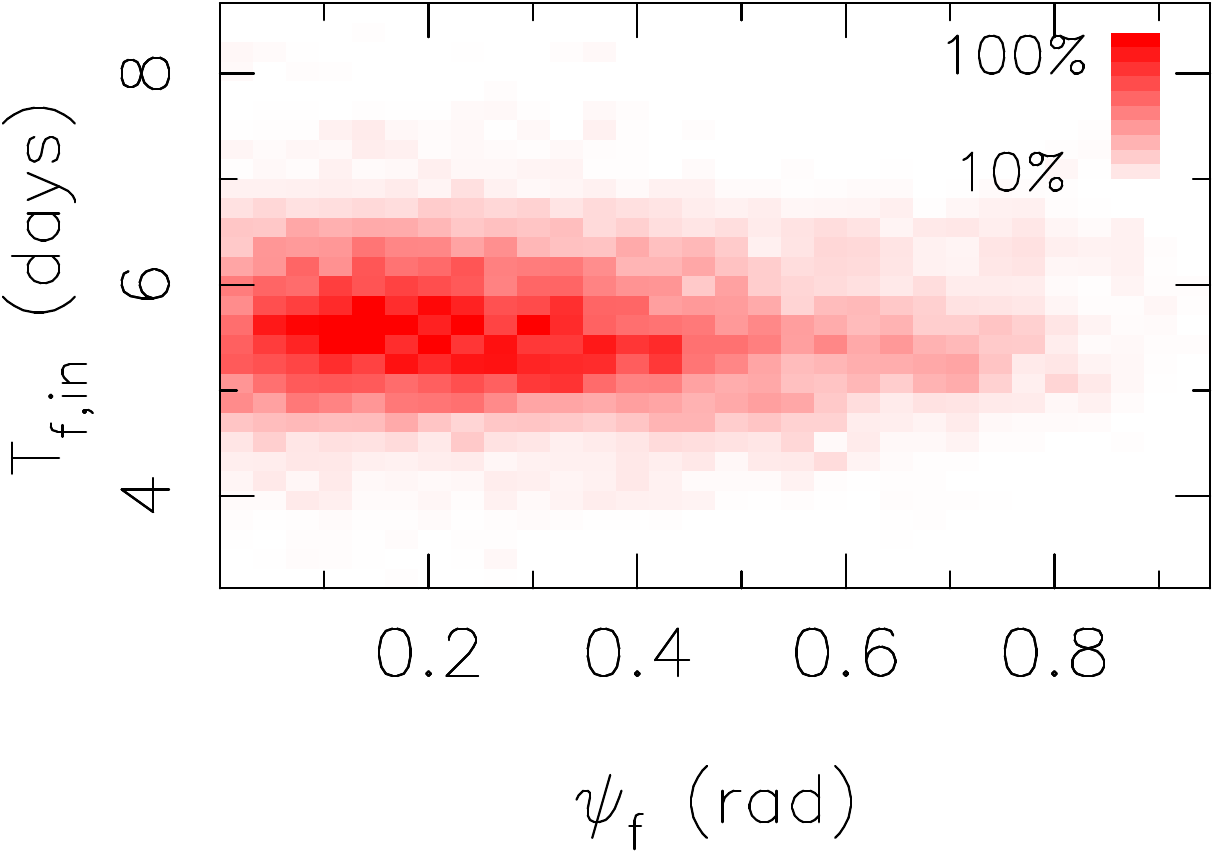}
\includegraphics[angle=270,width=0.135\textwidth,clip]{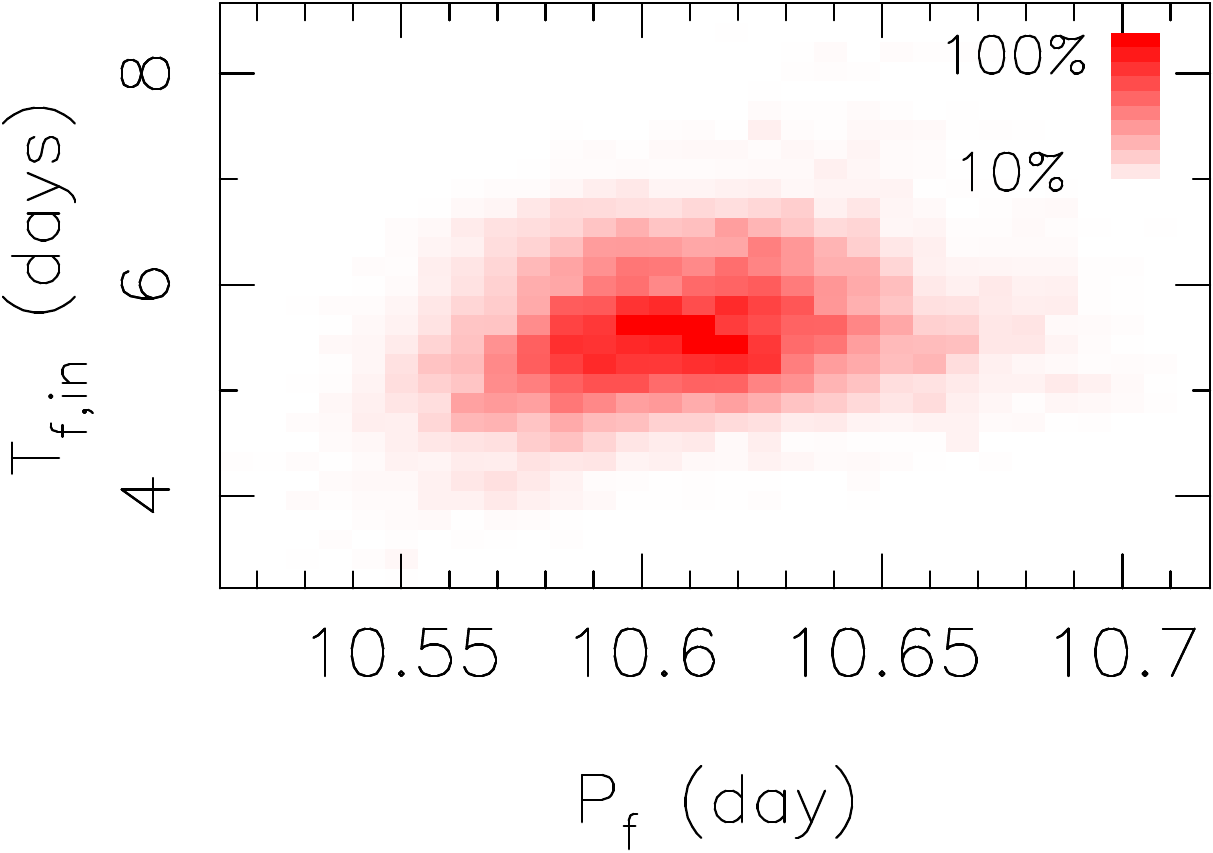}
\includegraphics[angle=270,width=0.135\textwidth,clip]{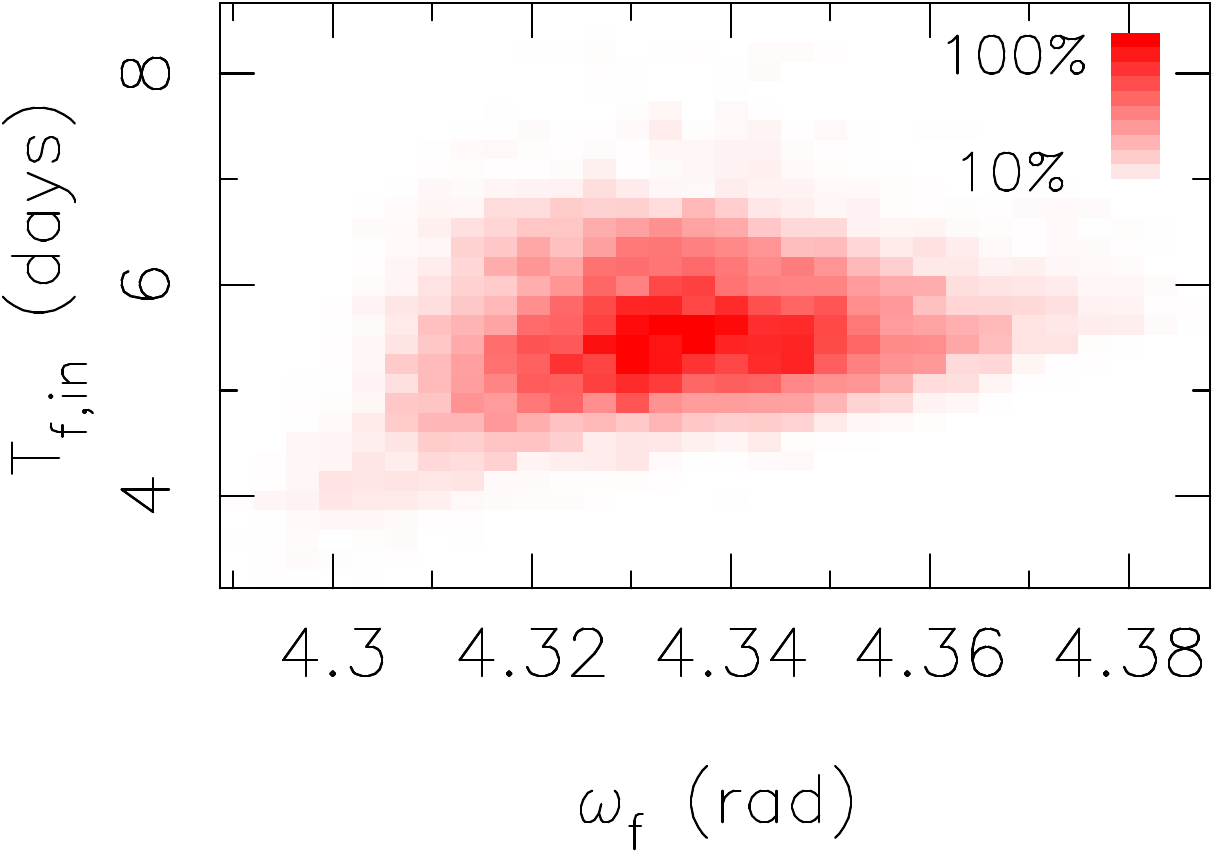}

\includegraphics[angle=270,width=0.135\textwidth,clip]{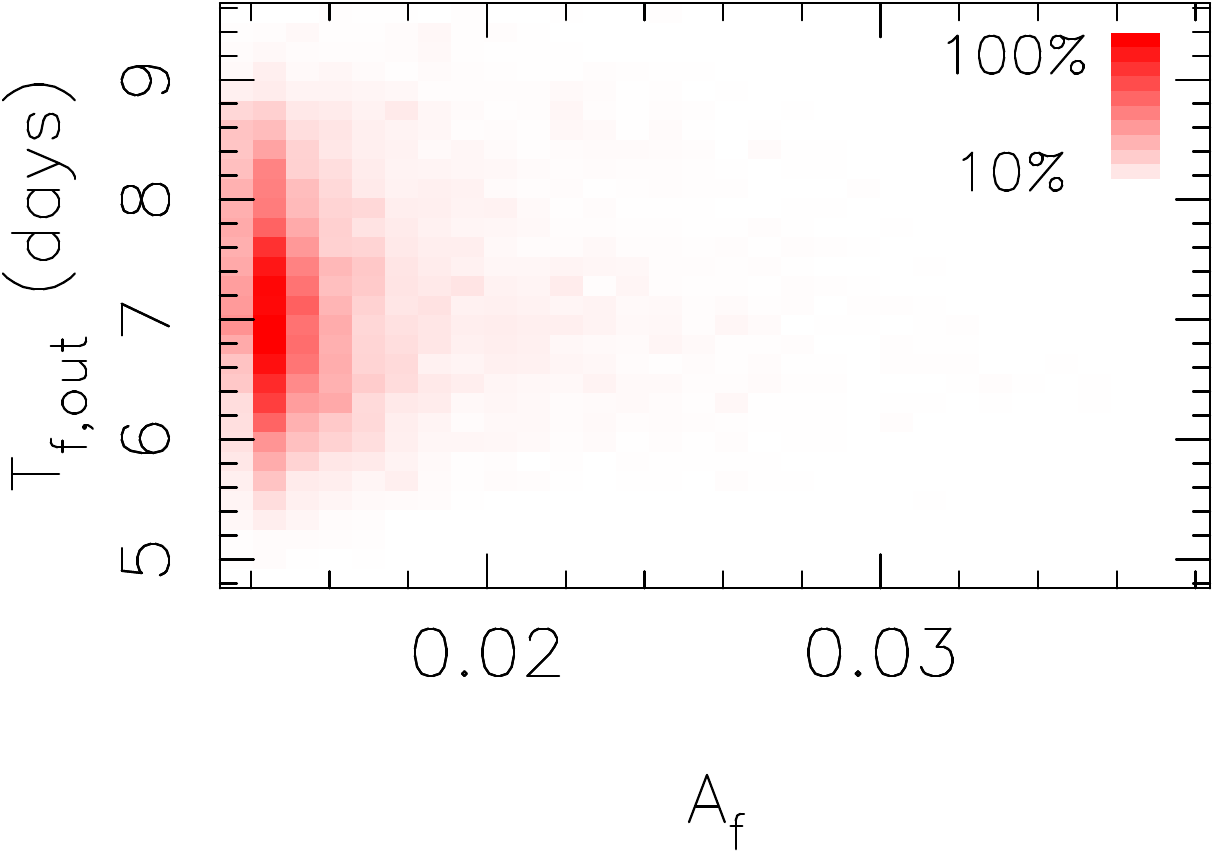}
\includegraphics[angle=270,width=0.135\textwidth,clip]{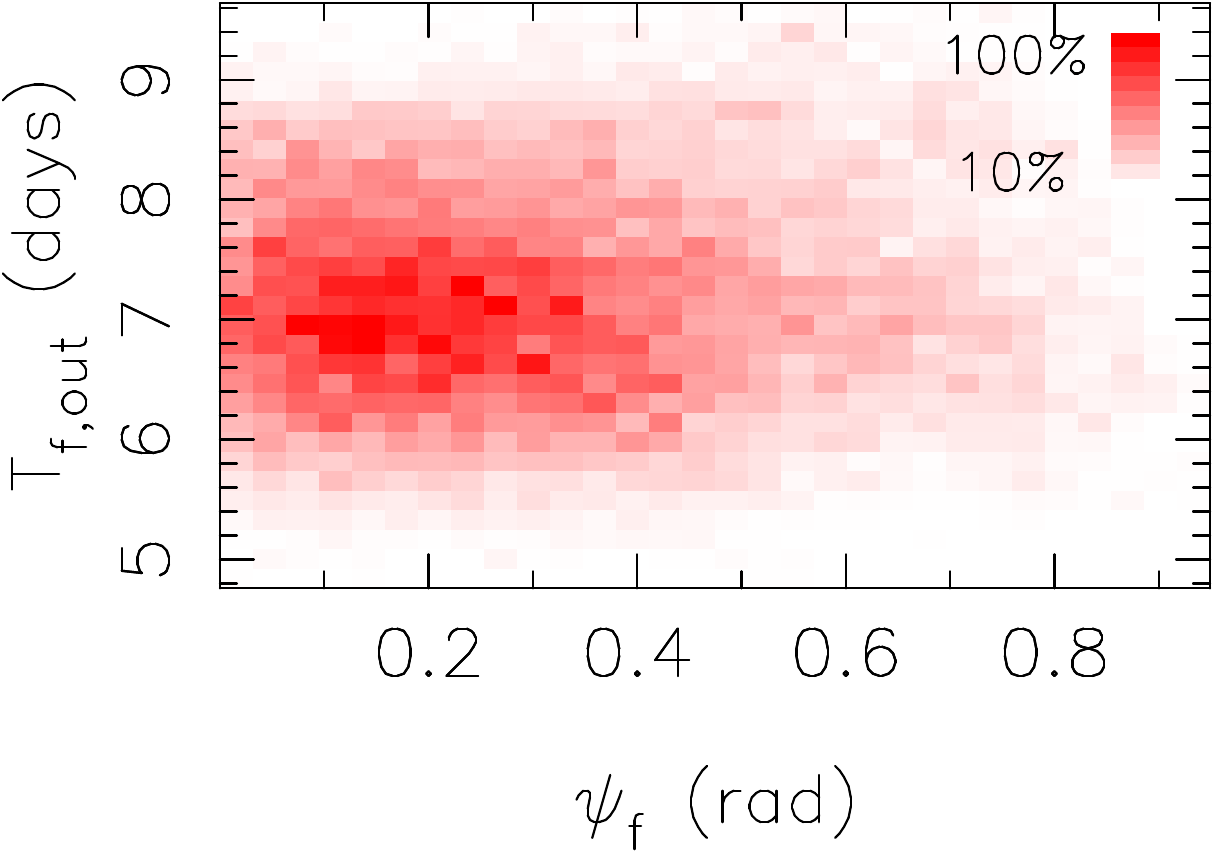}
\includegraphics[angle=270,width=0.135\textwidth,clip]{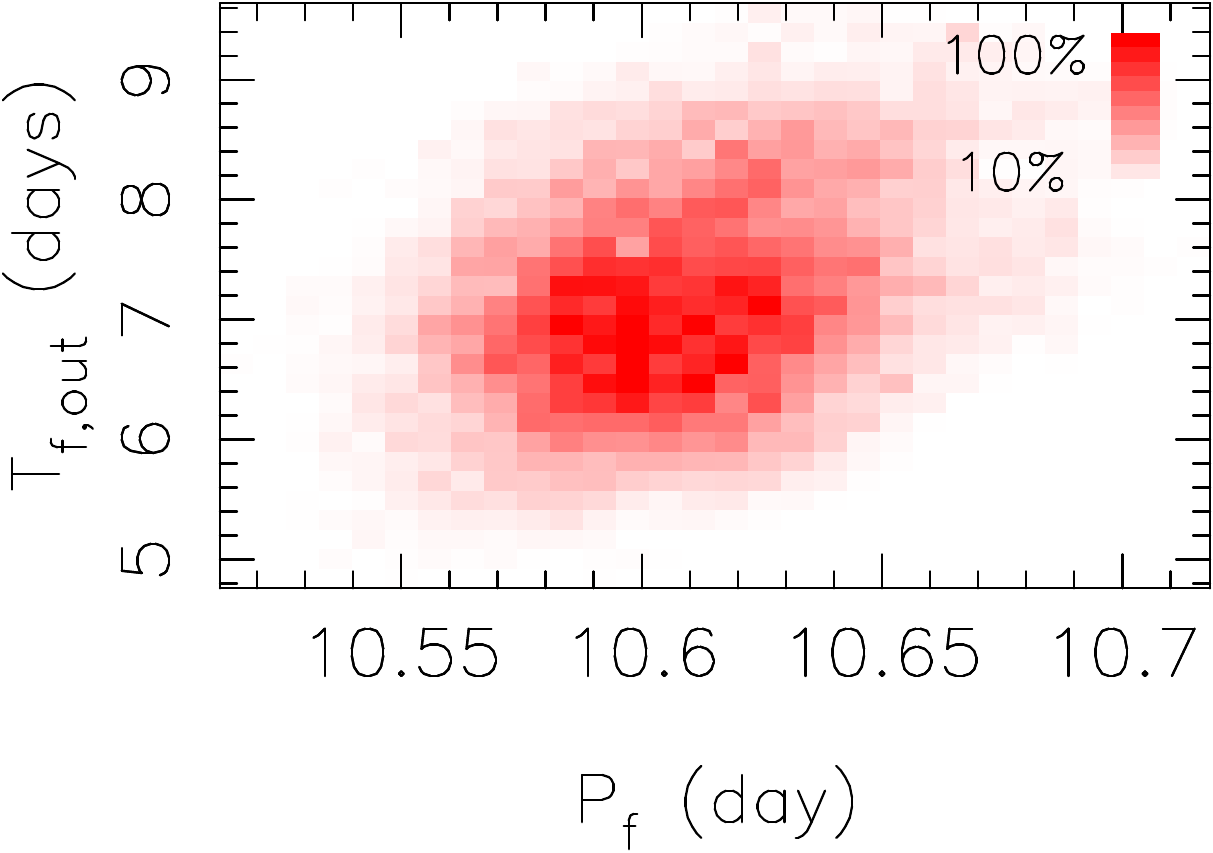}
\includegraphics[angle=270,width=0.135\textwidth,clip]{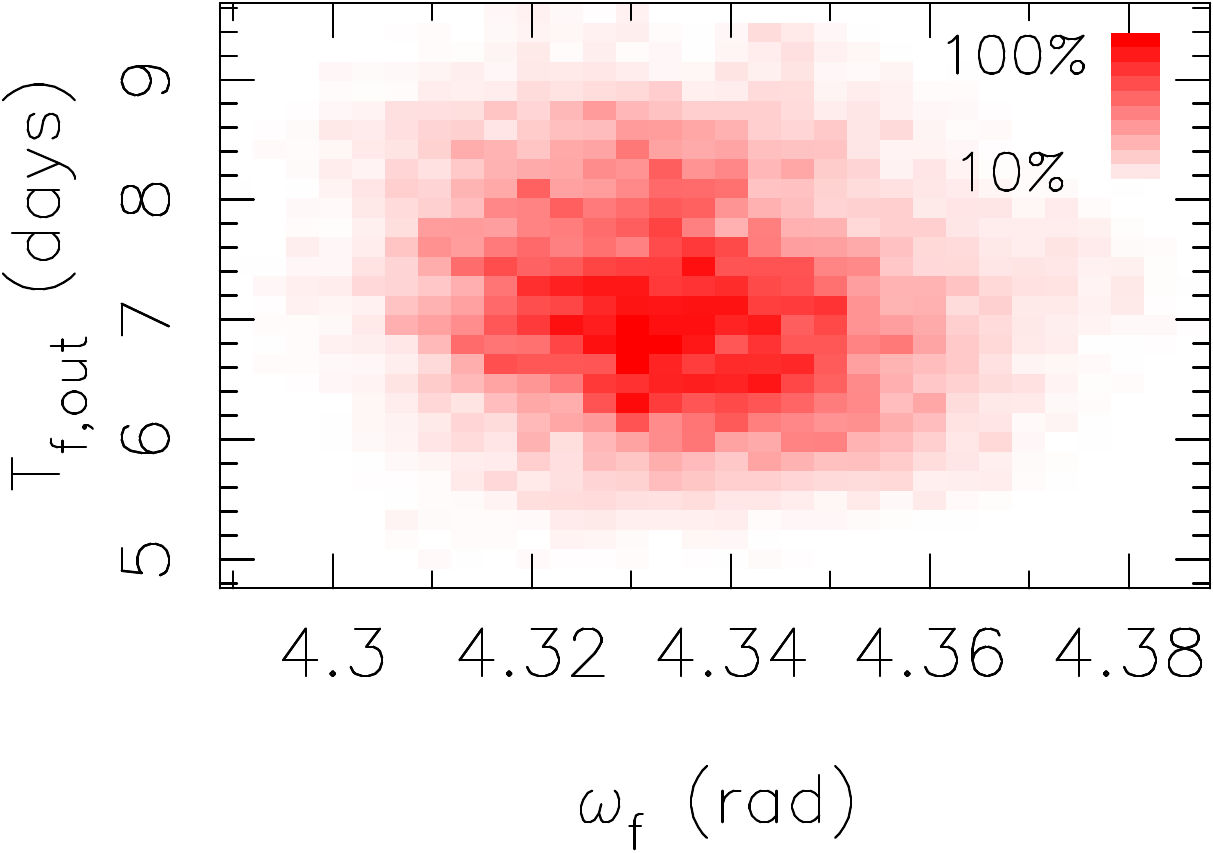}
\includegraphics[angle=270,width=0.135\textwidth,clip]{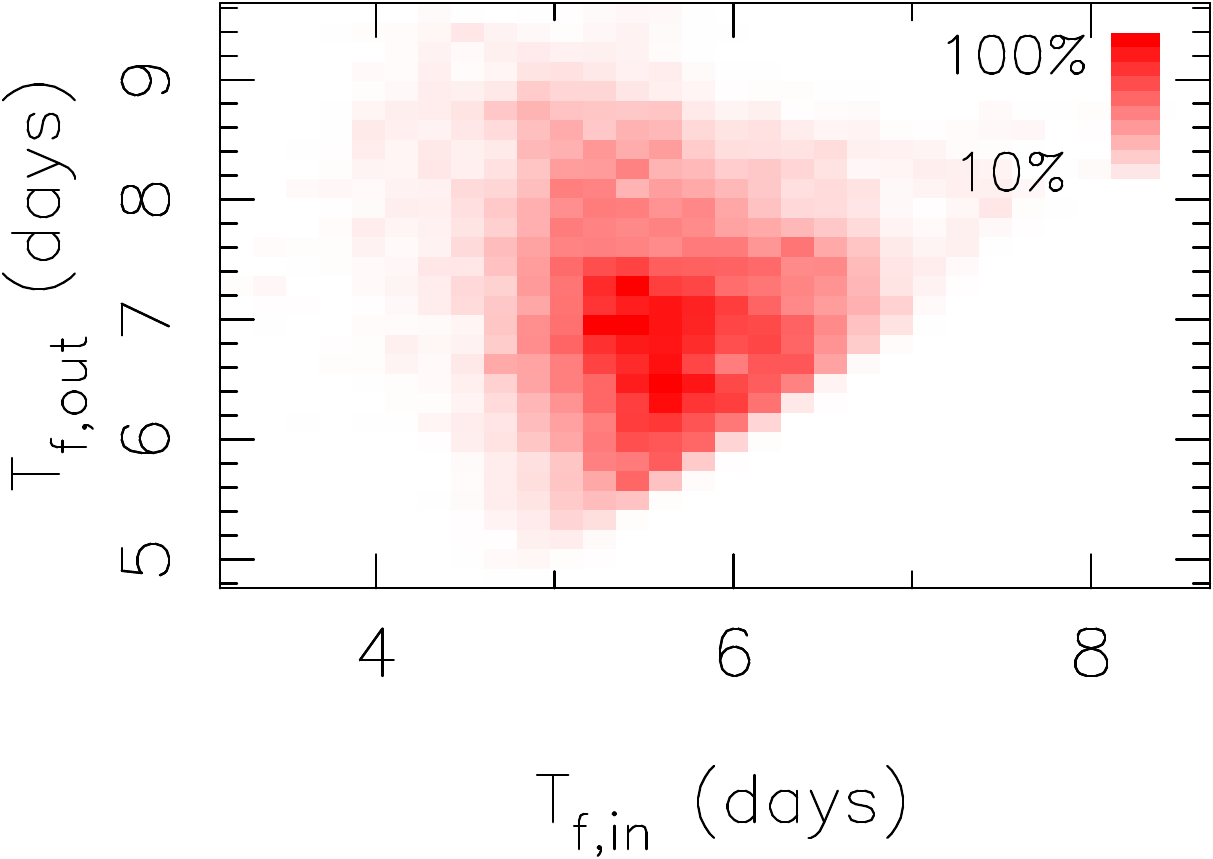}

\includegraphics[angle=270,width=0.135\textwidth,clip]{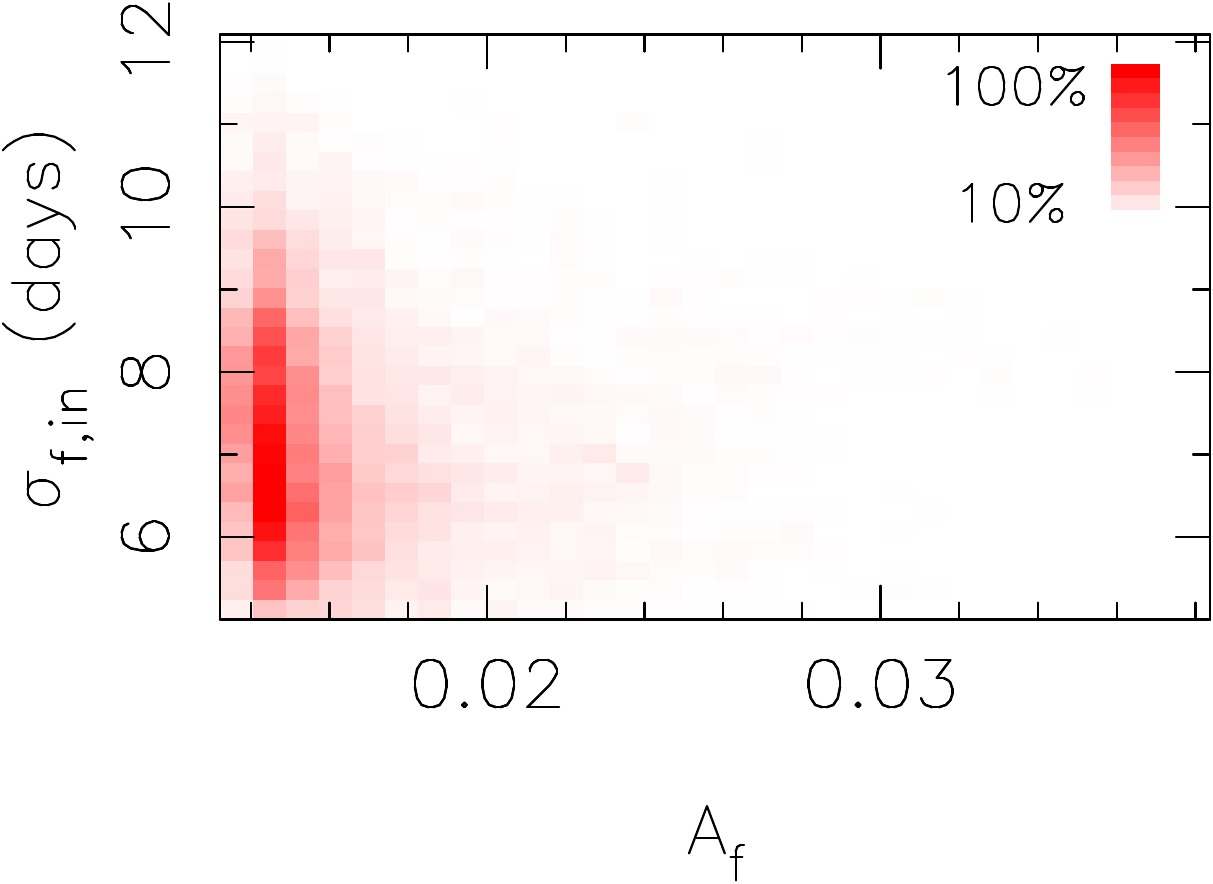}
\includegraphics[angle=270,width=0.135\textwidth,clip]{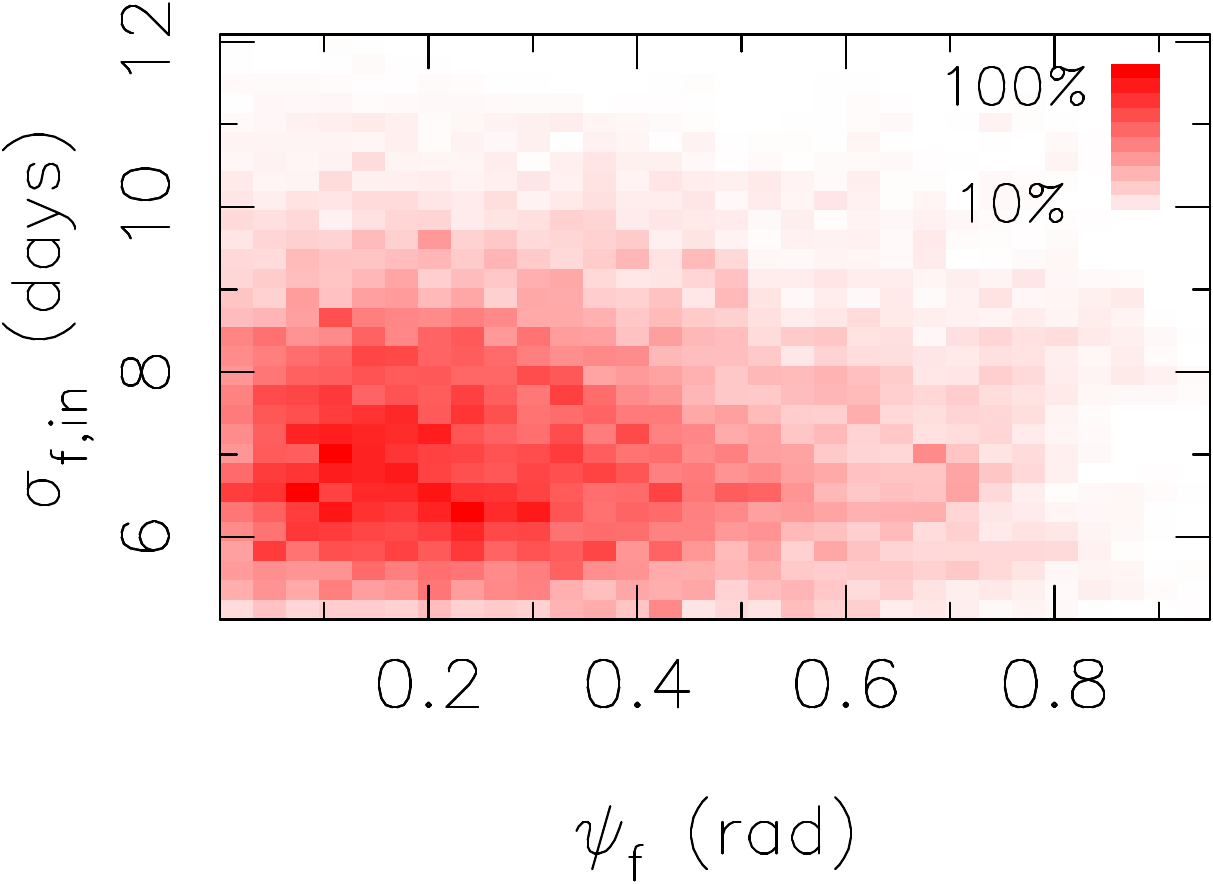}
\includegraphics[angle=270,width=0.135\textwidth,clip]{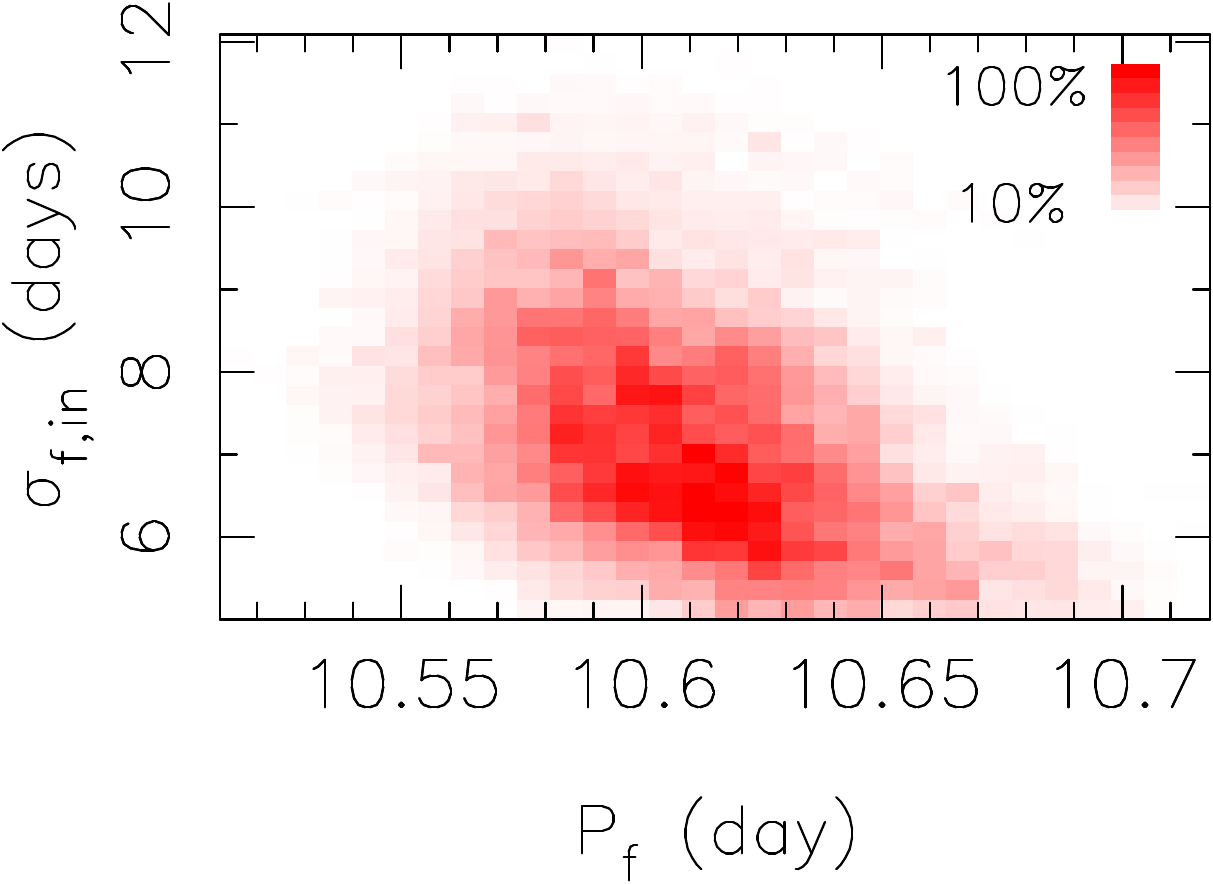}
\includegraphics[angle=270,width=0.135\textwidth,clip]{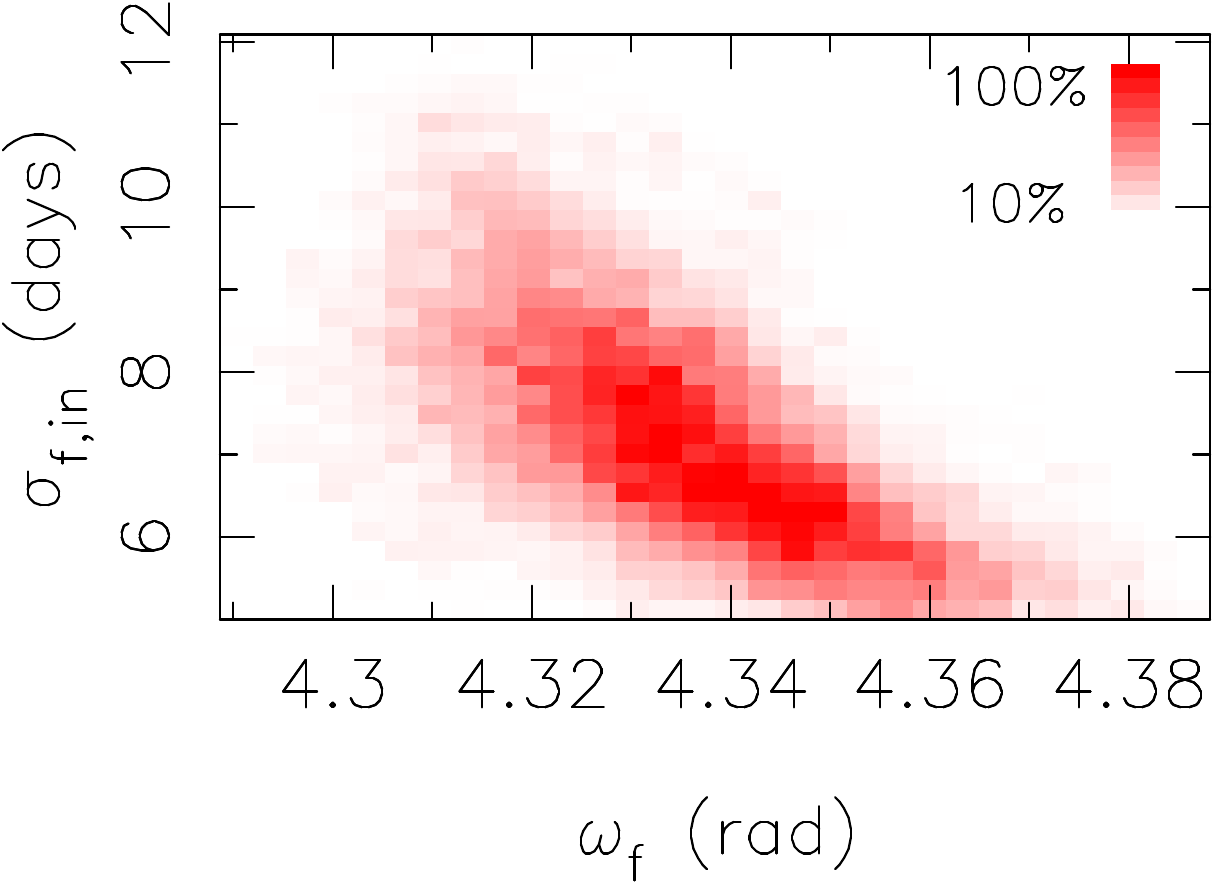}
\includegraphics[angle=270,width=0.135\textwidth,clip]{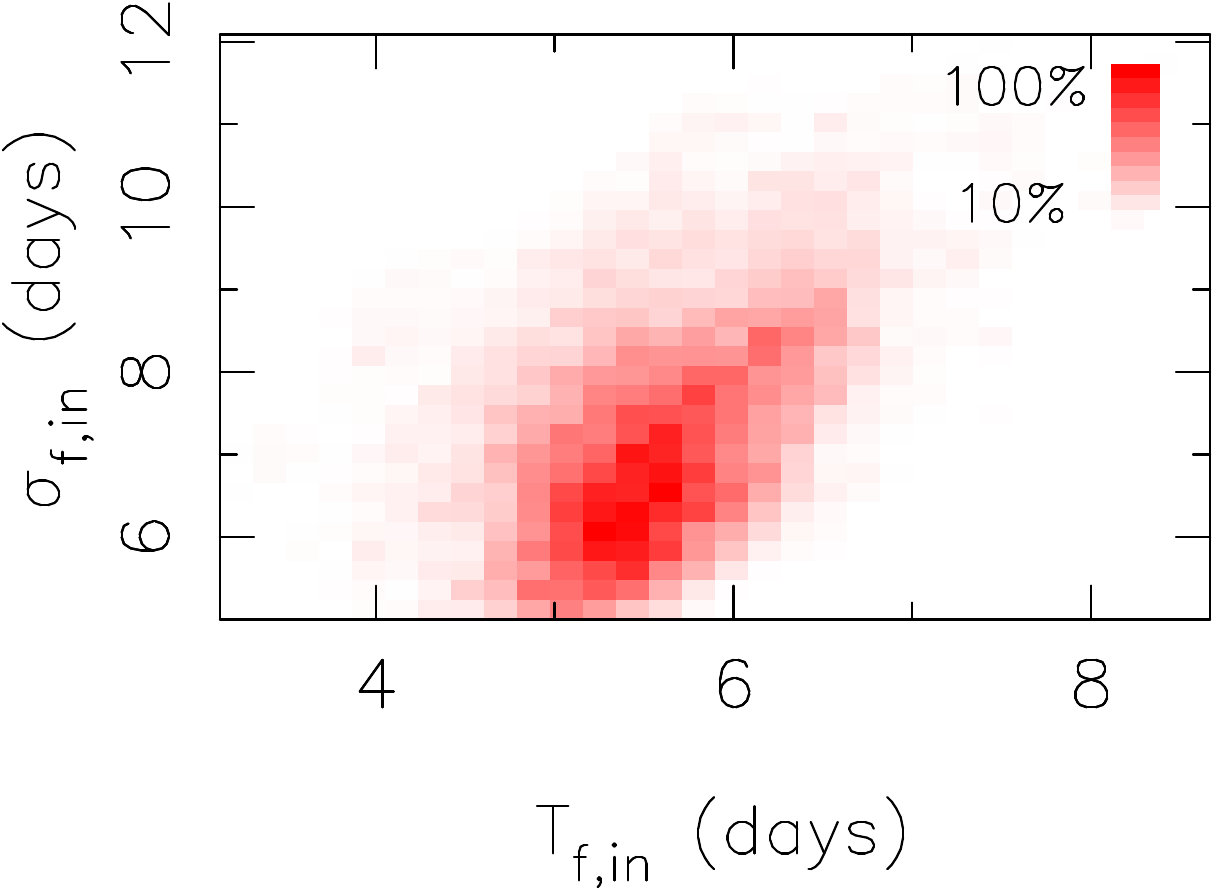}
\includegraphics[angle=270,width=0.135\textwidth,clip]{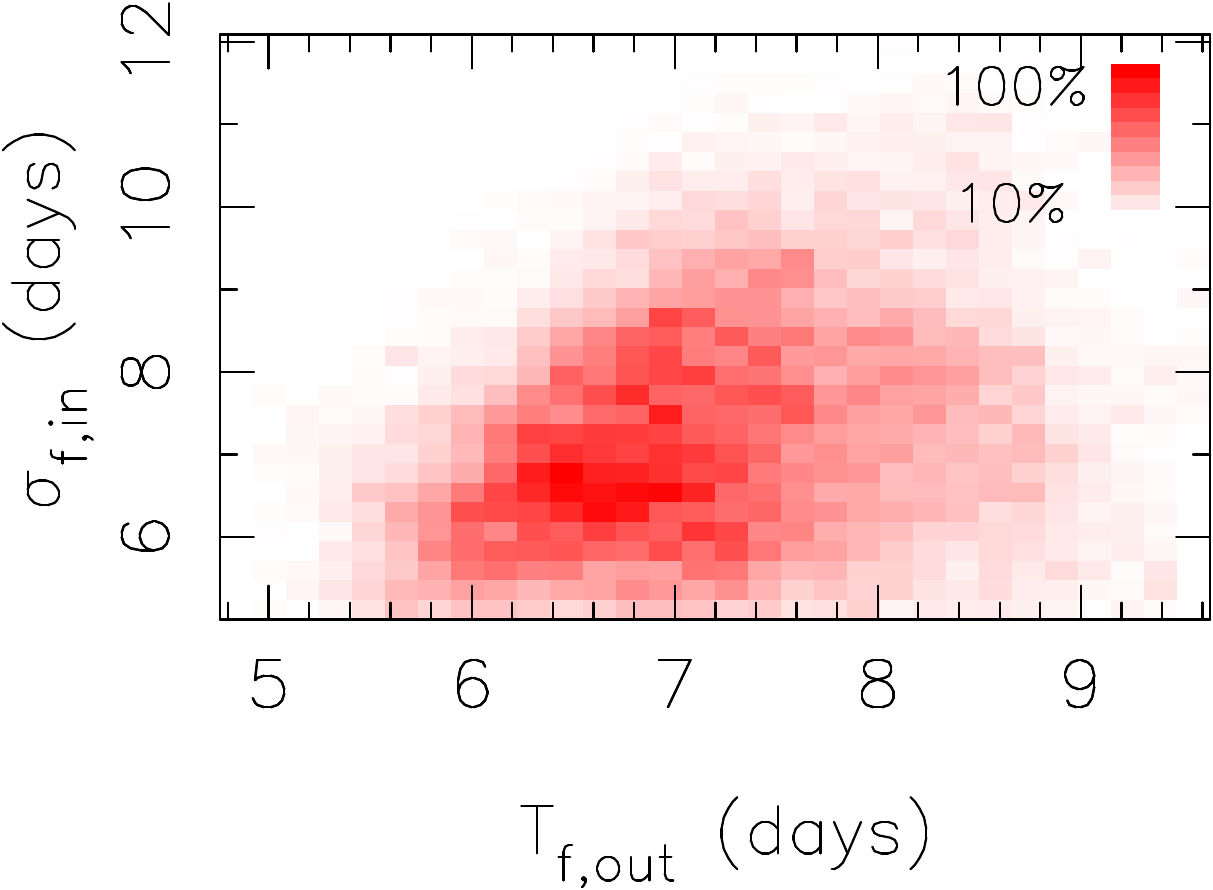}

\includegraphics[angle=270,width=0.135\textwidth,clip]{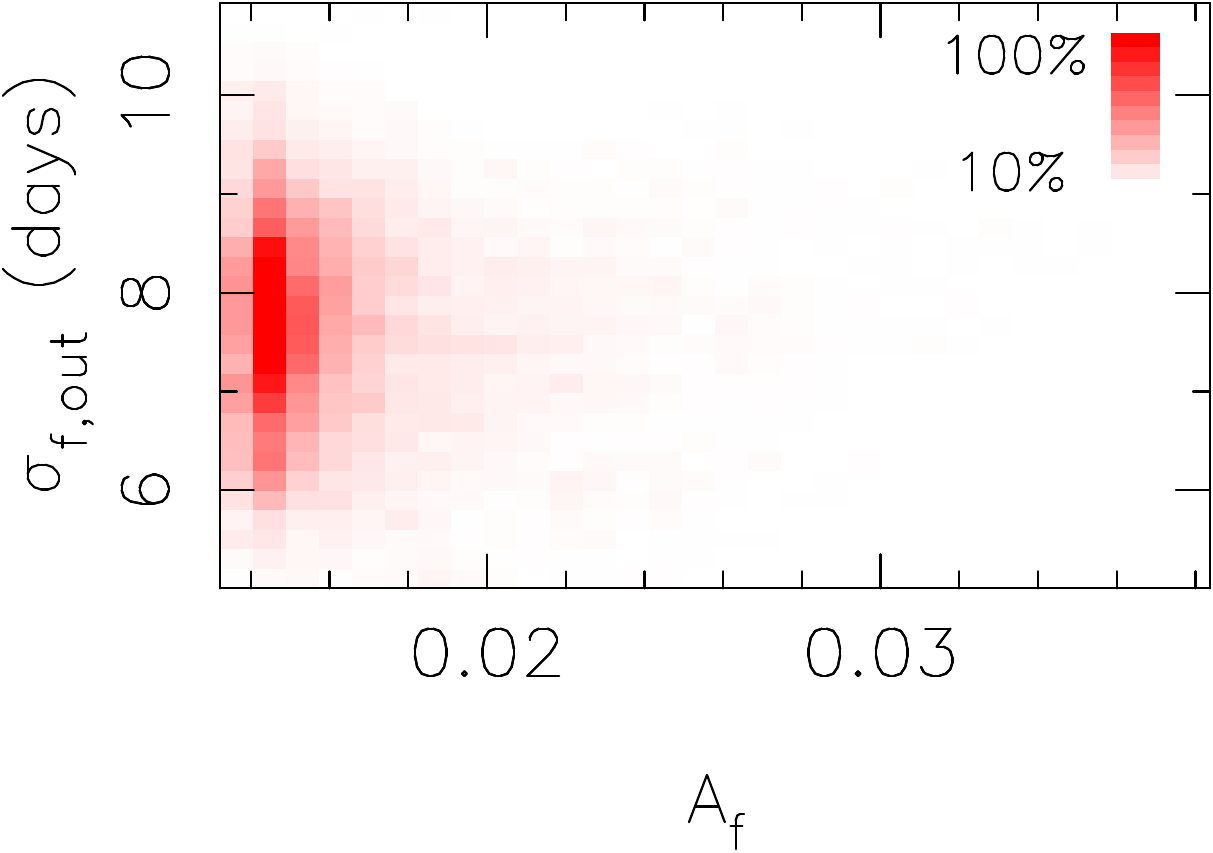}
\includegraphics[angle=270,width=0.135\textwidth,clip]{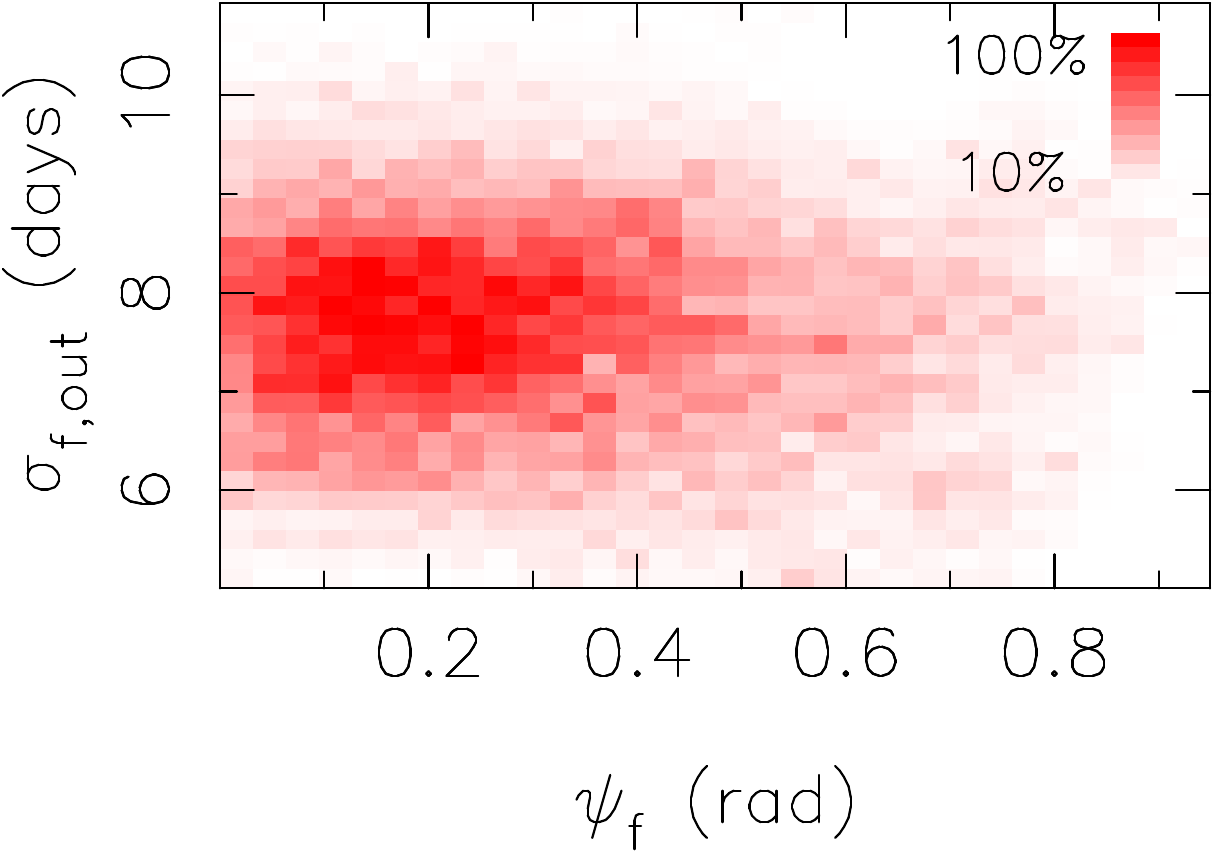}
\includegraphics[angle=270,width=0.135\textwidth,clip]{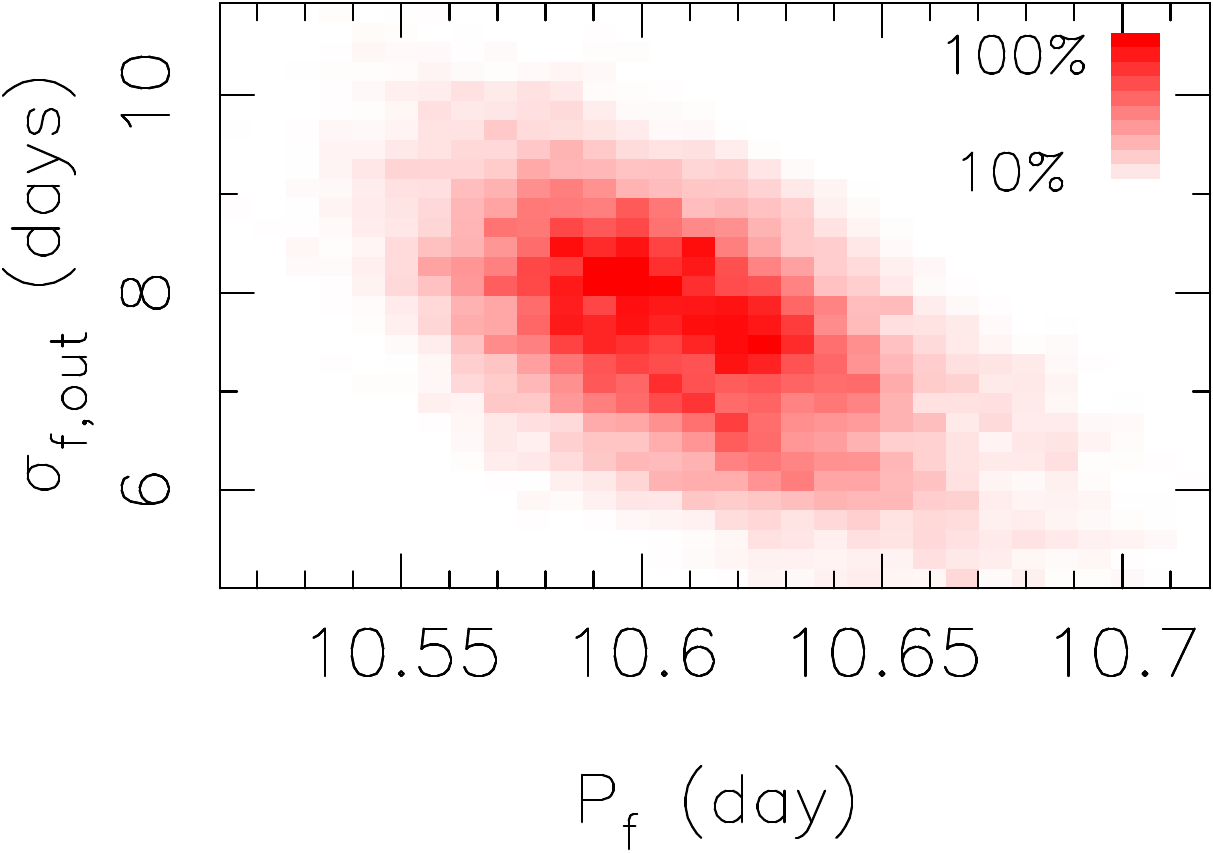}
\includegraphics[angle=270,width=0.135\textwidth,clip]{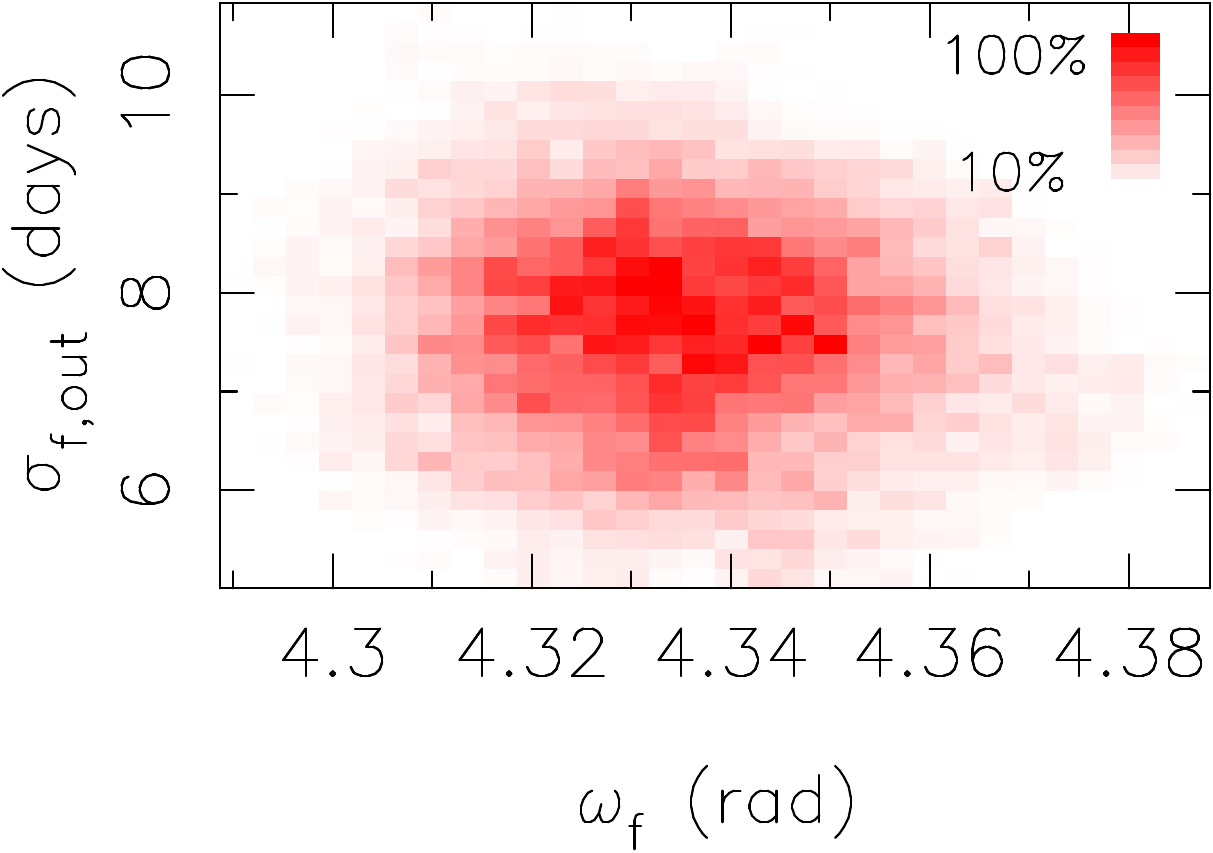}
\includegraphics[angle=270,width=0.135\textwidth,clip]{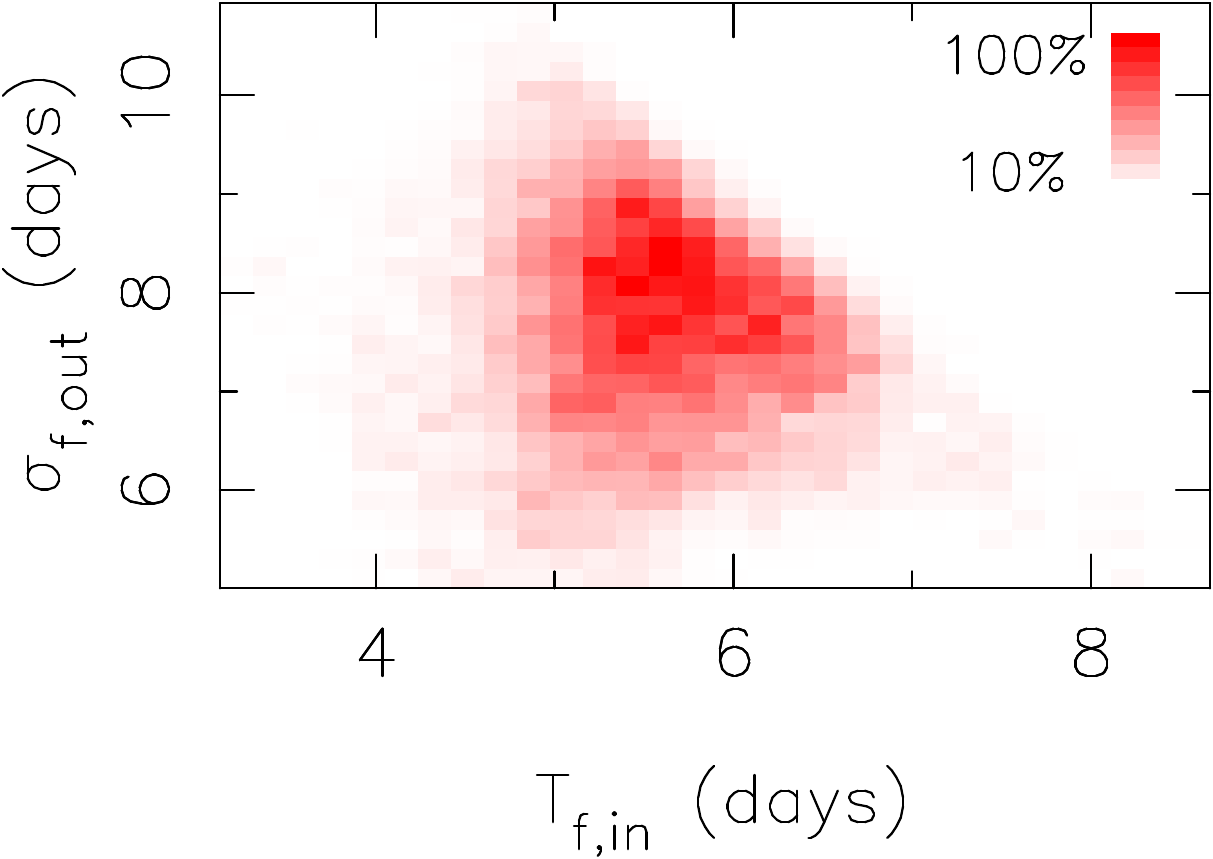}
\includegraphics[angle=270,width=0.135\textwidth,clip]{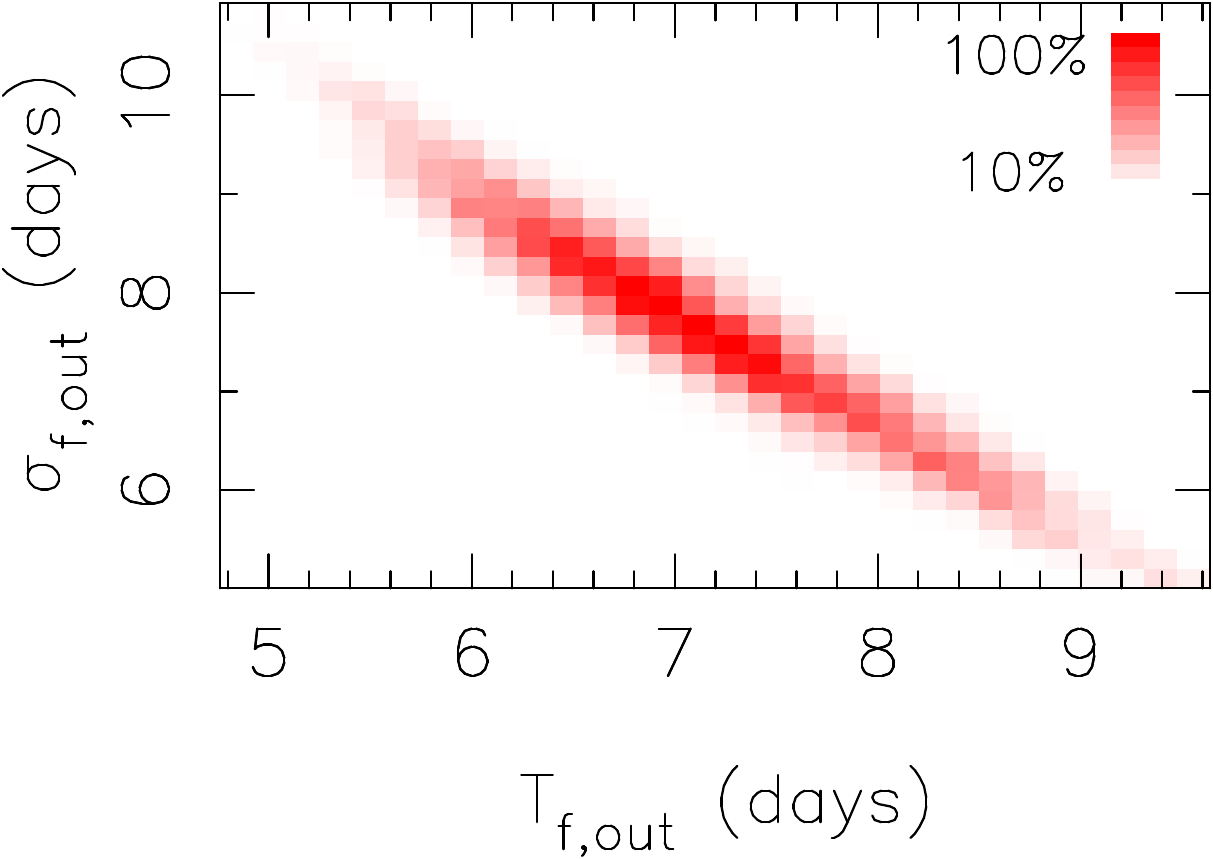}
\includegraphics[angle=270,width=0.135\textwidth,clip]{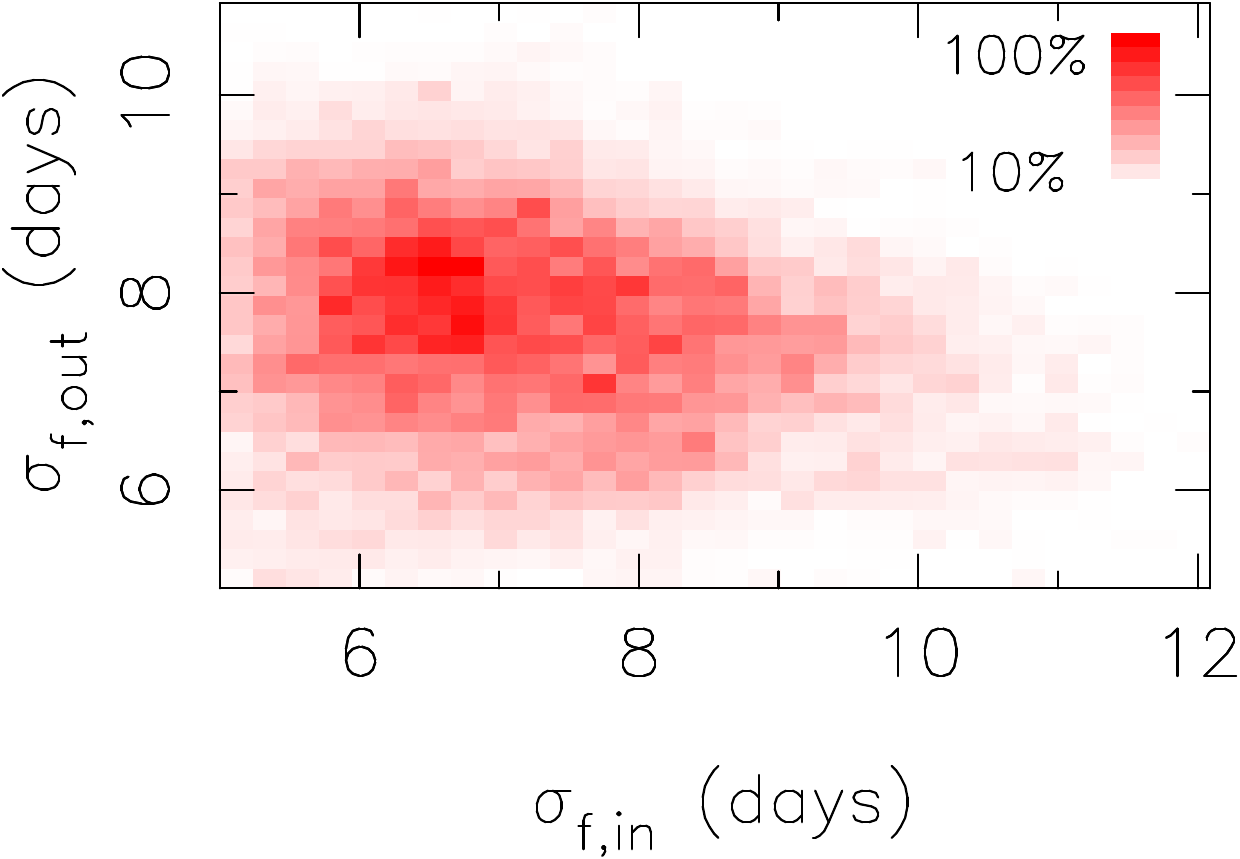}
\caption{Selected two-dimensional projections of sampled probability densities of parameters of spot 1 in data set S1034. The maxima are scaled to unity for each projection.}\label{fig:S1034_solution_contours}
\end{figure*}

\begin{figure*}
\includegraphics[angle=270,width=0.135\textwidth,clip]{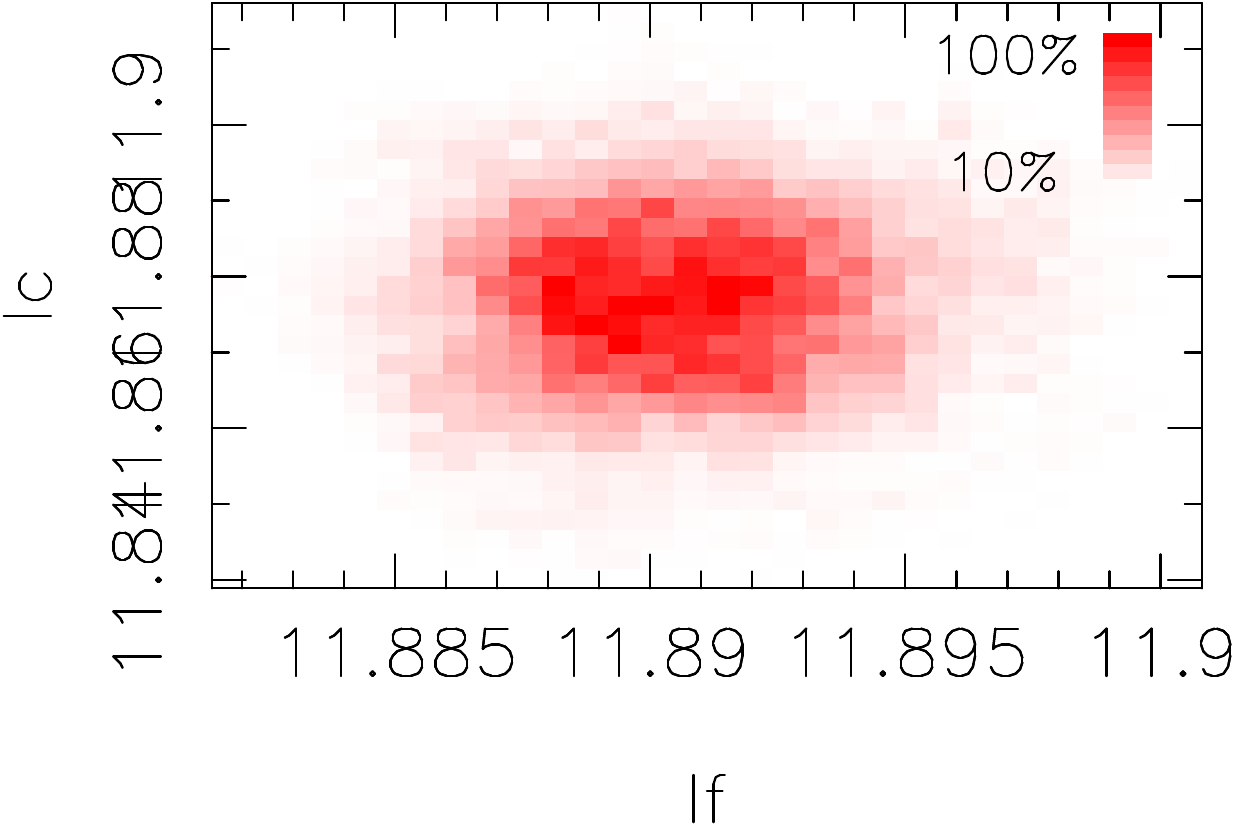}
\includegraphics[angle=270,width=0.135\textwidth,clip]{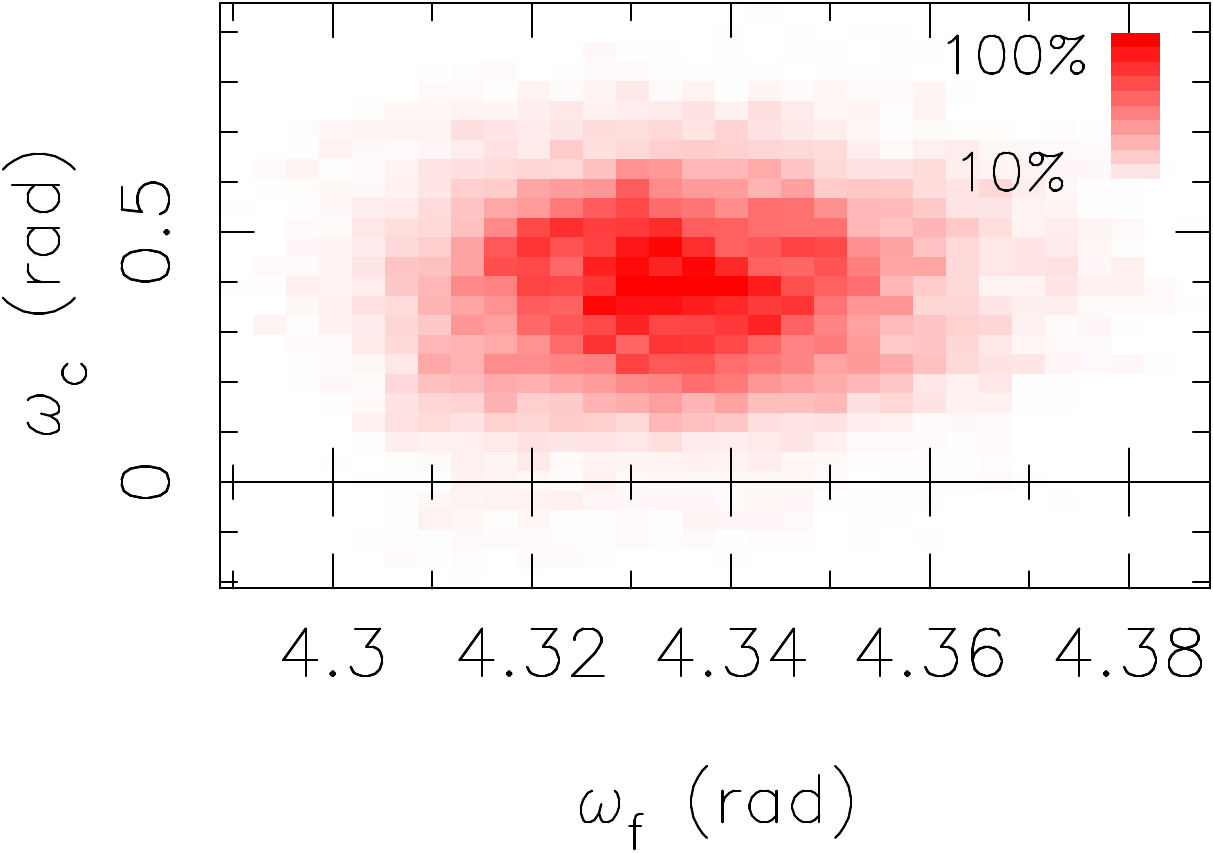}
\includegraphics[angle=270,width=0.135\textwidth,clip]{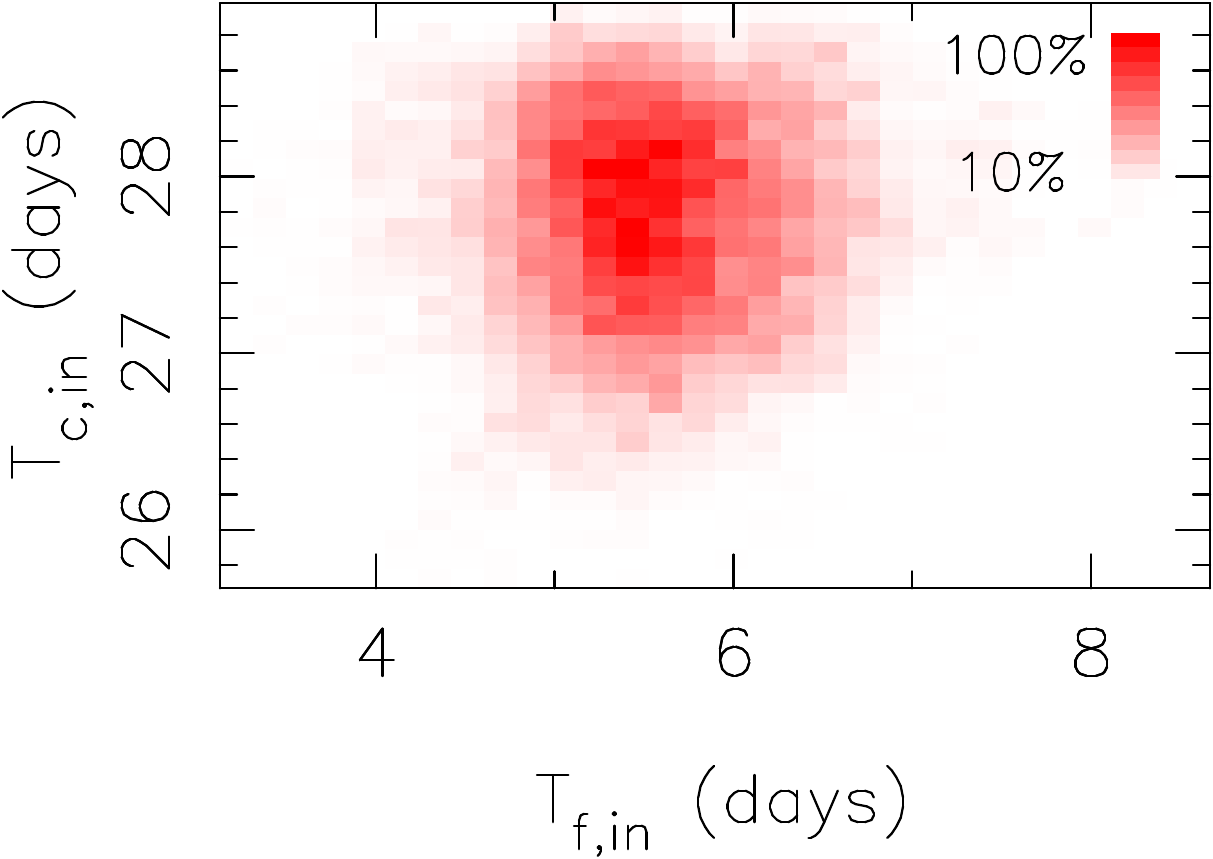}
\includegraphics[angle=270,width=0.135\textwidth,clip]{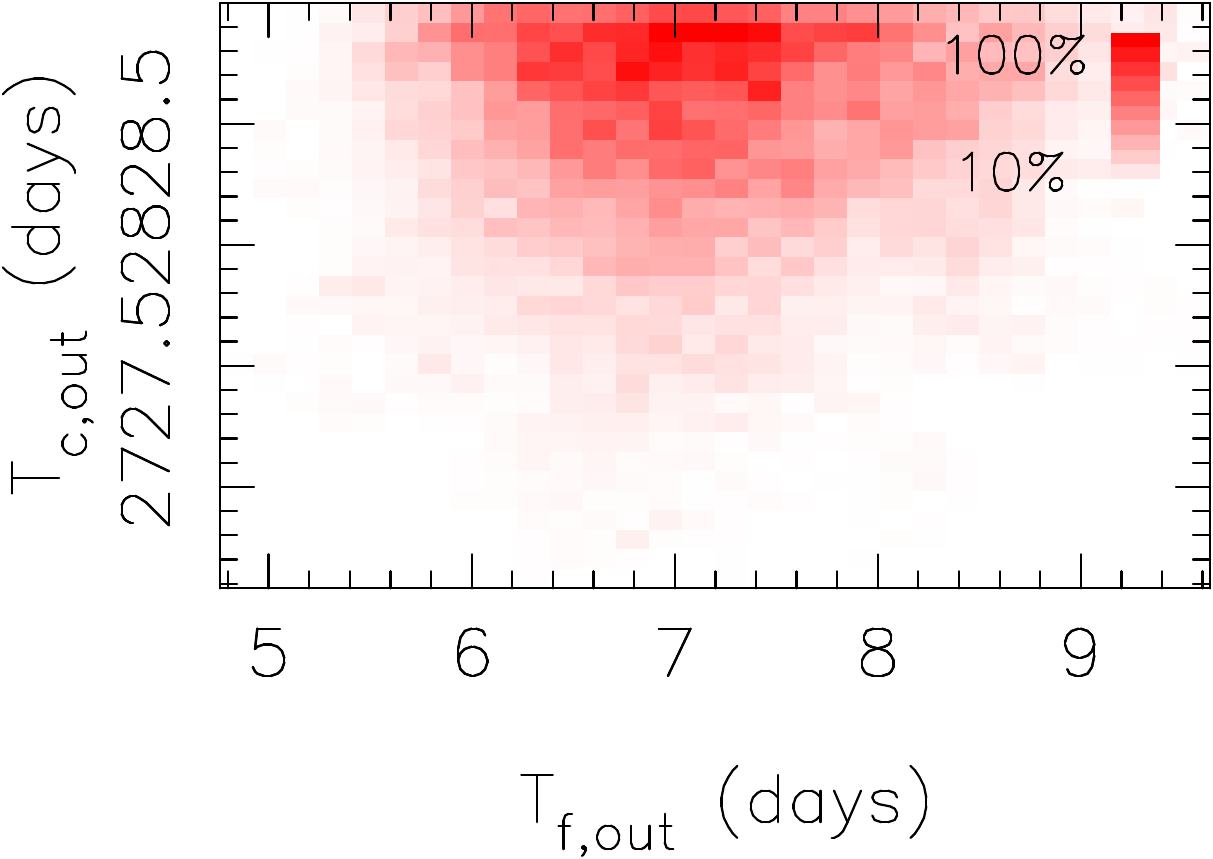}
\includegraphics[angle=270,width=0.135\textwidth,clip]{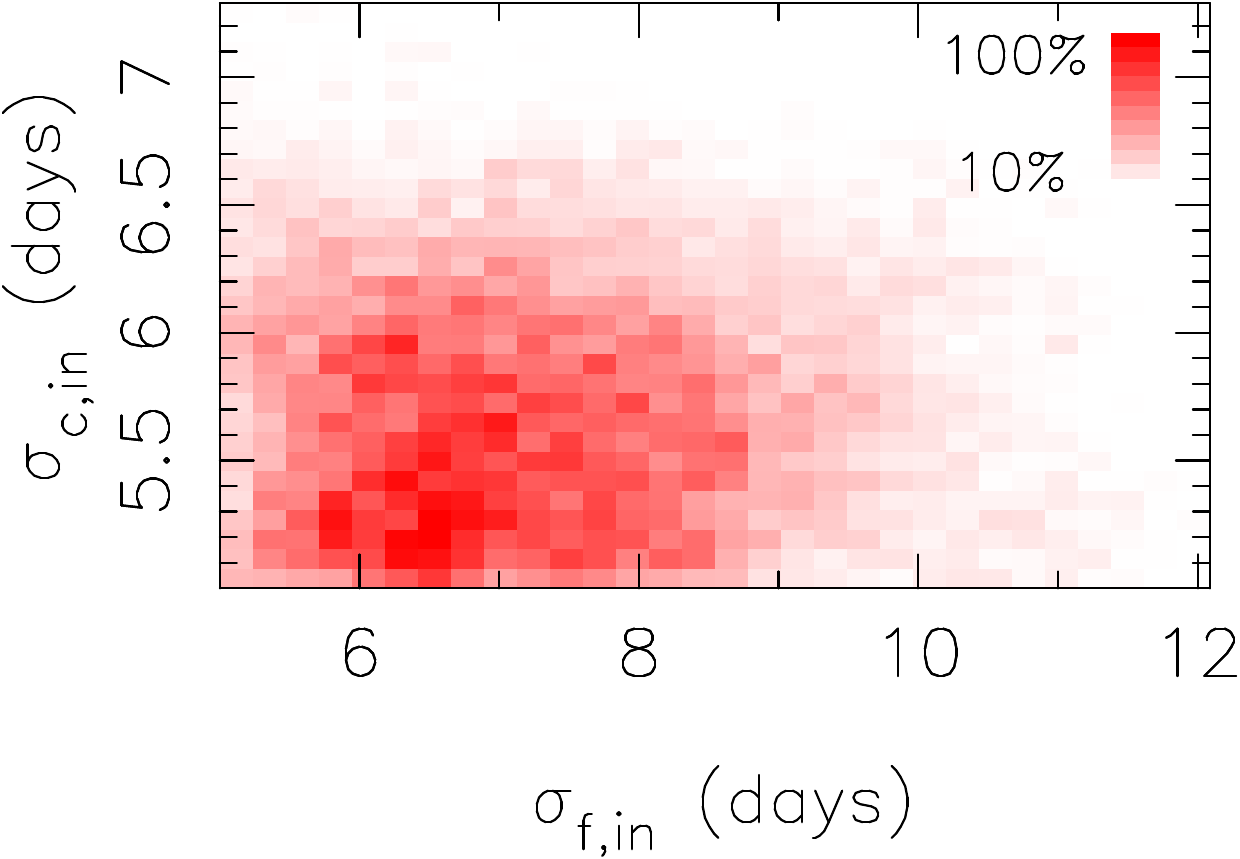}
\includegraphics[angle=270,width=0.135\textwidth,clip]{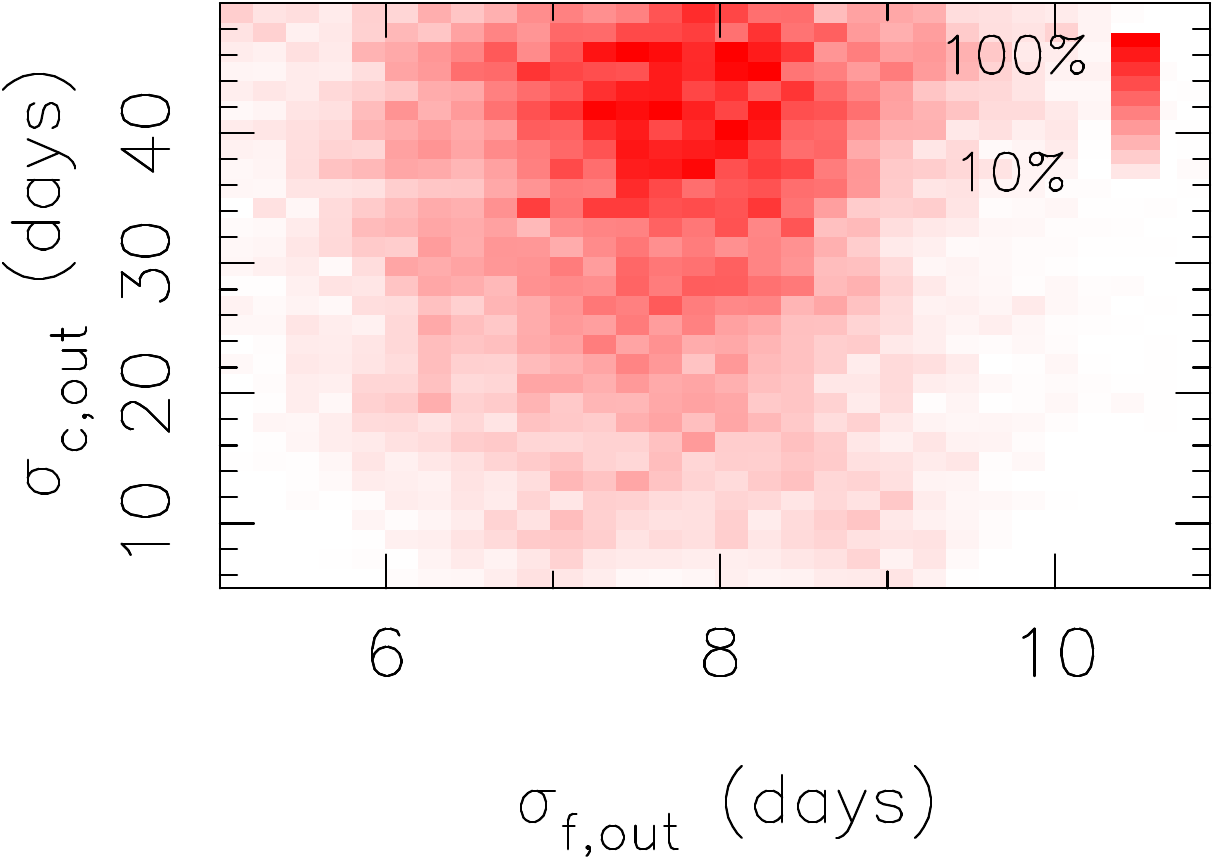}
\includegraphics[angle=270,width=0.135\textwidth,clip]{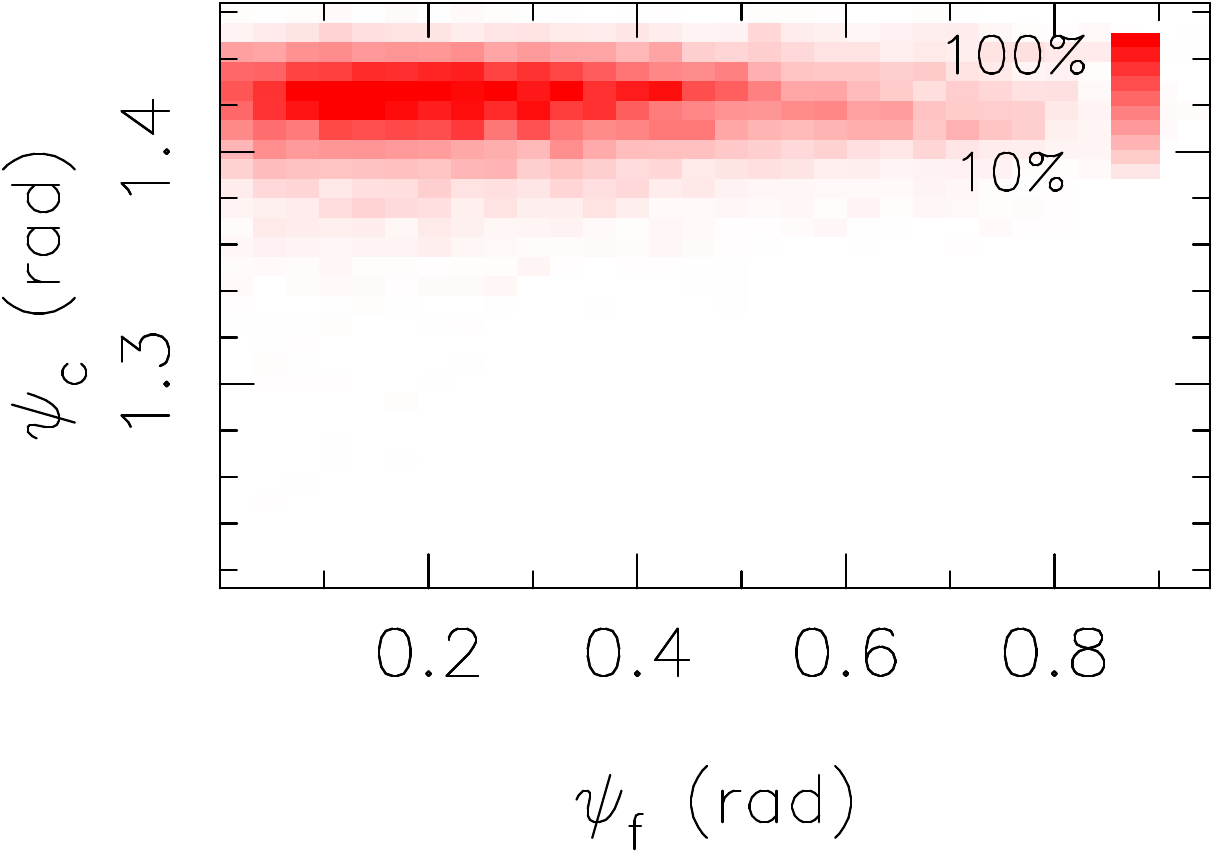}

\includegraphics[angle=270,width=0.135\textwidth,clip]{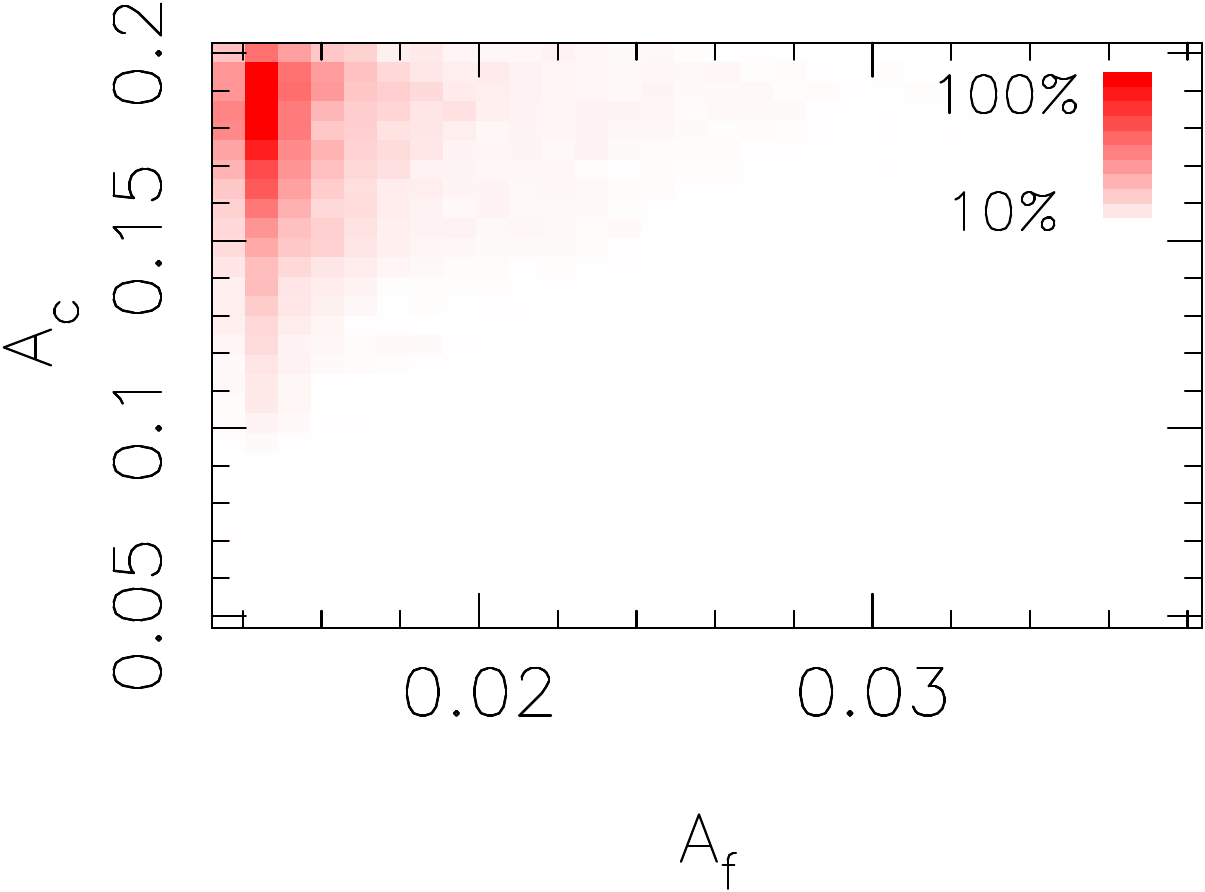}
\includegraphics[angle=270,width=0.135\textwidth,clip]{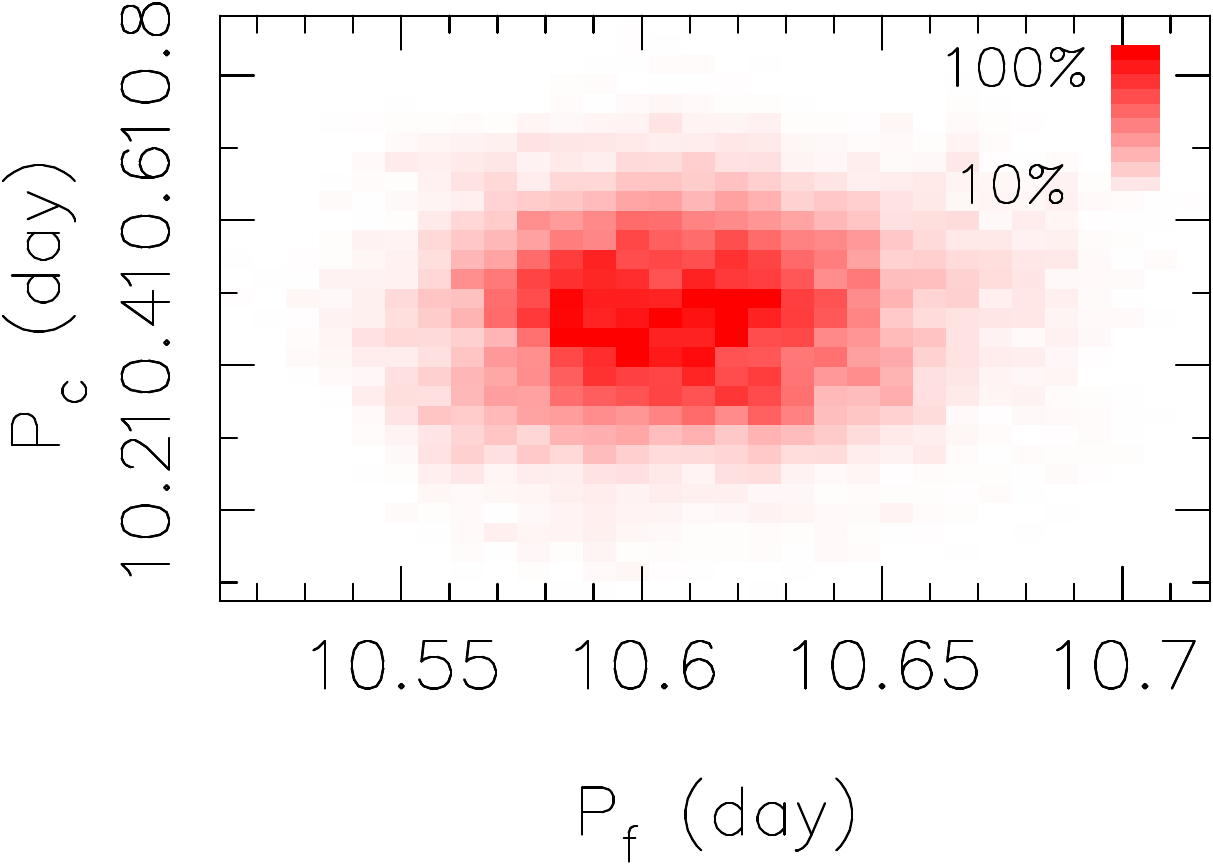}
\caption{As in Fig. \ref{fig:S1034_solution_contours} but between parameters of spots 1 and 2.}\label{fig:S1034_solution_contours2}
\end{figure*}

\begin{figure*}
\includegraphics[angle=270,width=0.135\textwidth,clip]{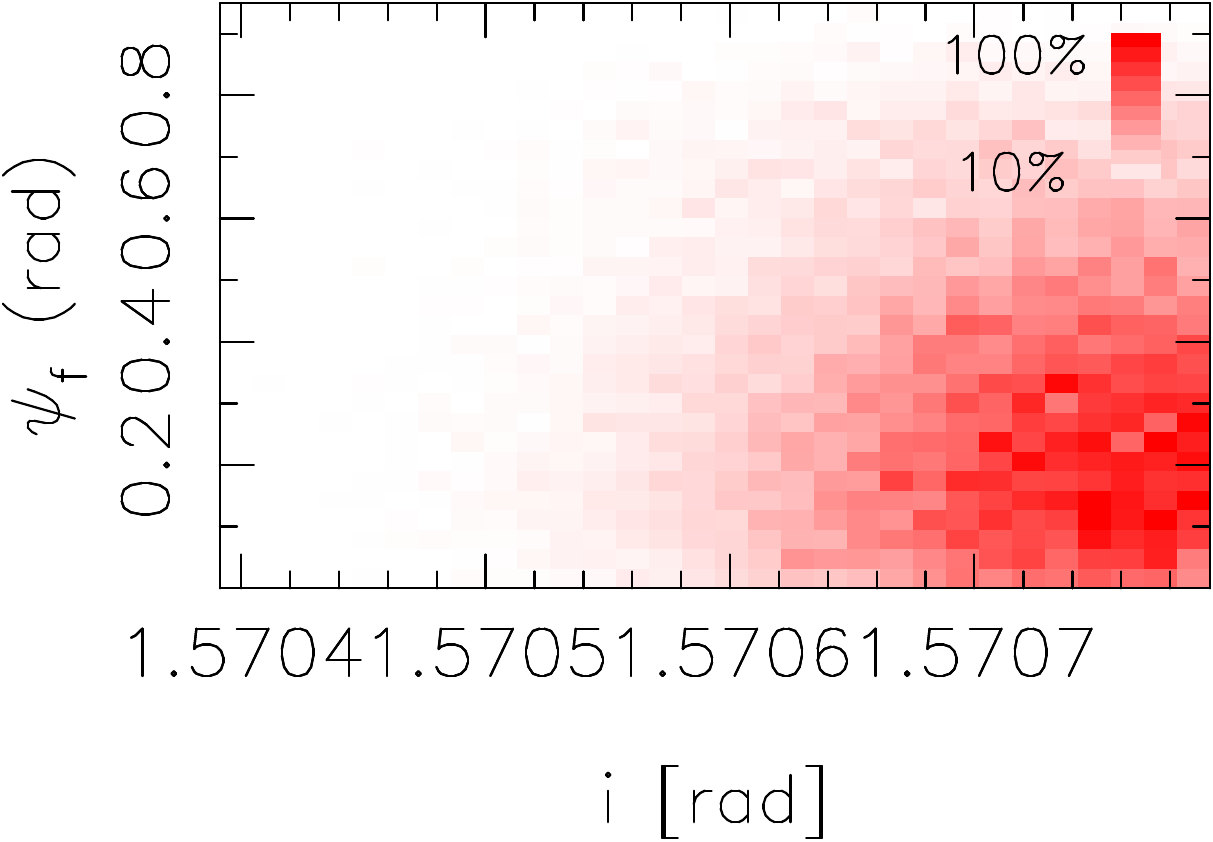}
\includegraphics[angle=270,width=0.135\textwidth,clip]{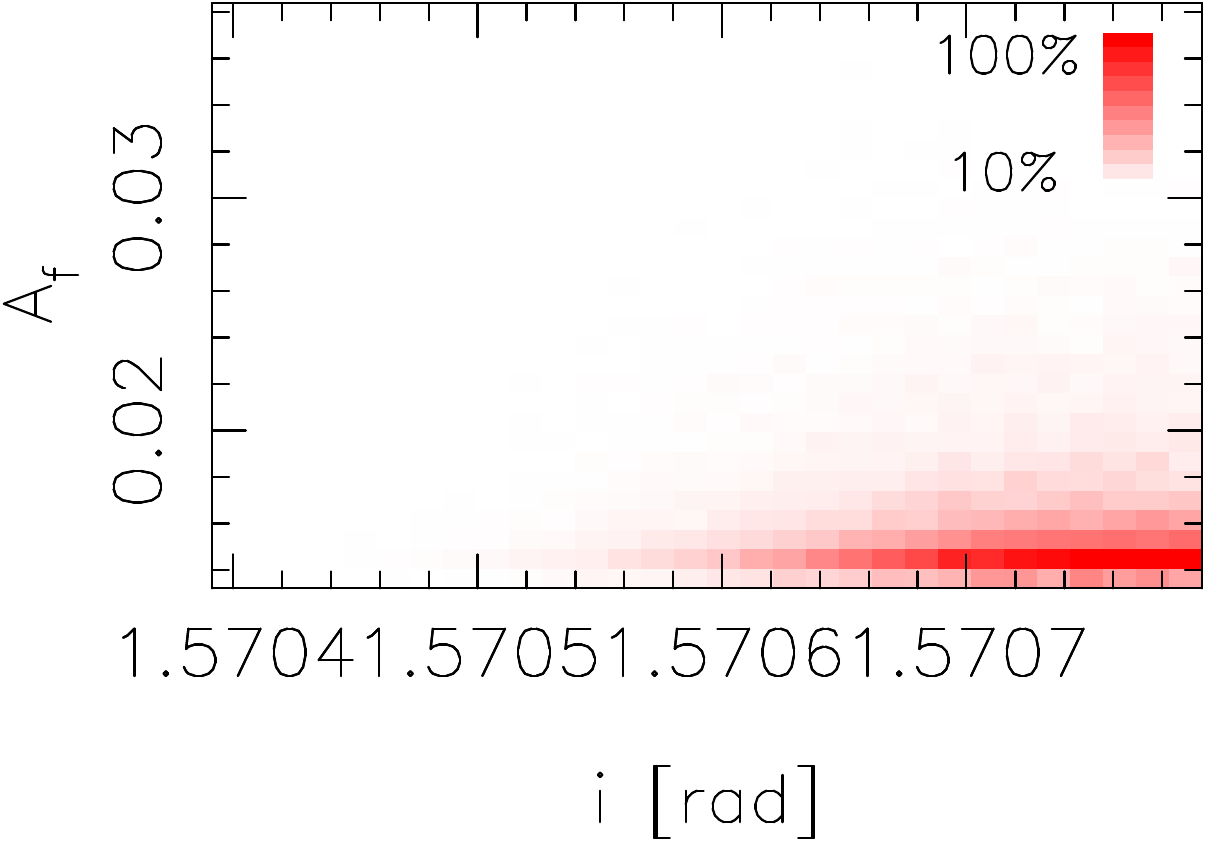}
\includegraphics[angle=270,width=0.135\textwidth,clip]{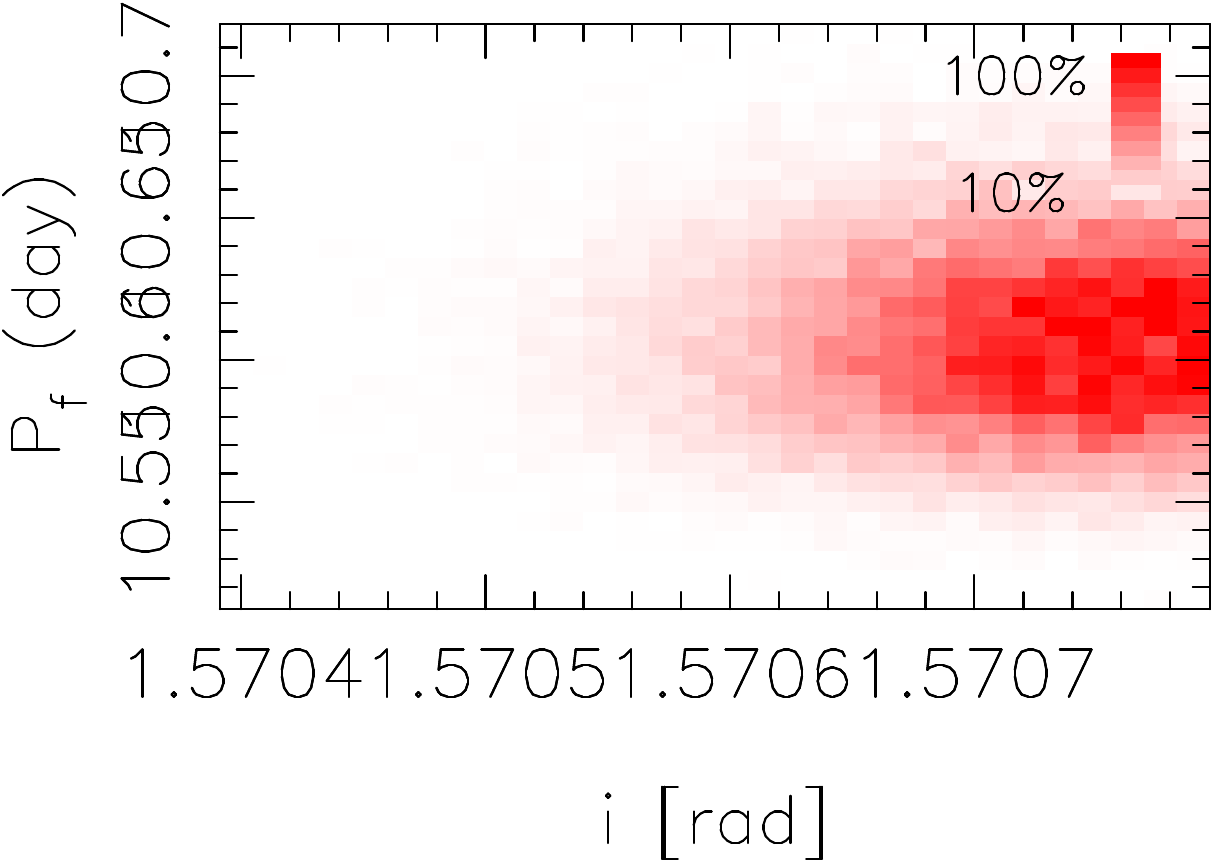}
\includegraphics[angle=270,width=0.135\textwidth,clip]{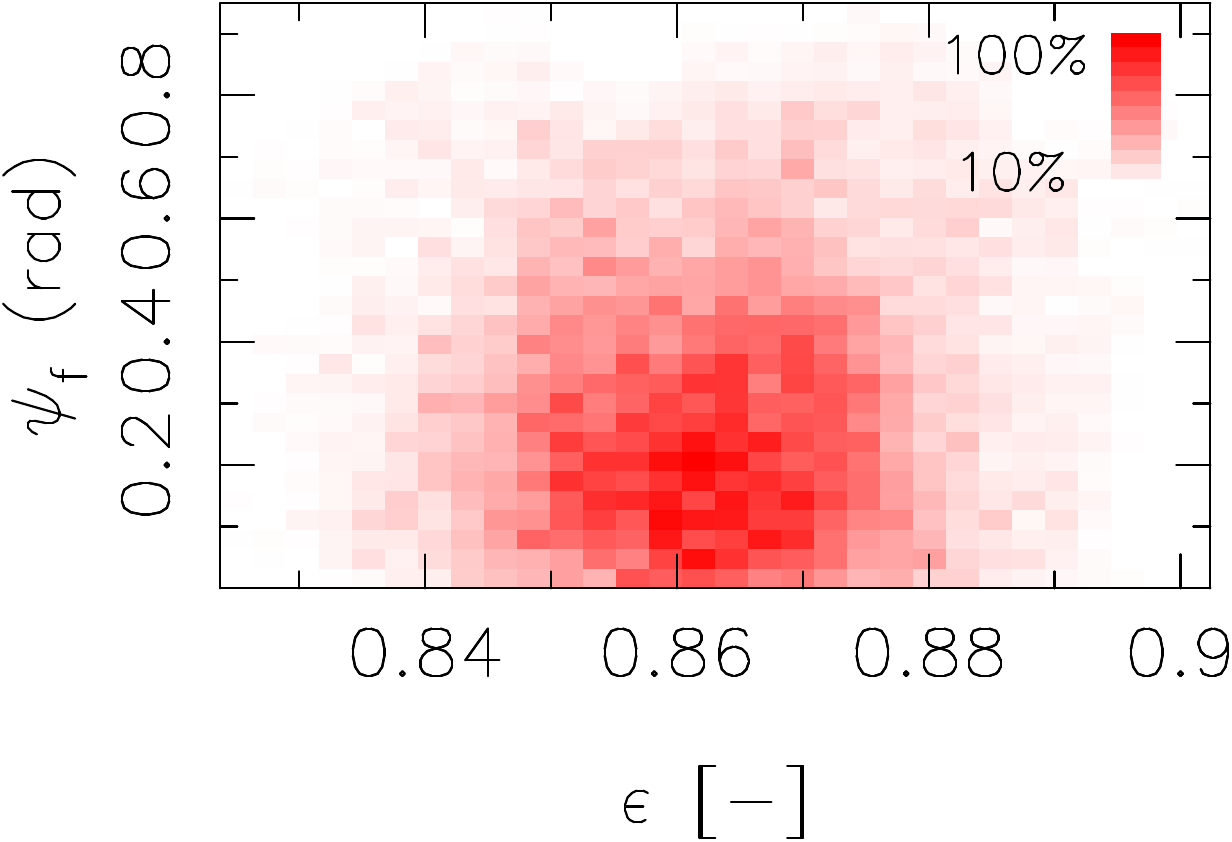}
\includegraphics[angle=270,width=0.135\textwidth,clip]{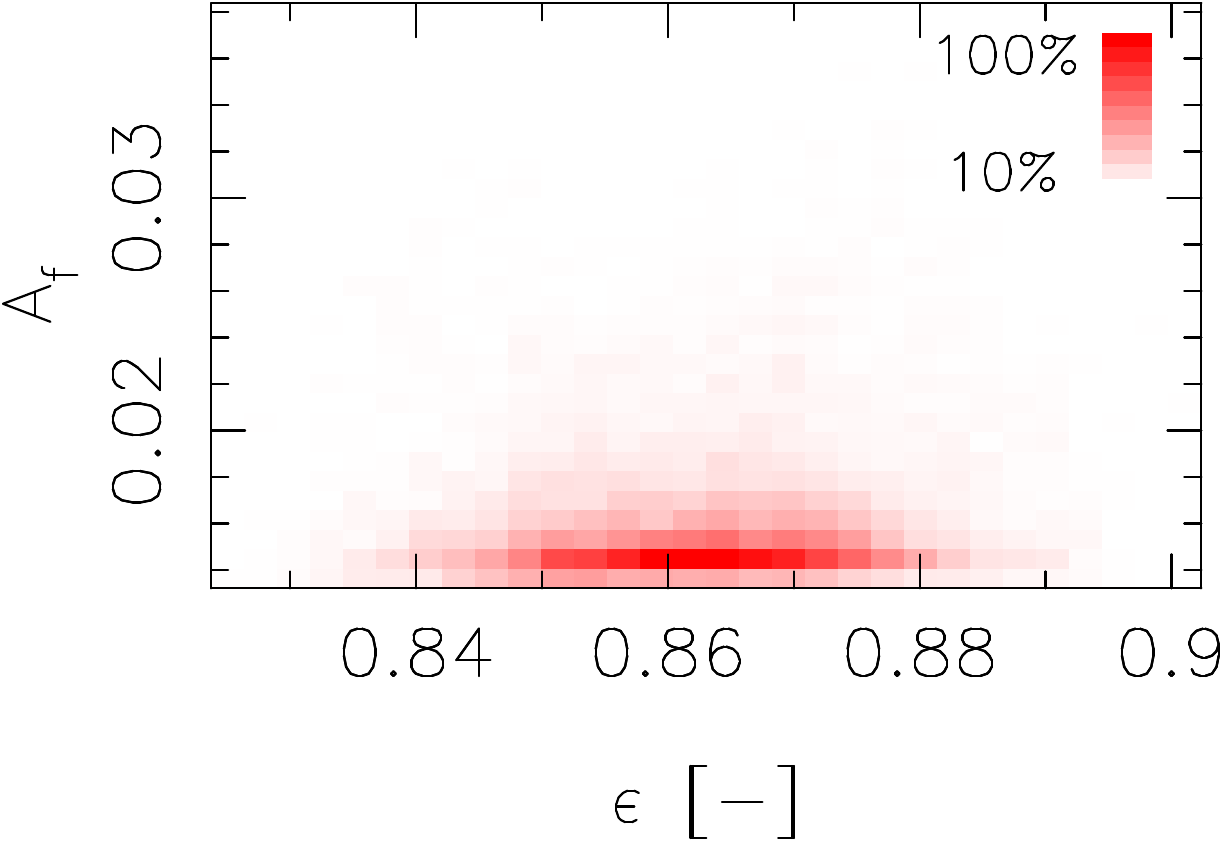}
\includegraphics[angle=270,width=0.135\textwidth,clip]{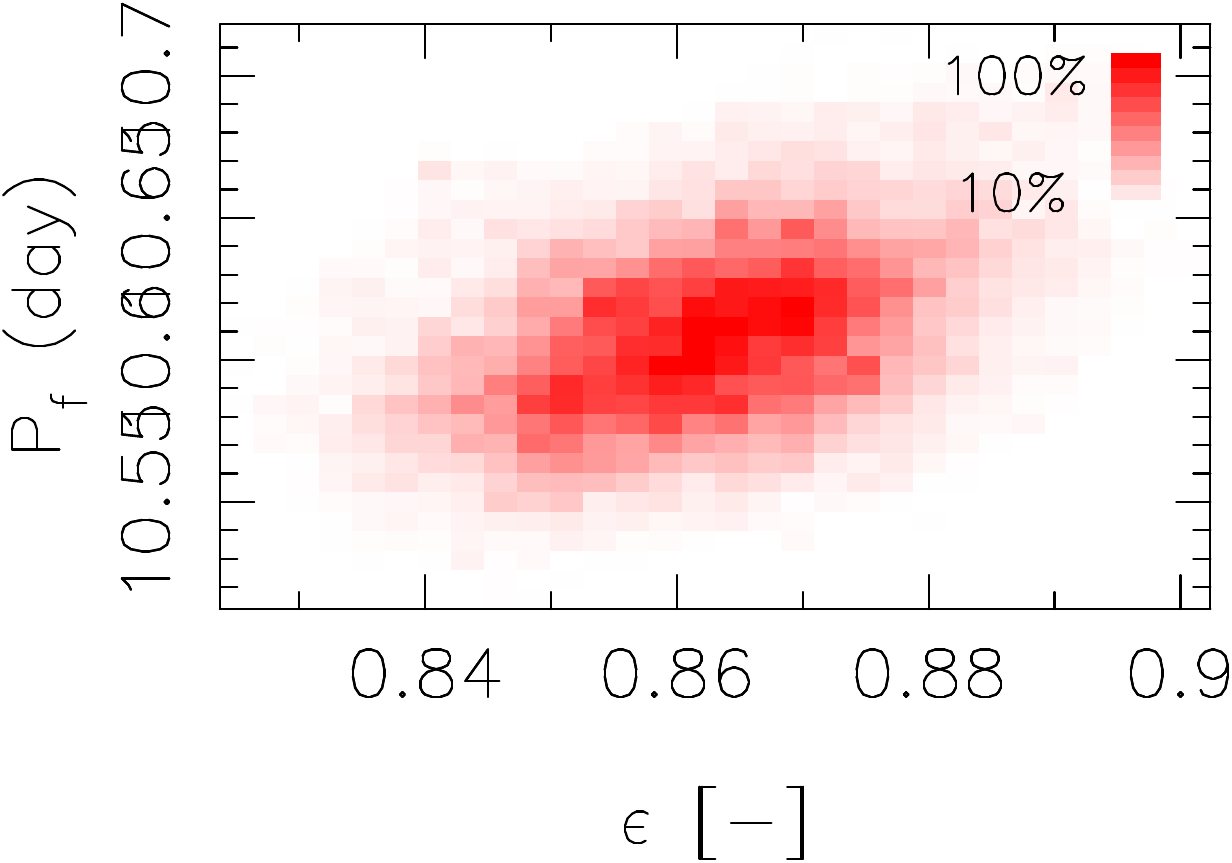}

\includegraphics[angle=270,width=0.135\textwidth,clip]{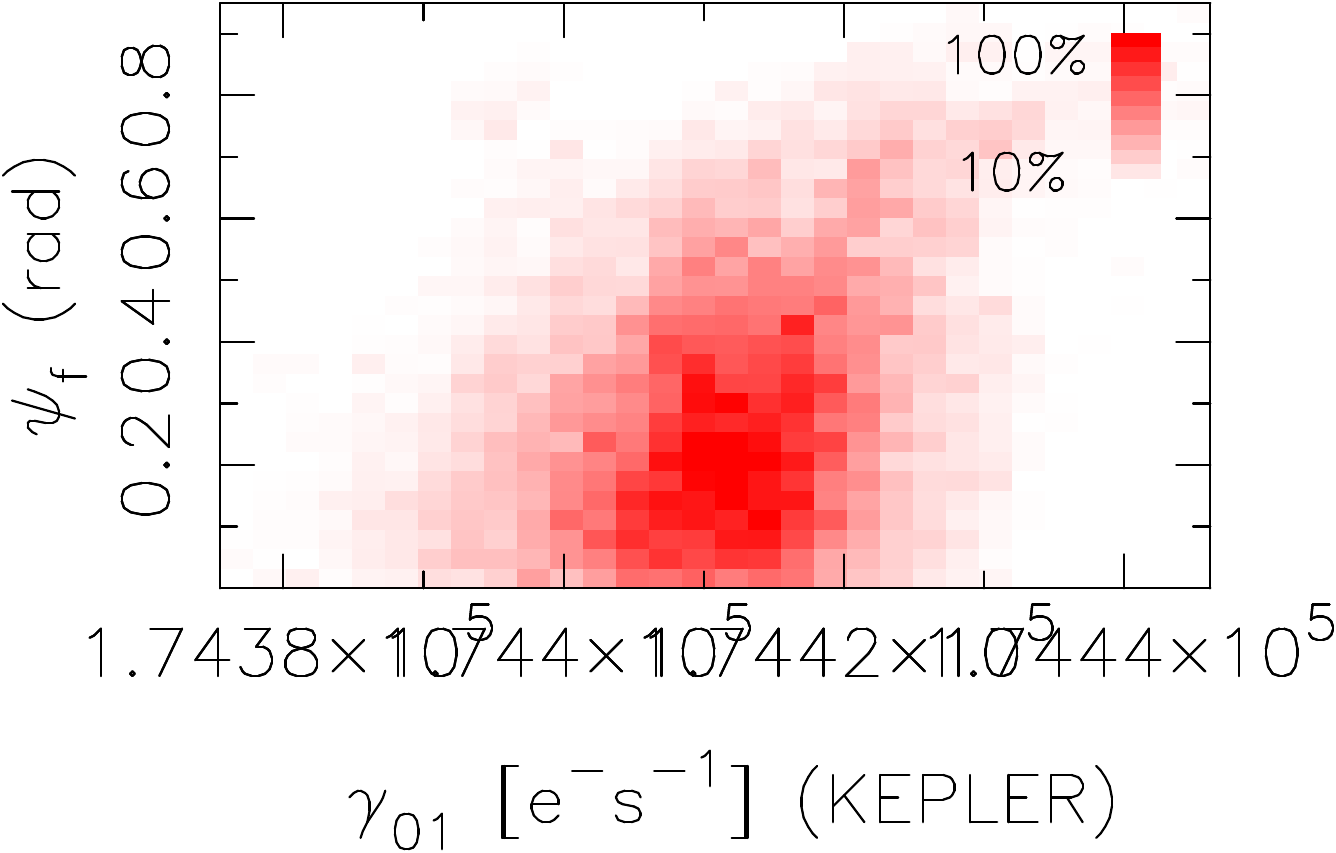}
\includegraphics[angle=270,width=0.135\textwidth,clip]{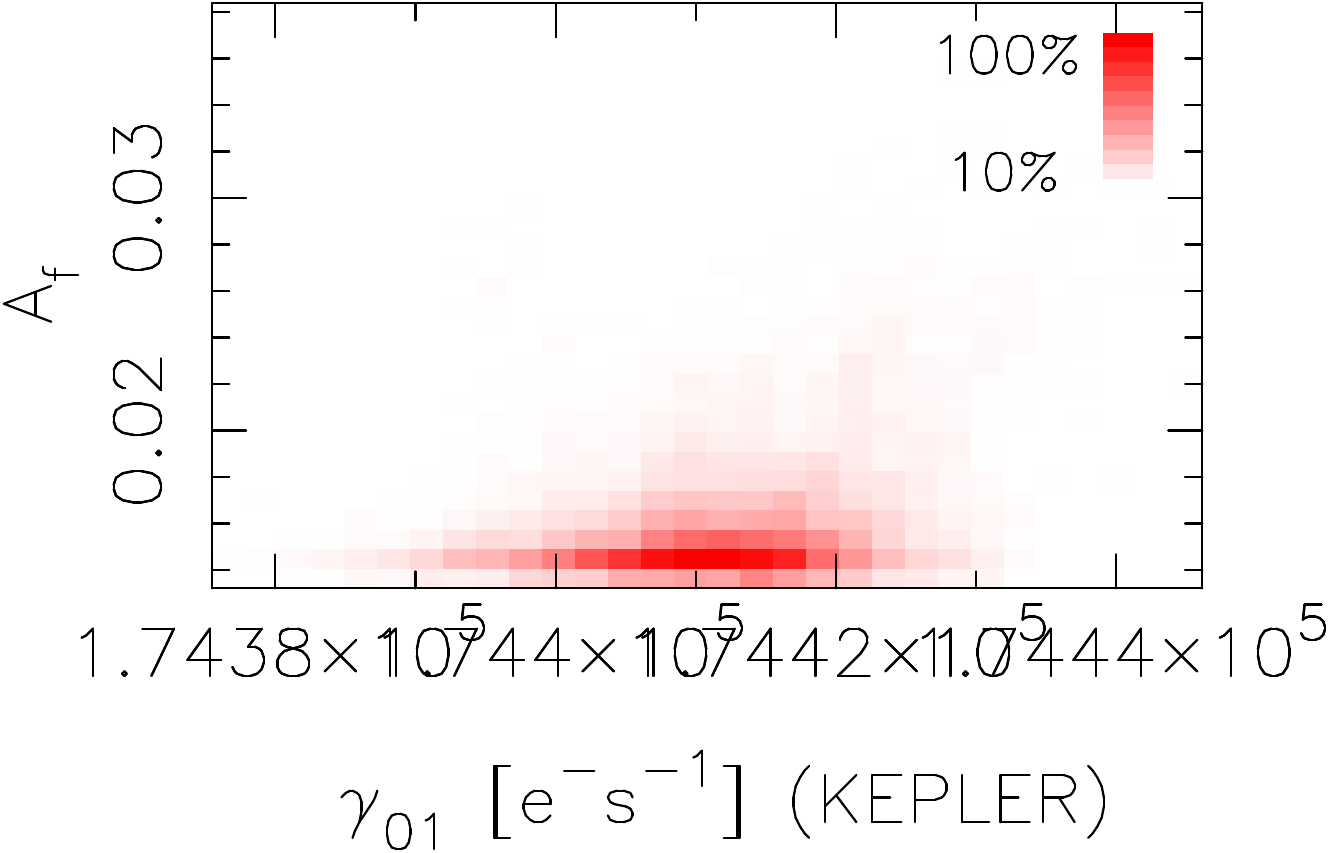}
\includegraphics[angle=270,width=0.135\textwidth,clip]{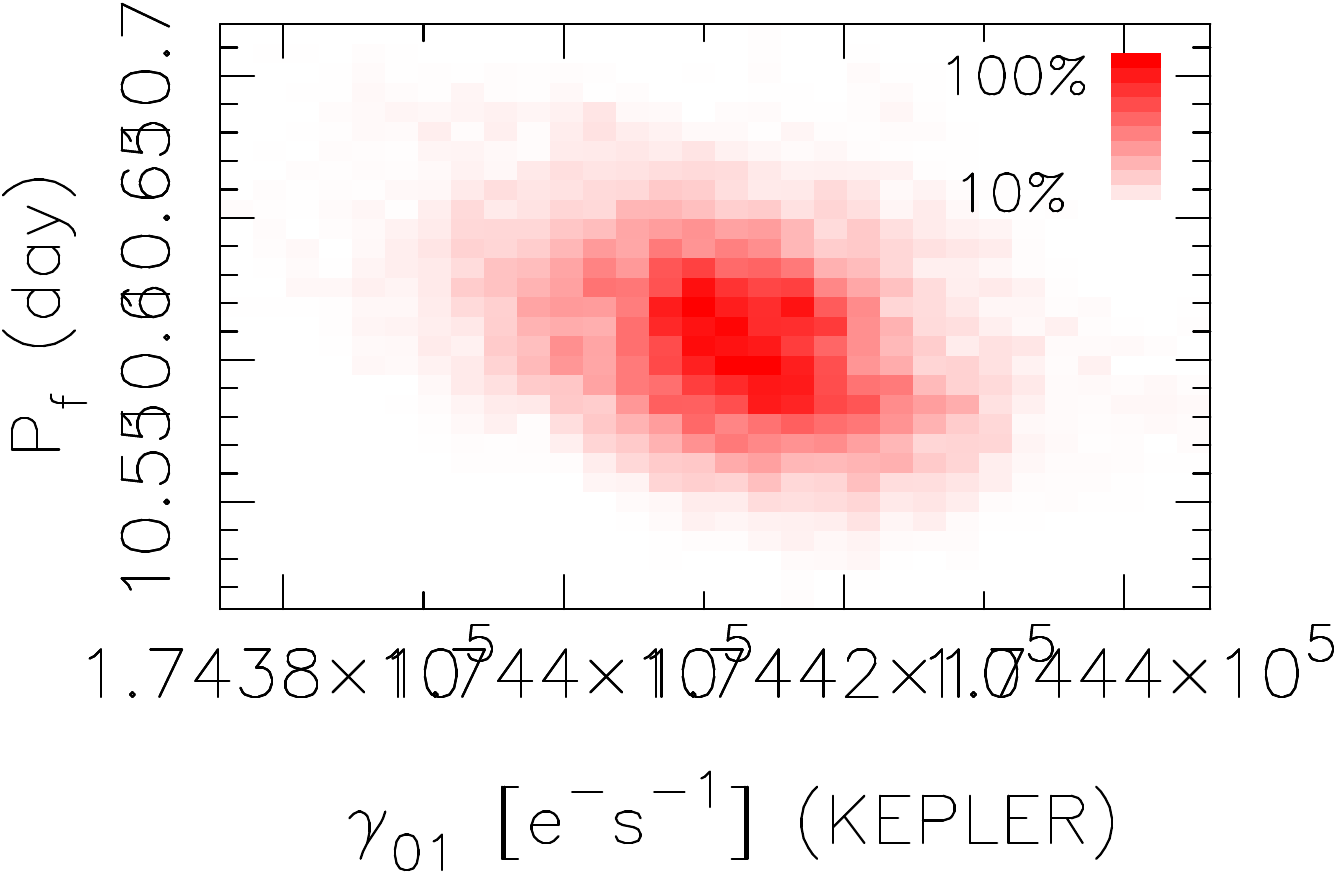}
\includegraphics[angle=270,width=0.135\textwidth,clip]{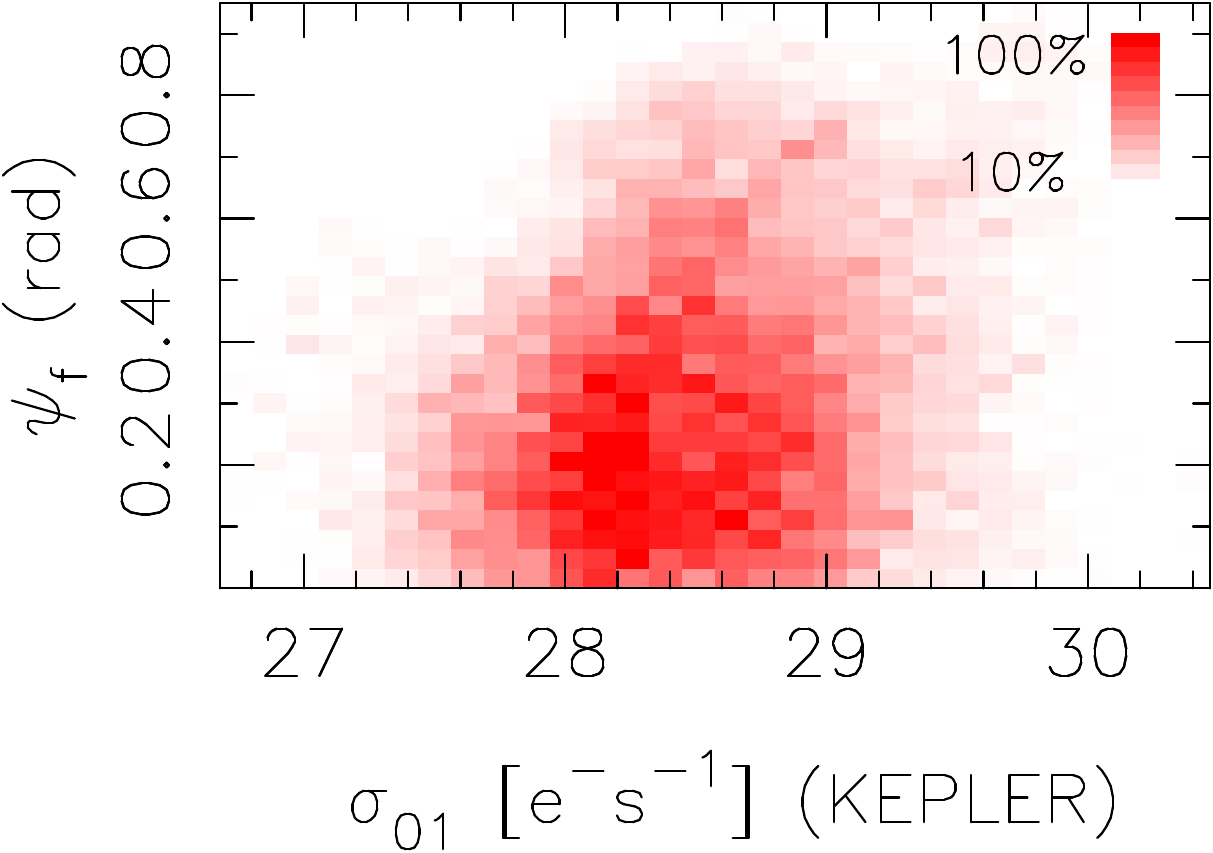}
\includegraphics[angle=270,width=0.135\textwidth,clip]{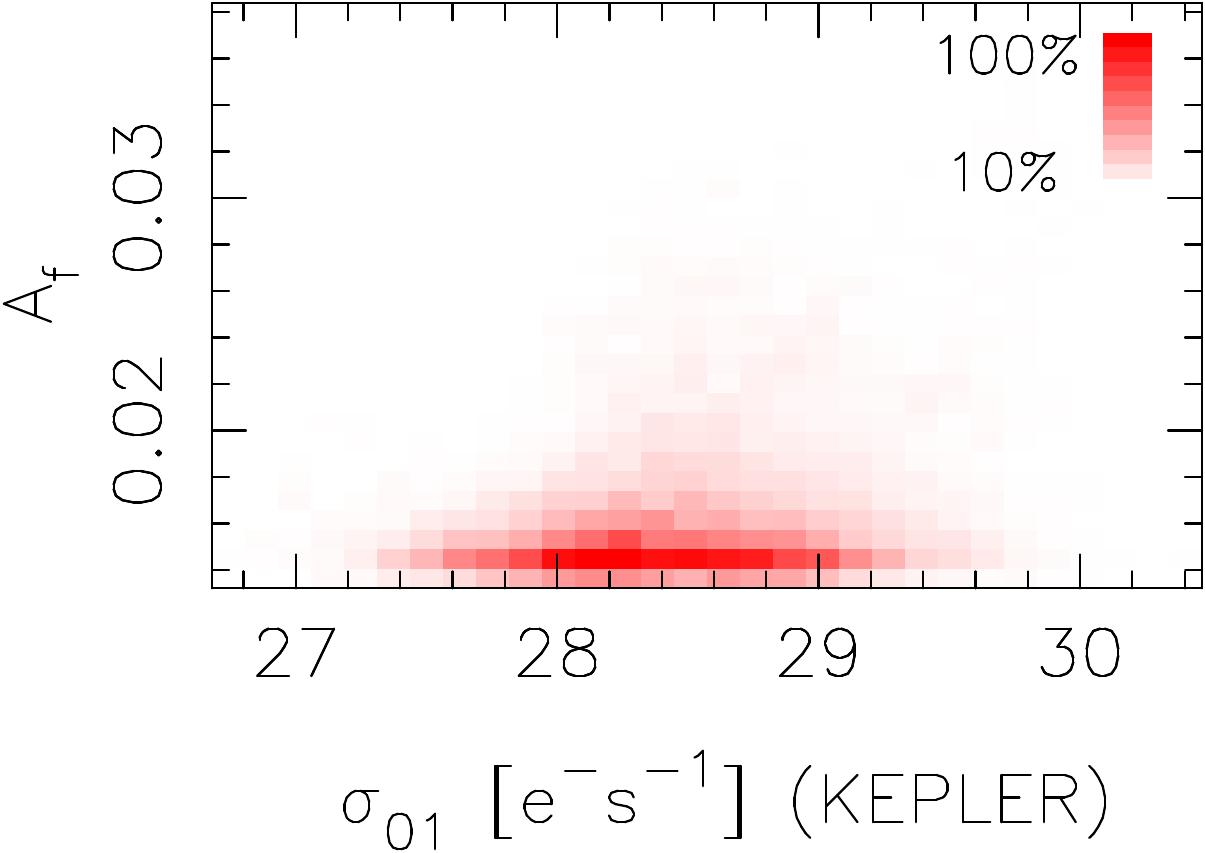}
\includegraphics[angle=270,width=0.135\textwidth,clip]{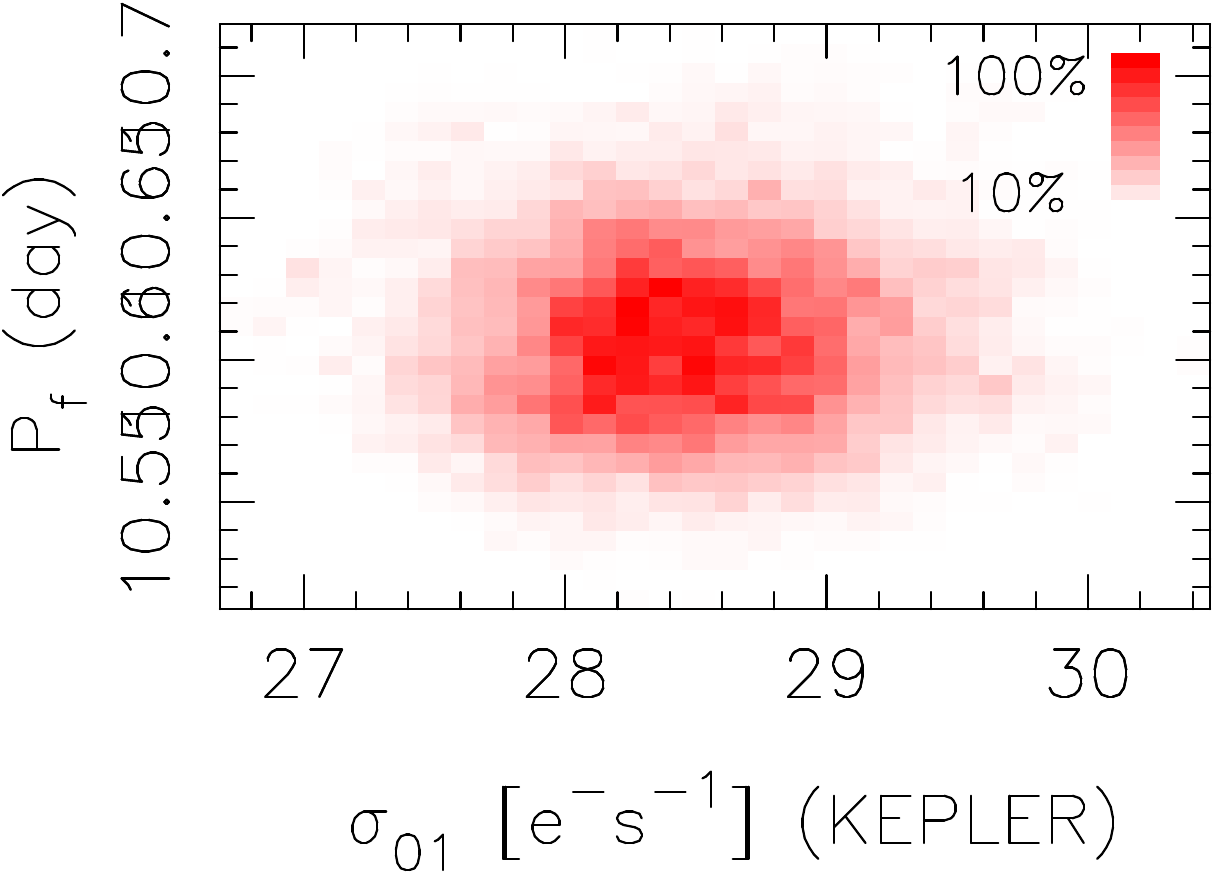}
\caption{As in Fig. \ref{fig:S1034_solution_contours} but between spot amplitude $A$, period $P$ and latitude $\psi$ and parameters: $i$, $\epsilon$, $\gamma$, and $\sigma$.}\label{fig:S1034_solution_contours3}
\end{figure*}

\begin{figure*}
\includegraphics[angle=270,width=0.135\textwidth,clip]{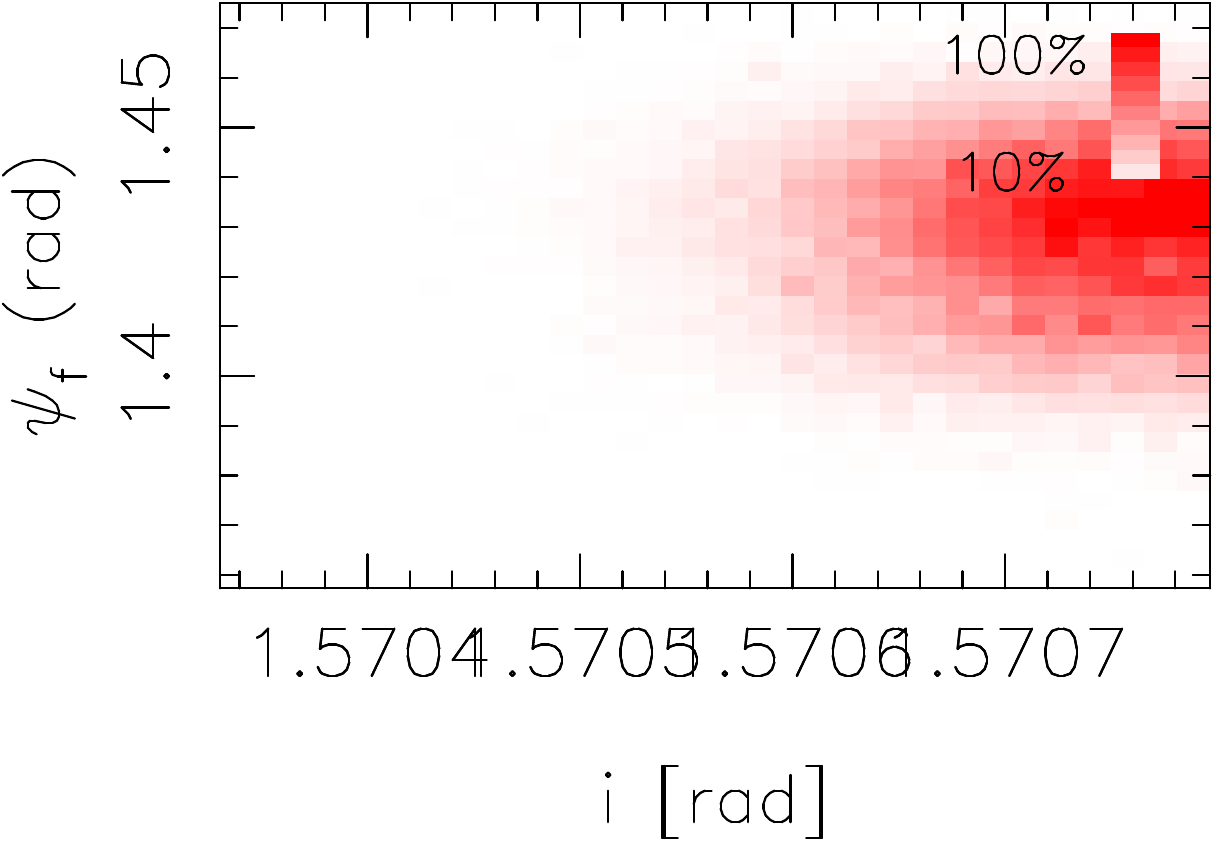}
\includegraphics[angle=270,width=0.135\textwidth,clip]{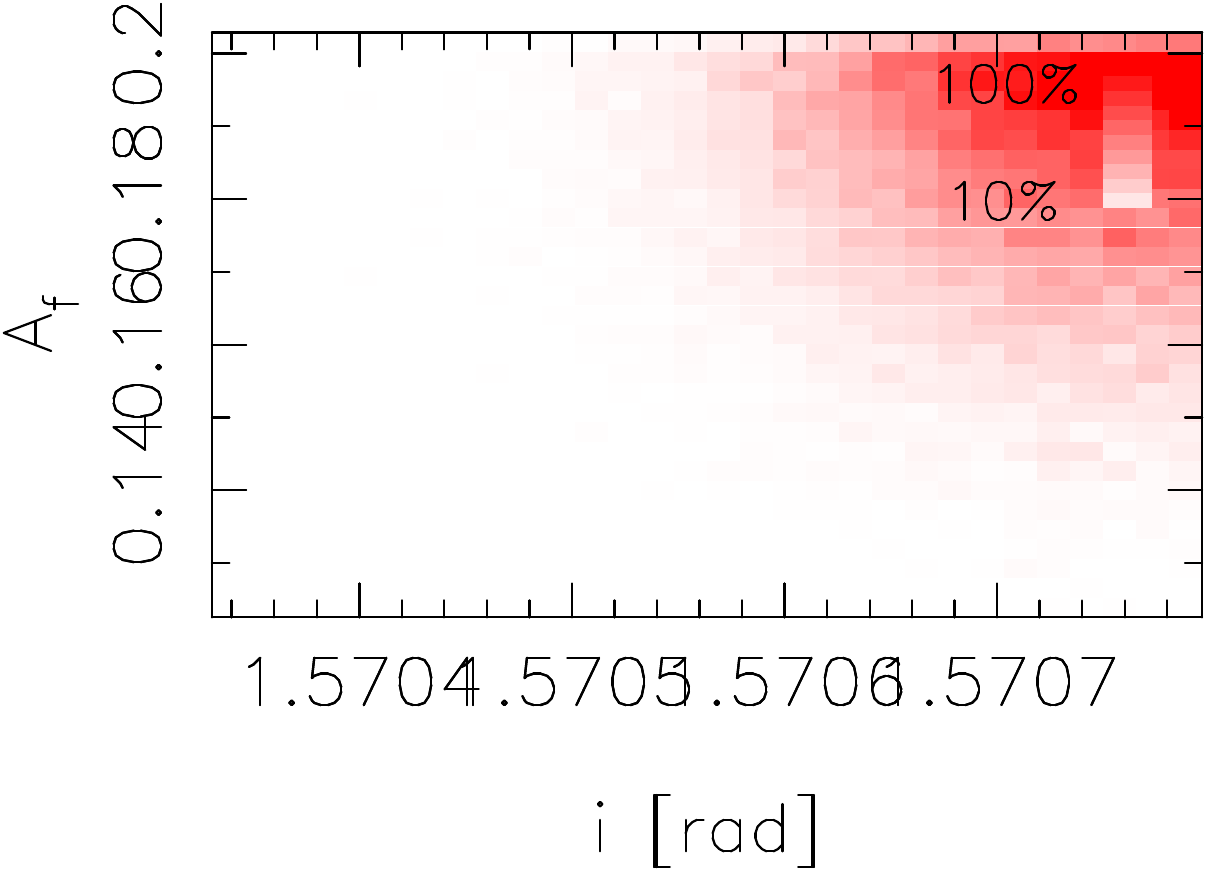}
\includegraphics[angle=270,width=0.135\textwidth,clip]{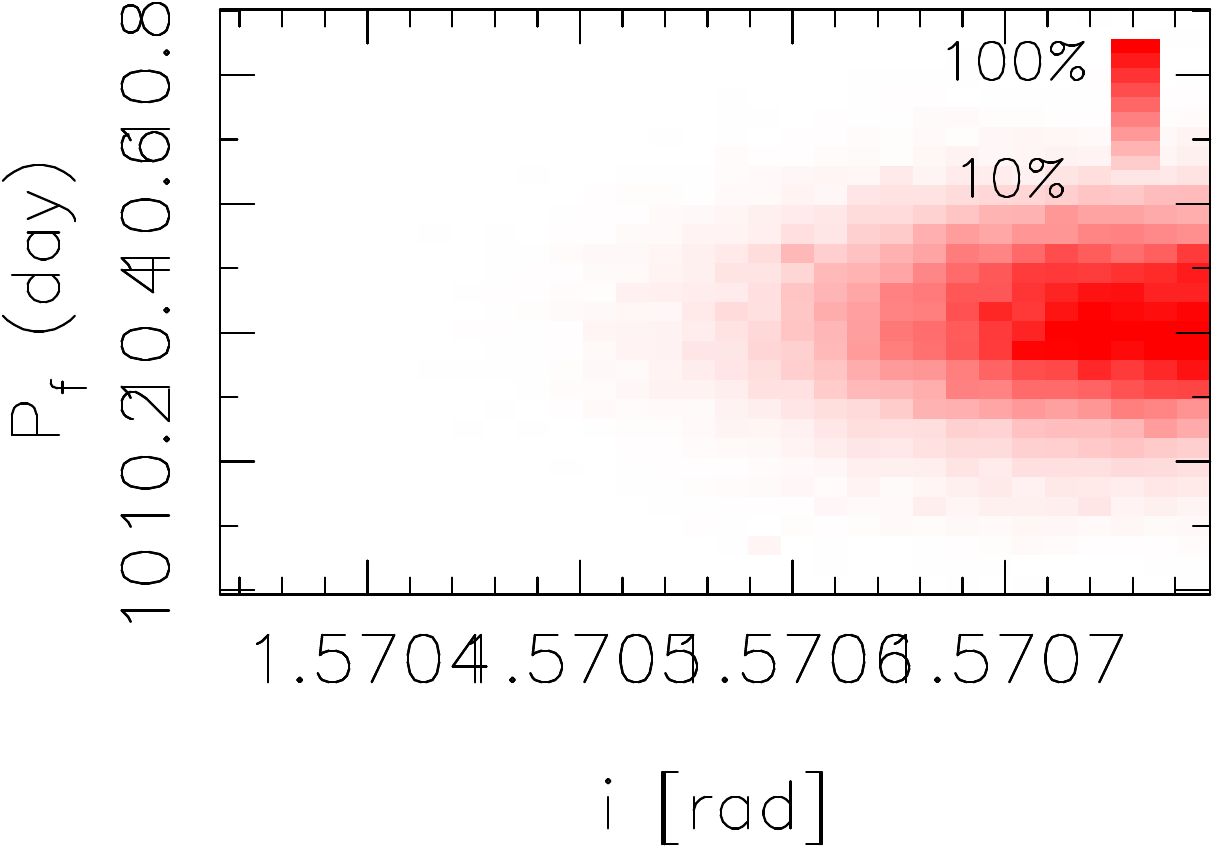}
\includegraphics[angle=270,width=0.135\textwidth,clip]{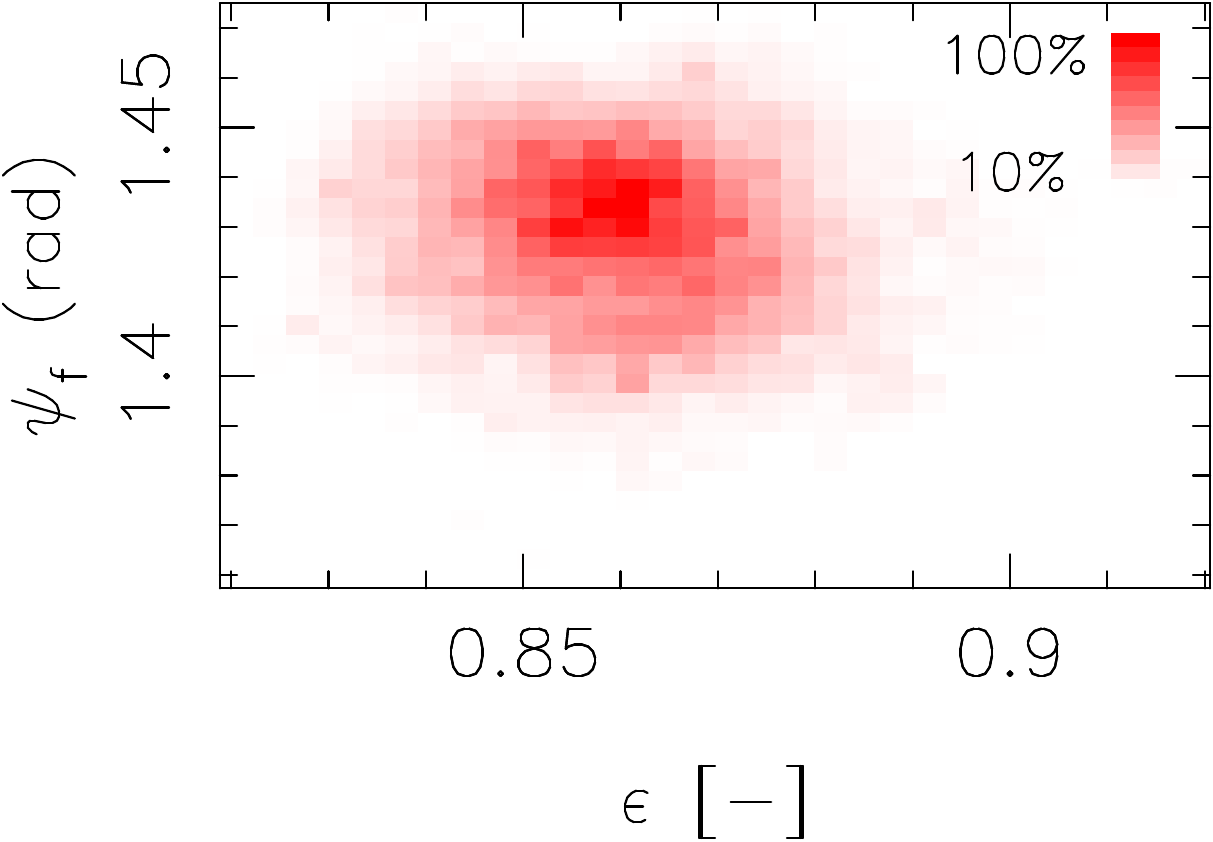}
\includegraphics[angle=270,width=0.135\textwidth,clip]{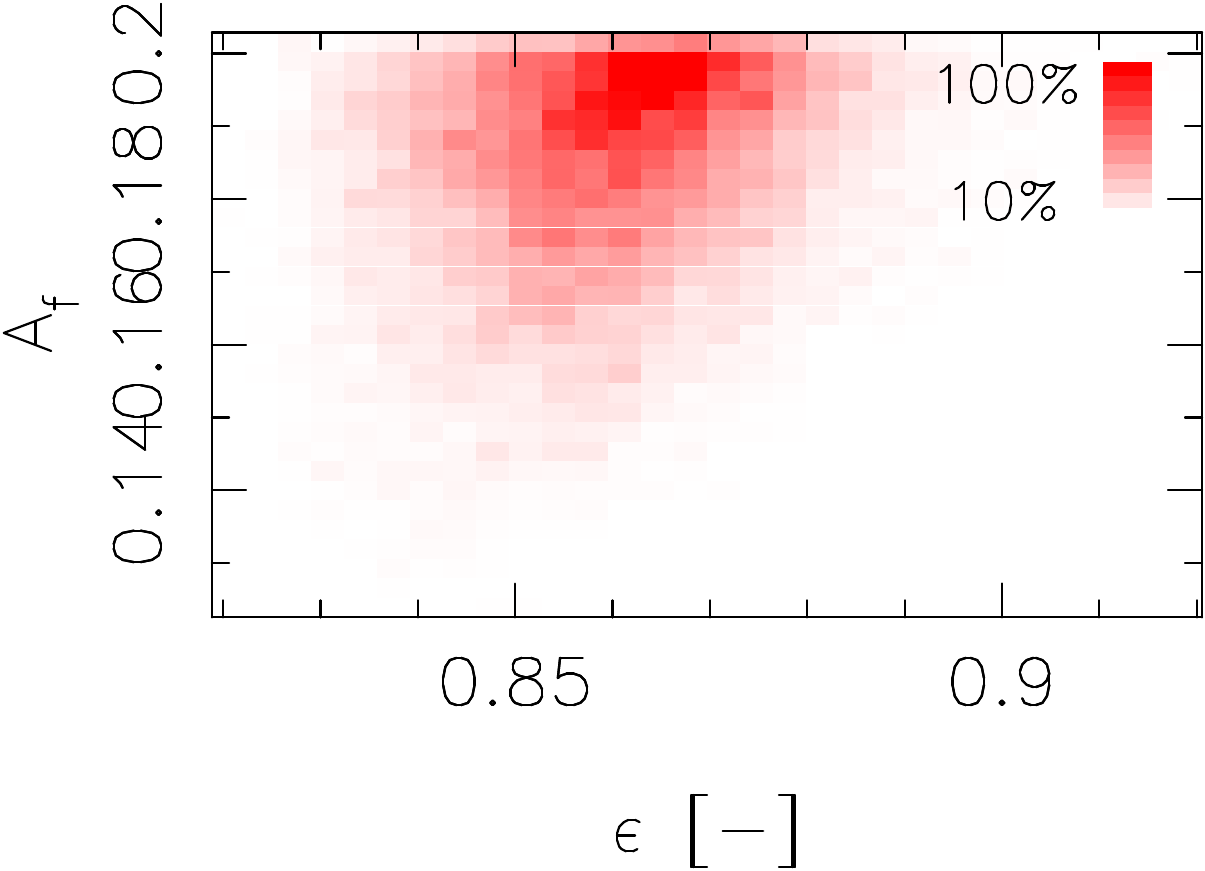}
\includegraphics[angle=270,width=0.135\textwidth,clip]{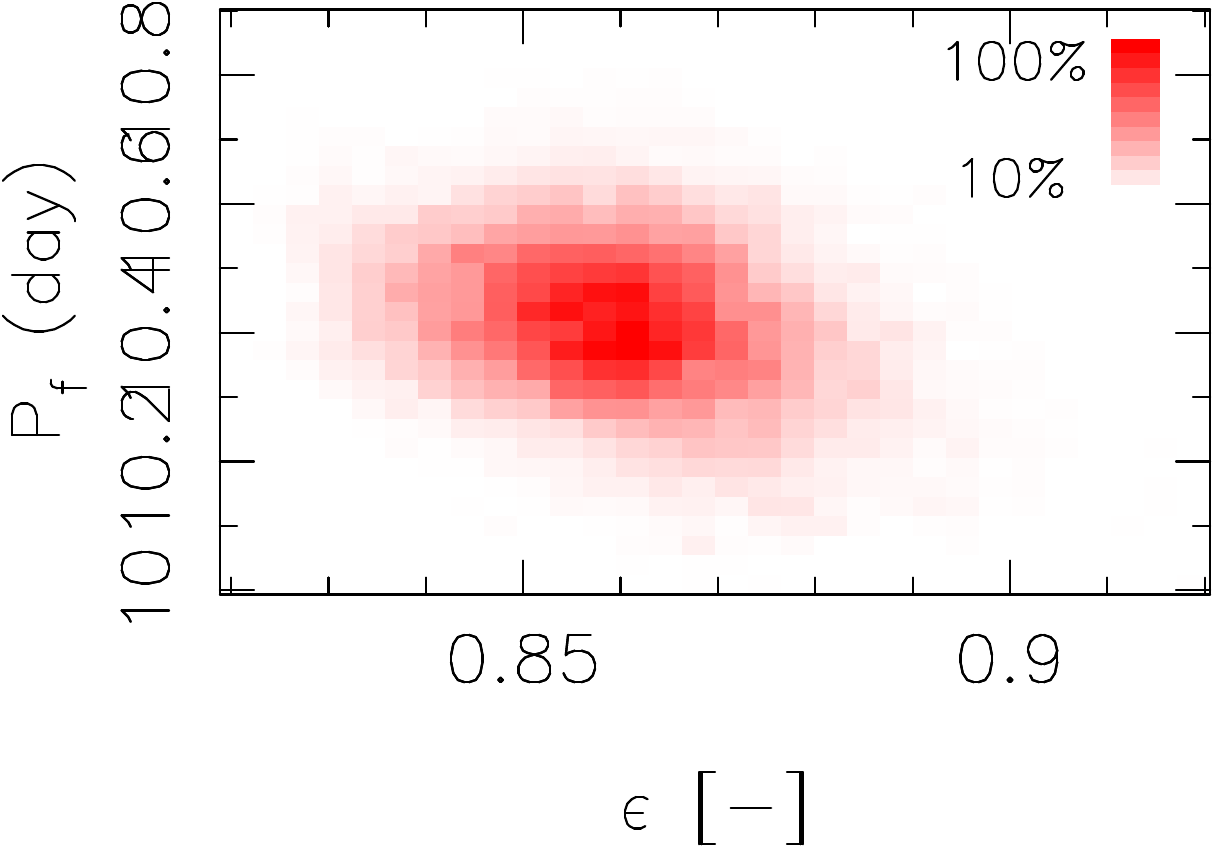}

\includegraphics[angle=270,width=0.135\textwidth,clip]{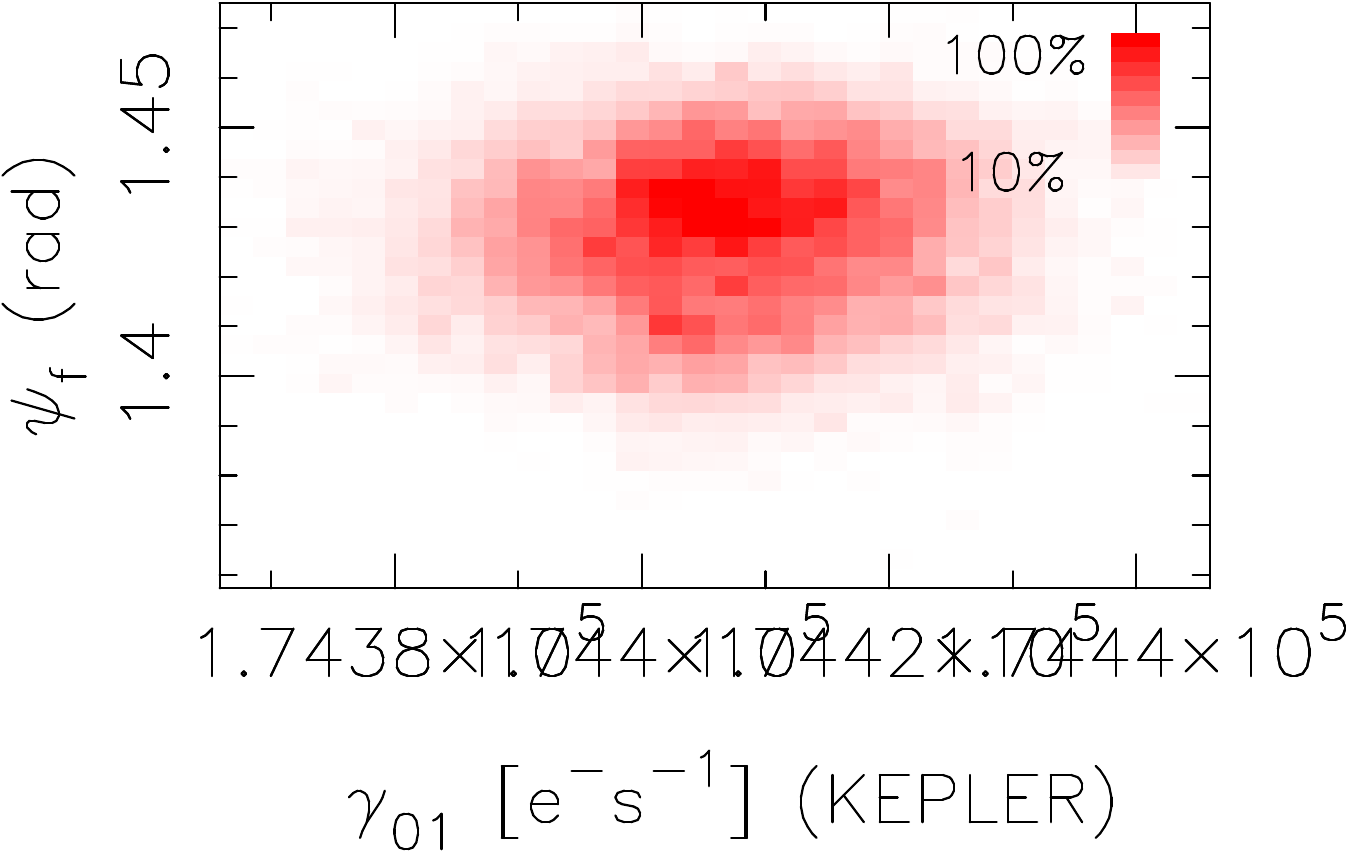}
\includegraphics[angle=270,width=0.135\textwidth,clip]{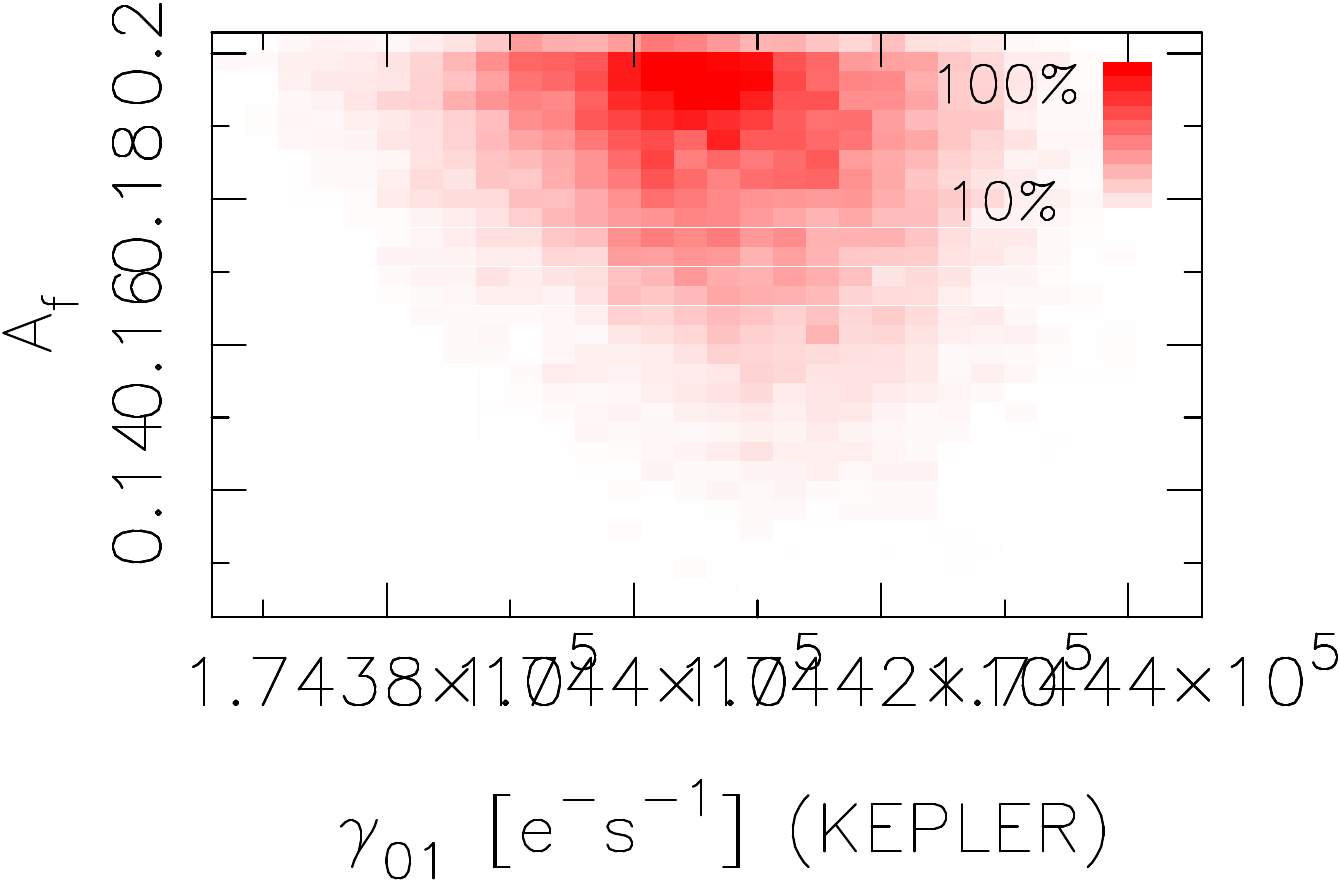}
\includegraphics[angle=270,width=0.135\textwidth,clip]{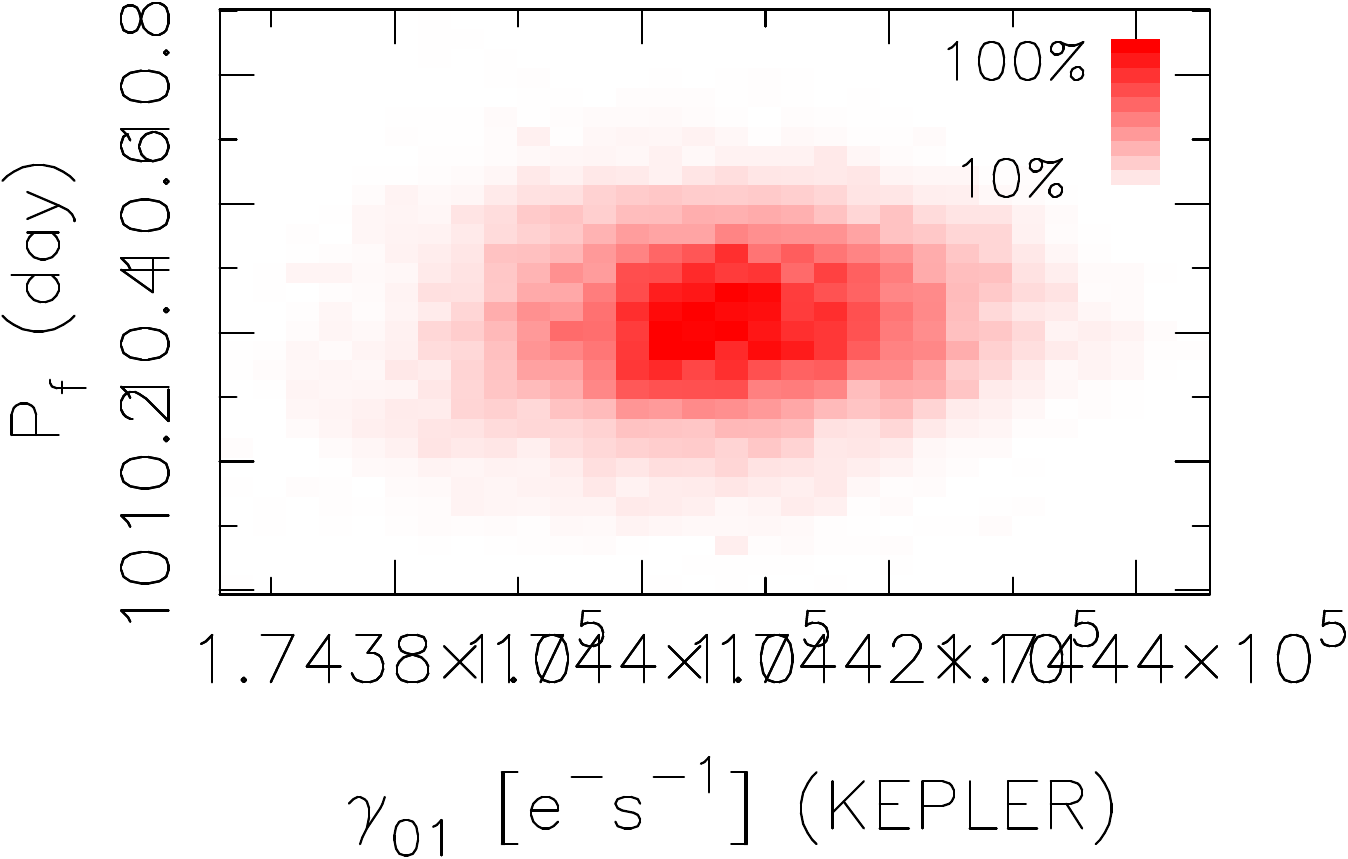}
\includegraphics[angle=270,width=0.135\textwidth,clip]{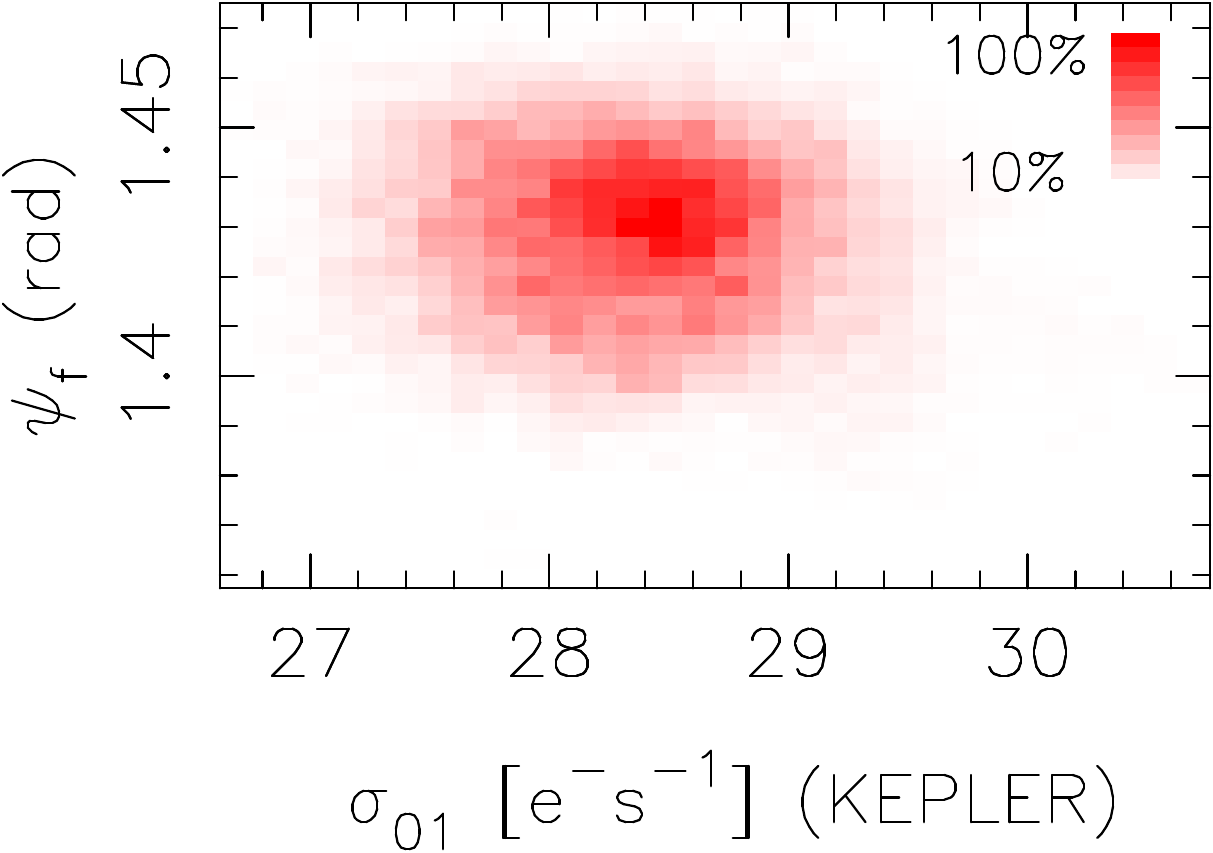}
\includegraphics[angle=270,width=0.135\textwidth,clip]{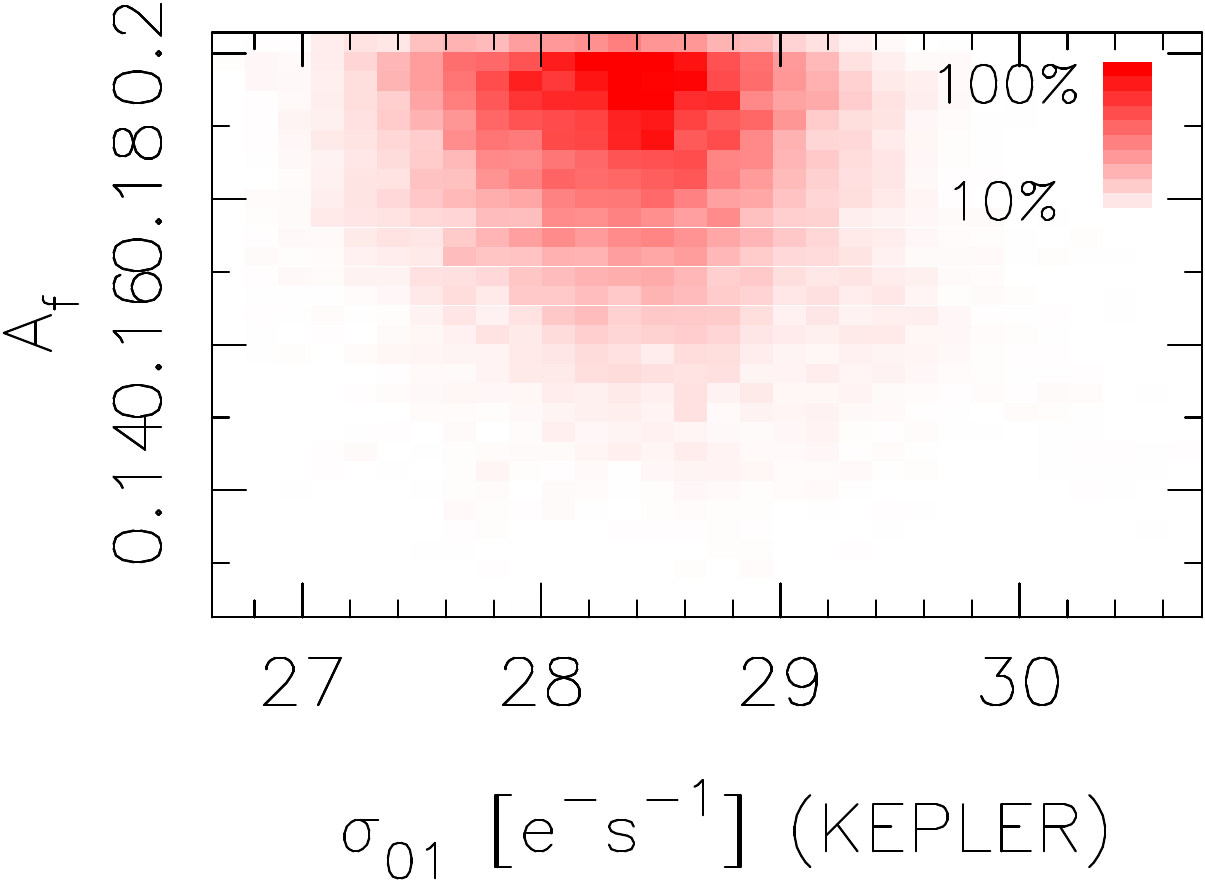}
\includegraphics[angle=270,width=0.135\textwidth,clip]{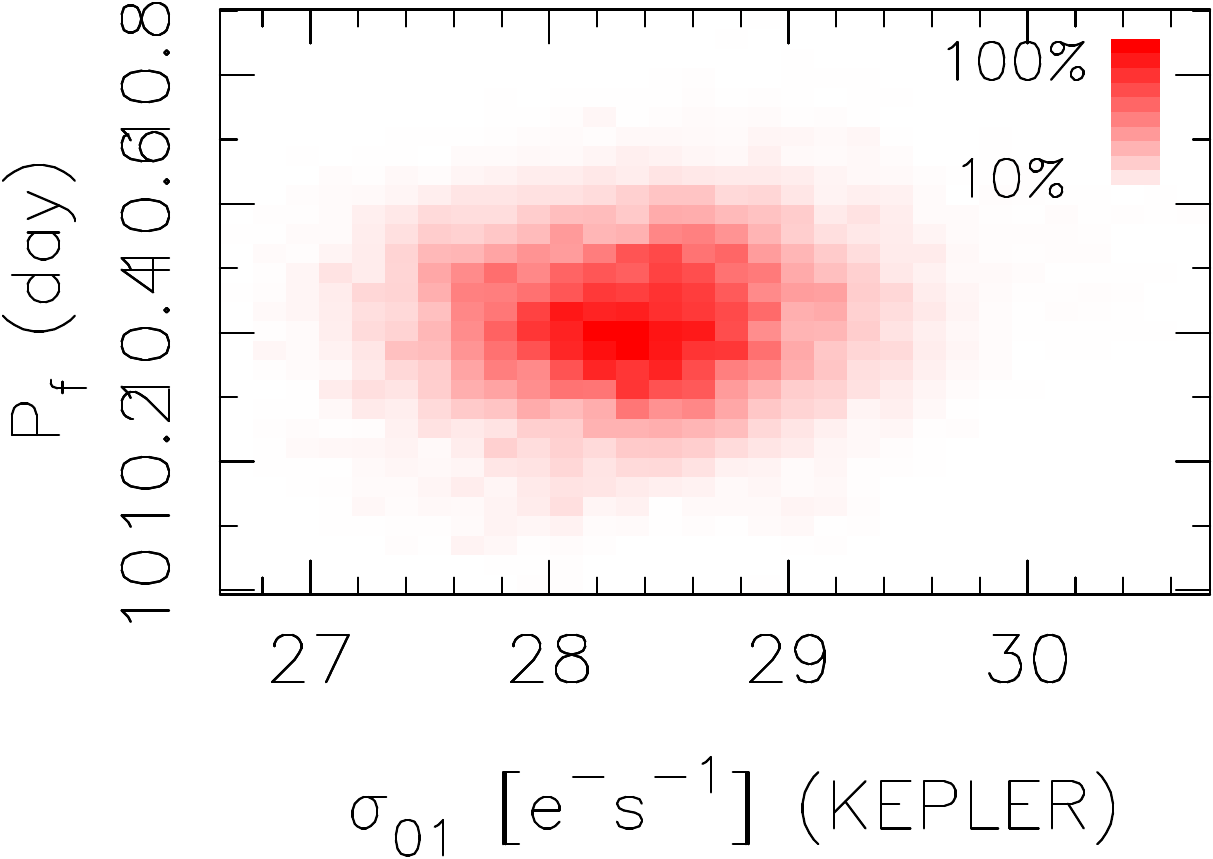}
\caption{As in Fig. \ref{fig:S1034_solution_contours3} but for spot 2 at higher latitude.}\label{fig:S1034_solution_contours4}
\end{figure*}


\begin{thebibliography}{100}\small
\bibitem[\protect\astroncite{Aigrain et al.}{2012}]{aigrain2012} Aigrain, S., Pont, F., \& Zucker, S. 2012, MNRAS, 419, 3147
\bibitem[\protect\astroncite{Ara\'ujo \& Valio}{2021}]{araujo2021a} Ara\'ujo, A. \& Valio, A. 2021a, ApJL, 907, L5
\bibitem[\protect\astroncite{Ara\'ujo \& Valio}{2021}]{araujo2021b} Ara\'ujo, A. \& Valio, A. 2021b, ApJL, 922, L23
\bibitem[\protect\astroncite{Ara\'ujo \& Valio}{2023}]{araujo2023} Ara\'ujo, A. \& Valio, A. 2023, MNRAS, 522, L16
\bibitem[\protect\astroncite{Aulanier et al.}{2012}]{aulanier2012} Aulanier, G., D\'emoulin, P., Schrijver, C. J. et al. 2012, A\&A, 549, A66
\bibitem[\protect\astroncite{Basri \& Shah}{2020}]{basri2020} Basri, G \& Shah R. 2020, ApJ, 901, 14
\bibitem[\protect\astroncite{Basri et al.}{2022}]{basri2022} Basri, G., Streichenberger, T., McWard, C., Edmond, L., et al. 2022, ApJ, 924, 31
\bibitem[\protect\astroncite{Breton et al.}{2024}]{breton2024} Breton, S. N., Lanza, A. F., \& Messina, S. 2024, \&A, 682, A67
\bibitem[\protect\astroncite{Claytor et al.}{2022}]{claytor2022} Claytor, Z. R., van Saders, J. L., Llama, J., Sadowski, P., et al. 2022, ApJ, 927, 219
\bibitem[\protect\astroncite{Claytor et al.}{2023}]{claytor2023} Claytor, Z. R., van Saders, J. L., Cao, L., Pinsonneault, M. H., et al. 2023, AAS submitted (arXiv:2307.05664)
\bibitem[\protect\astroncite{Feng et al.}{2016}]{feng2016} Feng, F., Tuomi, M., Jones, H. R. A., Butler, R. P., \& Vogt, S. S. 2016, MNRAS, 461, 2440
\bibitem[\protect\astroncite{Gaia Collaboration}{2018}]{gaia2018} Gaia Collaboration 2018, A\&A, 616, A1
\bibitem[\protect\astroncite{Gaia Collaboration}{2021}]{gaia2021} Gaia Collaboration 2021, A\&A, 649, A1
\bibitem[\protect\astroncite{Gelman \& Rubin}{1992}]{gelman1992} Gelman, A. \& Rubin, D. B. 1992, Stat. Sci., 7, 457
\bibitem[\protect\astroncite{G\"unther et al.}{2020}]{gunther2020} G\"unther, M. N., Zhan, Z., Seager, S., Rimmel, P. B., et al. 2020, AJ, 159, 60
\bibitem[\protect\astroncite{Haario et al.}{2001}]{haario2001} Haario, H., Saksman, E., \& Tamminen, J. 2001, Bernoulli, 7, 223
\bibitem[\protect\astroncite{Haris et al.}{2024}]{haris2024} Haris, A., Tuomi, M., \& Hackman, T. 2024, A\&A, 704, A102
\bibitem[\protect\astroncite{Harmon \& Crews}{2000}]{harmon2000} Harmon, R. O. \& Crews, L. J. 2000, AJ, 120, 3274
\bibitem[\protect\astroncite{Hastings}{1970}]{hastings1970} Hastings, W. 1970, Biometrika 57, 97
\bibitem[\protect\astroncite{Hirano et al.}{2014}]{hirano2014} Hirano, T., Sanchis-Ojeda, R., Takeda, et al. 2014, ApJ, 783, 9
\bibitem[\protect\astroncite{Howard \& MacGregor}{2022}]{howard2022} Howard, W. S. \& MacGregor, M., A., 2022, ApJ, 926, 204
\bibitem[\protect\astroncite{Ilin et al.}{2021}]{ilin2021} Ilin, E., Poppenhaeger, K., Schmidt, S. J., J\"arvinen, S. P., et al. 2021, MNRAS, 507, 1723
\bibitem[\protect\astroncite{Ioannidis \& Schmitt}{2020}]{ioannidis2020} Ioannidis, P. \& Schmitt, H. M. M. 2020, A\&A, 644, A26
\bibitem[\protect\astroncite{J\"arvinen et al.}{2008}]{jarvinen2008} J\"arvinen, S. P., Korhonen, H., Berdyugina, S. V., et al. 2008, A\&A, 488, 1047
\bibitem[\protect\astroncite{Kass \& Raftery}{1995}]{kass1995} Kass, R. E. \& Raftery, A. E. 1995, J. Am. Stat. Ass., 430, 773
\bibitem[\protect\astroncite{Kipping}{2012}]{kipping2012} Kipping, D. M. 2012, MNRAS, 427, 2487
\bibitem[\protect\astroncite{Koen}{2021}]{koen2021} Koen, C. 2021, A\&A, 647, L1
\bibitem[\protect\astroncite{Lamb}{2017}]{lamb2017} Lamb, D. A. 2017, ApJ, 836, 10
\bibitem[\protect\astroncite{Lanza et al.}{2019}]{lanza2019} Lanza, A. F., Netto, Y., Bonomo, A. S., Parviainen, H., et al. 2019, A\&A, 626, A38
\bibitem[\protect\astroncite{Lightkurve Collaboration}{2018}]{lightkurve2018} Lightkurve Collaboration 2018, Astrophysics Source Code Library, record ascl:1812.013
\bibitem[\protect\astroncite{Liddle}{2007}]{liddle2007} Liddle, A. R. 2007, MNRAS, 377, L74
\bibitem[\protect\astroncite{Luger et al.}{2021a}]{luger2021} Luger, R., Foreman-Mackey, D., Hedges, C., \& Hogg, D. W. 2021a, AJ, 162, 123
\bibitem[\protect\astroncite{Luger et al.}{2021b}]{luger2021b} Luger, R., Foreman-Mackey, D., \& Hedges, C. 2021b, AJ, 162, 124
\bibitem[\protect\astroncite{Luo et al.}{2019}]{luo2019} Luo, T., Liang, Y., \& IP, W.-H. 2019, AJ, 157, 238
\bibitem[\protect\astroncite{Mandel \& Agol}{2002}]{mandel2002} Mandel, K. \& Agol, E. 2002, ApJ, 580, L171
\bibitem[\protect\astroncite{Metropolis et al.}{1953}]{metropolis1953} Metropolis, N., Rosenbluth, A. W., Rosenbluth, M. N., et al. 1953, J. Chem. Phys., 21, 1087
\bibitem[\protect\astroncite{Morris et al.}{2017}]{morris2017} Morris, B. M., Hebb, L., Davenport, J. R. A., Rohn, G., \& Hawley, S. L. 2017, ApJ, 846, 99
\bibitem[\protect\astroncite{Morton et al.}{2016}]{morton2016} Morton, T. D., Bryson, S. T., Coughlin, J. L., Rowe, J. F., et al. 2016, ApJ, 822, 86
\bibitem[\protect\astroncite{Nielsen et al.}{2013}]{nielsen2013} Nielsen, M. B., Gizon, L., Schunker, H., \& Karoff, C. 2013, A\&A, 557, L10
\bibitem[\protect\astroncite{Reinhold et al.}{2013}]{reinhold2013} Reinhold, T., Reiners, A., \& Basri, G. 2013, A\&A, 560, A4
\bibitem[\protect\astroncite{Rodon\'o et al.}{1986}]{rodono1986} Rodon\'o, M., Cutispoto, G., Pazzani, V., et al. 1986, A\&A, 165, 135
\bibitem[\protect\astroncite{Stassun et al.}{2018}]{stassun2018} Stassun, K. G., Oelkers, R. J., Paegert, M., Torres, G., et al. 2018, AJ, 158, 138
\bibitem[\protect\astroncite{Sun et al.}{2019}]{sun2019} Sun, L., Ioannidis, P., Gu, S., Schmitt, J. H. M. M. et al. 2019, A\&A, 624, A15
\bibitem[\protect\astroncite{Silva-Valio}{2008}]{silva-valio2008} Silva-Valio, A. 2008, ApJ, 683, L179
\bibitem[\protect\astroncite{Silva-Valio et al.}{2010}]{silva-valio2010} Silva-Valio, A., Lanza, A. F., Alonso, R., \& Barge, P. 2010, A\&A, 510, A25
\bibitem[\protect\astroncite{Vida et al.}{2021}]{vida2021} Vida, K., B\'odi, A., Szklen\'ar, T., \& Seli, B. 2021, A\&A, 652, A107
\bibitem[\protect\astroncite{Walkowicz et al.}{2013}]{walkowicz2013} Walkowicz, L. M., Basri, G., \& Valenti, J. A. ApJS, 205, 17
\bibitem[\protect\astroncite{Wang et al.}{2014}]{wang2014} Wang, J., Xie, J.-W., Barclay, T., \& Fischer, D. A. 2014, ApJ, 783, 4
\bibitem[\protect\astroncite{Willamo et al.}{2022}]{willamo2022} Willamo, T., Lehtinen T. T., Hackman, T., K\"apyl\"a, M. J., et al. 2022, A\&A, 659, A71
\bibitem[\protect\astroncite{W\"ohl et al.}{2010}]{wohl2010} W\"ohl, H., Braj\u{s}a, R., Hanslmeier, A., \& Gissot, S. F. 2010, A\&A, 520, A29
\bibitem[\protect\astroncite{Xu et al.}{2021}]{xu2021} Xu, F., Gu, S., \& Ioannidis, P. 2021, MNRAS, 501, 1878
\bibitem[\protect\astroncite{Zeldes et al.}{2021}]{zeldes2021} Zeldes, J., Hinkle, J. T., Shappee, B. J., Avallone, E. A., et al. 2021, ApJ, submitted (arXiv:2109.04501)
\bibitem[\protect\astroncite{Zhan et al.}{2019}]{zhan2019} Zhan, Z., G\"unther, M. N., Rappaport, S., Ol\'ah, K., et al. 2019, ApJ, 876, 127
\end{thebibliography}
\end{document}